\documentclass[openacc]{rstransa}

\usepackage{ref_macros_sci}

\usepackage{anyfontsize}

\usepackage{graphicx}
\usepackage{multirow,tabularx}

\newcommand{\M}{{\rm M$_\odot$}}
\newcommand{\kms}{{\rm km~s$^{-1}$}}
\newcommand{\sqdeg}{{\rm deg}$^2$}
\newcommand{\cmg}{{\rm cm}$^2$~{\rm g}$^{-1}$}

\titlehead{MMA lessons}

\begin{document}

\title{Electromagnetic follow-up of gravitational waves: review and lessons learned}

\author{
M. Nicholl$^{1}$ and I. Andreoni$^{2,3,4}$}

\address{$^{1}$Astrophysics Research Centre, School of Mathematics and Physics, Queens University Belfast, Belfast BT7 1NN, UK
\\
$^{2}$Joint Space-Science Institute, University of Maryland, College Park, MD 20742, USA \\
$^{3}$Department of Astronomy, University of Maryland, College Park, MD 20742, USA \\
$^{4}$Astrophysics Science Division, NASA Goddard Space Flight Center, Mail Code 661, Greenbelt, MD 20771, USA}

\subject{Astrosphysics, Gravitational Waves}


\corres{Matt Nicholl\\
\email{matt.nicholl@qub.ac.uk}}

\begin{abstract}
The detection of gravitational waves (GWs) has provided a new tool to study the Universe, with the scientific return enriched when combined with established probes: electromagnetic (EM) radiation and energetic particles. Since the groundbreaking detection in 2017 of merging neutron stars producing GW emission, a gamma-ray burst and an optical `kilonova', the field has grown rapidly. At present, no additional neutron star mergers have been jointly detected in GW and EM radiation, but with upgrades in EM and GW facilities now is a chance to take stock of almost a decade of observations. We discuss the motivations for following up GW sources and the basic challenges of searching large areas for a rapidly-evolving EM signal. We examine how the kilonova counterpart to GW170817 was discovered and the association confirmed, and outline some of the key physics enabled by this discovery. We then review the status of EM searches since 2017, highlighting areas where more information (in GW alerts or catalogs) can improve efficiency, and discuss what we have learned about kilonovae despite the lack of further multi-messenger detections. We discuss upcoming facilities and the many lessons learned, considering also how these could inform searches for lensed mergers.
\end{abstract}


\begin{fmtext}
\end{fmtext}


\maketitle

\section{Introduction}
\label{sec:intro}


Gravitational waves (GWs) are propagating fluctuations in the curvature of spacetime, produced by time-varying gravitational fields (formally, variation in the quadrupole moment of the mass distribution). Their direct detection is enormously challenging: at cosmological distances, the signal is negligible unless the system involves massive objects moving at large fractions of the speed of light. These conditions are fulfilled mainly by compact objects in extremely tight binary orbits. Emission of GW radiation causes the orbit to shrink, and as it does so the amplitude (and frequency) of the emission increases. This eventually drives the binary components to merge, with a dramatic increase in the GW signal during the final few orbits. 

After decades of experimentation, the field of observational GW astronomy reached maturity on 14th September, 2015, when the Advanced Laser Interferometer Gravitational wave Observatory (Advanced LIGO) \cite{LIGOScientificCollaboration2015} detected the final inspiral and coalescence of two stellar mass black holes (BHs) \cite{Abbott2016}. This first detection, at the very beginning of the first GW Observing Run (O1), was a landmark event, and marked the first confirmation of binaries consisting of two BHs, and of BHs with masses $\approx 30$\,\M. 

Since then, the two LIGO detectors at Hanford and Livingston have undergone further upgrades in sensitivity, and Advanced Virgo \cite{Acernese2015} and the Kamioka GW detector (KAGRA) \cite{KagraCollaboration2019} have come online (in 2017 and 2020, respectively). Together, they constitute the International Gravitational-Wave Observatory Network (IGWN). After the second and the third Observing Runs (O2 and O3), the catalog of detected GW sources now includes 90 high-confidence compact object mergers \cite{Abbott2023} (with an additional $>100$ significant candidates so far in the ongoing O4). For most sources, both binary components have masses $>3$\,\M\ and are therefore very likely BHs. However, in the GW population studies conducted to date \cite{Abbott2021,Abbott2023,Abbott2024}, at least seven sources\footnote{GW170817 \cite{Abbott2017}, GW190425 \cite{Abbott2020a}, GW190426\_152155 \cite{Abbott2021}, GW190917\_114630 \cite{Abbott2024}, GW191219\_163120 \cite{Abbott2023}, GW200105 and GW200115 \cite{Abbott2021a}.} show evidence for one or both components having masses in the range $\approx1.1-2.5$\,\M, consistent with plausible neutron star (NS) masses.

GW source detection relies on interferometry using a pair of lasers in an L-shaped configuration. The passage of an oscillating GW signal through the detector changes the relative path length of these lasers, leading to oscillating constructive and destructive interference when the two lasers are recombined. This measures the time-dependent frequency and amplitude of the wave. For compact binary mergers, the frequency is proportional to the orbital frequency, thus this constrains the component masses. For a given mass, the amplitude is inversely proportional to the distance \cite{Finn1993}. However, with one detector, there is very little directional information. Directionality is best inferred using multiple detectors separated by large distances, and comparing the arrival times (as well as the relative phase and amplitude of the waveform) in each detector to triangulate the GW signal. For events detected by the two LIGO detectors and Virgo, the signal can typically be localized to a `skymap' with an area of tens to hundreds of square degrees (with higher signal-to-noise ratio corresponding to tighter localization), whereas with fewer detectors the skymaps span thousands of square degrees \cite{Abbott2023}.

One of the key goals since the first GW detection has been to identify electromagnetic (EM) radiation from the same source producing the GW signal. This is generally referred to as `multi-messenger' astronomy, where a `messenger' may refer to one of GW or EM radiation, or energetic particles such as neutrinos or cosmic rays. Each messenger carries different information about the physical conditions, and analysing them together can lead to major breakthroughs. An early example is the discovery of solar neutrinos \cite{Bahcall1989}. The first example of multi-messenger astronomy (MMA) applied to energetic transients is the detection of neutrinos from supernova SN~1987A, which confirmed the collapse of the progenitor stellar core to a NS \cite{Arnett1989}. In GW astrophysics, detecting EM radiation constrains several aspects that are difficult or impossible to measure from GW data alone: precise localization of the signal to a position within a specific galaxy to probe the environmental conditions; a precise redshift from the galaxy spectrum, enabling cosmological studies; the mass, velocity and composition of any matter ejected from the system, probing the resulting nucleosynthetic pattern; and the orientation of the merging binary relative to the observer, measurable if a jet is launched and detected. Measuring the orientation (and redshift) breaks degeneracies in analysing the GW signal, allowing better understanding of the component masses. Therefore MMA observations can be more than the sum of their parts, but often require searching large fractions of the sky to look for a signal that may be faint, short-lived, or (worse!) highly uncertain.

The merger between two BHs is not generally expected to produce EM emission, unless the event occurs in a gas-rich environment \cite{Perna2016,McKernan2019}. Not to be discouraged, astronomers conducted deep and wide EM follow-up campaigns of the early BH detections, in particular the first two sources GW150914 \cite{Evans2016,Kasliwal2016,Smartt2016,Soares-Santos2016} and GW151226 \cite{Cowperthwaite2016,Smartt2016a,Brocato2018} and the best localized sources such as GW170814 \cite{Doctor2019,Smith2019,Grado2020}. A possible gamma-ray counterpart (a flare with $2.9\sigma$ significance) was identified in temporal coincidence with GW150914 by the \textit{Fermi} satellite \cite{Connaughton2016}, but this was not detected by \textit{INTEGRAL} \cite{Savchenko2016}. No optical or radio emission was detected. To date, perhaps the most plausible EM counterpart to a binary BH merger is associated with GW190521, a source with component masses of $\approx85$ and $\approx66$\,\M\ inferred from the GW signal \cite{Abbott2020}. A subsequent optical flare lasting several months, from the centre of a galaxy within the GW localization volume, was identified by \cite{Graham2020}, suggesting a possible merger inside the accretion disk of a central supermassive BH. However, it is difficult to exclude that the flare was instead due to accretion onto the central supermassive BH, due to the possibility of chance coincidence within the skymap \cite{Palmese2021, Veronesi2024arXiv}.

In contrast, models of compact binary mergers involving at least one NS make clear predictions for observable signatures. NS mergers have long been suspected to be the cause of short-duration gamma-ray bursts (GRBs; \cite{Paczynski1986,Eichler1989,Berger2014}), producing a spike of hard gamma-ray emission which usually lasts for less than 2\,s (but see \S\ref{sec:o3}). The satellites that detect GRBs typically have a field of view that covers a large fraction of the sky (thousands of square degrees), but cannot usually localize the source to a specific galaxy (though the Burst Alert Telescope on-board the Neil Gehrels \textit{Swift} Observatory can localize sources to within a few arcminutes \cite{Barthelmy2005}). Most gamma-ray missions automatically cross-match any detected GRBs with GW alerts to search for temporal and spatial coincidence, as discussed previously in the case of GW150914. However, the prompt emission from these sources is visible only if the observer is within the $\sim10^\circ$ opening angle of the jet producing the gamma-rays \cite{Fong2015}, otherwise it is relativistically beamed out of our line of sight. The deceleration of the jet with the interstellar medium eventually leads to emission visible to an off-axis observer (and at all wavelengths, not just gamma-rays) as the shock spreads into their line of sight. However, this so-called `afterglow' emission can be very faint for events viewed far from the jet axis \cite{Metzger2012}.

Another, more isotropic signal from a NS merger has also been predicted. The decompression of dense, highly neutron-rich material is a promising site for rapid neutron capture, or the `r-process', in which neutrons are added to seed nuclei faster than the products can beta-decay \cite{Lattimer1974,Eichler1989,Freiburghaus1999}. This process must occur in nature, in order to match the observed abundance patterns of heavy elements, i.e.~those with atomic mass number $A\gtrsim80$ \cite{Burbidge1957}. In the NS merger scenario, subsequent radioactive decays of heavy r-process nuclei heat the ejected material, leading to an EM transient that has been termed a `kilonova' \cite{Li1998,Rosswog1999,Metzger2010}. This emission is thermal in character and peaks at optical or infrared (IR) wavelengths. Studying the kilonova provides a direct constraint on the mass of heavy elements produced, with deep implications for understanding cosmic chemical evolution. This strongly motivates EM follow-up observations of mergers between two NSs, or between a NS and a BH. The remainder of this paper will review the progress in (mainly) optical and near-infrared (NIR) imaging searches for kilonova emission from GW-detected compact binary mergers, and the lessons learned for future follow-up.

\section{Approaches to optical counterpart searches}
\label{sec:strategies}

\begin{figure}[!t]
\centering\includegraphics[width=\textwidth]{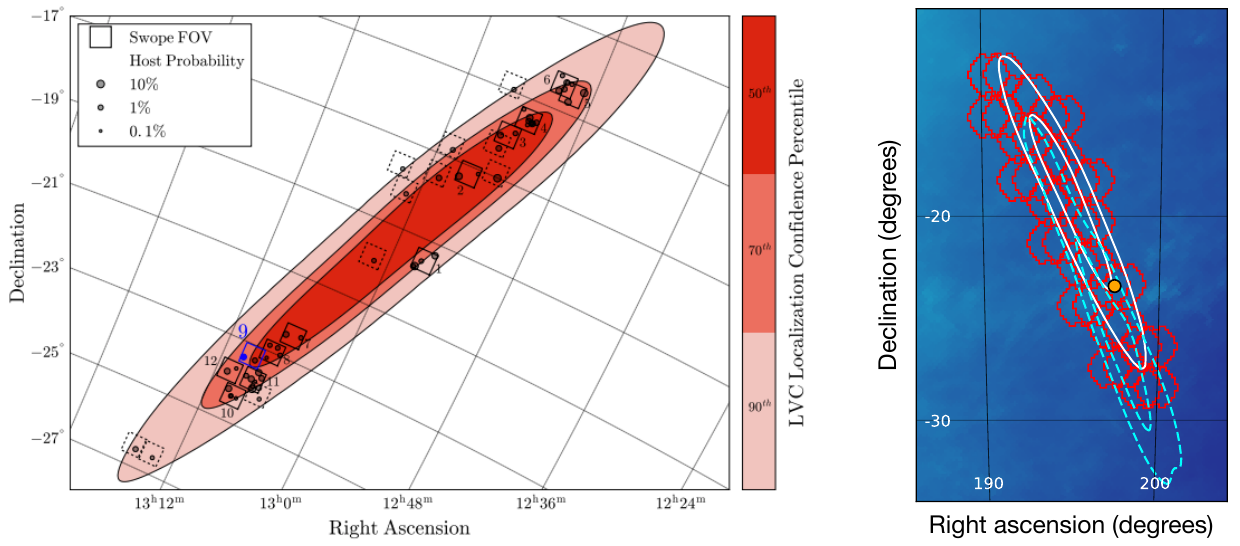}
\caption{Main EM search strategies used in GW follow-up. In both cases, the contours show the sky localization (skymap) from the LIGO-Virgo analysis of GW170817 (\S\ref{sec:170817}). Left: Example of galaxy-targeted follow-up, where telescope pointings are chosen to cover the most probable host galaxies. Each square shows one pointing with the Swope telescope, while the circles indicate known galaxies. Right: Example of tiled follow-up with a wide-field telescope, covering the full skymap. Each hexagon shows the footprint of one DECam pointing. The dashed blue lines show the shifted GW skymap after the final low-latency LIGO-Virgo analysis. Adapted from figures by D.~Coulter et al.~\cite{Coulter2017} and M.~Soares-Santos et al.~\cite{Soares-Santos2017}.}
\label{fig:strategies}
\end{figure}

Strategies to find optical counterparts to GW sources fall into two broad categories, which reflect the two techniques that have been used historically to search for other transients like supernovae. These two methods are illustrated in Figure \ref{fig:strategies}.

The first approach is to use telescopes with a wide field of view, and try to `tile' as much of the localization region as possible. This is analogous to the current generation of all-sky transient surveys (many of which now prioritise observing GW skymaps following an alert from the IGWN), such as the All-sky Automated Search for Supernovae (ASAS-SN) \cite{Shappee2014}, the Asteroid Terrestrial impact Last Alert System (ATLAS) \cite{Tonry2018}, the Gravitational Wave Optical Transient Observer (GOTO) \cite{Steeghs2022}, the Panoramic Survey Telescope and Rapid Response System (Pan-STARRS) \cite{Chambers2016}, and the Zwicky Transient Facility (ZTF) \cite{Bellm2019, Graham2019}.

The advantages of this `synoptic' method are that covering more area increases the chances of finding events that occur in unremarkable locations (e.g.\ at large projected distances from any bright galaxies) and more generally it does not require complete knowledge a priori of galaxy positions and distances within the skymap. Observing the full skymap also allows for robust inference in the case of a non-detection. The main disadvantage of this method is that wide-field telescopes often tend to be smaller in aperture (sacrificing depth for area) so searches may not be complete for faint (but still plausible) EM counterparts. There are exceptions to this: the Dark Energy Camera (DECam) \cite{Flaugher2015} can perform wide and deep searches using the 4\,m Blanco Telescope \cite{Kessler2015}, and in the near future the Vera C. Rubin Observatory will conduct the first wide-field time-domain survey with an 8\,m-class telescope \cite{Ivezic2019} (see \S\ref{sec:future}). Tiling the full skymap can also be slow (spending some fraction of observing time on regions of relatively low probability), but several strategies have been proposed to optimise the list of telescope pointings \cite{Ghosh2016,Rana2017,Coughlin2016,Coughlin2018} and even to observe sky regions at specific times when kilonova models are expected to be brightest \cite{Salafia2017}.

The alternative method, typically employed for telescopes with a narrower field of view, is to target the most likely host galaxies in which the merger could have occurred. This is more analogous to historical supernova searches that observed massive, nearby galaxies, e.g.\ the Lick Observatory Supernova Search \cite{Li2000}. In the GW follow-up case, galaxies can be prioritised for observation if they are massive (since the merger rate will scale with stellar mass), close to the highest probability regions of the skymap, and at a distance consistent with the GW inference. Several algorithms now exist to rank galaxies in the skymap within this formalism \cite{Gehrels2016,Arcavi2017a,Salmon2020,Ducoin2020, Coulter2024arXiv}. 

The advantages of this approach are that large telescopes capable of deep narrow-field observations can detect faint transients that may be beyond the reach of most survey telescopes, and it can be applied to essentially any telescope, enabling more astronomers to participate in the search. If the GW source does reside in a high probability galaxy, this approach can also be efficient for faster detection of a kilonova or luminous afterglow. However, since this approach will not cover the entire GW skymap, analysis of non-detections can be complicated. The construction of the target list also relies on the completeness of galaxy catalogs, a major issue for mergers outside of the very local Universe (see \S\ref{sec:o3}).

\section{GW170817, GRB\,170817A and AT2017gfo}
\label{sec:170817}

\begin{figure}[!t]
\centering\includegraphics[width=\textwidth]{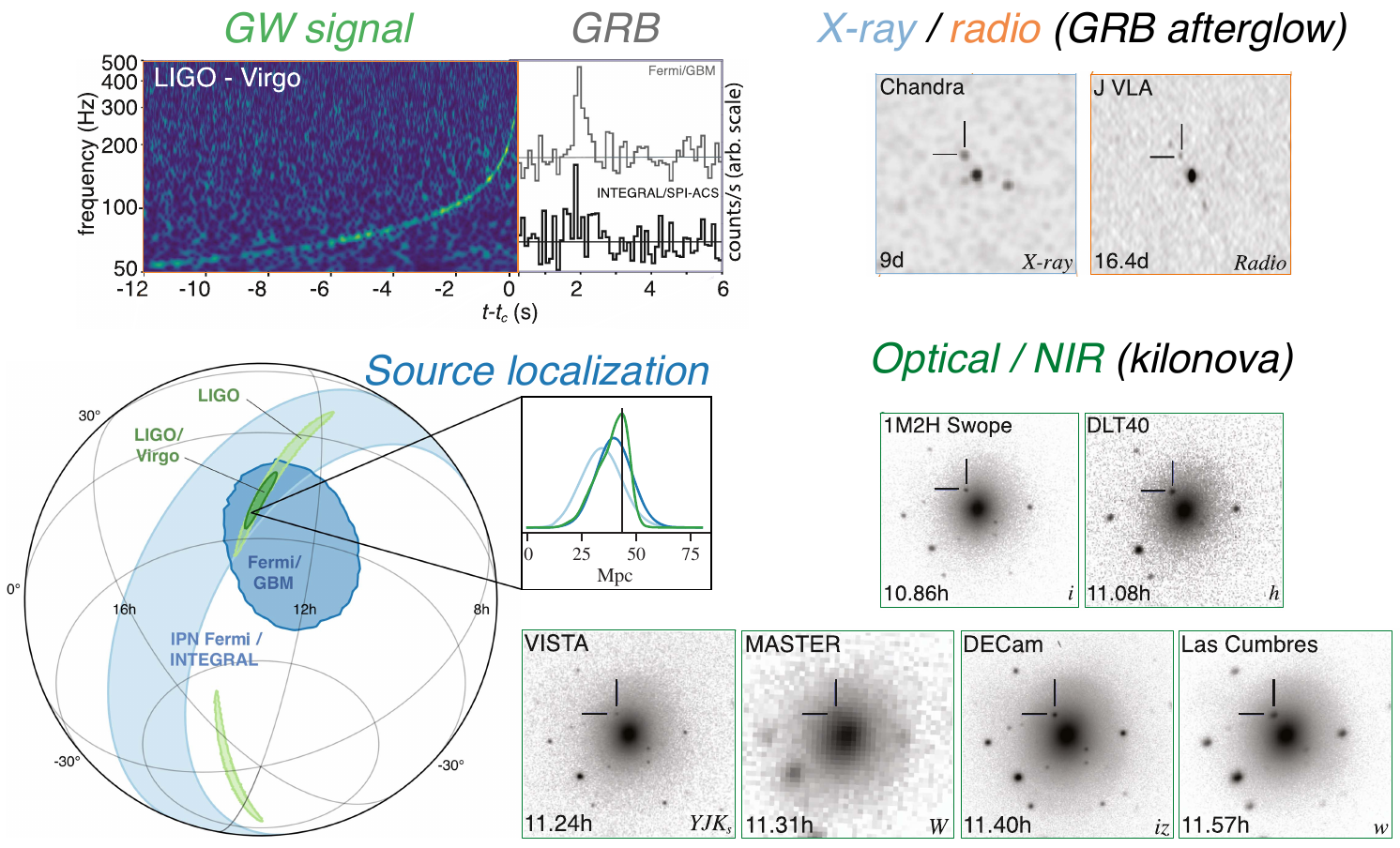}
\caption{The discovery and localization of GW170817 and the multi-wavelength emission with different messengers: GW emission, the GRB and its afterglow, and the kilonova. Adapted from original figures by the LIGO-Virgo Collaboration \cite{Abbott2017,Abbott2017mma}.}
\label{fig:170817-discovery}
\end{figure}

\subsection{GW discovery and GRB counterpart}

The first detection of GW emission from a merging NS binary occurred during the second GW observing run (O2) on August 17th, 2017, at 12:41:04 UTC \cite{Abbott2017}. GW170817 was a highly significant event initially identified by the LIGO Hanford detector. It was also detected at high signal-to-noise ratio by the LIGO Livingston detector, though a noise `glitch' in the detector (later modelled and removed) prevented immediate automatic identification of the signal \cite{Abbott2017}. In a stroke of good fortune, Virgo had recently joined the GW detector network on August 1st. Across the three detectors, the combined signal-to-noise ratio of 32.4 was at that time the highest yet observed for a GW event. This was largely due to the long inspiral time ($\sim100$\,s) spent in the LIGO/Virgo sensitive frequency band (BH mergers sweep through the band more rapidly, in $<1$\,s). The final GW analysis revealed a very typical double NS system, with both component masses in the range $1.17-1.60$\,\M\ \cite{Abbott2017}.

Although the signal was not significant in the Virgo data alone, this non-detection proved to be highly constraining for the sky localization: it indicated that the source must be in a direction of low-sensitivity, or `blind spot', of the Virgo antenna pattern at that time. The initial analysis from the two LIGO detectors localized GW170817 to a region of 190\,\sqdeg, whereas a few hours later the updated analysis including Virgo narrowed this to only 31\,\sqdeg\ (Figure \ref{fig:170817-discovery}). The final analysis conducted over subsequent days shrunk this further to 28\,\sqdeg. The distance inferred from the GW waveform was only $\approx 40$\,Mpc.

\begin{figure}[!t]
\centering\includegraphics[width=0.7\textwidth]{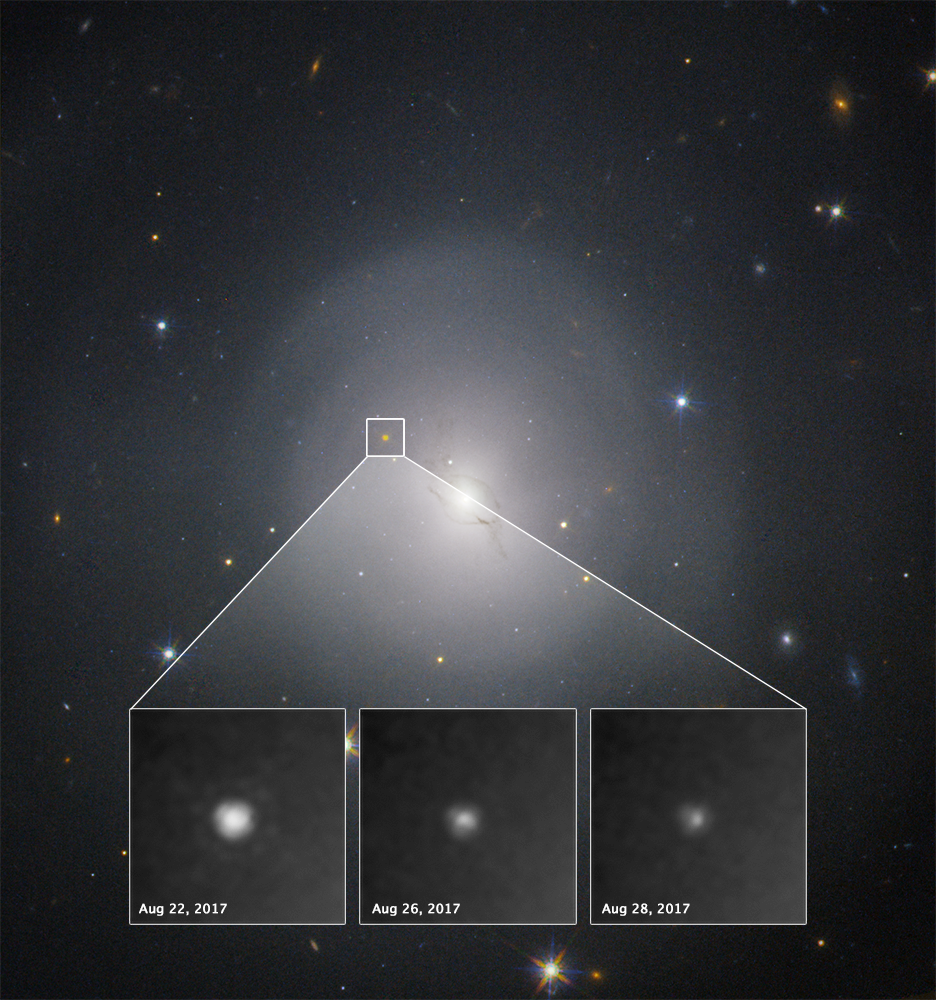}
\caption{\textit{Hubble Space Telescope} combined colour (F606W, F110W, F160W) image of NGC~4993, the host galaxy of GW170817 \cite{Blanchard2017,Levan2017,Pan2017}. Inset shows a zoom-in of the optical counterpart, AT2017gfo, fading over several nights. Image credit: A.~Levan, N.~Tanvir, A.~Fruchter and O.~Fox.}
\label{fig:ngc4993}
\end{figure}

As if the small skymap and proximity were not compelling enough, \textit{Fermi} and \textit{INTEGRAL} reported the detection of a short GRB (GRB\,170817A) arriving 1.7\,s after the GW signal, and with an overlapping localization region \cite{Goldstein2017,Savchenko2017}. The probability of temporal and spatial chance coincidence between the short GRB and GW signals was estimated to be $5\times10^{-8}$ \cite{Abbott2017grb}. At 40\,Mpc, this would be the closest short GRB (with a measured distance) ever observed. Its flux was surprisingly low, such would not have been detectable at a typical short GRB redshift of $z\sim0.5$. The likely explanation is that cosmological GRBs are viewed very close to the jet axis, whereas GRB\,170817A was viewed at $\sim20-30^\circ$ off-axis (see \S\ref{sec:phys}). Due to the steep fall-off in Lorentz factor with angle, most of the energy was beamed out of our line of sight. 

This is in some ways a very fortunate alignment. Being close enough to the jet axis to detect some gamma-ray emission had major implications for our understanding of fundamental physics: the difference of less than 2\,s between the arrival of the GW and EM signals constrains the fractional difference in speed between these two messengers to be $\lesssim 10^{-15}$ \cite{Abbott2017grb}. However, if we had been `too' on-axis, the GRB afterglow would dominate over the kilonova emission during the first few days after merger, making detection of heavy element signatures more complicated. From off-axis, the afterglow emission was initially faint, minimising contamination of the kilonova, but as it spread into our line of sight gradually over a few months it would enable detailed multi-wavelength observations and modeling of the angular structure of the jet \cite{Alexander2017,Alexander2018,Haggard2017,Hallinan2017,Margutti2017,Margutti2018,Ruan2018,Troja2017,Troja2018,Troja2019,DAvanzo2018,Dobie2018,Lyman2018,Mooley2018,Fong2019,Hajela2019,Lamb2019,Ghirlanda2019}. While arguably this was an ideal situation for the \emph{physical} analysis of GW170817 and its multi-wavelength emission, a more directly on-axis GRB and its afterglow would have been more luminous, which would be more helpful at higher redshifts if our goal is simply to \emph{detect} an EM counterpart (e.g. for source localization).

\subsection{Optical searches and confirming the kilonova}

As most of the localization probability was in the Southern Hemisphere, and at a Right Ascension close to evening twilight, most of the optical searches began immediately as the sun set in Chile on August 17th. Unfortunately, this was around 10 hours after the GW signal, meaning that crucial information about the early EM emission was lost. This highlights the need for telescopes, capable of rapid-response counterpart searches, \emph{spread over a range of longitudes}. 

Many groups and telescopes searched for the optical counterpart to GW170817/GRB\,170817A. The first detection and first report of an optical transient in the galaxy NGC~4993 (at a distance $38.9\pm1.3$\,Mpc; Figure \ref{fig:ngc4993}) came from the One Meter Two Hemisphere collaboration \cite{Coulter2017}, who used the Swope Telescope in Chile to perform a galaxy-targeted search. The source was named SSS17a (for Swope Supernova Survey) or AT2017gfo (following transient naming conventions of the International Astronomical Union). Within the next hour, and before the Swope detection was announced via a NASA General Coordinates Network\footnote{at that time known instead as the Gamma-ray Coordinates Network} (GCN) circular \cite{Coulter2017gcn}, five other telescopes independently detected the same optical transient. This included two other galaxy-targeted searches: Distance Less Than 40\,Mpc \cite{Yang2017gcn1,Valenti2017} and the Las Cumbres Observatory network \cite{Arcavi2017gcn1,Arcavi2017}. NGC~4993 was ranked as the 5th most likely host galaxy by the Las Cumbres team \cite{Arcavi2017} and the 12th by the Swope team \cite{Coulter2017}. The other three independent detections came from wide-field searches that tiled the skymap: VISTA \cite{Tanvir2017gcn,Tanvir2017}, MASTER \cite{Lipunov2017gcn,Lipunov2017}, and DECam \cite{Allam2017gcn,Soares-Santos2017}. The discovery data using different wavelengths and messengers are shown in Figure \ref{fig:170817-discovery}. Immediate optical and IR follow-up of the source was obtained by many groups, ultimately showing that AT2017gfo really was the optical counterpart \cite{Andreoni2017,Chornock2017,Covino2017,Cowperthwaite2017,Diaz2017,Drout2017,Evans2017,Hu2017,Kasliwal2017,McCully2017,Nicholl2017,Pian2017,Pozanenko2018,Shappee2017,Smartt2017,Troja2017,Utsumi2017}.

\begin{figure}[!t]
\centering\includegraphics[width=0.53\textwidth]{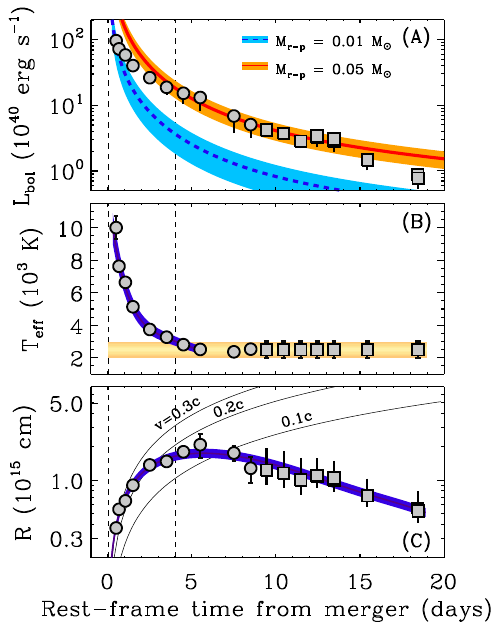}
\centering\includegraphics[width=0.42\textwidth]{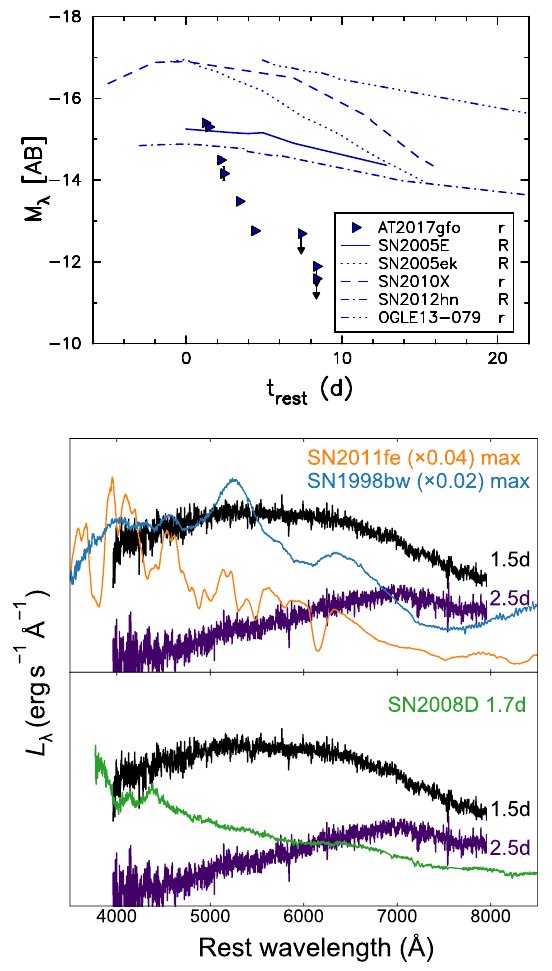}
\caption{Evidence that the source AT2017gfo was the optical counterpart to GW170817. Left: Luminosity, temperature and radius evolution. The expansion of the radius at early times is consistent with a mildly relativistic outflow; this was noted in the first night of observing. At later times, the luminosity evolution indicated that a few $\times0.01$\,\M\ of r-process elements were produced. Top right: From the second night onwards, the rapid decline of the optical light curve was faster than any historical supernova. The comparison shown is to some of the fastest-fading known supernovae. Lower right: The spectroscopic evolution of AT2017gfo was also unique. Compared to supernovae at the peak of the light curve, the early spectra of AT2017gfo are remarkably featureless. Compared to a supernova 1-2 days after explosion, AT2017gfo is already exceptionally red and cools extremely quickly. Adapted from figures by M.~Drout et al.~\cite{Drout2017}, S.~Smartt et al.~\cite{Smartt2017} and M.~Nicholl et al.~\cite{Nicholl2017}.}
\label{fig:170817-early}
\end{figure}

Some results from this detailed follow-up will be summarised in the next section. However, for the purpose of learning lessons applicable to future GW counterpart searches, it is instructive to examine how and when we arrived at the conclusion that AT2017gfo was the real deal. This is particularly important in the context of future GW detectors or lensed GW sources, where the counterparts of distant events are likely to appear faint and we may be required to identify them from relatively sparse EM data. For this, we turn to the GCN circulars archive, which provides a `time capsule' of how thinking evolved within the EM follow-up community in the days following GW170817. Aside from having a position consistent with the 3D sky localization, the first EM clue came from examining archival observations of NGC~4993. Astronomers were quick to check recent data from all-sky surveys, finding no evidence for a transient in ASAS-SN, SkyMapper or ATLAS data \cite{Cowperthwaite2017gcn1,Moller2017gcn,Smartt2017} prior to GW170817. The estimated explosion date was constrained to be $<2$\,days before the first optical detection. It was estimated that the probability of AT2017gfo being a coincidental core-collapse supernova, unrelated to GW170817, was $<10^{-4}$ \cite{Foley2017gcn}. A \textit{Hubble Space Telescope} image obtained on April 28, 2017 showed no source at the position of the transient to a limiting absolute magnitude $M>-7.2$ in $V$-band \cite{Foley2017gcn2}.

By the end of the first night, additional evidence emerged from the spectral energy distribution of the source. Using multi-band photometry, several groups inferred a blackbody temperature of $\approx 8000\,$K, and an expansion velocity of $\approx0.2c$ \cite{Cowperthwaite2017gcn2,Malesani2017gcn1,Cenko2017gcn} if the explosion occurred within the previous 2 days (or even faster if the explosion time was assumed to be the merger). This mildly relativistic expansion was consistent with kilonova model predictions, but faster than expected for any supernovae other than a few broad-lined SNe Ic. While this evidence obtained during the first night was certainly very promising, it remained circumstantial. Moreover, the first reports of relatively blue optical colours \cite{Nicholl2017gcn1} and a blue spectrum \cite{Drout2017gcn} appeared inconsistent with the expectations at that time that kilonova ejecta dominated by heavy elements would be distinctively red \cite{Barnes2013, Tanaka2013}.

On the second night of observing, it became possible to measure the time evolution of the optical transient. With this, compelling evidence emerged that AT2017gfo was a genuinely new transient unlike any supernova. Within an hour of sunset in Chile, reports emerged that the transient was declining rapidly, fading at optical wavelengths by around 0.5 magnitudes per day \cite{Yang2017gcn2,Nicholl2017gcn2,Chambers2017gcn}. Spaced-based UV observatories (see review by S.~Oates in this issue) or ground-based networks with a spread of latitudes confirmed this fast evolution at even higher cadence \cite{Arcavi2017gcn2,Cenko2017gcn}. This is much faster than any known supernova (Figure \ref{fig:170817-early}). While only one spectrum was obtained on the first night \cite{Drout2017gcn}, showing a blue continuum common in young transients, spectra obtained by many groups on night two showed a remarkable evolution. The first reports noted a colour temperature of $\approx 5000-6000$\,K, with a deficit of flux at bluer wavelengths even in comparison to such a cool blackbody \cite{Lyman2017gcn,Nicholl2017gcn3}. More direct evidence in favour of a kilonova was soon obtained: NIR photometry showed an increasing brightness at longer wavelengths \cite{Wiseman2017gcn,Malesani2017gcn2}, and the first spectrum from X-Shooter on the European Southern Observatory Very Large Telescope (VLT), covering the UV-NIR range, drew immediate comparisons to kilonova models \cite{Pian2017gcn}.

The fast fading, rapid photospheric expansion, smooth optical spectrum and the rapid shift of flux to longer wavelengths convinced most groups that AT2017gfo was the kilonova counterpart to GW170817, and subsequent observations focused almost exclusively on this source until the field entered solar conjunction around two weeks later. Two further lines of evidence confirmed this beyond any doubt. DECam re-imaged the majority of the GW skymap in $i$ and $z$ bands between August 31st and September 2nd, to test for any other rapidly fading transients in the field (any plausible counterpart would have faded significantly from August 17th to 31st). Only one source was detected in both bands, had a reliable point-spread function, and faded by at least $3\sigma$: AT2017gfo \cite{Soares-Santos2017}. And in the many detailed analyses of the full photometric and spectroscopic data sets, convincing matches were found with predictions from kilonova models \cite{Andreoni2017,Chornock2017,Covino2017,Cowperthwaite2017,Diaz2017,Drout2017,Evans2017,Hu2017,Kasliwal2017,McCully2017,Nicholl2017,Pian2017,Pozanenko2018,Shappee2017,Smartt2017,Troja2017,Utsumi2017,Arcavi2017,Tanvir2017,Kilpatrick2017,Tanaka2017,Kasen2017,Villar2017}.

While no doubt a landmark moment for astrophysics, the enormous success in finding and following up the counterpart of GW170817 at least partly reflects the fact that searching $\sim30$\,\sqdeg\ at $\sim40$\,Mpc is relatively easy for modern telescopes. As we will see in \S\ref{sec:o3}, the indications since 2017 are that `typical' cases are a lot more challenging.

\begin{figure}[!t]
\centering\includegraphics[width=0.47\textwidth]{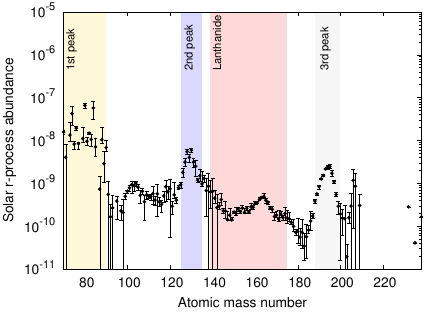}
\centering\includegraphics[width=0.47\textwidth]{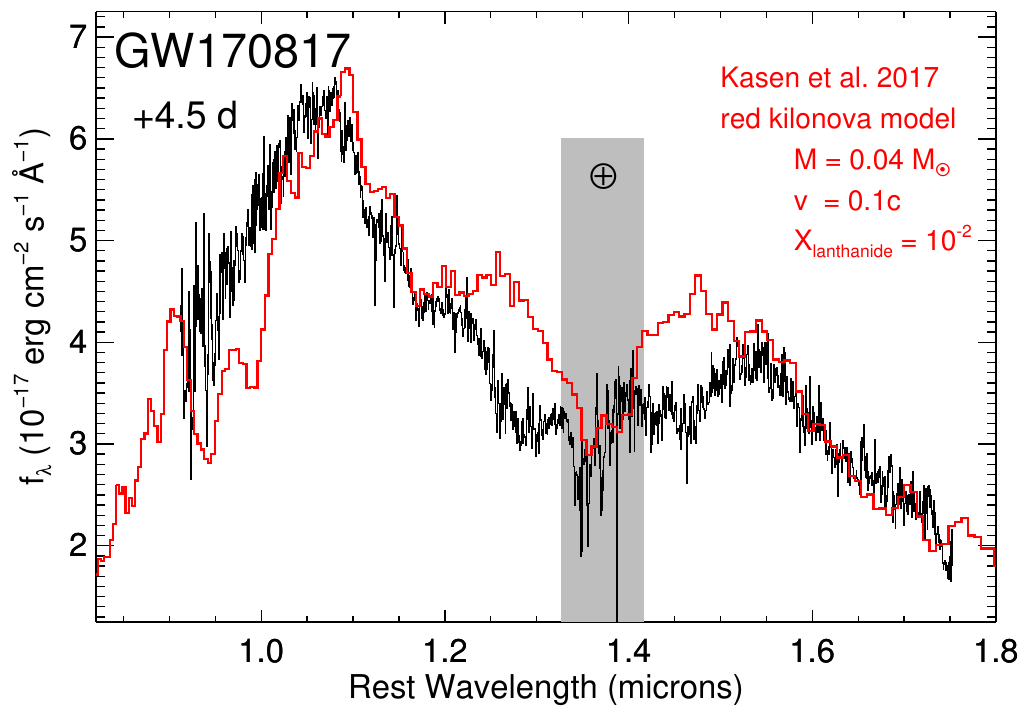}
\caption{Left: The solar r-process abundance pattern, showing the overall decline with mass number and the peaks at closed neutron shells. Highly neutron-rich environments are needed to synthesise elements beyond the second peak, but if this occurs the presence of lanthanides greatly affects the kilonova spectrum. Right: The NIR spectrum of AT2017gfo from the Gemini South telescope, showing an excellent match to a lanthanide-rich kilonova model. Adapted from figures by K.~Hotokezaka et al.~\cite{Hotokezaka2018} and R.~Chornock et al.~\cite{Chornock2017}.}
\label{fig:r-process}
\end{figure}

\subsection{Physical implications}
\label{sec:phys}

The exquisite data obtained for AT2017gfo had a huge impact on our understanding of NS physics, heavy element nucleosynthesis and cosmology. We will briefly summarise some of the major breakthroughs here. A much more detailed review of the multi-wavelength evolution and physical interpretation of GW170817/AT2017gfo is provided by \cite{margutti2021}.

One of the first goals was to understand the mass of r-process ejecta produced in the merger. While GW data can constrain whether mass was ejected from the system, only EM observations can probe its composition. The astrophysical r-process produces a distinct abundance pattern (Figure \ref{fig:r-process}), with peaks at atomic mass numbers $A\sim 80$, 130 and 195 (due to the increased stability of nuclei with full neutron shells having $N=50,82,126$ neutrons \cite{Burbidge1957}). Outside of these peaks, the abundances decline with mass number, due to the increasing demands on binding energy to keep adding neutrons. In order to produce elements beyond the second peak, an exceptionally high density of free neutrons is required. Whether these conditions are achieved in NS mergers is therefore a critical question for cosmic chemical evolution.

If produced, the elements with $A>140$ have a pronounced effect on the predicted kilonova spectrum. In particular, elements with open electron f-shells (lanthanides and actintides), have a huge number of possible electronic transitions, leading to large effective opacities at optical and UV wavelengths \cite{Barnes2013,Tanaka2013,Kasen2017}. A detailed review of kilonova spectral modelling is provided in this issue by C.~Collins; here we note only that model spectra (calculated even before 2017) for r-process material including lanthanides produced an excellent match to the spectrum of AT2017gfo (Figure \ref{fig:r-process}). This confirmed that the merger ejecta was composed of r-process material. Since the source of the optical and NIR luminosity is the radioactive decay of this material, the ejected mass can be inferred directly from the luminosity, with observations indicating a total ejecta mass of a few~$\times 0.01$\,\M. Multiplying this mass by the integrated NS merger rate suggests that these mergers could be responsible for the full cosmic budget of r-process elements \cite{Villar2017}.

\begin{figure}[!t]
\centering\includegraphics[width=\textwidth]{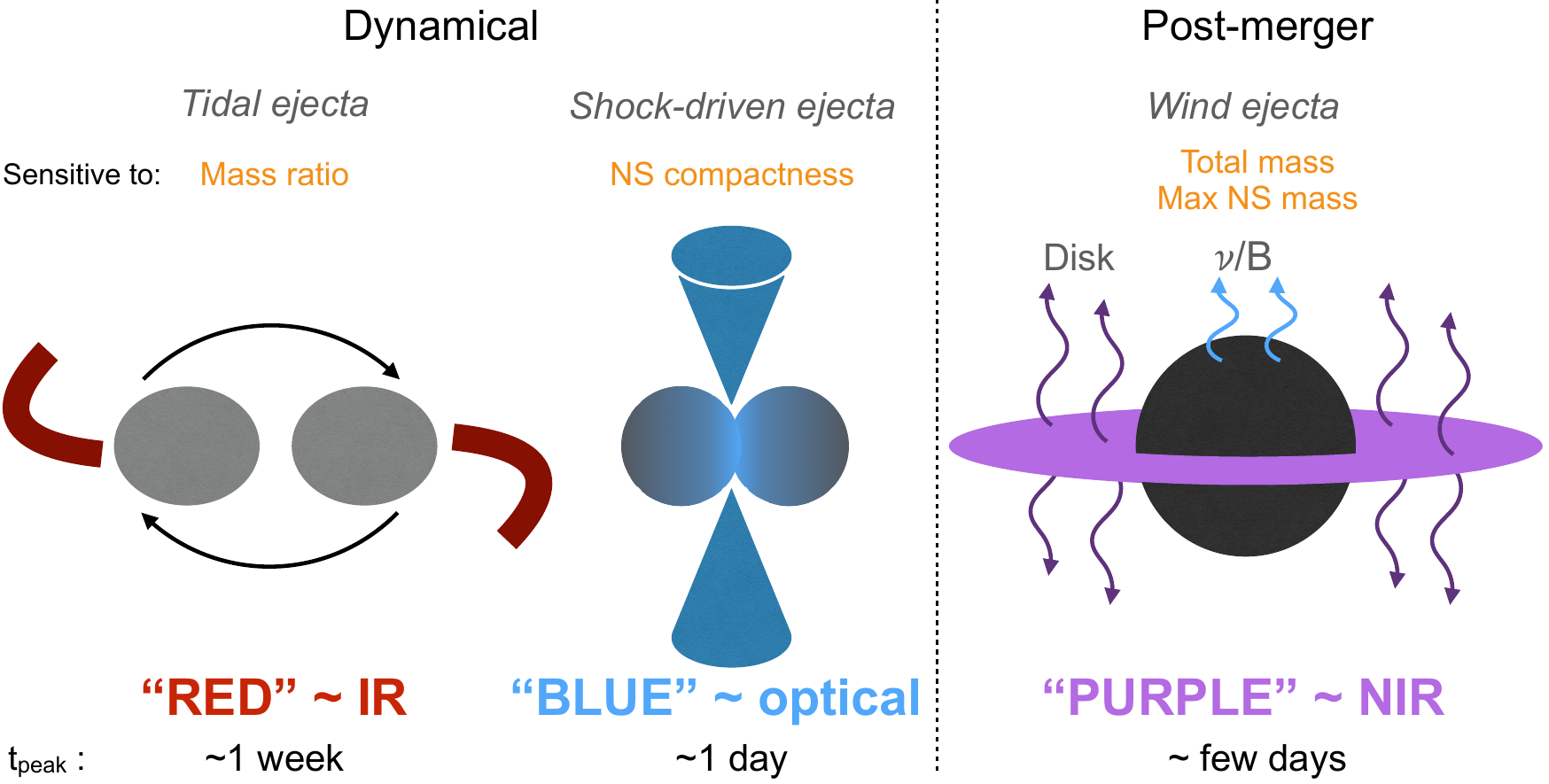}
\caption{Cartoon showing sources of ejecta in NS mergers. Tidal stripping of the less massive NS in a binary expels neutron-rich matter in the orbital plane, which can produce lanthanide-rich ejecta. More unequal mass ratios produce more of this `red' ejecta \cite{Rosswog1999,Sekiguchi2015}. If the lighter NS is not fully disrupted (expected if the binary mass ratio is $\gtrsim0.8$), the two will collide and further matter is ejected through shocks, primarily in the polar direction \cite{Hotokezaka2013,Bauswein2013}. This is strongly irradiated by neutrinos and likely does not make the heaviest r-process elements \cite{Wanajo2014,Goriely2015,Foucart2016}; these ejecta are considered `blue'. Models using a more compact (softer) equation of state predict more blue ejecta, as the collision occurs at smaller separation and hence higher orbital velocity. After the merger, an accretion disk forms around the remnant. Winds from the disk (or from the remnant surface) can dominate the ejecta mass budget, if the remnant is not too massive. In most cases the remnant is a rotationally-supported NS, but as it loses angular momentum it will collapse to a BH. Depending on how long the remnant survives as a NS and cools by neutrino emission, the wind may have a high, low or intermediate (`purple') opacity. The survival time of the remnant depends on the maximum mass of a NS allowed by the equation of state \cite{Metzger2014,Fernandez2016,Margalit2017}. In AT2017gfo we attribute the early blue emission mainly to shock-driven ejecta, and the long-lived redder emission to the disk wind ejecta \cite{Nicholl2021}.}
\label{fig:schematic}
\end{figure}

\begin{figure}[!t]
\centering\includegraphics[width=0.8\textwidth]{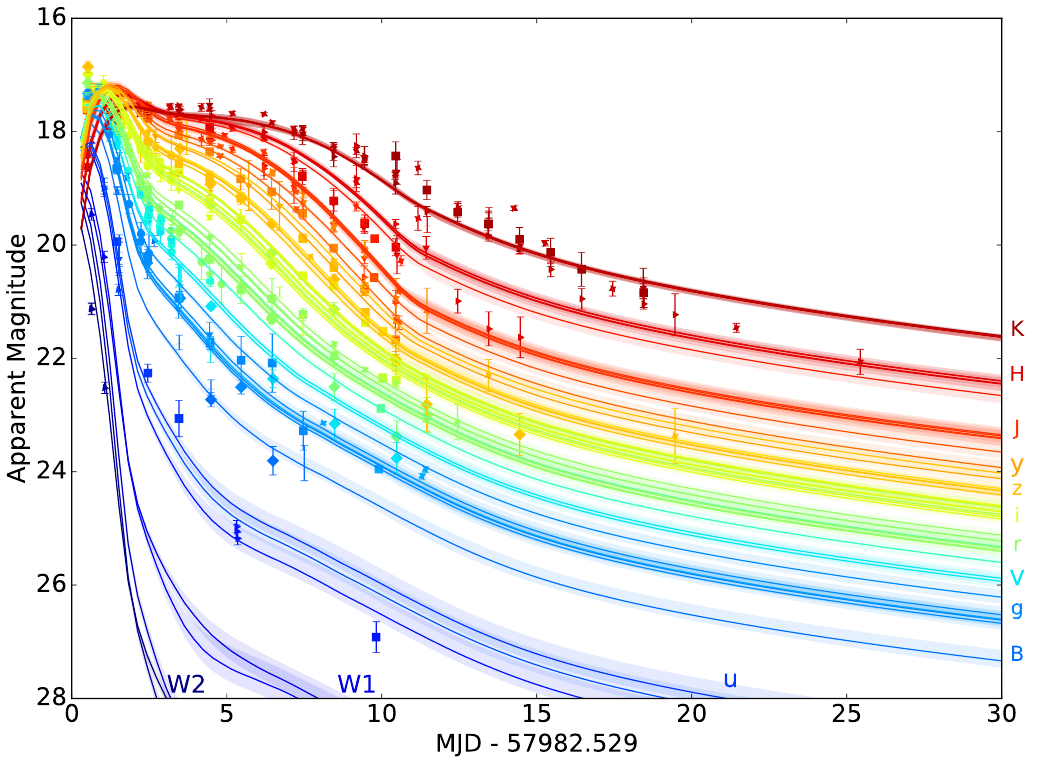}
\caption{Compilation by \cite{Villar2017} of the AT2017gfo photometry obtained by \cite{Andreoni2017,Arcavi2017,Coulter2017,Cowperthwaite2017,Diaz2017,Drout2017,Evans2017,Hu2017,Kasliwal2017,Pian2017,Pozanenko2018,Shappee2017,Smartt2017,Troja2017,Utsumi2017,Valenti2017,Tanvir2017,Lipunov2017}. The lines show fits with two- and three-component kilonova models. The peak on day one and the early UV emission is attributed to a low-opacity ($\kappa\approx0.5$\,\cmg) component with mass $M_{\rm b}\approx0.005-0.02$\,\M. The week-long peak in the $K$-band results from a high or intermediate opacity ($\kappa\approx3-10$\,\cmg) component with mass $M_{\rm r}\approx 0.02-0.04$\,\M. Reproduced from figure by V.~A.~Villar et al.~\cite{Villar2017}.}
\label{fig:villar}
\end{figure}

Given the robust detection of these heavy elements, two observed properties of AT2017gfo may initially seem surprising: the blue emission during the first $\sim$ day after merger, with a colour temperature of $\approx7000-11000$\,K \cite{Shappee2017} and detections even in the ultraviolet \cite{Evans2017}; and the immediate fast fade of the optical light curve. The light curve timescale of an expanding, internally heated transient is approximately 
\begin{align}\label{diffuse}
\begin{split}
t_{\rm peak} \sim \left(\frac{2 \kappa M}{\beta c v} \right)^{\frac{1}{2}},
\end{split}
\end{align}
where $\kappa$ is the average opacity, $M$ is the ejected mass with velocity $v$, and $\beta\approx13.7$ is a constant depending on the density profile \cite{Arnett1982}. For lanthanide-rich ejecta, $\kappa\approx10$\,\cmg\ \cite{Barnes2013,Tanaka2020}. For $v\approx0.2c$ and $M\approx0.02$\,\M, we find $t_{\rm peak}\approx 1$\,week. Ejecta expanding for 1 week at this velocity reach a radius $\sim10^{15}$\,cm, giving a temperature of a few $\times 1000$\,K and an overall `red' kilonova (even ignoring the forest of line absorption suppressing the flux at blue wavelengths yet further). On the other hand, lanthanide-poor ejecta have an effective opacity $\lesssim 1$\,\cmg, and can produce a `blue' kilonova \cite{Metzger2014} that peaks at $\sim 10,000$\,K within a day or two after merger. Simulations show that kilonovae can eject matter through different physical mechanisms, and that these ejecta components may be spatially distinct \cite{Rosswog1999,Sekiguchi2015,Bauswein2013,Hotokezaka2013}. Figure \ref{fig:schematic} shows the different sources of merger ejecta. These components can contain different fractions of lanthanides, depending on the extent to which they encounter neutrinos (which suppress the fraction of free neutrons through weak reactions). In short, it is possible to observe a `red' and `blue' kilonova simultaneously from the same merger. The ratio of their observed fluxes likely also depends on the observer viewing angle from the orbital axis.

\begin{figure}[!t]
\centering\includegraphics[width=\textwidth]{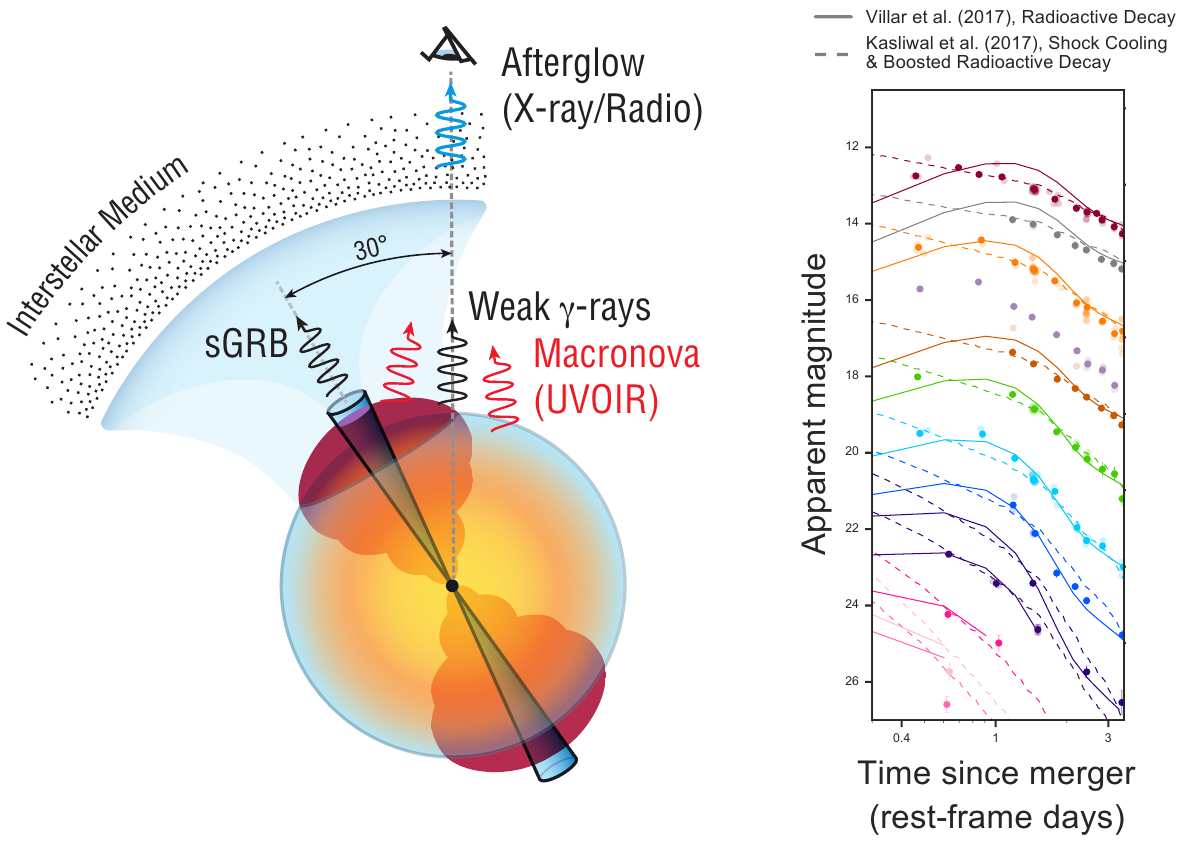}
\caption{Left: Schematic showing the possible contribution of a shock-heated `cocoon' to the early kilonova emission. Matter close to the polar axis, shown in dark red, is heated by the passage of the GRB jet, providing another luminosity source as it cools. Right: Models with shock cooling (from \cite{Kasliwal2017}; left) and neglecting shock cooling (from \cite{Villar2017}; Figure \ref{fig:villar}) can fit the data similarly well at $t\gtrsim 1$\,day after merger. Observations within the first hours are needed to distinguish between them. Adapted from figures by M.~Kasliwal et al.~\cite{Kasliwal2017} and I.~Arcavi et al.~\cite{Arcavi2018}.}
\label{fig:cocoon}
\end{figure}

Identifying distinct ejecta components in the data therefore provides important constraints on the merger physics. Modelling of the AT2017gfo light curve clearly required multiple emitting components to simultaneously match the early blue peak and the long-lived red emission. A model that includes a high, low and intermediate opacity component is shown in Figure \ref{fig:villar} \cite{Villar2017} (also highlighting the amount of optical and NIR imaging obtained by the numerous observing campaigns).
These models suggested that a ${\rm few}\times0.01$\,M$_\odot$ of blue ejecta were required to match the early colour and luminosity \cite{Cowperthwaite2017,Drout2017,McCully2017,Nicholl2017,Shappee2017,Villar2017,Bulla2019}. However, such a large quantity of low-opacity material is difficult to produce in merger simulations \cite{Bauswein2013}. This has led some authors to suggest that at least part of the early blue luminosity may have arisen instead from the interaction of the GRB jet with the polar ejecta, forming a `cocoon' of shock-heated material \cite{Kasliwal2017,Piro2018,Gottlieb2018}. Confirming which energy source dominates the early light curve requires observations within the first few hours after merger, which were not obtained for GW170817 (Figure \ref{fig:cocoon}). This clearly shows the importance of \emph{rapid response} follow-up. Radioactive decay of free neutrons can also heat the ejecta during the first $\sim$~hour \cite{Metzger2015}.

On the other hand, most studies agree that the longer-lived red component is primarily from disk winds, which can provide the required ejecta mass of $\approx 0.02-0.04$\,M$_\odot$. The masses of both the blue and redder ejecta have been used to place constraints on the NS equation of state (EoS). The EoS is the relation between pressure and density for nuclear matter, which determines the maximum stable mass of NS (the Tolman-Oppenheimer-Volkoff mass or $M_{\rm TOV}$) and the radius for a given mass. It was pointed out early on that the relatively large mass of blue (shock-driven) ejecta favours a fairly compact radius $R_{\rm NS}\lesssim12$\,km, or a `soft' EoS \cite{Nicholl2017}. Analysis of the energy extracted from the post-merger disk was used to constrain the lifetime of the NS remnant, indicating a maximum mass $M_{\rm TOV}<2.17$\,\M \cite{Margalit2017}. Several studies have tried to combine the EM constraints with those from the GW signal, either by modelling both simultaneously or by including the posteriors from GW waveform modelling in their EM model priors \cite{Coughlin2019,Dietrich2020,Nicholl2021,Breschi2021,Raaijmakers2021}. Further constraints can be folded in based on nuclear physics theory \cite{Dietrich2020} and known massive pulsars \cite{Raaijmakers2021}. Including multiple messengers can break model degeneracies, and tighten constraints on the equation of state. These models appear to be converging towards a NS radius of approximately $R_{\rm NS}\approx12$\,km.

The multi-messenger observations of GW170817 have been similarly impactful for cosmology. The EM detection localizes the merger to a specific galaxy, enabling an accurate measurement of its redshift. At the same time, the amplitude of the GW signal provides a measure of the distance \emph{independent of the local distance ladder}. This so-called `standard siren' approach \cite{Schutz1986} was used to measure the Hubble constant as $H_0=70^{+12}_{-8}$\,\kms\ \cite{Abbott2017H0}, consistent with both the local distance ladder estimate \cite{Riess2016} and the cosmic microwave background estimate \cite{Planck2016}. Although GW170817 did not resolve the tension between these two measurements, it is nonetheless impressive to constrain $H_0$ to $\sim15\%$ using only a single source. 

The EM detection also helps the cosmological analysis in another way. The largest degeneracy in the GW distance measurement is the unknown inclination of the binary orbit. Modelling of the GRB afterglow \cite{Guidorzi2017,Palmese2024} and kilonova emission \cite{Dhawan2020,Bulla2022} constrains our viewing angle and reduces the uncertainty on $H_0$. The gold standard in this regard is Very-Long Baseline Interferometry (VLBI) observations of the radio jet. Measurements of the jet motion in GW170817 constrained the viewing angle to between $14-28^\circ$ off-axis \cite{Mooley2018a,Ghirlanda2019}, leading to $H_0$ measurements with uncertainties of $\lesssim5$\,\kms\ \cite{Hotokezaka2019,Gianfagna2024}. Superluminal motion was also observed in the optical, thanks to the fine spatial resolution of the {\it Hubble Space Telescope}. Coupled with radio VLBI observations, this further narrowed down the viewing angle to $19-25^\circ$ \cite{Mooley2022}.  While the kilonova emission is more isotropic and subject to more modelling assumptions than the GRB afterglow emission (and is therefore less constraining overall on the viewing angle for a given event), it has the advantage of being easier to detect for events far off-axis, enabling future population studies if the kilonova emission can be accurately modelled \cite{Doctor2020}. It has also been suggested that the luminosity of the kilonova emission itself might be standardizable in the future, potentially opening another avenue for cosmological exploration \cite{Coughlin2020PhRvR, Coughlin2020NatCo}.

\section{GW follow-up after 2017}
\label{sec:o3}

\subsection{O3 alerts and parameters informing follow-up}
\label{sec:gcns}

After O2 and beginning with O3, GW alerts became available in real-time to the public (previously, these alerts were only available to those who had signed a memorandum of understanding with the LIGO-Virgo Collaboration). A large number of teams conducted EM follow-up of GW events, but at the time of writing no confirmed EM counterparts have been detected since GW170817 (though one claim exists for the binary BH merger, GW190521 \cite{Graham2020}).

In the age of open GW events, the alert format is standardized and made available over Kafka and via GCN notices or SCiMMA\footnote{\url{https://scimma.org/}} (the Scalable Cyberinfrastructure for Multi-messenger Astrophysics). The content of IGWN alerts is explained in a regularly maintained online Users Guide\footnote{\url{https://emfollow.docs.ligo.org/userguide/}}.  Plain text GCN circulars also remain important, as they explain and summarise the alert content in human-readable form. Alerts include the localization and distance posteriors, plus a limited number of additional parameters derived from waveform template matching, but do not currently include fundamental physical properties of the sources such as component masses. The alerts report the source significance via the False Alarm Rate (FAR), expressed as the inverse of the time one would have to wait on average to find a comparably large `signal' produced by detector noise. They also report which GW instruments participated in the detection. Typically several alerts are issued for a given source, which may update the skymap or source parameters following a more detailed and time-consuming waveform analysis, or retract a spurious event after human vetting. A key goal of this section is to ask how to decide in real time which GW alerts are most likely to be genuine, to avoid wasting telescope time.

Preliminary source classification is provided in the alerts, via the probabilities of a source being a binary black hole (BBH), binary neutron star (BNS), neutron star - black hole system (NSBH), or noise (terrestrial). The terrestrial probability is $p_{\rm terr} = 1-p_{\rm astro}$ (the latter being the total probability that a source is astrophysical). \emph{Under the assumption that the source is astrophysical}, we are also provided with the probability that at least one of the binary components has a mass consistent with a NS (\texttt{HasNS}), the probability that at least one component lies in the `mass gap'\footnote{The range above any plausible NS mass but below observed stellar mass BHs. Note that during O3, `MassGap' was reported as a separate class; this has been changed for O4.} between $3-5$\,\M (\texttt{HasMassGap}), and the probability that there is some mass outside of the event horizon of the merger remnant and therefore a possibility of EM emission (\texttt{HasRemnant}). This last parameter can be crucial in assessing whether to follow up NSBH sources, as we will discuss. That these probabilities are conditional means that it is not a contradiction for an alert to have both a large $p_{\rm terr}$ and a large \texttt{HasNS}, for example, but in this case we should interpret the latter with caution.

The public GW alerts are preserved in the Gravitational-Wave Candidate Event Database (GraceDB\footnote{\url{https://gracedb.ligo.org/}}). This makes it possible to revisit the events reported in real time and compare them to the set of GW events found by the full GWTC analyses \cite{Abbott2021,Abbott2023,Abbott2024}. These two sets of events do not have a one-to-one correspondence, as some events `detected' by the low-latency pipeline (even in addition to those retracted via GCN) are no longer significant in the final analysis, and conversely the GWTC analysis picks out additional signals that were not detected in low latency. We can therefore investigate whether any of the parameters reported in the alerts have predictive power for selecting events that have are recovered in the GWTC analysis.

We find 77 events from O3 in GraceDB, of which 45 are in GWTC-3. Of the remainder, 24 events were retracted, mostly within minutes (but occasionally within hours or days). The others were not retracted but are not recovered in the GWTC analyses; we assume such sources were also spurious. We label each source with its event type, taking care to update the source classifications from GraceDB if GWTC-2, GWTC-3 or another published IGWN paper classified a source as being a BNS, NSBH or in the mass gap (since a merger containing a NS may be capable of producing a kilonova). These particular events are listed in Table \ref{tab:o3-sources}. We will discuss the EM follow-up efforts for some of these sources in more detail later.

\begin{table}[!t]
\caption{BNS, NSBH and Mass gap sources reported in real-time during O3 and persisting in GWTC-2 or GWTC-3.}
 \label{tab:o3-sources}
 \begin{tabular}{lcc}
 \hline
Name & Class & Reference \\
\hline
GW190425 & BNS & \cite{Abbott2020a} \\
GW190426\_152155 & NSBH & \cite{Abbott2021} \\
GW200105\_162426 & NSBH & \cite{Abbott2021a} \\
GW200115\_042309 & NSBH & \cite{Abbott2021a} \\
GW190814  & Mass gap &  \cite{Abbott2020b} \\
GW190924\_021846 & Mass gap & \cite{Abbott2023} \\
GW190930\_133541 & Mass gap &  \cite{Abbott2023} \\
GW200316\_215756 & Mass gap &  \cite{Abbott2023} \\
\hline
 \end{tabular}
 \vspace*{-4pt}
 \end{table}

The vast majority of GraceDB sources that turned out to be spurious also belong to the BNS and NSBH classes. If a source is marginal, it is more important to release it early to the community if it is likely to contain a NS. The IGWN therefore conduct early warning searches for these events, which in turn leads to a larger number of early candidates that are retracted after vetting. Conversely, most BBH events in GraceDB remain in GWTC. In our analysis we therefore consider both the full set of events and the set of NS-bearing events separately, as their statistics (completeness and purity) under different selection criteria are quite different.

\begin{figure}[!t]
\centering\includegraphics[width=0.9\textwidth]{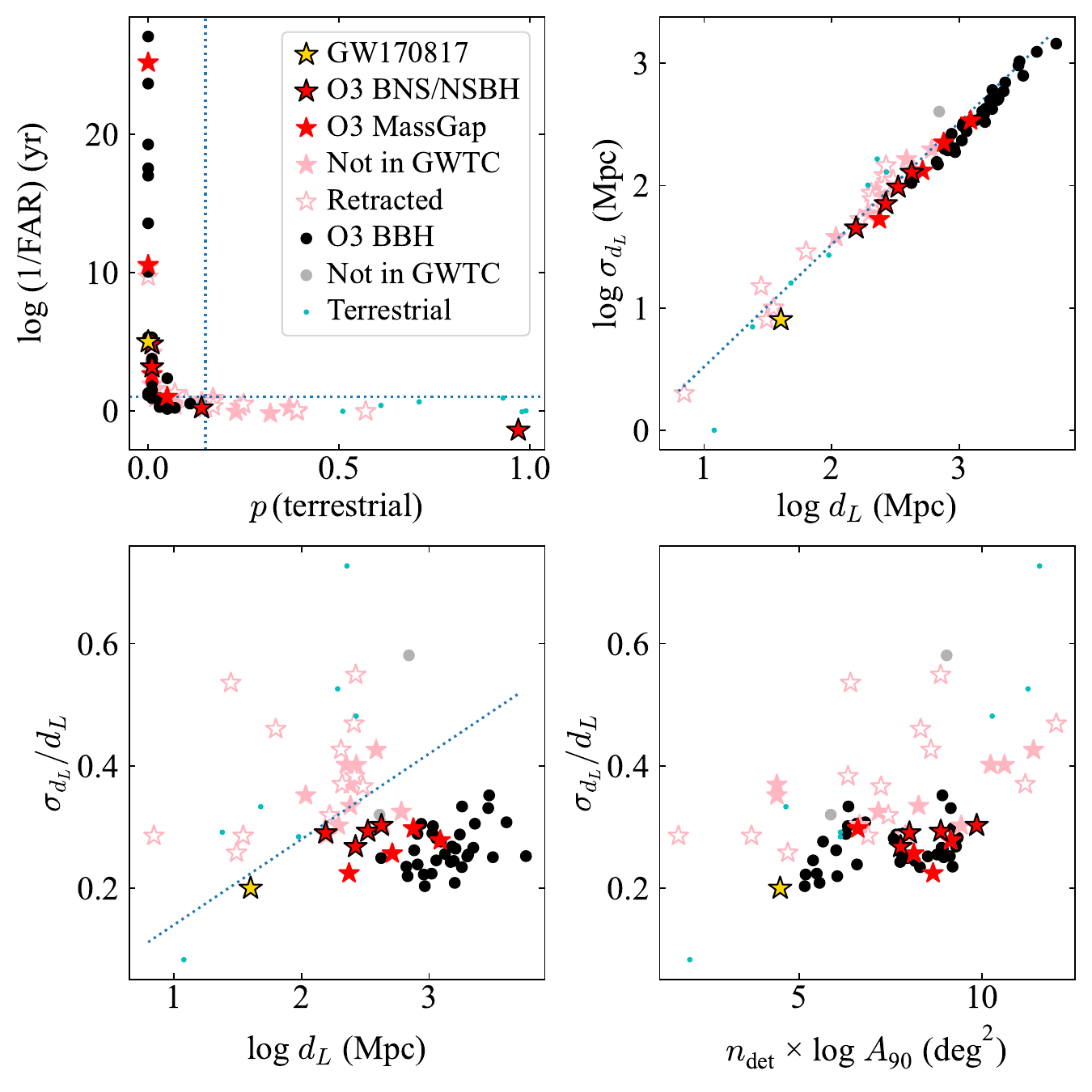}
\caption{Selected parameters reported in real-time GCN circulars for GW events in O3, obtained from GraceDB. Events are divided into BBH, BNS, NSBH and Mass gap classes, as well as those classified as likely terrestrial even in real time. Events that did not make it into GWTC-3 are shown in lighter colours. Empty symbols indicate that a GCN retraction notice was issued. Dotted lines indicate the cuts used to select sub-samples in Table \ref{tab:gwcuts}.}
\label{fig:gwparams}
\end{figure}

We use the parameter values from the \textsc{bayestar} pipeline \cite{Singer2016}, as these are generally available at the time of the earliest circulars, when the first decisions on follow-up are made. Specifically, we consider the median of the GW-inferred luminosity distance distribution ($d_L$) and its standard deviation ($\sigma_{d_L}$), the area of the 90\% localization region ($A_{90}$), the number of detectors participating in the detection ($n_{\rm det}$), the inverse FAR, and the probability that an event is a BBH, BNS, NSBH, Mass-gap or Terrestrial source. We show the combinations of parameters that appear to be informative in Figure \ref{fig:gwparams}, and examine the effectiveness of different selection cuts on the O3 sample in Table \ref{tab:gwcuts}.

The majority of events with low FAR ($1/{\rm FAR}\gtrsim10$\,yr) are real, but many other real events have higher FARs similar to those of the spurious sources. Setting 10\,yr as the threshold in 1/FAR yields an 80\% pure sample when including all alerts, but the purity is much lower (45\%) when looking at only the NS-bearing events. This cut also misses $\gtrsim25\%$ of real events. A cut on $p({\rm terrestrial})<0.15$ returns a more complete sample (98\% overall and 88\% for NS events), but with lower purity.

The quantity with the most discriminating power appears to be the fractional uncertainty in $d_L$. While both real and spurious events appear to follow a relation $\sigma_{d_L}\propto d_L$, there is a clear offset between the two groups. Real events almost always have $\sigma_{d_L} \lesssim 0.33 d_L$, whereas spurious ones lie above this line. Applying this cut yields a sample that is 98\% complete (100\% for NS events) and 75\% (44\%) pure. 

A relation of this form could perhaps be anticipated because the S/N or amplitude, $\mathcal{A}$, of the GW detection is proportional to $1/d_L$, and the uncertainty is proportional to $1/\mathcal{A}$, suggesting that $\sigma_{d_L}/d_L$ should be approximately constant \cite{Finn1993}. While the fractional uncertainty in amplitude is expected to be $\sigma_\mathcal{A}/\mathcal{A}\approx1/8$ \cite{Finn1993}, we expect $\sigma_{d_L}/d_L>\sigma_\mathcal{A}/\mathcal{A}$, because the distance is also degenerate with the source inclination for a given $\mathcal{A}$ \cite{Berry2015}. We find that by applying an empirical cut selecting events with $\sigma_{d_L}/d_L$ < 0.14 $\log d_L$, we can obtain a remarkable 100\% complete and 88\% pure sample of events (67\% pure for NS events).

\begin{table}[!ht]
\caption{Application of different cuts to the O3 alert sample to select reliable GW events in real time. These cuts are shown pictorially in Figure \ref{fig:gwparams}. Completeness is the number of real events passing each cut divided by the total number of real events; purity is the number of real events passing each cut divided by the total number of events passing the cut; the F1 score is the harmonic mean of completeness and purity.}
 \label{tab:gwcuts}
 \begin{tabular}{lcccc}
 \hline
Cut & 1/FAR $>10$~yr & $p({\rm terr})<0.15$ & $\sigma_{d_L}/d_L<0.33$ & $\sigma_{d_L}/d_L<0.14 \log{d_L}$ \\
\hline
All sources \\
\hline
Completeness & 0.73 & 0.98 & 0.98 & 1.0 \\
Purity & 0.80 & 0.77 & 0.75 & 0.88 \\
F1 score & 0.77 & 0.86 & 0.85 & 0.94 \\
\hline
\multicolumn{3}{l}{BNS / NSBH / Mass-gap} \\
\hline
Completeness & 0.75 & 0.88 & 1.0 & 1.0 \\
Purity & 0.46 & 0.39 & 0.44 & 0.67 \\
F1 score & 0.57 & 0.54 & 0.62 & 0.80 \\
\hline
 \end{tabular}
 \vspace*{-4pt}
 \end{table}
 
However, we caution that the overall number of events used in our analysis is small, and any automated application of cuts that are too strict may lead to important events being missed (including sources that are unusually nearby!). Our suggestion is therefore simply that the location of a new GW alert in the multi-dimensional parameter space of Figure \ref{fig:gwparams} may provide clues as to its reliability. Another diagnostic we find here is that real events cluster much more tightly than spurious ones in the quantity $n_{\rm det}\times \log A_{90}$.

Another important property, not used in our analysis here, is the probability that a signal is coherent between multiple GW detectors. A strategy suggested by \cite{Kiendrebeogo2023} is to follow up events with a Bayes Factor for coherence $\log({\rm BCI})>4$. Machine learning can similarly help to differentiate between real events and noise. Taking a similar approach of using only information available from the real-time alerts, \cite{Cabero2020} trained a convolutional neural network on two- and three-dimensional skymaps from O3 alerts, achieving a completeness of 92\% and a purity of 97\%.

\subsection{Individual events}

\subsubsection{GW190425}
\label{sec:190425}

The second BNS merger detected by LIGO occurred early in O3. GW190425 was flagged by the Livingston detector with a FAR of 1 per 69,000\,yr \cite{Abbott2020a}. While not public information initially, the final GW analysis showed this to be a remarkable source, with an estimated total mass of $3.4^{+0.3}_{-0.1}$\,\M. This is heavier than any Galactic NS binary, and requires one or both components is above the canonical $\approx 1.4$\,\M\ for a NS. As this was virtually a single-detector event (the LIGO Hanford detector was offline, and the S/N ratio in Virgo was only $\approx2.5$), the final 90\% credible region produced by the \textsc{LALinference} pipeline spanned 8,284\,\sqdeg\ -- about 20\% of the entire sky, and almost 300 times larger than that for GW170817. At a distance of $159^{+69}_{-71}$\,Mpc, this merger was also about 4 times further away and therefore likely much fainter. 

Despite the daunting task of searching a volume approaching $10^7$\,Mpc$^3$, many EM groups conducted follow-up campaigns using at least 24 telescopes. In addition to hundreds of GCNs, several papers emerged presenting detailed accounts of the searches \cite{Hosseinzadeh2019,Lundquist2019,Coughlin2019b,Antier2020,Gompertz2020,Paterson2020,Oates2021,Smartt2024}. These included both wide-field and galaxy-targeted searches. A summary of community follow-up efforts was compiled by \cite{Hosseinzadeh2019}, and updated by \cite{Paterson2020} and \cite{Rastinejad2022b}. In total, 69 transients were reported in real time \cite{Hosseinzadeh2019}, from searches by ZTF, ATLAS, Pan-STARRS, \textit{Swift} and \textit{Gaia}. Although a few of these sources appeared briefly to be promising candidates, these were ruled out as potential kilonovae. Some could be excluded immediately without further follow-up: ZTF reported serendipitous \emph{pre-merger} detections of many candidates during their regular survey, while other sources had cataloged host galaxy redshifts that were inconsistent with the GW distance estimate.
Such cuts can be very effective: a later meta-analysis of the full sample of O3 transients from follow-up of many GW events showed that $\approx30\%$ of candidates can typically be ruled out by pre-merger detections, and $\approx20\%$ by host galaxy distance measurements \cite{Rastinejad2022b}.

\begin{figure}[!t]
\centering\includegraphics[width=0.7\textwidth]{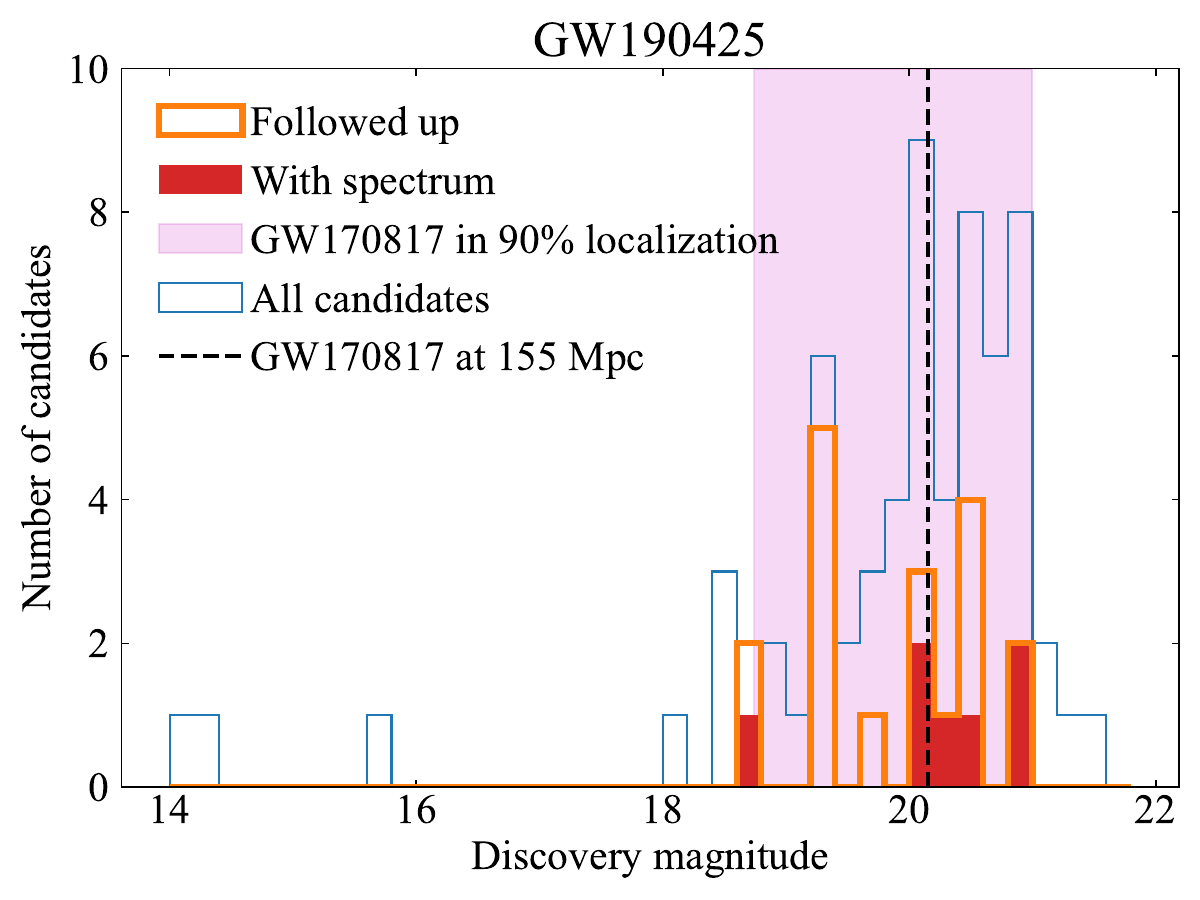}
\caption{Summary of follow-up of GW190425. The histogram shows the magnitudes of all transients reported via GCNs by EM searches, and highlights those that received follow-up imaging and spectroscopic classifications. None proved to be the counterpart to the GW event. The shaded region shows the expected magnitude range of an AT2017gfo-like kilonova within the localization volume. Reproduced from figure by G.~Hosseinzadeh et al.~\cite{Hosseinzadeh2019}.}
\label{fig:190425}
\end{figure}

In the nights that followed GW190425, targeted follow-up was conducted for 18 sources that were not excluded by simple cuts. Typically these were selected because their host galaxies had secure redshift measurements that were consistent with the GW distance. Where distances were not available, a consistency check could be applied: if the transient was within the 90\% distance range of GW190425, would it have an absolute magnitude similar to AT2017gfo ($\approx-16$\,mag)? This consistency check can be seen in the magnitude distribution of those transients that were followed up, shown in Figure \ref{fig:190425}. However, it also transpired that the peak apparent magnitude of AT2017gfo in this distance range would have been $\approx19-21$\,mag, comparable to the limiting magnitudes of the deeper follow-up surveys, so unfortunately this criterion excludes only a small fraction of the reported candidates.

Follow-up observations included seven spectroscopic classifications, which were all supernovae \cite{gcn24204,gcn24205,gcn24215,gcn24230,gcn24233,gcn24269,gcn24321}. Several of these sources were observed spectroscopically by more than one group (usually before a classification had been publicised). A detailed meta-analysis of all sources detected in O3 follow-up found that events that received multiple spectra were generally detected early, at a distance consistent with GW constraints, and showed evidence for red colours, suggesting that teams were successful in allocating spectroscopic time to promising candidates \cite{Rastinejad2022b}. For events that did not receive spectroscopic observations, photometric follow-up either failed to confirm the existence of a few marginal candidates, or showed evolution too slow to be a kilonova. An additional few sources were excluded by later detections during routine survey operations by Pan-STARRS and the Catalina Sky Survey (via the SAGUARO collaboration; \cite{Paterson2020}), when a plausible kilonova would have long since faded. A final list of 20 sources were not detected in further follow-up \cite{Paterson2020}, and cannot be definitively excluded as counterparts. However, even this likely represents an incomplete list of transients within the localization volume.


Assessing the fraction of the localization volume covered by EM follow-up searches is not trivial, as the distance reached by observations depends on the assumed luminosity of the transient. A meta-analysis of community follow-up of GW190425 was instructive here \cite{Hosseinzadeh2019}. Galaxy-targeted searches were relatively deep: 304 out of the 373 observed galaxies brighter than $-19$\,mag (where galaxy catalogs appeared to be fairly complete at this distance) were imaged to a limiting magnitude that would have detected an AT2017gfo-like kilonova. However, given the size of the localization volume, this constitutes only $\sim 1.2\%$ of the expected galaxy counts, so detecting the counterpart with this approach would have required a substantial degree of good fortune.

Wide-field searches, on the other hand, had mixed success. Most of the searches were estimated to cover $\approx 0-8\%$ of the integrated probability volume \cite{Hosseinzadeh2019,Paterson2020,Gompertz2020}, though notably Pan-STARRS covered $\approx25\%$ \cite{Smartt2024} and ZTF covered $\approx40\%$ \cite{Coughlin2019b}. The searches that covered 0\% were too shallow to detect an AT2017gfo-like kilonova at the distance of GW190425. A GRB afterglow could provide a brighter EM target, however this is expected to be brighter than the kilonova only if we are fortunate enough to have the jet directly along our line of sight, and even then only during the first $\lesssim 1$\,day after merger \cite{Hosseinzadeh2019}. In this case we would also expect a GRB coincident with the GW signal. This indicates that deep, wide-field surveys are the only efficient way to find the counterparts of poorly localized GW events.

In fact, the assumption of an AT2017gfo-like kilonova is probably optimistic in this case. The more massive merger that caused GW190425 is likely to collapse more quickly to a BH than in GW170817, and therefore to produce a fainter and likely redder kilonova. Models now exist that can predict the kilonova light curve directly from the binary parameters accessible to GW detectors, such as the chirp mass and viewing angle \cite{Coughlin2018,Barbieri2019,Kruger2020,Nicholl2021,Mochkovitch2021,Nedora2022,Gompertz2023a,Setzer2023}. If the component masses and inclinations inferred from GW observations could be released to the community in real time, this could help to optimise the observing strategy, and also to alleviate the substantial telescope time that could otherwise be wasted by follow-up efforts that are too shallow. At later times, models that directly link GW and ejecta properties can also be very useful in interpreting EM non-detections \cite{Foley2020,Sagues2021,Raaijmakers2021a,Barbieri2021} (though it has also been pointed out that many models are highly sensitive to simplifications in the radiative transfer \cite{Bulla2023}).

Although wide-field searches can cover more of the probability, observations targeting specific regions or galaxies can be motivated by other information. This could be a coincident multi-wavelength signal such as a GRB or fast radio burst (FRB), or a particular sight line of interest. For this special issue, a relevant example could be a known strong-lensing cluster within the skymap.  In such cases, a specific physical question (e.g., is a GW source lensed?) can potentially be addressed without needing to cover the entire skymap \cite{Smith2019,Smith2023,Bianconi2023}. Such science cases may require observing to greater depth than most wide-field surveys can provide \cite{Smith2023}.

Although only realised retrospectively, GW190425 provides a case study in doing multi-messenger science with particular galaxies of interest. The Canadian Hydrogen Intensity Mapping Experiment (CHIME) detected an FRB within the skymap of GW190425, 2.5 hours after the GW signal -- close enough in time that the probability of chance coincidence was only $\approx 0.005$ \cite{Moroianu2023}. Only one galaxy (UGC10667) is consistent with the positions of both the FRB and GW190425 \cite{Panther2023}. If the FRB was physically related to the NS merger that caused GW190425, the merger remnant would have needed to survive for 2.5 hours before collapsing to a BH (in contrast to expectations for such a massive binary). Such a long-lived NS remnant would likely produce an overly luminous kilonova: the NS will be rotating maximally and can be highly magnetised (a `magnetar'), and continually injects energy while it spins down \cite{Price2006,Yu2013,Zrake2013,Metzger2019,Sarin2022,Kiuchi2023}. A kilonova in this galaxy was ruled out by observations from ATLAS, Pan-STARRS and ZTF, disfavouring a physical association between the FRB and GW190425 \cite{Smartt2024}. The association has also been questioned on theoretical grounds, based on the required equation of state \cite{MaganaHernandez2024} and the high optical depth to radio emission in NS merger ejecta \cite{Radice2024}. We also note that no FRB was detected during the follow-up of GW170817 with the Australian Square Kilometre Array Pathfinder (ASKAP) and Parkes radio telescopes \cite{Andreoni2017}.

\subsubsection{GW190814}
\label{sec:190814}

GW190814 is another well localized event found during O3, this time in the mass gap. The IGWN pipelines originally placed it at a distance of $267 \pm 52$\,Mpc, with the 90\% integrated localization probability enclosed in a 23\,deg$^2$ area. These values were revised to $241^{+41}_{-45}$\,Mpc and 18\,deg$^2$ in GWTC-2 \cite{Abbott2020b, Abbott2021}. GW190814 was followed up extensively due to the tight localization, and the possibility that as a mass gap source the lighter object could be a NS. The refined GW analysis later revealed the system to have a 22.2--24.3\,M$_\odot$ BH merging with a compact object with a mass of 2.50--2.67\,M$_\odot$ (90\% credible level) \cite{Abbott2020b}, making the latter either the lightest BH or heaviest NS found in a compact object binary. The GW source was studied in an attempt to gain new insights on the NS equation of state \cite{Fattoyev2020, Tsokaros2020} and to measure the Hubble constant via the standard siren method \cite{Palmese2020, Vasylyev2020}.

The approximately equatorial location of GW190814 made it feasible for ground-based telescopes in both hemispheres to search for possible EM counterparts. The relatively small volume to probe made it possible to use the synoptic approach of tiling the skymap \cite{Dobie2019, Andreoni2020bv, Gompertz2020, Morgan2020, Kasliwal2020, Vieira2020, Watson2020, Becerra2021MNRAS, Chang2021, deWet2021, Dobie2022}, a galaxy-targeted approach based on available catalogs \cite{Gomez2019, Page2020, Alexander2021, Oates2021} or even a combination of the two \cite{Ackley2020, Thakur2020, Kilpatrick2021}. More than 100 EM counterpart candidates were identified during the searches, especially in the optical, but all the transients were ruled out as counterparts to GW190814 via spectroscopic classification or photometric evolution incompatible with models. This result is compatible with later analyses of the GW signal, which indicated that the lighter object is likely a BH rather than a NS \cite{Tews2021}. The high mass ratio between the components of the binary means that even if the lighter component was a NS, it would have been tidally disrupted inside the innermost stable circular orbit (ISCO) of the more massive component, and therefore the mass of ejecta would be negligible \cite{Foucart2012, Kawaguchi2015} (see \S\ref{sec:nsbh}).

\begin{figure}[!t]
\centering\includegraphics[width=0.47\textwidth]{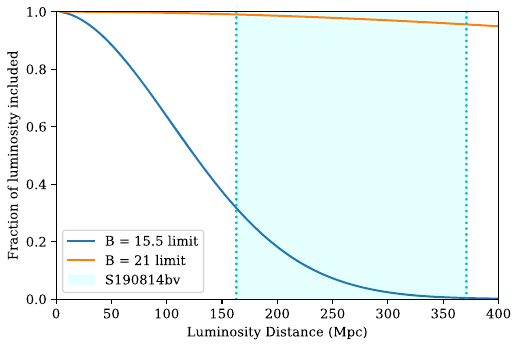}
\centering\includegraphics[width=0.47\textwidth]{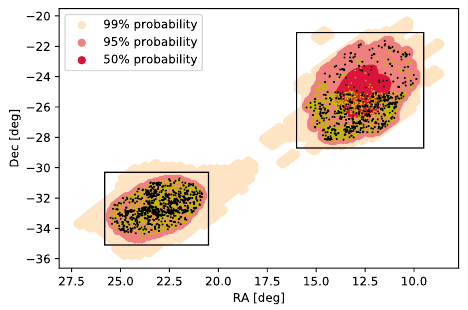}
\caption{Left: Estimated completeness of galaxy catalogs as a function of distance, assuming a catalog limiting magnitude of $B=15.5$ (approximately representative of current catalogs) or $B=21$. The shallow catalog is highly incomplete beyond $\approx 100$\,Mpc, but the deeper catalog is achievable with multi-plexed spectroscopic surveys. Right: The inhomogeneous footprints of different catalogs. The black points show photometric galaxies and the yellow points show spectroscopic galaxies within the skymap of GW190814. Adapted from figures by I.~Andreoni et al.~\cite{Andreoni2020bv} and the ENGRAVE collaboration (K.~Ackley et al.) \cite{Ackley2020}.}
\label{fig:catalogs}
\end{figure}

One clear lesson that emerged during the follow-up of GW190814 is the need for more complete galaxy catalogs beyond 200\,Mpc.  Catalogs were used by teams with small field-of-view instruments to conduct galaxy-targeted searches, and were also extremely important during wide-field searches to prioritize candidates for spectroscopic and multi-wavelength characterization. Photometric redshift catalogs are generally not reliable at low redshift, but several proved helpful to exclude very luminous transients likely associated with distant galaxies, where the relative uncertainties of photometric redshifts are smaller (at least for bright sources \cite{Zhou2021}). Photometric catalogs used in the search included the 2MASS Photometric
Redshift catalog \cite{Bilicki2014}, the Legacy Survey \cite{Dey2019,Zhou2021} and PS1-STRM \cite{Beck2021}.

Spectroscopic galaxy catalogs are much more powerful, as they have definitive and precise distance information. Within the skymap of GW190814, the 2dF Galaxy Redshift Survey \cite{Colless2001} was particularly rich in spectroscopic redshifts. Spectroscopic catalogs are reasonably complete for very nearby sources (dozens of Mpc, e.g. \cite{Nissanke2013, Gehrels2016, Singer2016}), but much less so for sources at the distance of GW190814. As well as the rapid decrease with distance, catalog completeness is not uniform over the sky (Figure \ref{fig:catalogs}). The issue of catalog incompleteness within the skymap of GW190814 was highlighted by several studies (e.g.~\cite{Andreoni2020bv,Ackley2020,Kilpatrick2021}). 

Given a Schechter luminosity function \cite{Gehrels2016}, if we could obtain spectra for all galaxies down to $B = 21$\,mag, redshift catalogs could reach $>97\%$ completeness within the $2\sigma$ distance range of GW190814 \cite{Andreoni2020bv}. If they can achieve this depth, the current and upcoming massively multiplexed spectroscopic surveys such as the Dark Energy Spectroscopic Instrument (DESI) \cite{DESI2016}, the 4-metre Multi-Object Spectroscopic Telescope (4MOST) \cite{deJong2019} and the Subaru Prime Focus Spectrograph (PFS) \cite{Takada2014} can be game-changers in this regard. In the more immediate term, the NASA Extragalactic Database (NED) has already undergone several recent updates and improvements. It is now the most complete local-volume catalog beyond $\approx80$\,Mpc, and provides automated cross-matching against GW alerts \cite{Cook2023}, making it an increasingly valuable tool for multi-messenger astronomy.


\subsubsection{NSBH events}
\label{sec:nsbh}

While the mass of the lighter component in GW190814 is probably more consistent with a BH than with a NS, several other events in O3 data contained a secondary where the GW posteriors indicated a mass $\lesssim2$\,\M\ (i.e.~allowed by current estimates of the NS equation of state; e.g.~\cite{Antoniadis2013,Margalit2017,Cromartie2020,Fonseca2021,Dietrich2020}), alongside a BH primary.

\begin{figure}[!t]
\centering\includegraphics[width=0.45\textwidth]{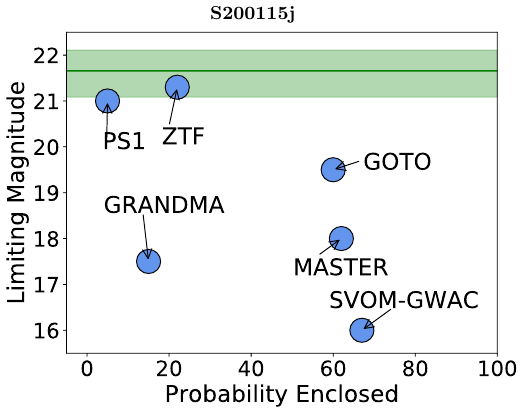}
\centering\includegraphics[width=0.5\textwidth]{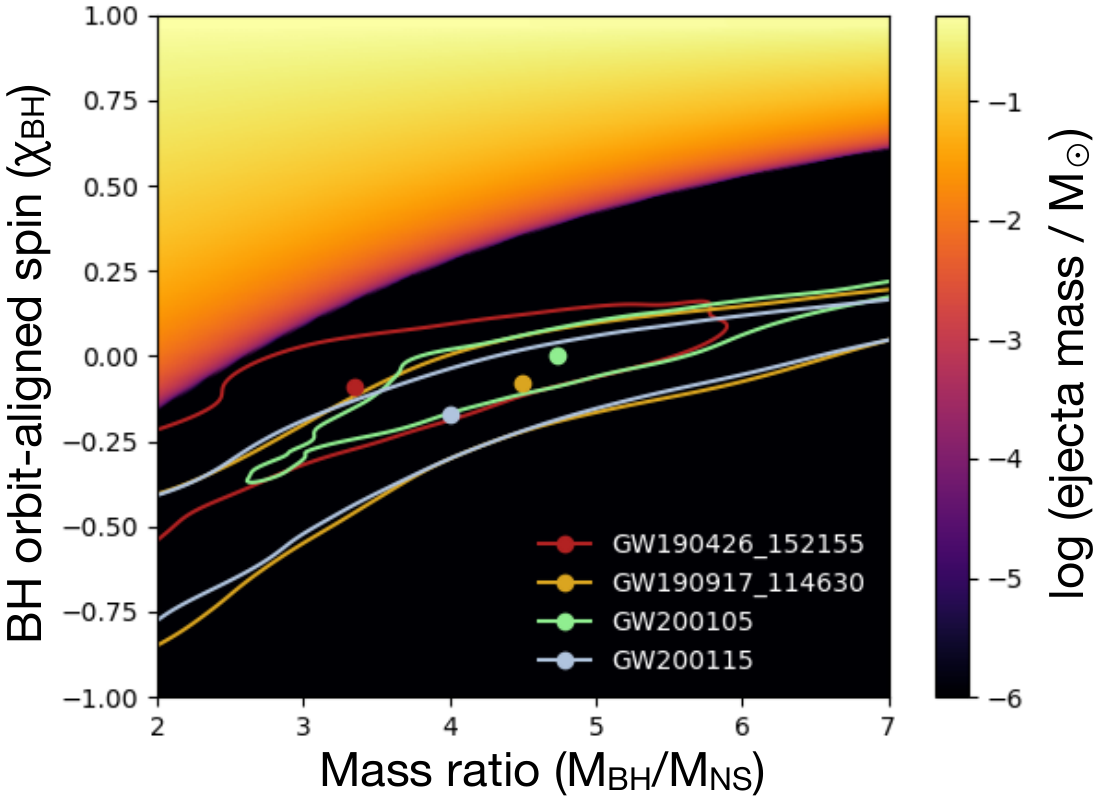}
\caption{Left: Summary of EM follow-up for NSBH merger GW200115. The green shaded region shows the apparent magnitude of an AT2017gfo-like kilonova within the distance range of GW200115. Only ZTF and Pan-STARRS are sensitive to such a kilonova; the surveys that covered a wider fraction of the two-dimensional skymap (``Probability Enclosed'') are not constraining for the luminosity of a potential counterpart.
Right: Ejecta mass from NSBH binaries as a function of the mass ratio, defined here as $M_{\rm BH}/M_{\rm NS}$, and the component of the BH dimensionless spin projected along the orbital angular momentum vector, using the prescription from \cite{Foucart2018}. Large ejecta masses are possible, but only if the BH spin is aligned with the binary orbit -- otherwise the NS crosses the ISCO before it is tidally disrupted, resulting in minimal ejecta. Contours show NSBH candidates from O3. Despite the modest mass ratios for most observed systems, none of these sources were likely to have had a bright EM counterpart. Adapted from figures by M.~Coughlin et al.~\cite{Coughlin2020c} and B.~Gompertz et al.~\cite{Gompertz2022}.}
\label{fig:nsbh}
\end{figure}

Such events identified in real-time alerts are listed in Table \ref{tab:o3-sources}. GW190426\_152155 was identified just one day after the BNS GW190425, and also received substantial follow-up observations \cite{Hosseinzadeh2019,Lundquist2019,Kasliwal2020,Gompertz2020,Paterson2020,Goldstein2019,Kumar2022,Oates2021,Gourdji2023}. No EM counterpart was identified. This source was reported in GWTC-2 at rather low significance \cite{Abbott2021}, and is not recovered in the analyses used in GWTC-2.1 or GWTC-3, making its veracity unclear. The other two real-time events were GW200105 and GW200115 (names abbreviated following \cite{Abbott2021a}). As the first high-significance NSBH mergers identified by the IGWN, their GW data were analysed in detail \cite{Abbott2021a}. They were
also followed up by EM telescopes, but with no counterpart detections \cite{Anand2021,Dichiara2021}. It was pointed out however, that most searches conducted were not sensitive to a kilonova comparable to AT2017gfo, and the deeper searches covered only a small fraction of the skymaps (Figure \ref{fig:nsbh}) \cite{Coughlin2020c}. Completing the O3 NSBH sample, GW190917\_114630 was not detected in real time, but was found in the GWTC-2.1 analysis \cite{Abbott2024}, and similarly GW191219\_163120 was only identified in the GWTC-3 analysis \cite{Abbott2023}.

The detailed waveform modeling of these events reported BH masses in the range $\approx6-30$\,\M, and without significant BH spins in the direction of the orbital angular momentum. Population synthesis studies \cite{Broekgaarden2021,Chattopadhyay2022} found that the typical masses and the implied NSBH merger rate were consistent with progenitors formed via isolated binary evolution, expected to be the dominant channel to make NSBH mergers\footnote{See \cite{Mandel2022} for a review of the different formation channels and constraints on their relative rates.}. In contrast, the very asymmetric GW191219\_163120 may be an example of a system that formed dynamically \cite{Gompertz2022}. 

While the GW event rate is in line with expectations, the distributions of masses and spins do not favour detectable EM counterparts. In order for a NSBH merger to eject a substantial amount of matter, the NS must be tidally disrupted outside of the ISCO. A smaller BH mass or a larger orbit-aligned spin would both act to decrease the ISCO, bringing it inside the radius where tidal disruption occurs \cite{Foucart2012, Kawaguchi2015}. For a non-spinning BH and a fiducial NS, this condition is fulfilled only for a BH mass $\lesssim3.5$\,\M, but aligned rotation can increase this limiting mass by a factor of several. It was noted that GW200115 (and possibly two lower significance events) appeared to show a BH spin \emph{anti}-aligned with the orbit  \cite{Fragione2021,Gompertz2022}, though this is sensitive to the priors assumed in the GW analysis \cite{Mandel2021}. 

We show the joint constraints on the binary mass ratio and projected BH spin for the O3 NSBH candidates in Figure \ref{fig:nsbh}; in all of these events, the NS crosses the ISCO of the BH before it is disrupted \cite{Gompertz2022}. EM counterparts are therefore largely excluded by the GW data for all of the O3 events \cite{Abbott2021a,Zhu2021}. Recent analyses estimate that only $\sim1-10\%$ of NSBH mergers detected with current GW facilities will have detectable EM emission \cite{Fragione2021a,Drozda2022,Biscoveanu2023,Colombo2023}. This is based primarily on the mass ratios predicted by population synthesis and the assumption that most BHs are born with low spins \cite{Fuller2019}. Several authors have suggested mechanisms by which BHs in these binaries can retain (or attain) faster spins, which would promote EM detectability (e.g.~\cite{Hu2022,Steinle2023,Wang2024}) -- but whether these are common in nature remains to be seen. 

The expectation that many NSBH mergers will not be EM sources suggests (somewhat obviously) that particular attention should be paid to the parameters \texttt{HasRemnant} and possibly the combination of \texttt{HasNS} and \texttt{HasMassgap}. A source with high probability of containing both a NS and a mass gap object could indicate that the BH is of relatively low mass, and therefore more capable of disrupting a NS before it crosses the ISCO. Alternatively, a large \texttt{HasMassgap} could also indicate a gravitationally-lensed BNS merger, which may require a different follow-up strategy \cite{Smith2023}. 

In principle, NSBH systems can eject more matter than BNS mergers, reaching up to $\sim0.1$\,\M\ \cite{Tanaka2014,Kyutoku2015,Kawaguchi2016}. However, these systems do not produce shock-driven ejecta, and the remnant is always a BH (reducing neutrino irradiation of the disk wind; see Figure \ref{fig:schematic}). Therefore the emission from NSBH kilonovae is expected to be much redder than BNS kilonovae \cite{Fernandez2017,Kyutoku2018}, with predicted $g-z\sim 1$\,mag and $g-K\gtrsim2$\,mag for fiducial models \cite{Gompertz2023a}. This may make optical searches challenging, even for events that eject significant mass.

\subsubsection{BBH follow-up}

O3 was rich in loud BBH mergers, generally much more tightly localized than BNS and NSBH events, albeit typically at much larger distances. Theoretical studies on possible EM transient emission from BBH mergers are less developed than those of NS mergers, but the emergence of candidate counterparts in various form may motivate further exploration here. The proposed emission mechanisms typically involve the interaction of the merging binary or the compact remnant with a surrounding medium such as the accretion disk of an active galactic nucleus (AGN). The interaction may happen for example in the form of accretion \cite{Bartos2017},  
breakout emission from a post-merger jet \cite{Tagawa2023}, jetted Bondi accretion \cite{McKernan2019} and ram-pressure stripping of gas surrounding the remnant. AGN in particular have been identified as promising sites to search for EM counterparts to high-mass BH mergers \cite{McKernan2012, McKernan2014}. Hierarchical mergers (where one or both BHs in the merger are themselves the product of a previous merger) are expected to be found more frequently in deep gravitational potentials, such as in galactic nuclei \cite{Gerosa2019}. 

The most massive source found during O3 was GW190521, which had a total mass of $\sim 150$\,M$_\odot$ \cite{Abbott2021, Abbott2023}. A few weeks after this event, a luminous flare coincident with a known AGN, J124942.3+344929, was detected by ZTF \cite{Graham2020}. This generated substantial interest as the AGN is within the 90\% GW probability area and at a distance consistent with the GW signal \cite{Graham2020}. The flare could plausibly have been generated by the interaction of the high-mass merger with the AGN disk \cite{McKernan2019}, and therefore represented the first candidate optical counterpart to a BBH merger. However, subsequent analysis of the probability of chance coincidence within the localization volume reduced the statistical significance of the possible EM-GW association \cite{Palmese2021, Veronesi2024arXiv}.

A ZTF search for EM counterparts to all BBH mergers observable from Palomar revealed 9 candidate counterparts coincident with AGNs \cite{Graham2023}. This number is higher than theoretically predicted, and robust methods will need to be developed to remove false positives as multi-messenger searches expand. 
Other searches for EM counterparts to BBH mergers conducted in O3 focused on probing potential early jetted emission \cite{Gompertz2020, Paterson2020}, but this has yet to reveal compelling candidates.

\subsubsection{O4 events}
The LIGO-Virgo-KAGRA fourth Observing run (O4) started on May 24, 2023. By the mid-point of O4, detector upgrades have enabled sensitivity to BNS mergers within $\sim 160$\,Mpc. The Virgo detector was offline during the first half of O4, but re-joined the network when the second part of O4 began on April 10, 2024. At the time of writing, three GW events that likely included at least one NS have so far been deemed `Significant' by the analysis pipelines, namely S230518h \cite{2023GCN.33813_S230518h_disc, 2023GCN.33816_S230518h_update, 2023GCN.33884_S230518h_update2}, S230529ay \cite{2023GCN.33813_S230518h_disc, 2023GCN.33891_S230529ay_update, 2023GCN.34148_S230529ay_update2}, and S240422ed \cite{2024GCN.36236_S240422ed_disc, 2024GCN.36240_S240422ed_update} (the prefix `S', rather than `GW', is given to events that have not yet been confirmed by a detailed offline analysis from the IGWN). All of them were classified as likely NSBH mergers. Several more NSBH alerts were issued automatically during O4, but were retracted after vetting.

The main properties of these sources from the GW alerts are listed in Table\,\ref{tab:nsbh}. S230518h was the least likely to have ejected material (\texttt{HasRemnant}\,$<1\%$) \cite{2023GCN.33813_S230518h_disc, 2023GCN.33816_S230518h_update} according to the parameter estimation pipelines \cite{Ashton2019bilby, Chatterjee2020}. S230529ay had a greater probability of ejected matter (\texttt{HasRemnant}\,$=7\%$), and a high probability that one of its binary components was in the mass gap between 3\,M$_\odot$ to 5\,M$_\odot$ \cite{2023GCN.34148_S230529ay_update2}. While this was already more encouraging than any of the O3 NSBH events, S240422ed initially appeared even more promising, with \texttt{HasRemnant} $>99\%$ \cite{2024GCN.36236....1L}. However, while this paper was under review, a re-analysis of S240422ed was announced in July 2024, reducing the FAR to 1 in 35 days, and re-classifying it as terrestrial with 93\% probability \cite{2024GCN.36812....1L}. The initial GCN \cite{2024GCN.36236....1L} noted that both LIGO Hanford and Livingston data were affected by glitches (noise) at the time of observing, prompting these additional offline analyses. This should serve as a cautionary tale for any GW event contaminate by glitches, even if the real-time significance appears very high.

On the other hand, the final analysis of S230529ay (now updated to GW230529) published in August 2024 confirmed it to be a real signal, resulting from the merger of a 2.5-4.5\,M$_\odot$ mass gap object (presumably a low-mass BH) with a NS secondary \cite{Abac2024}. Unfortunately, the poor localization makes it challenging to put constraints on the EM emission from this system, but it provides encouragement that NSBH mergers capable of producing an EM signal do exist in nature.

 \begin{table}[!ht]
 \caption{Significant GW events found until May 23, 2024 during O4, which likely include at least one NS. The first column reports the event name, the second the false alarm rate (FAR), the third the area included in the 90\% integrated probability contour, the last the distance to the source. Skymaps and posteriors were obtained using the \texttt{Bilby} analysis framework. All these events were classified as likely NSBH mergers, though S240422ed (marked with $*$) was later downgraded to likely terrestrial \cite{2024GCN.36812....1L}. As detailed follow-up observations were conducted prior to retraction, we report its parameters here for completeness.
 }
 \label{tab:nsbh}
 \begin{tabular}{lccccccl}
 \hline
 Event & 1/FAR (yr) & \texttt{HasNs} & \texttt{HasRemnant} & \texttt{HasMassgap} & A$_{90}$ (deg$^2$) & D (Mpc) \\
 \hline
S230518h & 98.5 & $>99\%$ & $<1\%$ & $<1\%$ & 460 & $204 \pm 57$ \\
S230529ay & 160.4 & $98\%$ & $ 7\%$ & 73\% & 24,534 & $197 \pm 62$ \\
S240422ed$^*$  & $10^5$ &  $>99\%$ &  $>99\%$ & $34\%$ & 259 & $188 \pm 43$ \\
\hline
 \end{tabular}
 \vspace*{-4pt}
 \end{table}

The astronomical community carried out follow-up observations across the spectrum for these events, from the radio to gamma-rays. As in previous observing runs, details about the observations, lists of possible host galaxies, counterpart candidates and their photometric and spectroscopic classification were communicated primarily via GCN circulars. Within months of each event, 25, 12, and 89 circulars were published for S230518h, S230529ay, and S240422ed, respectively\footnote{These numbers include the GCN circulars issued by the LIGO-Virgo-KAGRA collaboration}. Each of these events presented observers with challenges: the low likelihood of ejecta in S230518h; the large area (much of it in solar conjunction) for S230529ay; and, in the case of S240422ed, the GW localization was directly through the Galactic plane, making it difficult to find extragalactic transients behind the high extinction and crowded stellar fields.  
All the candidate EM counterparts found during these observing campaigns, once characterized, were deemed unlikely to be related to the GW sources. 

Across O2, O3 and O4, the role of Virgo (and in the future, KAGRA and potentially LIGO-India) has become very evident in reducing the area of the high-probability localizations. Reducing the size of skymaps even with a Virgo non-detection, as in the case of S240422ed (and GW170817), enables substantially better targeting of EM observations, more feasible vetting of counterparts, and more constraining limits in the result of an EM non-detection. This is true especially for optical and NIR telescopes. For events that are poorly localized, for example events similar to S230529ay that are detected by only one GW interferometer, neutrino detectors and very wide-field gamma-ray monitors in space may still provide hope of detecting a multi-messenger counterpart.

We also note that optical follow-up observations of some well-localized BBH mergers are still on-going in O4 (e.g. \cite{Sherman2024GCN.36288}; Cabrera et al., in prep) with the primary objective to search for flares from mergers inside AGN disks \cite{McKernan2019,Graham2023}.

\subsection{Further constraints on the kilonova population}
\label{sec:stats}

\subsubsection{Statistical constraints from multi-messenger searches}


Since GW170817, no BNS or NSBH event discovered by the IGWN  has been detected in all-sky neutrino and gamma-ray detectors, or in targeted follow-up observations. For some events with detailed follow-up, non-detections of EM emission have been used in attempts to place some constraints on ejecta properties such as masses, velocities, compositions, opacities, and viewing angles (see for example references in \S\ref{sec:190425}--\ref{sec:190814}). However, given the many challenges outlined above, the limits derived from the EM efforts are not especially constraining for the kilonova source population. Another approach is to perform a statistical analysis, putting together optical upper limits for all the NS-bearing mergers into a common framework (e.g.~NIMBUS \cite{Mohite2022}). Such an analysis has been performed using ZTF data \cite{Bellm2019, Graham2019} from the O3 and O4a searches \cite{Kasliwal2020, Ahumada2024arXiv}. 

In this case the luminosity function of optical kilonovae was obtained following the equation:
\begin{align}\label{lumfunc}
\begin{split}
(1 - {\rm CL}) = \prod_{i=1}^{N} (1-f_{b}\cdot p_{i}\cdot(1-t_{i})),
\end{split}
\end{align}
where CL is the confidence level, $f_{\rm b}$ is the maximum allowed fraction of kilonovae brighter than a given peak absolute magnitude, $p_{\rm i}$ is the probability of a kilonova detection within a given GW event skymap, and $t_{\rm i}$ is the probability that a GW trigger was of terrestrial origin, defined as $1 - p_\mathrm{astro}$. The luminosity function can then be derived by solving for $f_{\rm b}$ at 90\% confidence in each luminosity bin. Combining ZTF observations for those events which were confirmed as likely astrophysical in GWTC3 \cite{Abbott2023} and five additional events from O4 (S230518h,
S230529ay, S230627c, S230731an, S231113bw; note that these are not yet confirmed astrophysical), it was found that a maximum fraction of 76\% of kilonovae can be brighter than $-17.5$\,mag \cite{Ahumada2024arXiv}. 

This result is not surprising, given that kilonovae are expected to be fainter than most supernova classes. Applying realistic filtering criteria necessary for real transient identification, the maximum fraction of kilonovae brighter than $-17.5$\,mag and fading by 1\,mag\,day$^{-1}$
became 92\%, i.e.~current observations can barely constrain the optical emission. AT2017gfo peaked at around $-16$\,mag in optical bands; more kilonovae in this magnitude range cannot be excluded by searches so far. The ability to place deeper limits is affected by the difficulty of detecting faint, fast-fading transients: the efficiency in recovering an AT2017gfo-like kilonova in the combined ZTF searches was estimated at around 36\% \cite{Ahumada2024arXiv}.

This statistical approach is likely to become more constraining over time due to the combination of: i) a larger number of follow-ups prompted by additional significant BNS or NSBH alerts; ii) better completeness over the localization volumes observed by more wide-field surveys; iii) deeper observations that can probe the presence of a faint ($M < -16$\,mag) kilonova. Of course, actual \emph{detections} of more counterparts besides AT2017gfo still remains the most desirable way forward to constrain the luminosity function.

\subsubsection{Kilonovae from GRB follow-up and optical surveys}

\begin{figure}[!t]
\centering\includegraphics[width=\textwidth]{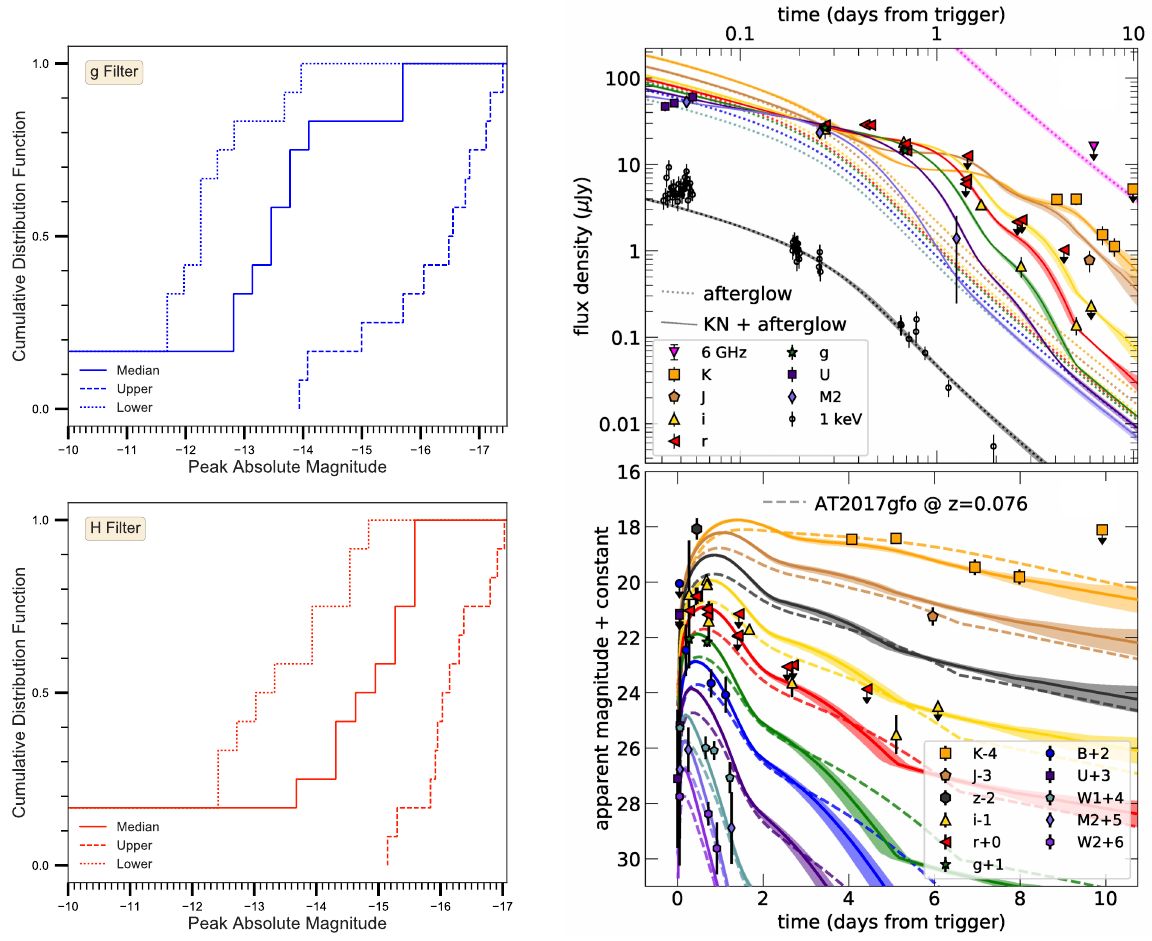}
\caption{Constraints on kilonovae from GRB observations. Left: Distributions of peak absolute magnitudes in the $g$ (top) and $H$ (bottom) bands. The luminosities have been estimated by fitting a combined afterglow plus kilonova model \cite{Troja2018,Kasen2017}. The median optical luminosity is likely between $-12$ and $-16$\,mag. Right: The well-observed kilonova in GRB\,211211A. The top panel shows the observed light curve data compared to a combined afterglow plus kilonova model \cite{Lamb2019,Nicholl2021}; the contrast between the kilonova and afterglow is greatest in the NIR $K$-band. The bottom panel shows the data and kilonova model after subtracting the afterglow, compared to AT2017gfo. The long-term NIR evolution is extremely similar, though GRB\,211211A initially appears somewhat brighter in the optical. Adapted from figures by figure by S.~Ascenzi et al.~\cite{Ascenzi2019} and J.~Rastinejad et al.~\cite{Rastinejad2022}.}
\label{fig:grbs}
\end{figure}

Fortunately, GW follow-up is not the only way to discover kilonovae. The first kilonova to be identified was discovered following the short burst GRB\,130603B \cite{Tanvir2013,Berger2013}. The GRB follow-up revealed luminous NIR emission, relative to the expected fading of the afterglow, in \textit{Hubble Space Telescope} imaging at a phase of $\approx 1$\,week. Several other kilonova candidates have been identified in \emph{archival} data following a similar approach. These include GRB\,060614 \cite{Yang2015,Jin2015}, GRB\,050709 \cite{Jin2016}, GRB\,150101B \cite{Troja2018a}, GRB\,070707 \cite{Zhu2023}, and GRB\,080503 \cite{Gao2015,Zhou2023}. Additionally, a few `over-luminous' candidates have been identified, which may be attributable to kilonovae with extra energy injection from a magnetar remnant \cite{Price2006,Yu2013,Zrake2013,Metzger2019,Sarin2022,Kiuchi2023}: GRBs 050724, 070714B, and 061006
\cite{Gao2017}, and GRB\,200522A \cite{Fong2020,OConnor2021}. Most of these candidates consist of only one or a few data points where the observed emission in an optical band (usually a redder band) lies above expectations for a fading GRB afterglow. The exception was GRB\,060821B \cite{Lamb2019,Troja2019}, where careful afterglow modelling allowed for the kilonova to be extracted from data at several epochs and wavelengths. 

Since GW170817 provided us with a clear empirical template for how the kilonova emission can evolve, \emph{real-time} kilonova searches following GRBs have been quite successful. Kilonova signatures were identified early in GRB\,211211A \cite{Rastinejad2022,Gompertz2023a} and GRB\,230307A \cite{Levan2024}, and extensive follow-up observations were obtained for both sources \cite{Rastinejad2022,Troja2019a,Levan2024,Yang2024,Gillanders2023}, including the only spectra of a kilonova other than AT2017gfo. These spectra were obtained for GRB\,230307A using the \textit{James Webb Space Telescope} \cite{Levan2024}. The late-time NIR spectra show an emission line matching AT2017gfo and possibly associated with [Te III] \cite{Hotokezaka2023}. Remarkably, neither of these events was a canonical short-duration GRB. Both belonged to the `extended emission' sub-class, with a short-hard spike of gamma-ray emission followed by a longer, softer component lasting a few tens of seconds \cite{Norris2006,Norris2010}. The kilonova detections proved that extended emission GRBs can originate from NS mergers, challenging the phenomenological division of GRBs into `long' and `short' in favour of a physical division into `merger' and `collapsar' \cite{Gompertz2023a}. The physical mechanism behind the extended emission is still not fully understood, but may be related to a magnetised central engine \cite{Metzger2008,Bucciantini2012,Giacomazzo2013,Gao2022} or fallback accretion in a NSBH merger \cite{Rosswog2007,Desai2019,Gompertz2020a,Zhu2022}. The merger of a NS and a white dwarf has also been suggested, but it is not clear that this would produce the observed kilonova emission \cite{Yang2022}.

With the larger population of kilonovae detected using GRBs, the community have begun to investigate the diversity of kilonova emission. GRB\,130603B clearly exhibits a higher NIR luminosity than AT2017gfo, implying a larger red ejecta mass, whereas GRB\,211211A appears brighter at bluer wavelengths. GRB\,160821B is overall fainter than the other well-observed events. From statistical studies \cite{Gompertz2018,Ascenzi2019,Rossi2020,Rastinejad2021}, the population appears to have at least a couple of magnitudes of spread in peak luminosity in both the optical and the NIR. Current estimates find that the optical ($g$-band) peak spans $\approx$[$-12,-17$]\,mag, while the NIR ($H$-band) peak spans $\approx$[$-13,-16$]\,mag \cite{Ascenzi2019}. For several additional GRBs, deep limits suggest kilonovae much fainter than AT2017gfo, while others have bright afterglows that could mask even a luminous kilonova \cite{Gompertz2018}. However, we are a long way from fully characterising kilonova diversity. From the largest sample of 85 merger GRBs \cite{Rastinejad2021}, it was found that only $\approx14\%$ of events had observations that were sensitive to a blue ejecta component of similar mass to AT2017gfo, while most observations are completely unconstraining for red ejecta components.

We also note that observations of the GRB-kilonova population may be biased by our viewing angle. First, the need to model out the GRB afterglow introduces degeneracies in the analysis at early times. Second, since all GRB-discovered kilonovae are viewed close to the jet (orbital) axis, we are much more sensitive to the properties of the polar ejecta, compared to the equatorial ejecta (Figure \ref{fig:schematic}). Therefore interpreting observations of GRB-kilonovae requires models that carefully consider the effect of viewing angle (e.g.~\cite{{Kasen2015,Kawaguchi2018,Wollaeger2018,Bulla2019,Korobkin2020}}). The structure of the ejecta can also be modified close to the pole by the passage of the GRB jet. In addition to the shock-heating noted in \S\ref{sec:170817}, the jet may also `push' matter out of the way to reveal inner, hotter regions with a lower lanthanide content, resulting in brighter or bluer emission for the same total ejecta mass \cite{Nativi2021,Klion2021}.

While GRB and GW triggers have been responsible for all events identified to date, finding kilonovae in optical or IR surveys independently of GW or GRB triggers would be extremely valuable. Unveiling such a population would inform us on the distribution of kilonova ejecta masses, velocities, and compositions without the viewing-angle biased imposed by the coincidence with a GRB\footnote{Note however that independent of any GRB detection, kilonova emission may be subject to its own angle-dependent detection bias if these sources are brighter in optical light closer to the pole.}. In principle, deep optical surveys may be sensitive to kilonovae beyond the horizon of IGWN detectors, and can therefore provide insights on possible trends as a function of redshift and test whether kilonovae can be utilised as an independent probe of $H_0$ \cite{Coughlin2020PhRvR, Coughlin2020NatCo}. Despite numerous searches in wide-field surveys, kilonovae have stubbornly eluded serendipitous discovery \cite{Doctor2017, Kasliwal2017, Smartt2017, Yang2017, Andreoni2020, Andreoni2021ztfrest, McBrien2021, Li2023}. In the volume-limited search for faint and fast-fading transients in Pan-STARRS  \cite{McBrien2021}, two candidates passed all selection criteria, but their classifications remain uncertain. The Transiting Exoplanet Survey Satellite (TESS) could detect a handful of kilonovae independent of GW triggers, owing to its wide field of view and continuous, high-cadence coverage \cite{Mo2023}. Prospects for kilonova detectability in deep, wide-field future surveys will largely depend on the survey strategy employed (i.e. the choice of cadence and filters), which we discuss further in the next section.

\section{Future outlook}
\label{sec:future}

So far in the first decade of GW follow-up, we have experienced tremendous success with GW170817, followed mainly by challenges in following up other sources with larger distances, wider localization areas, and likely lower intrinsic luminosities. The fundamental challenge is well understood: \emph{kilonovae fade on timescales of days, which is comparable to the amount of time it takes to search a few thousand square degrees to reasonable depth}. In such large skymaps, targeted searches are limited by catalog completeness. As GW detectors become more sensitive, more of the volume moves towards larger distances, there will be many more galaxies in the skymap, and both targeted and synoptic searches will have even more contaminating supernovae to weed out.

However, there are reasons to be optimistic: our experience with GW170817 and AT2017gfo shows that \emph{detecting the uniquely fast rate of fading compared to any other extragalactic transient (and ideally measuring a red colour) is probably sufficient to at least identify that a source is the likely EM counterpart to a GW event}. This requires covering the region of interest\footnote{Note that this does not necessarily have to be the entire skymap; some science cases like lensed kilonova searches may need to target only sight lines with a high optical depth to lensing \cite{Smith2019,Smith2023,Bianconi2023}.} at least $2-3$ times within a few days after merger. This is feasible with current wide-field telescopes, but existing surveys are shallow (reaching $\sim20$\,mag), and will struggle to measure reliably the rate of fading for targets that peak close to their detection limits. Most areas of the sky also lack deep reference images, leading to false detections of `new' sources just below the depths cataloged by surveys like Pan-STARRS \cite{Flewelling2020} or the Legacy Surveys \cite{Dey2019}. For the required combination of depth and the ability to search wide areas, the next generation of telescopes will be game-changing.

\subsection{Vera C. Rubin Observatory}

From this perspective, perhaps the most significant new facility to come online will be
the Vera C. Rubin Observatory \cite{Ivezic2019}, equipped with a wide-field 9.6\,deg$^2$ camera that will scan the optical sky in $u$-$g$-$r$-$i$-$z$-$y$ bands to remarkable depths. The mean effective aperture of the telescope is 6.4m which, coupled with the high sensitivity of the detectors, will make it possible to achieve a limiting magnitude of $r \sim 24$ mag in 30s exposures. The main scope of the Rubin Observatory is to complete the 10-year Legacy Survey of Space and Time (LSST) across a $\sim 18,000$ deg$^2$ footprint of Southern sky starting in late 2025\footnote{The Vera Rubin Observatory project timeline is regularly updated at: \url{ls.st/dates}}. This will establish a deep and continuous light curve history over the Southern sky, which will be enormously useful for excluding faint variable sources as contaminants in multi-messenger searches.

Due to the modest cadence of this baseline survey strategy, with observations of the same field every $\approx3$\,days on average\footnote{\url{https://survey-strategy.lsst.io}}, it will be challenging to find a large number of kilonovae serendipitously. While hundreds may be present in the data, most will be found at large distance and with only a few sparse detections, and may therefore remain unidentified and unclassified \cite{Andreoni2019PASP, Setzer2019MNRAS,Andreoni2022serendip, Setzer2023}. However, the possibility to interrupt the planned LSST observing sequence to perform target-of-opportunity (ToO) observations has been incorporated in the scheduling system. This will be crucial to find the faint and distant optical counterparts of GW sources. The Survey Cadence Optimization Committee (SCOC) recommended that up to 3\% of Rubin time may be dedicated to ToOs of all kinds, a significant fraction of which may be devoted to GW follow-up \cite{SCOCtoo}. 

The exact amount of time that will be designated for ToO observations and specifically for GW follow-up is still to be determined as of June 2024. 
The number of GW events that can be observed by Rubin will depend on many factors including intrinsic merger rates, GW detector sensitivity, width of the localization regions, visibility, ToO activation criteria based on the GW alert content, and the strategy that will be adopted (i.e. the trade-off between the number of observations of each field, the exposure time, the number of filters, etc.). A report was prepared in March--April 2024 by a large fraction of the Rubin community interested in ToO observations, outlining the expected number of triggers and options for the strategy to adopt \cite{rubintoo}. Based on the official IGWN observing scenario simulations \cite{Kiendrebeogo2023} for the Fifth Observing Run (O5), it was estimated that 18 triggers may have FAR\,$<1$\,yr$^{-1}$ and be localized to an area smaller than 100 deg$^2$. Given the observability constraints, Rubin is expected to trigger follow-up observations for 6 BNS and up to 2 NSBH events per year in O5 \cite{rubintoo}.

    \begin{figure}[!t]
\renewcommand{\tabularxcolumn}[1]{m{#1}}
\setlength\tabcolsep{3pt}
\begin{tabularx}{\textwidth}{XX}
\multirow{2}*[15ex]{\includegraphics[width=0.62\columnwidth]{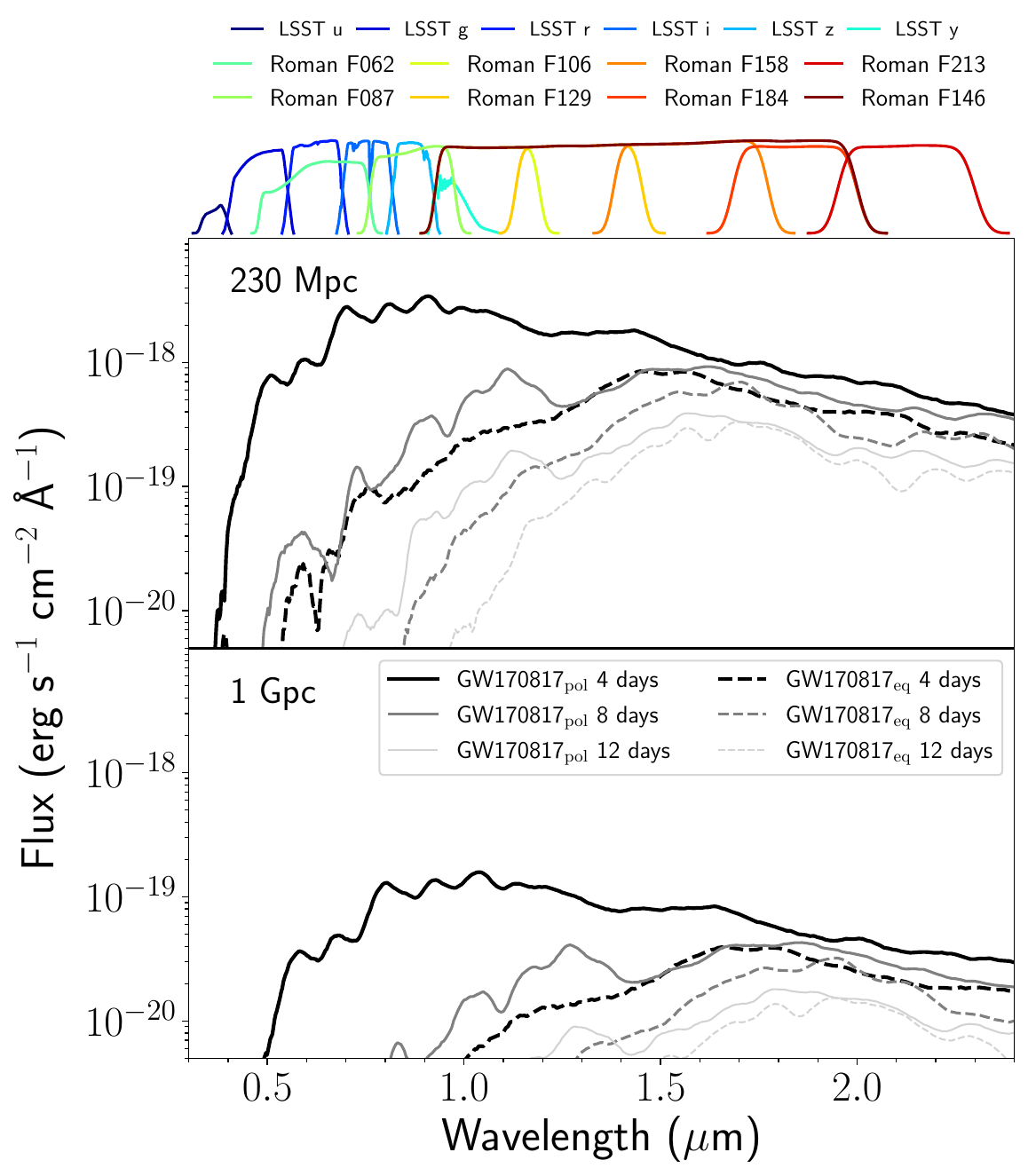}}   
        & \hspace{0.66in}  \includegraphics[width=0.365\columnwidth]{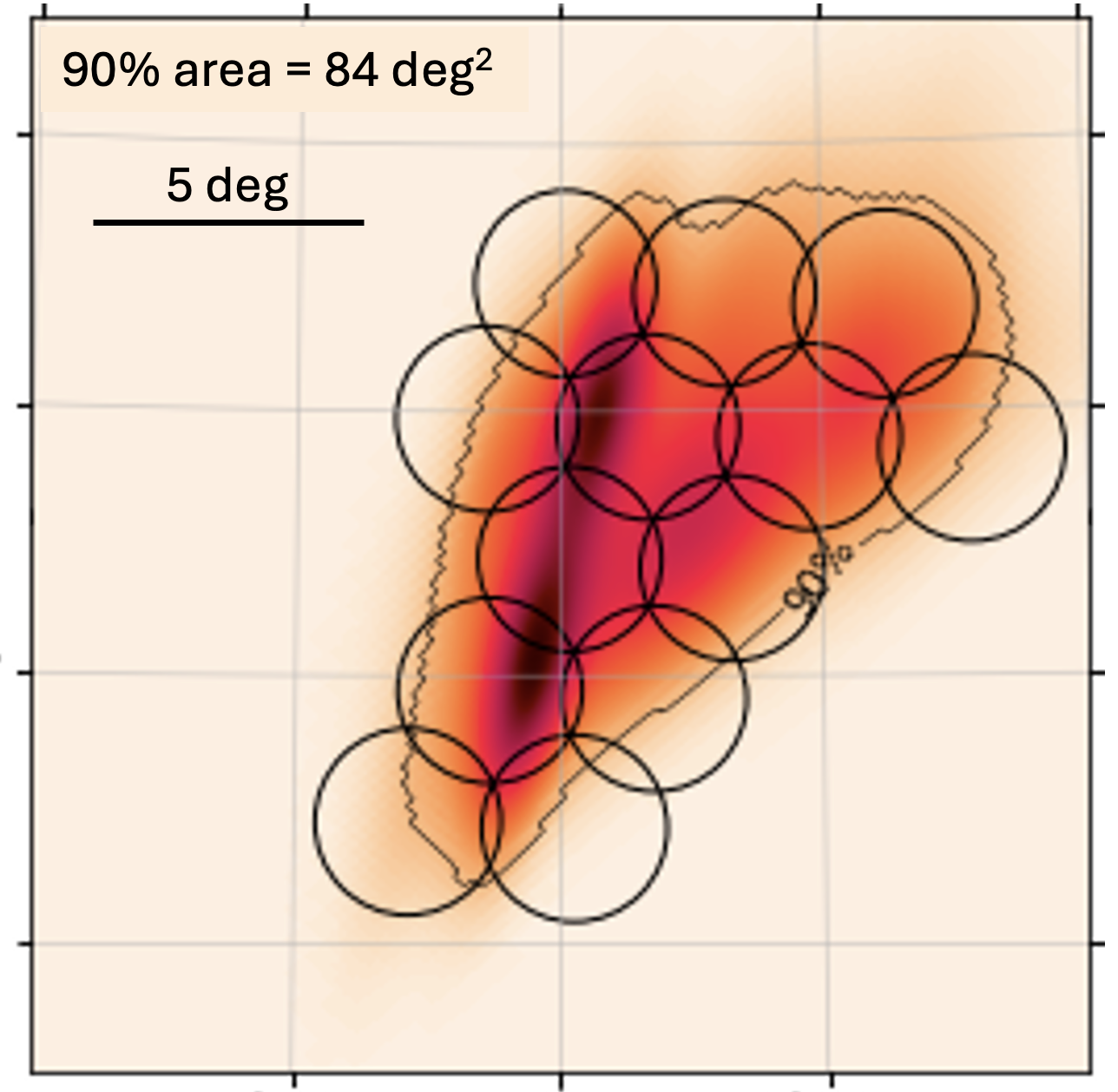}    \\
        & \hspace{0.7in}  \includegraphics[width=0.35\columnwidth]{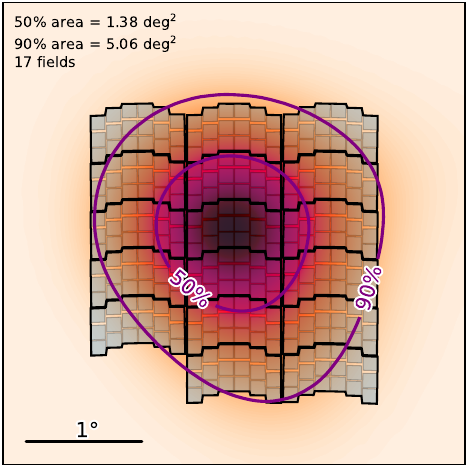}    \\
\end{tabularx}
\caption{Left: Kilonova model spectra at 230\,Mpc (top) and 1\,Gpc (bottom), obtained using \texttt{possis} \cite{Bulla2019, Bulla2023}. Spectra
are shown at three epochs (4, 8 and 12 days after the
merger) both for the best-fit model to AT2017gfo (solid lines) and for the same model viewed from an equatorial
viewing angle (dashed lines). The Rubin and {\it Roman} passbands are shown
at the top. Right: Example IGWN skymaps for reasonably well-localized
events tiled with Rubin (top, $A_{90} = 84$\,deg$^2$) and {\it Roman} ($A_{90} = 5$\,deg$^2$). The Rubin footprint has been approximated to a circle for simplicity. These figures were taken and modified from I.~Andreoni et al.~\cite{Andreoni2022ToO, Andreoni2024APh}.}
\label{fig:rubin_roman}
    \end{figure}

The detectable kilonova or GRB afterglow counterparts are expected to be fast-evolving and red (or rapidly reddening) in the optical bands. Proposed strategies for the follow-up of BNS and NSBH mergers with Rubin typically envision multi-band, repeated observations in the first few days from the GW trigger \cite{Margutti2018WP, Cowperthwaite2019, Andreoni2022ToO, Smith2023} (see also \cite{Scolnic2018, Andreoni2019PASP, Bianco2019, Setzer2019MNRAS, Almualla2021, Sagues2021, Andreoni2022serendip, Gompertz2023, Hlozek2023, Setzer2023, Ragosta2024} for other studies of kilonova detectability in Rubin LSST). The number of bands ranges from two (for very deep lensed kilonova searches \cite{Smith2023}) up to five (the maximum number that can be accommodated in the Rubin filter wheel). For well-localised events, multiple visits with three bands per epoch ($g+r+i$) on the first night, followed by exposures in two bands on subsequent nights, appears to be a reasonable compromise to catch a possible early blue component, measure the temperature in the early phase (even before targeted follow-up), and monitor the luminosity and color evolution to identify viable kilonova candidates. Coarsely localized events can be followed up using only two filters ($g+i$ or $g+z$) in order to cover a wider area, but still ensuring broad wavelength coverage to measure the colors of candidates \cite{rubintoo}. Another good option for coarsely localized events is to ``re-weight" the regular LSST survey to cover most of the region of interest in the nights following a GW trigger, with minimal disruption to the main survey. This strategy has already been implemented successfully by other wide-field surveys such as ZTF during O4 \cite{Ahumada2024arXiv}. 

\subsection{Nancy Grace Roman Space Telescope}

While Rubin is optimal for finding kilonova counterparts in the local Universe, another planned facility will extend our reach to detect more distant sources. The {\it Nancy Grace Roman} space mission, planned to launch in 2027, will be equipped with a Wide Field Instrument (WFI) with an effective field of view of 0.281\,deg$^2$ and a resolution of 0.1 arcsec per pixel in the optical and near infrared. The WFI is equipped with 8 filters covering the 0.48--2.3 $\mu$m wavelength range (Figure \ref{fig:rubin_roman}). According to the instrument technical webpage\footnote{\url{https://roman.gsfc.nasa.gov/science/WFI_technical.html}}, the instrument will be able to detect $\sim 25$\,mag point sources in most filters in $< 1$\,minute of exposure time. The infrared sensitivity will enable discovery of very faint or obscured kilonovae, including those without a `blue' ejecta component and those seen from equatorial viewing angles.

{\it Roman} will be capable of finding kilonovae all the way out to $z \sim 1$ \cite{Chase2022}. Several kilonovae may be detected by {\it Roman} during the High Latitude Time-Domain Survey \cite{Scolnic2018, Chase2022, Andreoni2024APh} and possibly as part of the Roman High Latitude Wide Area Survey, if the ``Transient Exploration in the high latitude Wide Area Survey" (TEWAS) strategy is implemented \cite{Rest2024}. It is expected that {\it Roman} will be able to perform 1--6 ToO observations to follow up well-localized ($A_{90} < 10$\,deg$^2$) BNS or NSBH mergers over 1.5 years in O5, and 4--21 events over 1.5 years in O6 \cite{Andreoni2024APh}.
Also relevant for this issue, it was estimated that {\it Roman} could detect $\sim 3$ lensed kilonovae during follow-up observations of Cosmic Explorer triggers of lensed binary NSs \cite{Ma2023}. We emphasise that synergies between Roman and Rubin could be extremely valuable. Their pass bands are largely complementary, and both observatories will be able to carry out deep, synoptic coverage of the best localized events from O6, as shown in Figure \ref{fig:rubin_roman}.

\subsection{The roles of coordination and redundancy in observing strategies}


The low detection rate of well-localized GW events bearing a NS in O3 and O4a shows the critical importance of following up these high-value events as comprehensively and efficiently as possible. We discuss here two aspects relevant to how we as a community can maximise the chances of a positive kilonova identification in O4b and beyond: {\it collaboration} between groups and {\it redundancy} in search data. 

From the very beginning of IGWN operations, it was obvious that collaborative efforts could help to guarantee coverage in many time zones, both hemispheres, and at multiple wavelengths. Quoting the GROWTH collaboration webpage\footnote{\url{https://www.growth.caltech.edu/}}, there were evident advantages in creating ``a network of telescopes to continuously observe the transient sky unbeaten by sunrise". Such networks have typically engaged several institutions coming together and often collaborating in the publication of scientific results. While operated by a single team, the Las Cumbres Observatory network also conducts searches that benefit from the geographical spread of its observing nodes. Even in the era of Rubin, observatories at different longitudes will be essential, as there is no guarantee it will be night in Chile when a well-localized BNS merger occurs.

The dispersed approach increases the capacity of a follow-up team (and potentially expands the team itself), but another motivation for coordinating between groups is to increase the efficiency of using a particularly important resource. While many groups obtained spectra of AT2017gfo, the high signal-to-noise ratio and broad wavelength coverage obtained with the X-Shooter spectrograph on the 8\,m European Southern Observatory Very Large Telescope (VLT) clearly demonstrated that this is an ideal instrument for kilonova studies \cite{Pian2017,Smartt2017}. From O3 onwards, most European teams have joined forces within the ENGRAVE consortium to submit a single proposal for VLT time, in order to ensure that time is spent observing the best kilonova candidates at the optimal times, rather than being rushed to trigger due to competition\footnote{For other observatories with multiple approved follow-up programs, even when data sharing between teams is mandatory it has been advantageous to trigger first in order to set the observing strategy.}. This has also motivated ENGRAVE members to cooperate in sharing their results from photometric searches, as this can speed up the identification of promising targets for VLT \cite{Ackley2020,Agudo2023}. Similar approaches can (and have) been employed for other highly competitive resources such as \textit{HST} and \textit{JWST}. In the near future, Rubin will become another critical resource, and signs are positive that the community are working together to maximise its efficient use in GW follow-up \cite{Andreoni2022ToO,SCOCtoo}.

The coarse localizations of distant GW events has also encouraged the community to investigate tools to facilitate multi-telescope coordination to more effectively cover large skymaps. One of these tools is \texttt{TreasureMap}\footnote{\url{https://treasuremap.space/}} \cite{Wyatt2020}, an online platform where people can programmatically upload tables of their planned or executed observations. Collaborating (or competing) teams can keep track of when, were, in which band, and to what depth the GW skymap has been observed, and factor these aspects into their own follow-up plans. As a case study, it was recently proposed that if observatories worldwide had coordinated their follow-up, they could have covered all of the visible part of the skymap of GW190425 (the BNS merger in O3), corresponding to 75\% of the total $A_{90}=9,881$\,\sqdeg\ \cite{Keinan2024arXiv}. In comparison, only a total of $\approx50\%$ was covered by teams pursuing independent strategies. This shows that for very poorly localized events, reacting to observations already performed can enable total coverage over a larger area.

The flipside of the coin is that data redundancy is often necessary to achieve completeness. Instruments can differ significantly in wavelength coverage, achievable depth, pixel scale, linearity of the detector response, uniformity of the point spread function, etc, and chip gaps can leave holes in coverage even for patches that have been observed. Equally importantly, practical challenges in transient identification (encompassing image processing, image subtraction, real/bogus classification) ensure that the efficiency of recovering sources is always $<1$. Completeness decreases especially near complex galaxy structures, galactic nuclei and bright sources \cite{Frohmaier2017}, but also the edge of the detectors. Single-image transient detections are often discarded\footnote{Though new methods such as convolutional neural networks can aid in finding supernova candidates in individual images with reasonable confidence, e.g.~\cite{Carrasco-Davis2021}.}, but multi-instrument observations can provide unbiased confirmation of such events, and its photometric evolution can be probed if cross-instrument photometric calibration can be performed reliably. Reporting of measurements in a machine-readable form, which can be ingested by transient alert brokers, would enable effectively higher cadence light curves from the combination of multiple data sources, so that fast-fading candidates can be identified efficiently.

In summary, whether to pursue a strategy of coordination between many groups or to ensure reasonable data redundancy should be evaluated on a case-by-case basis. When Rubin comes online, it may be preferable to build coordinated strategies around the community-led Rubin ToO observations.
Prompt observations with both small and large telescopes remain valuable when a bright counterpart is expected, for example when the sources are nearby (e.g.~GW170817) or when a coincident GRB detection indicates that the afterglow may be detectable in the first few hours after merger. Deep wide-field observations with a facility such as Rubin can serve as an anchor for other telescopes, which can then choose to observe areas unreachable from Cerro Pach\'{o}n, or to supplement the Rubin observations at different cadences or wavelengths. 

With comprehensive coverage of the skymap, efficient source classification is then of paramount importance in identifying the true EM counterpart among potentially hundreds of contaminants. Therefore rapid acquisition, reduction, analysis and dissemination of follow-up data are key. Coordinated transient characterization currently makes use of GCN circulars to report what has been done. Classification spectra (at least those of unrelated supernovae in the skymap) should be published immediately via the IAU Transient Name Server\footnote{\url{www.wis-tns.org}}, to prevent further wasted follow-up. Reassuringly, $\approx 90\%$ of sources reported in GCN circulars during O3 were also reported to the TNS (and in fact, a much larger number of sources were reported to the TNS with no accompanying GCN) \cite{Rastinejad2022b}. 

\section{Conclusion}

In this paper we have attempted to review the key developments in observational follow-up of GW sources, with a particular focus on how GW170817/AT2017gfo was discovered during O2 and why it has been so difficult to replicate this success during O3 and the early part of O4. Despite the lack of EM counterparts, this remains an incredibly active and fast-moving field. We conclude by summarising some of the key lessons from this review that may be useful going forward. Some of these lessons overlap with previous assessments of the follow-up conducted during O3 \cite{Coughlin2020c}.

\begin{enumerate}
    \item The majority of GRBs associated with NS mergers will be off-axis; therefore kilonovae are in general more likely to be the observable counterpart. However, if a GRB is detected from a distant merger, its afterglow is likely to be the most luminous optical signature at early times.

    \item AT2017gfo, the kilonova following GW170817, would most likely have been reliably identified based on its decline rate, colours, and lack of previous variability. This is promising for future high-redshift searches, where spectroscopy may not always be possible (but is still highly desirable for physical analysis). 

    \item Detecting kilonovae unlocks a wide range of physics: finding more sources will answer important questions in neutron star structure, nucleosynthesis and cosmology, but it is important to understand their diversity.

    \item Rapid dissemination of GW-inferred source parameters from the IGWN, both in low-latency and via regular updates, is essential in prioritising follow-up, especially as detectors become more sensitive and the event rate increases. Some of these parameters can be used to reject false triggers (in particular the fractional uncertainty in distance).
    
    \item The probabilities of different source classes, and the quantities \texttt{HasNS}, \texttt{HasMassgap} and  \texttt{HasRemnant}, can help to motivate follow-up, but their interpretation is not always straightforward. Most NSBH sources will likely be EM-faint, and the kilonova luminosity function even for BNS mergers is not well constrained observationally.

    \item Wide-field and galaxy targeted searches are both regularly employed in follow-up, and have distinct advantages and disadvantages, though for many GW events only wide-field telescopes reaching limiting magnitudes of $\gtrsim 20$ can cover a substantial fraction of the three-dimensional probability volume.

    \item All searches benefit from combining telescopes at a range of latitudes and longitudes, to mitigate for the effects of daylight, weather, and hemisphere.

    \item A lot of telescope time is spent on searches that are too shallow to detect a physically motivated kilonova. Access to component masses and inclinations would enable the use of physical models that predict the EM luminosity, which could greatly improve the efficiency of follow-up.

    \item Observations need not always be uniform over the skymap. In particular, contextual information such as a GRB, FRB or neutrino alert, or the presence of a known strong gravitational lens in the skymap, may motivate deeper, targeted observations of a particular line of sight.

    \item However, interpretation of any targeted search relies on the completeness of galaxy catalogs. Spectroscopic redshift catalogs are highly incomplete at the typical distances of GW sources, but this may be mitigated in the near future by multiplexed spectroscopic surveys.

    \item The Vera Rubin Observatory will be a game-changer in this field, through the use of deep ToO observations and the excellent light curve history at every point in the Southern sky. The combination of light curve history and improved galaxy redshift information is highly effective for contaminant rejection. The Nancy Grace Roman Space Telescope will be highly complementary in extending kilonova detection to higher redshift.

    \item The size of the GW skymap is one of the most important factors determining whether follow-up observations are constraining. The impact of Virgo (and soon KAGRA) in shrinking these skymaps as sensitivity increases will be critical for covering the full probability and performing meaningful inference.

    \item In the era of Rubin and upgraded GW detectors, coordination between groups will be essential to optimise follow-up efficiency and contaminant rejection. However, observations of the same sky with multiple facilities will remain important, providing some degree of redundancy to mitigate for (e.g.) weather, data artefacts or uncertainties in image subtraction. Prompt reporting of candidates in a standardised, machine readable format will streamline this process.

\end{enumerate}

\vskip6pt

\ack{
We thank the organisers of the Royal Society Specialist Discussion Meeting, `Multi-messenger Gravitational Lensing'.
We also thank two anonymous referees, and Mattia Bulla and Michael Coughlin for providing comments on the manuscript. We are grateful to Michael Fulton and Stephen Smartt for compiling the data in Figure \ref{fig:gwparams}.
We acknowledge the many EM-GW groups working hard to make this science possible. While we have tried to be as thorough as possible in reviewing the literature, doubtless there is more that could have been included. Any omissions are entirely accidental, and we apologise for any contribution that we may have overlooked.
MN is supported by the European Research Council (ERC) under the European Union’s Horizon 2020 research and innovation programme (grant agreement No.~948381) and by UK Space Agency Grant No.~ST/Y000692/1. 
}


\bibliographystyle{RS}
\bibliography{refs}

\begin{thebibliography}{99}

\bibitem{LIGOScientificCollaboration2015}
{LIGO Scientific Collaboration}. 2015  {Advanced LIGO}. {\em Classical and Quantum Gravity} \textbf{32}, 074001.
(\href{http://dx.doi.org/10.1088/0264-9381/32/7/074001}{10.1088/0264-9381/32/7/074001})

\bibitem{Abbott2016}
{LIGO Scientific Collaboration}, {Virgo Collaboration}. 2016  {Observation of Gravitational Waves from a Binary Black Hole Merger}. {\em \prl} \textbf{116}, 061102.
(\href{http://dx.doi.org/10.1103/PhysRevLett.116.061102}{10.1103/PhysRevLett.116.061102})

\bibitem{Acernese2015}
{Acernese} F, {Agathos} M, {Agatsuma} K, {Aisa} D, {Allemandou} N, {Allocca} A, {Amarni} J, {Astone} P, {Balestri} G, {Ballardin} G, {Barone} F, {Baronick} JP, {Barsuglia} M, {Basti} A, {Basti} F, {Bauer} TS, {Bavigadda} V, {Bejger} M, {Beker} MG, {Belczynski} C, {Bersanetti} D, {Bertolini} A, {Bitossi} M, {Bizouard} MA, {Bloemen} S, {Blom} M, {Boer} M, {Bogaert} G, {Bondi} D, {Bondu} F, {Bonelli} L, {Bonnand} R, {Boschi} V, {Bosi} L, {Bouedo} T, {Bradaschia} C, {Branchesi} M, {Briant} T, {Brillet} A, {Brisson} V, {Bulik} T, {Bulten} HJ, {Buskulic} D, {Buy} C, {Cagnoli} G, {Calloni} E, {Campeggi} C, {Canuel} B, {Carbognani} F, {Cavalier} F, {Cavalieri} R, {Cella} G, {Cesarini} E, {Mottin} EC, {Chincarini} A, {Chiummo} A, {Chua} S, {Cleva} F, {Coccia} E, {Cohadon} PF, {Colla} A, {Colombini} M, {Conte} A, {Coulon} JP, {Cuoco} E, {Dalmaz} A, {D'Antonio} S, {Dattilo} V, {Davier} M, {Day} R, {Debreczeni} G, {Degallaix} J, {Del{\'e}glise} S, {Pozzo} WD, {Dereli} H, {Rosa} RD, {Fiore} LD, {Lieto} AD, {Virgilio} AD,
  {Doets} M, {Dolique} V, {Drago} M, {Ducrot} M, {Endr{\H{o}}czi} G, {Fafone} V, {Farinon} S, {Ferrante} I, {Ferrini} F, {Fidecaro} F, {Fiori} I, {Flaminio} R, {Fournier} JD, {Franco} S, {Frasca} S, {Frasconi} F, {Gammaitoni} L, {Garufi} F, {Gaspard} M, {Gatto} A, {Gemme} G, {Gendre} B, {Genin} E, {Gennai} A, {Ghosh} S, {Giacobone} L, {Giazotto} A, {Gouaty} R, {Granata} M, {Greco} G, {Groot} P, {Guidi} GM, {Harms} J, {Heidmann} A, {Heitmann} H, {Hello} P, {Hemming} G, {Hennes} E, {Hofman} D, {Jaranowski} P, {Jonker} RJG, {Kasprzack} M, {K{\'e}f{\'e}lian} F, {Kowalska} I, {Kraan} M, {Kr{\'o}lak} A, {Kutynia} A, {Lazzaro} C, {Leonardi} M, {Leroy} N, {Letendre} N, {Li} TGF, {Lieunard} B, {Lorenzini} M, {Loriette} V, {Losurdo} G, {Magazz{\`u}} C, {Majorana} E, {Maksimovic} I, {Malvezzi} V, {Man} N, {Mangano} V, {Mantovani} M, {Marchesoni} F, {Marion} F, {Marque} J, {Martelli} F, {Martellini} L, {Masserot} A, {Meacher} D, {Meidam} J, {Mezzani} F, {Michel} C, {Milano} L, {Minenkov} Y, {Moggi} A, {Mohan} M,
  {Montani} M, {Morgado} N, {Mours} B, {Mul} F, {Nagy} MF, {Nardecchia} I, {Naticchioni} L, {Nelemans} G, {Neri} I, {Neri} M, {Nocera} F, {Pacaud} E, {Palomba} C, {Paoletti} F, {Paoli} A, {Pasqualetti} A, {Passaquieti} R, {Passuello} D, {Perciballi} M, {Petit} S, {Pichot} M, {Piergiovanni} F, {Pillant} G, {Piluso} A, {Pinard} L, {Poggiani} R, {Prijatelj} M, {Prodi} GA, {Punturo} M, {Puppo} P, {Rabeling} DS, {R{\'a}cz} I, {Rapagnani} P, {Razzano} M, {Re} V, {Regimbau} T, {Ricci} F, {Robinet} F, {Rocchi} A, {Rolland} L, {Romano} R, {Rosi{\'n}ska} D, {Ruggi} P, {Saracco} E, {Sassolas} B, {Schimmel} F, {Sentenac} D, {Sequino} V, {Shah} S, {Siellez} K, {Straniero} N, {Swinkels} B, {Tacca} M, {Tonelli} M, {Travasso} F, {Turconi} M, {Vajente} G, {van Bakel} N, {van Beuzekom} M, {van den Brand} JFJ, {Van Den Broeck} C, {van der Sluys} MV, {van Heijningen} J, {Vas{\'u}th} M, {Vedovato} G, {Veitch} J, {Verkindt} D, {Vetrano} F, {Vicer{\'e}} A, {Vinet} JY, {Visser} G, {Vocca} H, {Ward} R, {Was} M, {Wei} LW, {Yvert} M,
  {{\.z}ny} AZ, {Zendri} JP. 2015  {Advanced Virgo: a second-generation interferometric gravitational wave detector}. {\em Classical and Quantum Gravity} \textbf{32}, 024001.
(\href{http://dx.doi.org/10.1088/0264-9381/32/2/024001}{10.1088/0264-9381/32/2/024001})

\bibitem{KagraCollaboration2019}
{Kagra Collaboration}. 2019  {KAGRA: 2.5 generation interferometric gravitational wave detector}. {\em Nature Astronomy} \textbf{3}, 35--40.
(\href{http://dx.doi.org/10.1038/s41550-018-0658-y}{10.1038/s41550-018-0658-y})

\bibitem{Abbott2023}
{Ligo Scientific Collaboration}, {Virgo Collaboration}, {Kagra Collaboration}. 2023  {GWTC-3: Compact Binary Coalescences Observed by LIGO and Virgo during the Second Part of the Third Observing Run}. {\em Physical Review X} \textbf{13}, 041039.
(\href{http://dx.doi.org/10.1103/PhysRevX.13.041039}{10.1103/PhysRevX.13.041039})

\bibitem{Abbott2021}
{LIGO Scientific Collaboration}, {Virgo Collaboration}. 2021  {GWTC-2: Compact Binary Coalescences Observed by LIGO and Virgo during the First Half of the Third Observing Run}. {\em Physical Review X} \textbf{11}, 021053.
(\href{http://dx.doi.org/10.1103/PhysRevX.11.021053}{10.1103/PhysRevX.11.021053})

\bibitem{Abbott2024}
{LIGO Scientific Collaboration}, {Virgo Collaboration}. 2024  {GWTC-2.1: Deep extended catalog of compact binary coalescences observed by LIGO and Virgo during the first half of the third observing run}. {\em \prd} \textbf{109}, 022001.
(\href{http://dx.doi.org/10.1103/PhysRevD.109.022001}{10.1103/PhysRevD.109.022001})

\bibitem{Abbott2017}
{LIGO Scientific Collaboration}, {Virgo Collaboration}. 2017  {GW170817: Observation of Gravitational Waves from a Binary Neutron Star Inspiral}. {\em \prl} \textbf{119}, 161101.
(\href{http://dx.doi.org/10.1103/PhysRevLett.119.161101}{10.1103/PhysRevLett.119.161101})

\bibitem{Abbott2020a}
{LIGO Scientific Collaboration}, {Virgo Collaboration}. 2020  {GW190425: Observation of a Compact Binary Coalescence with Total Mass {\ensuremath{\sim}} 3.4 M$_{{\ensuremath{\odot}}}$}. {\em \apjl} \textbf{892}, L3.
(\href{http://dx.doi.org/10.3847/2041-8213/ab75f5}{10.3847/2041-8213/ab75f5})

\bibitem{Abbott2021a}
{Ligo Scientific Collaboration}, {VIRGO Collaboration}, {KAGRA Collaboration}. 2021  {Observation of Gravitational Waves from Two Neutron Star-Black Hole Coalescences}. {\em \apjl} \textbf{915}, L5.
(\href{http://dx.doi.org/10.3847/2041-8213/ac082e}{10.3847/2041-8213/ac082e})

\bibitem{Finn1993}
{Finn} LS, {Chernoff} DF. 1993  {Observing binary inspiral in gravitational radiation: One interferometer}. {\em \prd} \textbf{47}, 2198--2219.
(\href{http://dx.doi.org/10.1103/PhysRevD.47.2198}{10.1103/PhysRevD.47.2198})

\bibitem{Bahcall1989}
{Bahcall} JN. 1989 {\em {Neutrino Astrophysics}}.

\bibitem{Arnett1989}
{Arnett} WD, {Bahcall} JN, {Kirshner} RP, {Woosley} SE. 1989  {Supernova 1987A.}. {\em \araa} \textbf{27}, 629--700.
(\href{http://dx.doi.org/10.1146/annurev.aa.27.090189.003213}{10.1146/annurev.aa.27.090189.003213})

\bibitem{Perna2016}
{Perna} R, {Lazzati} D, {Giacomazzo} B. 2016  {Short Gamma-Ray Bursts from the Merger of Two Black Holes}. {\em \apjl} \textbf{821}, L18.
(\href{http://dx.doi.org/10.3847/2041-8205/821/1/L18}{10.3847/2041-8205/821/1/L18})

\bibitem{McKernan2019}
{McKernan} B, {Ford} KES, {Bartos} I, {Graham} MJ, {Lyra} W, {Marka} S, {Marka} Z, {Ross} NP, {Stern} D, {Yang} Y. 2019  {Ram-pressure Stripping of a Kicked Hill Sphere: Prompt Electromagnetic Emission from the Merger of Stellar Mass Black Holes in an AGN Accretion Disk}. {\em \apjl} \textbf{884}, L50.
(\href{http://dx.doi.org/10.3847/2041-8213/ab4886}{10.3847/2041-8213/ab4886})

\bibitem{Evans2016}
{Evans} PA, {Kennea} JA, {Barthelmy} SD, {Beardmore} AP, {Burrows} DN, {Campana} S, {Cenko} SB, {Gehrels} N, {Giommi} P, {Gronwall} C, {Marshall} FE, {Malesani} D, {Markwardt} CB, {Mingo} B, {Nousek} JA, {O'Brien} PT, {Osborne} JP, {Pagani} C, {Page} KL, {Palmer} DM, {Perri} M, {Racusin} JL, {Siegel} MH, {Sbarufatti} B, {Tagliaferri} G. 2016  {Swift follow-up of the gravitational wave source GW150914}. {\em \mnras} \textbf{460}, L40--L44.
(\href{http://dx.doi.org/10.1093/mnrasl/slw065}{10.1093/mnrasl/slw065})

\bibitem{Kasliwal2016}
{Kasliwal} MM, {Cenko} SB, {Singer} LP, {Corsi} A, {Cao} Y, {Barlow} T, {Bhalerao} V, {Bellm} E, {Cook} D, {Duggan} GE, {Ferretti} R, {Frail} DA, {Horesh} A, {Kendrick} R, {Kulkarni} SR, {Lunnan} R, {Palliyaguru} N, {Laher} R, {Masci} F, {Manulis} I, {Miller} AA, {Nugent} PE, {Perley} D, {Prince} TA, {Quimby} RM, {Rana} J, {Rebbapragada} U, {Sesar} B, {Singhal} A, {Surace} J, {Van Sistine} A. 2016  {iPTF Search for an Optical Counterpart to Gravitational-wave Transient GW150914}. {\em \apjl} \textbf{824}, L24.
(\href{http://dx.doi.org/10.3847/2041-8205/824/2/L24}{10.3847/2041-8205/824/2/L24})

\bibitem{Smartt2016}
{Smartt} SJ, {Chambers} KC, {Smith} KW, {Huber} ME, {Young} DR, {Cappellaro} E, {Wright} DE, {Coughlin} M, {Schultz} ASB, {Denneau} L, {Flewelling} H, {Heinze} A, {Magnier} EA, {Primak} N, {Rest} A, {Sherstyuk} A, {Stalder} B, {Stubbs} CW, {Tonry} J, {Waters} C, {Willman} M, {Anderson} JP, {Baltay} C, {Botticella} MT, {Campbell} H, {Dennefeld} M, {Chen} TW, {Della Valle} M, {Elias-Rosa} N, {Fraser} M, {Inserra} C, {Kankare} E, {Kotak} R, {Kupfer} T, {Harmanen} J, {Galbany} L, {Gal-Yam} A, {Le Guillou} L, {Lyman} JD, {Maguire} K, {Mitra} A, {Nicholl} M, {Olivares E} F, {Rabinowitz} D, {Razza} A, {Sollerman} J, {Smith} M, {Terreran} G, {Valenti} S, {Gibson} B, {Goggia} T. 2016  {Pan-STARRS and PESSTO search for an optical counterpart to the LIGO gravitational-wave source GW150914}. {\em \mnras} \textbf{462}, 4094--4116.
(\href{http://dx.doi.org/10.1093/mnras/stw1893}{10.1093/mnras/stw1893})

\bibitem{Soares-Santos2016}
{Soares-Santos} M, {Kessler} R, {Berger} E, {Annis} J, {Brout} D, {Buckley-Geer} E, {Chen} H, {Cowperthwaite} PS, {Diehl} HT, {Doctor} Z, {Drlica-Wagner} A, {Farr} B, {Finley} DA, {Flaugher} B, {Foley} RJ, {Frieman} J, {Gruendl} RA, {Herner} K, {Holz} D, {Lin} H, {Marriner} J, {Neilsen} E, {Rest} A, {Sako} M, {Scolnic} D, {Sobreira} F, {Walker} AR, {Wester} W, {Yanny} B, {Abbott} TMC, {Abdalla} FB, {Allam} S, {Armstrong} R, {Banerji} M, {Benoit-L{\'e}vy} A, {Bernstein} RA, {Bertin} E, {Brown} DA, {Burke} DL, {Capozzi} D, {Carnero Rosell} A, {Carrasco Kind} M, {Carretero} J, {Castander} FJ, {Cenko} SB, {Chornock} R, {Crocce} M, {D'Andrea} CB, {da Costa} LN, {Desai} S, {Dietrich} JP, {Drout} MR, {Eifler} TF, {Estrada} J, {Evrard} AE, {Fairhurst} S, {Fernandez} E, {Fischer} J, {Fong} W, {Fosalba} P, {Fox} DB, {Fryer} CL, {Garcia-Bellido} J, {Gaztanaga} E, {Gerdes} DW, {Goldstein} DA, {Gruen} D, {Gutierrez} G, {Honscheid} K, {James} DJ, {Karliner} I, {Kasen} D, {Kent} S, {Kuropatkin} N, {Kuehn} K, {Lahav} O, {Li}
  TS, {Lima} M, {Maia} MAG, {Margutti} R, {Martini} P, {Matheson} T, {McMahon} RG, {Metzger} BD, {Miller} CJ, {Miquel} R, {Mohr} JJ, {Nichol} RC, {Nord} B, {Ogando} R, {Peoples} J, {Plazas} AA, {Quataert} E, {Romer} AK, {Roodman} A, {Rykoff} ES, {Sanchez} E, {Scarpine} V, {Schindler} R, {Schubnell} M, {Sevilla-Noarbe} I, {Sheldon} E, {Smith} M, {Smith} N, {Smith} RC, {Stebbins} A, {Sutton} PJ, {Swanson} MEC, {Tarle} G, {Thaler} J, {Thomas} RC, {Tucker} DL, {Vikram} V, {Wechsler} RH, {Weller} J, {DES Collaboration}. 2016  {A Dark Energy Camera Search for an Optical Counterpart to the First Advanced LIGO Gravitational Wave Event GW150914}. {\em \apjl} \textbf{823}, L33.
(\href{http://dx.doi.org/10.3847/2041-8205/823/2/L33}{10.3847/2041-8205/823/2/L33})

\bibitem{Cowperthwaite2016}
{Cowperthwaite} PS, {Berger} E, {Soares-Santos} M, {Annis} J, {Brout} D, {Brown} DA, {Buckley-Geer} E, {Cenko} SB, {Chen} HY, {Chornock} R, {Diehl} HT, {Doctor} Z, {Drlica-Wagner} A, {Drout} MR, {Farr} B, {Finley} DA, {Foley} RJ, {Fong} W, {Fox} DB, {Frieman} J, {Garcia-Bellido} J, {Gill} MSS, {Gruendl} RA, {Herner} K, {Holz} DE, {Kasen} D, {Kessler} R, {Lin} H, {Margutti} R, {Marriner} J, {Matheson} T, {Metzger} BD, {Neilsen}, E.~H. J, {Quataert} E, {Rest} A, {Sako} M, {Scolnic} D, {Smith} N, {Sobreira} F, {Strampelli} GM, {Villar} VA, {Walker} AR, {Wester} W, {Williams} PKG, {Yanny} B, {Abbott} TMC, {Abdalla} FB, {Allam} S, {Armstrong} R, {Bechtol} K, {Benoit-L{\'e}vy} A, {Bertin} E, {Brooks} D, {Burke} DL, {Carnero Rosell} A, {Carrasco Kind} M, {Carretero} J, {Castander} FJ, {Cunha} CE, {D'Andrea} CB, {da Costa} LN, {Desai} S, {Dietrich} JP, {Evrard} AE, {Fausti Neto} A, {Fosalba} P, {Gerdes} DW, {Giannantonio} T, {Goldstein} DA, {Gruen} D, {Gutierrez} G, {Honscheid} K, {James} DJ, {Johnson} MWG, {Johnson}
  MD, {Krause} E, {Kuehn} K, {Kuropatkin} N, {Lima} M, {Maia} MAG, {Marshall} JL, {Menanteau} F, {Miquel} R, {Mohr} JJ, {Nichol} RC, {Nord} B, {Ogando} R, {Plazas} AA, {Reil} K, {Romer} AK, {Sanchez} E, {Scarpine} V, {Sevilla-Noarbe} I, {Smith} RC, {Suchyta} E, {Tarle} G, {Thomas} D, {Thomas} RC, {Tucker} DL, {Weller} J, {DES Collaboration}. 2016  {A DECam Search for an Optical Counterpart to the LIGO Gravitational-wave Event GW151226}. {\em \apjl} \textbf{826}, L29.
(\href{http://dx.doi.org/10.3847/2041-8205/826/2/L29}{10.3847/2041-8205/826/2/L29})

\bibitem{Smartt2016a}
{Smartt} SJ, {Chambers} KC, {Smith} KW, {Huber} ME, {Young} DR, {Chen} TW, {Inserra} C, {Wright} DE, {Coughlin} M, {Denneau} L, {Flewelling} H, {Heinze} A, {Jerkstrand} A, {Magnier} EA, {Maguire} K, {Mueller} B, {Rest} A, {Sherstyuk} A, {Stalder} B, {Schultz} ASB, {Stubbs} CW, {Tonry} J, {Waters} C, {Wainscoat} RJ, {Della Valle} M, {Dennefeld} M, {Dimitriadis} G, {Firth} RE, {Fraser} M, {Frohmaier} C, {Gal-Yam} A, {Harmanen} J, {Kankare} E, {Kotak} R, {Kromer} M, {Mandel} I, {Sollerman} J, {Gibson} B, {Primak} N, {Willman} M. 2016  {A Search for an Optical Counterpart to the Gravitational-wave Event GW151226}. {\em \apjl} \textbf{827}, L40.
(\href{http://dx.doi.org/10.3847/2041-8205/827/2/L40}{10.3847/2041-8205/827/2/L40})

\bibitem{Brocato2018}
{Brocato} E, {Branchesi} M, {Cappellaro} E, {Covino} S, {Grado} A, {Greco} G, {Limatola} L, {Stratta} G, {Yang} S, {Campana} S, {D'Avanzo} P, {Getman} F, {Melandri} A, {Nicastro} L, {Palazzi} E, {Pian} E, {Piranomonte} S, {Pulone} L, {Rossi} A, {Tomasella} L, {Amati} L, {Antonelli} LA, {Ascenzi} S, {Benetti} S, {Bulgarelli} A, {Capaccioli} M, {Cella} G, {Dadina} M, {De Cesare} G, {D'Elia} V, {Ghirlanda} G, {Ghisellini} G, {Giuffrida} G, {Iannicola} G, {Israel} G, {Lisi} M, {Longo} F, {Mapelli} M, {Marinoni} S, {Marrese} P, {Masetti} N, {Patricelli} B, {Possenti} A, {Radovich} M, {Razzano} M, {Salvaterra} R, {Schipani} P, {Spera} M, {Stamerra} A, {Stella} L, {Tagliaferri} G, {Testa} V, {Grawita-Gravitational Wave Inaf Team}. 2018  {GRAWITA: VLT Survey Telescope observations of the gravitational wave sources GW150914 and GW151226}. {\em \mnras} \textbf{474}, 411--426.
(\href{http://dx.doi.org/10.1093/mnras/stx2730}{10.1093/mnras/stx2730})

\bibitem{Doctor2019}
{Doctor} Z, {Kessler} R, {Herner} K, {Palmese} A, {Soares-Santos} M, {Annis} J, {Brout} D, {Holz} DE, {Sako} M, {Rest} A, {Cowperthwaite} P, {Berger} E, {Foley} RJ, {Conselice} CJ, {Gill} MSS, {Allam} S, {Balbinot} E, {Butler} RE, {Chen} HY, {Chornock} R, {Cook} E, {Diehl} HT, {Farr} B, {Fong} W, {Frieman} J, {Fryer} C, {Garc{\'\i}a-Bellido} J, {Margutti} R, {Marshall} JL, {Matheson} T, {Metzger} BD, {Nicholl} M, {Paz-Chinch{\'o}n} F, {Salim} S, {Sauseda} M, {Secco} LF, {Smith} RC, {Smith} N, {Vivas} AK, {Tucker} DL, {Abbott} TMC, {Avila} S, {Bechtol} K, {Bertin} E, {Brooks} D, {Buckley-Geer} E, {Burke} DL, {Carnero Rosell} A, {Carrasco Kind} M, {Carretero} J, {Castander} FJ, {D'Andrea} CB, {da Costa} LN, {De Vicente} J, {Desai} S, {Doel} P, {Flaugher} B, {Fosalba} P, {Gaztanaga} E, {Gerdes} DW, {Goldstein} DA, {Gruen} D, {Gruendl} RA, {Gutierrez} G, {Hartley} WG, {Hollowood} DL, {Honscheid} K, {Hoyle} B, {James} DJ, {Jeltema} T, {Kent} S, {Kuehn} K, {Kuropatkin} N, {Lahav} O, {Lima} M, {Maia} MAG, {March} M,
  {Menanteau} F, {Miller} CJ, {Miquel} R, {Neilsen} E, {Nord} B, {Ogando} RLC, {Plazas} AA, {Roodman} A, {Sanchez} E, {Scarpine} V, {Schindler} R, {Schubnell} M, {Serrano} S, {Sevilla-Noarbe} I, {Smith} M, {Sobreira} F, {Suchyta} E, {Swanson} MEC, {Tarle} G, {Thomas} D, {Walker} AR, {Wester} W, {DES Collaboration}. 2019  {A Search for Optical Emission from Binary Black Hole Merger GW170814 with the Dark Energy Camera}. {\em \apjl} \textbf{873}, L24.
(\href{http://dx.doi.org/10.3847/2041-8213/ab08a3}{10.3847/2041-8213/ab08a3})

\bibitem{Smith2019}
{Smith} GP, {Bianconi} M, {Jauzac} M, {Richard} J, {Robertson} A, {Berry} CPL, {Massey} R, {Sharon} K, {Farr} WM, {Veitch} J. 2019  {Deep and rapid observations of strong-lensing galaxy clusters within the sky localization of GW170814}. {\em \mnras} \textbf{485}, 5180--5191.
(\href{http://dx.doi.org/10.1093/mnras/stz675}{10.1093/mnras/stz675})

\bibitem{Grado2020}
{Grado} A, {Cappellaro} E, {Covino} S, {Getman} F, {Greco} G, {Limatola} L, {Yang} S, {Amati} L, {Benetti} S, {Branchesi} M, {Brocato} E, {Botticella} M, {Campana} S, {Cantiello} M, {Dadina} M, {D'Ammando} F, {De Cesare} G, {D'Elia} V, {Della Valle} M, {Iodice} E, {Longo} G, {Mapelli} M, {Masetti} N, {Nicastro} L, {Palazzi} E, {Possenti} A, {Radovich} M, {Rossi} A, {Salvaterra} R, {Stella} L, {Stratta} G, {Testa} V, {Tomasella} L. 2020  {Search for the optical counterpart of the GW170814 gravitational wave event with the VLT Survey Telescope}. {\em \mnras} \textbf{492}, 1731--1754.
(\href{http://dx.doi.org/10.1093/mnras/stz3536}{10.1093/mnras/stz3536})

\bibitem{Connaughton2016}
{Connaughton} V, {Burns} E, {Goldstein} A, {Blackburn} L, {Briggs} MS, {Zhang} BB, {Camp} J, {Christensen} N, {Hui} CM, {Jenke} P, {Littenberg} T, {McEnery} JE, {Racusin} J, {Shawhan} P, {Singer} L, {Veitch} J, {Wilson-Hodge} CA, {Bhat} PN, {Bissaldi} E, {Cleveland} W, {Fitzpatrick} G, {Giles} MM, {Gibby} MH, {von Kienlin} A, {Kippen} RM, {McBreen} S, {Mailyan} B, {Meegan} CA, {Paciesas} WS, {Preece} RD, {Roberts} OJ, {Sparke} L, {Stanbro} M, {Toelge} K, {Veres} P. 2016  {Fermi GBM Observations of LIGO Gravitational-wave Event GW150914}. {\em \apjl} \textbf{826}, L6.
(\href{http://dx.doi.org/10.3847/2041-8205/826/1/L6}{10.3847/2041-8205/826/1/L6})

\bibitem{Savchenko2016}
{Savchenko} V, {Ferrigno} C, {Mereghetti} S, {Natalucci} L, {Bazzano} A, {Bozzo} E, {Brandt} S, {Courvoisier} TJL, {Diehl} R, {Hanlon} L, {von Kienlin} A, {Kuulkers} E, {Laurent} P, {Lebrun} F, {Roques} JP, {Ubertini} P, {Weidenspointner} G. 2016  {INTEGRAL Upper Limits on Gamma-Ray Emission Associated with the Gravitational Wave Event GW150914}. {\em \apjl} \textbf{820}, L36.
(\href{http://dx.doi.org/10.3847/2041-8205/820/2/L36}{10.3847/2041-8205/820/2/L36})

\bibitem{Abbott2020}
{LIGO Scientific Collaboration}, {Virgo Collaboration}. 2020  {GW190521: A Binary Black Hole Merger with a Total Mass of 150 M$_{{\ensuremath{\odot}}}$}. {\em \prl} \textbf{125}, 101102.
(\href{http://dx.doi.org/10.1103/PhysRevLett.125.101102}{10.1103/PhysRevLett.125.101102})

\bibitem{Graham2020}
{Graham} MJ, {Ford} KES, {McKernan} B, {Ross} NP, {Stern} D, {Burdge} K, {Coughlin} M, {Djorgovski} SG, {Drake} AJ, {Duev} D, {Kasliwal} M, {Mahabal} AA, {van Velzen} S, {Belecki} J, {Bellm} EC, {Burruss} R, {Cenko} SB, {Cunningham} V, {Helou} G, {Kulkarni} SR, {Masci} FJ, {Prince} T, {Reiley} D, {Rodriguez} H, {Rusholme} B, {Smith} RM, {Soumagnac} MT. 2020  {Candidate Electromagnetic Counterpart to the Binary Black Hole Merger Gravitational-Wave Event S190521g$^{*}$}. {\em \prl} \textbf{124}, 251102.
(\href{http://dx.doi.org/10.1103/PhysRevLett.124.251102}{10.1103/PhysRevLett.124.251102})

\bibitem{Palmese2021}
{Palmese} A, {Fishbach} M, {Burke} CJ, {Annis} J, {Liu} X. 2021  {Do LIGO/Virgo Black Hole Mergers Produce AGN Flares? The Case of GW190521 and Prospects for Reaching a Confident Association}. {\em \apjl} \textbf{914}, L34.
(\href{http://dx.doi.org/10.3847/2041-8213/ac0883}{10.3847/2041-8213/ac0883})

\bibitem{Veronesi2024arXiv}
{Veronesi} N, {van Velzen} S, {Rossi} EM. 2024  {AGN flares as counterparts to the mergers detected by LIGO and Virgo: a novel spatial correlation analysis}. {\em arXiv e-prints} p. arXiv:2405.05318.
(\href{http://dx.doi.org/10.48550/arXiv.2405.05318}{10.48550/arXiv.2405.05318})

\bibitem{Paczynski1986}
{Paczynski} B. 1986  {Gamma-ray bursters at cosmological distances}. {\em \apjl} \textbf{308}, L43--L46.
(\href{http://dx.doi.org/10.1086/184740}{10.1086/184740})

\bibitem{Eichler1989}
{Eichler} D, {Livio} M, {Piran} T, {Schramm} DN. 1989  {Nucleosynthesis, neutrino bursts and {\ensuremath{\gamma}}-rays from coalescing neutron stars}. {\em \nat} \textbf{340}, 126--128.
(\href{http://dx.doi.org/10.1038/340126a0}{10.1038/340126a0})

\bibitem{Berger2014}
{Berger} E. 2014  {Short-Duration Gamma-Ray Bursts}. {\em \araa} \textbf{52}, 43--105.
(\href{http://dx.doi.org/10.1146/annurev-astro-081913-035926}{10.1146/annurev-astro-081913-035926})

\bibitem{Barthelmy2005}
{Barthelmy} SD, {Barbier} LM, {Cummings} JR, {Fenimore} EE, {Gehrels} N, {Hullinger} D, {Krimm} HA, {Markwardt} CB, {Palmer} DM, {Parsons} A, {Sato} G, {Suzuki} M, {Takahashi} T, {Tashiro} M, {Tueller} J. 2005  {The Burst Alert Telescope (BAT) on the SWIFT Midex Mission}. {\em \ssr} \textbf{120}, 143--164.
(\href{http://dx.doi.org/10.1007/s11214-005-5096-3}{10.1007/s11214-005-5096-3})

\bibitem{Fong2015}
{Fong} W, {Berger} E, {Margutti} R, {Zauderer} BA. 2015  {A Decade of Short-duration Gamma-Ray Burst Broadband Afterglows: Energetics, Circumburst Densities, and Jet Opening Angles}. {\em \apj} \textbf{815}, 102.
(\href{http://dx.doi.org/10.1088/0004-637X/815/2/102}{10.1088/0004-637X/815/2/102})

\bibitem{Metzger2012}
{Metzger} BD, {Berger} E. 2012  {What is the Most Promising Electromagnetic Counterpart of a Neutron Star Binary Merger?}. {\em \apj} \textbf{746}, 48.
(\href{http://dx.doi.org/10.1088/0004-637X/746/1/48}{10.1088/0004-637X/746/1/48})

\bibitem{Lattimer1974}
{Lattimer} JM, {Schramm} DN. 1974  {Black-Hole-Neutron-Star Collisions}. {\em \apjl} \textbf{192}, L145.
(\href{http://dx.doi.org/10.1086/181612}{10.1086/181612})

\bibitem{Freiburghaus1999}
{Freiburghaus} C, {Rosswog} S, {Thielemann} FK. 1999  {R-Process in Neutron Star Mergers}. {\em \apjl} \textbf{525}, L121--L124.
(\href{http://dx.doi.org/10.1086/312343}{10.1086/312343})

\bibitem{Burbidge1957}
{Burbidge} EM, {Burbidge} GR, {Fowler} WA, {Hoyle} F. 1957  {Synthesis of the Elements in Stars}. {\em Reviews of Modern Physics} \textbf{29}, 547--650.
(\href{http://dx.doi.org/10.1103/RevModPhys.29.547}{10.1103/RevModPhys.29.547})

\bibitem{Li1998}
{Li} LX, {Paczy{\'n}ski} B. 1998  {Transient Events from Neutron Star Mergers}. {\em \apjl} \textbf{507}, L59--L62.
(\href{http://dx.doi.org/10.1086/311680}{10.1086/311680})

\bibitem{Rosswog1999}
{Rosswog} S, {Liebend{\"o}rfer} M, {Thielemann} FK, {Davies} MB, {Benz} W, {Piran} T. 1999  {Mass ejection in neutron star mergers}. {\em \aap} \textbf{341}, 499--526.
(\href{http://dx.doi.org/10.48550/arXiv.astro-ph/9811367}{10.48550/arXiv.astro-ph/9811367})

\bibitem{Metzger2010}
{Metzger} BD, {Mart{\'\i}nez-Pinedo} G, {Darbha} S, {Quataert} E, {Arcones} A, {Kasen} D, {Thomas} R, {Nugent} P, {Panov} IV, {Zinner} NT. 2010  {Electromagnetic counterparts of compact object mergers powered by the radioactive decay of r-process nuclei}. {\em \mnras} \textbf{406}, 2650--2662.
(\href{http://dx.doi.org/10.1111/j.1365-2966.2010.16864.x}{10.1111/j.1365-2966.2010.16864.x})

\bibitem{Shappee2014}
{Shappee} BJ, {Prieto} JL, {Grupe} D, {Kochanek} CS, {Stanek} KZ, {De Rosa} G, {Mathur} S, {Zu} Y, {Peterson} BM, {Pogge} RW, {Komossa} S, {Im} M, {Jencson} J, {Holoien} TWS, {Basu} U, {Beacom} JF, {Szczygie{\l}} DM, {Brimacombe} J, {Adams} S, {Campillay} A, {Choi} C, {Contreras} C, {Dietrich} M, {Dubberley} M, {Elphick} M, {Foale} S, {Giustini} M, {Gonzalez} C, {Hawkins} E, {Howell} DA, {Hsiao} EY, {Koss} M, {Leighly} KM, {Morrell} N, {Mudd} D, {Mullins} D, {Nugent} JM, {Parrent} J, {Phillips} MM, {Pojmanski} G, {Rosing} W, {Ross} R, {Sand} D, {Terndrup} DM, {Valenti} S, {Walker} Z, {Yoon} Y. 2014  {The Man behind the Curtain: X-Rays Drive the UV through NIR Variability in the 2013 Active Galactic Nucleus Outburst in NGC 2617}. {\em \apj} \textbf{788}, 48.
(\href{http://dx.doi.org/10.1088/0004-637X/788/1/48}{10.1088/0004-637X/788/1/48})

\bibitem{Tonry2018}
{Tonry} JL, {Denneau} L, {Heinze} AN, {Stalder} B, {Smith} KW, {Smartt} SJ, {Stubbs} CW, {Weiland} HJ, {Rest} A. 2018  {ATLAS: A High-cadence All-sky Survey System}. {\em \pasp} \textbf{130}, 064505.
(\href{http://dx.doi.org/10.1088/1538-3873/aabadf}{10.1088/1538-3873/aabadf})

\bibitem{Steeghs2022}
{Steeghs} D, {Galloway} DK, {Ackley} K, {Dyer} MJ, {Lyman} J, {Ulaczyk} K, {Cutter} R, {Mong} YL, {Dhillon} V, {O'Brien} P, {Ramsay} G, {Poshyachinda} S, {Kotak} R, {Nuttall} LK, {Pall{\'e}} E, {Breton} RP, {Pollacco} D, {Thrane} E, {Aukkaravittayapun} S, {Awiphan} S, {Burhanudin} U, {Chote} P, {Chrimes} A, {Daw} E, {Duffy} C, {Eyles-Ferris} R, {Gompertz} B, {Heikkil{\"a}} T, {Irawati} P, {Kennedy} MR, {Killestein} T, {Kuncarayakti} H, {Levan} AJ, {Littlefair} S, {Makrygianni} L, {Marsh} T, {Mata-Sanchez} D, {Mattila} S, {Maund} J, {McCormac} J, {Mkrtichian} D, {Mullaney} J, {Noysena} K, {Patel} M, {Rol} E, {Sawangwit} U, {Stanway} ER, {Starling} R, {Str{\o}m} P, {Tooke} S, {West} R, {White} DJ, {Wiersema} K. 2022  {The Gravitational-wave Optical Transient Observer (GOTO): prototype performance and prospects for transient science}. {\em \mnras} \textbf{511}, 2405--2422.
(\href{http://dx.doi.org/10.1093/mnras/stac013}{10.1093/mnras/stac013})

\bibitem{Chambers2016}
{Chambers} KC, {Magnier} EA, {Metcalfe} N, {Flewelling} HA, {Huber} ME, {Waters} CZ, {Denneau} L, {Draper} PW, {Farrow} D, {Finkbeiner} DP, {Holmberg} C, {Koppenhoefer} J, {Price} PA, {Rest} A, {Saglia} RP, {Schlafly} EF, {Smartt} SJ, {Sweeney} W, {Wainscoat} RJ, {Burgett} WS, {Chastel} S, {Grav} T, {Heasley} JN, {Hodapp} KW, {Jedicke} R, {Kaiser} N, {Kudritzki} RP, {Luppino} GA, {Lupton} RH, {Monet} DG, {Morgan} JS, {Onaka} PM, {Shiao} B, {Stubbs} CW, {Tonry} JL, {White} R, {Ba{\~n}ados} E, {Bell} EF, {Bender} R, {Bernard} EJ, {Boegner} M, {Boffi} F, {Botticella} MT, {Calamida} A, {Casertano} S, {Chen} WP, {Chen} X, {Cole} S, {Deacon} N, {Frenk} C, {Fitzsimmons} A, {Gezari} S, {Gibbs} V, {Goessl} C, {Goggia} T, {Gourgue} R, {Goldman} B, {Grant} P, {Grebel} EK, {Hambly} NC, {Hasinger} G, {Heavens} AF, {Heckman} TM, {Henderson} R, {Henning} T, {Holman} M, {Hopp} U, {Ip} WH, {Isani} S, {Jackson} M, {Keyes} CD, {Koekemoer} AM, {Kotak} R, {Le} D, {Liska} D, {Long} KS, {Lucey} JR, {Liu} M, {Martin} NF, {Masci} G,
  {McLean} B, {Mindel} E, {Misra} P, {Morganson} E, {Murphy} DNA, {Obaika} A, {Narayan} G, {Nieto-Santisteban} MA, {Norberg} P, {Peacock} JA, {Pier} EA, {Postman} M, {Primak} N, {Rae} C, {Rai} A, {Riess} A, {Riffeser} A, {Rix} HW, {R{\"o}ser} S, {Russel} R, {Rutz} L, {Schilbach} E, {Schultz} ASB, {Scolnic} D, {Strolger} L, {Szalay} A, {Seitz} S, {Small} E, {Smith} KW, {Soderblom} DR, {Taylor} P, {Thomson} R, {Taylor} AN, {Thakar} AR, {Thiel} J, {Thilker} D, {Unger} D, {Urata} Y, {Valenti} J, {Wagner} J, {Walder} T, {Walter} F, {Watters} SP, {Werner} S, {Wood-Vasey} WM, {Wyse} R. 2016  {The Pan-STARRS1 Surveys}. {\em arXiv e-prints} p. arXiv:1612.05560.
(\href{http://dx.doi.org/10.48550/arXiv.1612.05560}{10.48550/arXiv.1612.05560})

\bibitem{Bellm2019}
{Bellm} EC, {Kulkarni} SR, {Graham} MJ, {Dekany} R, {Smith} RM, {Riddle} R, {Masci} FJ, {Helou} G, {Prince} TA, {Adams} SM, {Barbarino} C, {Barlow} T, {Bauer} J, {Beck} R, {Belicki} J, {Biswas} R, {Blagorodnova} N, {Bodewits} D, {Bolin} B, {Brinnel} V, {Brooke} T, {Bue} B, {Bulla} M, {Burruss} R, {Cenko} SB, {Chang} CK, {Connolly} A, {Coughlin} M, {Cromer} J, {Cunningham} V, {De} K, {Delacroix} A, {Desai} V, {Duev} DA, {Eadie} G, {Farnham} TL, {Feeney} M, {Feindt} U, {Flynn} D, {Franckowiak} A, {Frederick} S, {Fremling} C, {Gal-Yam} A, {Gezari} S, {Giomi} M, {Goldstein} DA, {Golkhou} VZ, {Goobar} A, {Groom} S, {Hacopians} E, {Hale} D, {Henning} J, {Ho} AYQ, {Hover} D, {Howell} J, {Hung} T, {Huppenkothen} D, {Imel} D, {Ip} WH, {Ivezi{\'c}} {\v{Z}}, {Jackson} E, {Jones} L, {Juric} M, {Kasliwal} MM, {Kaspi} S, {Kaye} S, {Kelley} MSP, {Kowalski} M, {Kramer} E, {Kupfer} T, {Landry} W, {Laher} RR, {Lee} CD, {Lin} HW, {Lin} ZY, {Lunnan} R, {Giomi} M, {Mahabal} A, {Mao} P, {Miller} AA, {Monkewitz} S, {Murphy} P,
  {Ngeow} CC, {Nordin} J, {Nugent} P, {Ofek} E, {Patterson} MT, {Penprase} B, {Porter} M, {Rauch} L, {Rebbapragada} U, {Reiley} D, {Rigault} M, {Rodriguez} H, {van Roestel} J, {Rusholme} B, {van Santen} J, {Schulze} S, {Shupe} DL, {Singer} LP, {Soumagnac} MT, {Stein} R, {Surace} J, {Sollerman} J, {Szkody} P, {Taddia} F, {Terek} S, {Van Sistine} A, {van Velzen} S, {Vestrand} WT, {Walters} R, {Ward} C, {Ye} QZ, {Yu} PC, {Yan} L, {Zolkower} J. 2019  {The Zwicky Transient Facility: System Overview, Performance, and First Results}. {\em \pasp} \textbf{131}, 018002.
(\href{http://dx.doi.org/10.1088/1538-3873/aaecbe}{10.1088/1538-3873/aaecbe})

\bibitem{Graham2019}
{Graham} MJ, {Kulkarni} SR, {Bellm} EC, {Adams} SM, {Barbarino} C, {Blagorodnova} N, {Bodewits} D, {Bolin} B, {Brady} PR, {Cenko} SB, {Chang} CK, {Coughlin} MW, {De} K, {Eadie} G, {Farnham} TL, {Feindt} U, {Franckowiak} A, {Fremling} C, {Gezari} S, {Ghosh} S, {Goldstein} DA, {Golkhou} VZ, {Goobar} A, {Ho} AYQ, {Huppenkothen} D, {Ivezi{\'c}} {\v{Z}}, {Jones} RL, {Juric} M, {Kaplan} DL, {Kasliwal} MM, {Kelley} MSP, {Kupfer} T, {Lee} CD, {Lin} HW, {Lunnan} R, {Mahabal} AA, {Miller} AA, {Ngeow} CC, {Nugent} P, {Ofek} EO, {Prince} TA, {Rauch} L, {van Roestel} J, {Schulze} S, {Singer} LP, {Sollerman} J, {Taddia} F, {Yan} L, {Ye} QZ, {Yu} PC, {Barlow} T, {Bauer} J, {Beck} R, {Belicki} J, {Biswas} R, {Brinnel} V, {Brooke} T, {Bue} B, {Bulla} M, {Burruss} R, {Connolly} A, {Cromer} J, {Cunningham} V, {Dekany} R, {Delacroix} A, {Desai} V, {Duev} DA, {Feeney} M, {Flynn} D, {Frederick} S, {Gal-Yam} A, {Giomi} M, {Groom} S, {Hacopians} E, {Hale} D, {Helou} G, {Henning} J, {Hover} D, {Hillenbrand} LA, {Howell} J, {Hung} T,
  {Imel} D, {Ip} WH, {Jackson} E, {Kaspi} S, {Kaye} S, {Kowalski} M, {Kramer} E, {Kuhn} M, {Landry} W, {Laher} RR, {Mao} P, {Masci} FJ, {Monkewitz} S, {Murphy} P, {Nordin} J, {Patterson} MT, {Penprase} B, {Porter} M, {Rebbapragada} U, {Reiley} D, {Riddle} R, {Rigault} M, {Rodriguez} H, {Rusholme} B, {van Santen} J, {Shupe} DL, {Smith} RM, {Soumagnac} MT, {Stein} R, {Surace} J, {Szkody} P, {Terek} S, {Van Sistine} A, {van Velzen} S, {Vestrand} WT, {Walters} R, {Ward} C, {Zhang} C, {Zolkower} J. 2019  {The Zwicky Transient Facility: Science Objectives}. {\em \pasp} \textbf{131}, 078001.
(\href{http://dx.doi.org/10.1088/1538-3873/ab006c}{10.1088/1538-3873/ab006c})

\bibitem{Flaugher2015}
{Flaugher} B, {Diehl} HT, {Honscheid} K, {Abbott} TMC, {Alvarez} O, {Angstadt} R, {Annis} JT, {Antonik} M, {Ballester} O, {Beaufore} L, {Bernstein} GM, {Bernstein} RA, {Bigelow} B, {Bonati} M, {Boprie} D, {Brooks} D, {Buckley-Geer} EJ, {Campa} J, {Cardiel-Sas} L, {Castander} FJ, {Castilla} J, {Cease} H, {Cela-Ruiz} JM, {Chappa} S, {Chi} E, {Cooper} C, {da Costa} LN, {Dede} E, {Derylo} G, {DePoy} DL, {de Vicente} J, {Doel} P, {Drlica-Wagner} A, {Eiting} J, {Elliott} AE, {Emes} J, {Estrada} J, {Fausti Neto} A, {Finley} DA, {Flores} R, {Frieman} J, {Gerdes} D, {Gladders} MD, {Gregory} B, {Gutierrez} GR, {Hao} J, {Holland} SE, {Holm} S, {Huffman} D, {Jackson} C, {James} DJ, {Jonas} M, {Karcher} A, {Karliner} I, {Kent} S, {Kessler} R, {Kozlovsky} M, {Kron} RG, {Kubik} D, {Kuehn} K, {Kuhlmann} S, {Kuk} K, {Lahav} O, {Lathrop} A, {Lee} J, {Levi} ME, {Lewis} P, {Li} TS, {Mandrichenko} I, {Marshall} JL, {Martinez} G, {Merritt} KW, {Miquel} R, {Mu{\~n}oz} F, {Neilsen} EH, {Nichol} RC, {Nord} B, {Ogando} R, {Olsen} J,
  {Palaio} N, {Patton} K, {Peoples} J, {Plazas} AA, {Rauch} J, {Reil} K, {Rheault} JP, {Roe} NA, {Rogers} H, {Roodman} A, {Sanchez} E, {Scarpine} V, {Schindler} RH, {Schmidt} R, {Schmitt} R, {Schubnell} M, {Schultz} K, {Schurter} P, {Scott} L, {Serrano} S, {Shaw} TM, {Smith} RC, {Soares-Santos} M, {Stefanik} A, {Stuermer} W, {Suchyta} E, {Sypniewski} A, {Tarle} G, {Thaler} J, {Tighe} R, {Tran} C, {Tucker} D, {Walker} AR, {Wang} G, {Watson} M, {Weaverdyck} C, {Wester} W, {Woods} R, {Yanny} B, {DES Collaboration}. 2015  {The Dark Energy Camera}. {\em \aj} \textbf{150}, 150.
(\href{http://dx.doi.org/10.1088/0004-6256/150/5/150}{10.1088/0004-6256/150/5/150})

\bibitem{Kessler2015}
{Kessler} R, {Marriner} J, {Childress} M, {Covarrubias} R, {D'Andrea} CB, {Finley} DA, {Fischer} J, {Foley} RJ, {Goldstein} D, {Gupta} RR, {Kuehn} K, {Marcha} M, {Nichol} RC, {Papadopoulos} A, {Sako} M, {Scolnic} D, {Smith} M, {Sullivan} M, {Wester} W, {Yuan} F, {Abbott} T, {Abdalla} FB, {Allam} S, {Benoit-L{\'e}vy} A, {Bernstein} GM, {Bertin} E, {Brooks} D, {Carnero Rosell} A, {Carrasco Kind} M, {Castander} FJ, {Crocce} M, {da Costa} LN, {Desai} S, {Diehl} HT, {Eifler} TF, {Fausti Neto} A, {Flaugher} B, {Frieman} J, {Gerdes} DW, {Gruen} D, {Gruendl} RA, {Honscheid} K, {James} DJ, {Kuropatkin} N, {Li} TS, {Maia} MAG, {Marshall} JL, {Martini} P, {Miller} CJ, {Miquel} R, {Nord} B, {Ogando} R, {Plazas} AA, {Reil} K, {Romer} AK, {Roodman} A, {Sanchez} E, {Sevilla-Noarbe} I, {Smith} RC, {Soares-Santos} M, {Sobreira} F, {Tarle} G, {Thaler} J, {Thomas} RC, {Tucker} D, {Walker} AR, {DES Collaboration}. 2015  {The Difference Imaging Pipeline for the Transient Search in the Dark Energy Survey}. {\em \aj} \textbf{150},
  172.
(\href{http://dx.doi.org/10.1088/0004-6256/150/6/172}{10.1088/0004-6256/150/6/172})

\bibitem{Ivezic2019}
{Ivezi{\'c}} {\v{Z}}, {Kahn} SM, {Tyson} JA, {LSST Science Collaboration}. 2019  {LSST: From Science Drivers to Reference Design and Anticipated Data Products}. {\em \apj} \textbf{873}, 111.
(\href{http://dx.doi.org/10.3847/1538-4357/ab042c}{10.3847/1538-4357/ab042c})

\bibitem{Ghosh2016}
{Ghosh} S, {Bloemen} S, {Nelemans} G, {Groot} PJ, {Price} LR. 2016  {Tiling strategies for optical follow-up of gravitational-wave triggers by telescopes with a wide field of view}. {\em \aap} \textbf{592}, A82.
(\href{http://dx.doi.org/10.1051/0004-6361/201527712}{10.1051/0004-6361/201527712})

\bibitem{Rana2017}
{Rana} J, {Singhal} A, {Gadre} B, {Bhalerao} V, {Bose} S. 2017  {An Enhanced Method for Scheduling Observations of Large Sky Error Regions for Finding Optical Counterparts to Transients}. {\em \apj} \textbf{838}, 108.
(\href{http://dx.doi.org/10.3847/1538-4357/838/2/108}{10.3847/1538-4357/838/2/108})

\bibitem{Coughlin2016}
{Coughlin} M, {Stubbs} C. 2016  {Maximizing the probability of detecting an electromagnetic counterpart of gravitational-wave events}. {\em Experimental Astronomy} \textbf{42}, 165--178.
(\href{http://dx.doi.org/10.1007/s10686-016-9503-4}{10.1007/s10686-016-9503-4})

\bibitem{Coughlin2018}
{Coughlin} MW, {Tao} D, {Chan} ML, {Chatterjee} D, {Christensen} N, {Ghosh} S, {Greco} G, {Hu} Y, {Kapadia} S, {Rana} J, {Salafia} OS, {Stubbs} CW. 2018  {Optimizing searches for electromagnetic counterparts of gravitational wave triggers}. {\em \mnras} \textbf{478}, 692--702.
(\href{http://dx.doi.org/10.1093/mnras/sty1066}{10.1093/mnras/sty1066})

\bibitem{Salafia2017}
{Salafia} OS, {Colpi} M, {Branchesi} M, {Chassande-Mottin} E, {Ghirlanda} G, {Ghisellini} G, {Vergani} SD. 2017  {Where and When: Optimal Scheduling of the Electromagnetic Follow-up of Gravitational-wave Events Based on Counterpart Light-curve Models}. {\em \apj} \textbf{846}, 62.
(\href{http://dx.doi.org/10.3847/1538-4357/aa850e}{10.3847/1538-4357/aa850e})

\bibitem{Coulter2017}
{Coulter} DA, {Foley} RJ, {Kilpatrick} CD, {Drout} MR, {Piro} AL, {Shappee} BJ, {Siebert} MR, {Simon} JD, {Ulloa} N, {Kasen} D, {Madore} BF, {Murguia-Berthier} A, {Pan} YC, {Prochaska} JX, {Ramirez-Ruiz} E, {Rest} A, {Rojas-Bravo} C. 2017  {Swope Supernova Survey 2017a (SSS17a), the optical counterpart to a gravitational wave source}. {\em Science} \textbf{358}, 1556--1558.
(\href{http://dx.doi.org/10.1126/science.aap9811}{10.1126/science.aap9811})

\bibitem{Soares-Santos2017}
{Soares-Santos} M, {Holz} DE, {Annis} J, {Chornock} R, {Herner} K, {Berger} E, {Brout} D, {Chen} HY, {Kessler} R, {Sako} M, {Allam} S, {Tucker} DL, {Butler} RE, {Palmese} A, {Doctor} Z, {Diehl} HT, {Frieman} J, {Yanny} B, {Lin} H, {Scolnic} D, {Cowperthwaite} P, {Neilsen} E, {Marriner} J, {Kuropatkin} N, {Hartley} WG, {Paz-Chinch{\'o}n} F, {Alexander} KD, {Balbinot} E, {Blanchard} P, {Brown} DA, {Carlin} JL, {Conselice} C, {Cook} ER, {Drlica-Wagner} A, {Drout} MR, {Durret} F, {Eftekhari} T, {Farr} B, {Finley} DA, {Foley} RJ, {Fong} W, {Fryer} CL, {Garc{\'{\i}}a-Bellido} J, {Gill} MSS, {Gruendl} RA, {Hanna} C, {Kasen} D, {Li} TS, {Lopes} PAA, {Louren{\c c}o} ACC, {Margutti} R, {Marshall} JL, {Matheson} T, {Medina} GE, {Metzger} BD, {Mu{\~n}oz} RR, {Muir} J, {Nicholl} M, {Quataert} E, {Rest} A, {Sauseda} M, {Schlegel} DJ, {Secco} LF, {Sobreira} F, {Stebbins} A, {Villar} VA, {Vivas} K, {Walker} AR, {Wester} W, {Williams} PKG, {Zenteno} A, {Zhang} Y, {Abbott} TMC, {Abdalla} FB, {Banerji} M, {Bechtol} K,
  {Benoit-L{\'e}vy} A, {Bertin} E, {Brooks} D, {Buckley-Geer} E, {Burke} DL, {Carnero Rosell} A, {Carrasco Kind} M, {Carretero} J, {Castander} FJ, {Crocce} M, {Cunha} CE, {D'Andrea} CB, {da Costa} LN, {Davis} C, {Desai} S, {Dietrich} JP, {Doel} P, {Eifler} TF, {Fernandez} E, {Flaugher} B, {Fosalba} P, {Gaztanaga} E, {Gerdes} DW, {Giannantonio} T, {Goldstein} DA, {Gruen} D, {Gschwend} J, {Gutierrez} G, {Honscheid} K, {Jain} B, {James} DJ, {Jeltema} T, {Johnson} MWG, {Johnson} MD, {Kent} S, {Krause} E, {Kron} R, {Kuehn} K, {Kuhlmann} S, {Lahav} O, {Lima} M, {Maia} MAG, {March} M, {McMahon} RG, {Menanteau} F, {Miquel} R, {Mohr} JJ, {Nichol} RC, {Nord} B, {Ogando} RLC, {Petravick} D, {Plazas} AA, {Romer} AK, {Roodman} A, {Rykoff} ES, {Sanchez} E, {Scarpine} V, {Schubnell} M, {Sevilla-Noarbe} I, {Smith} M, {Smith} RC, {Suchyta} E, {Swanson} MEC, {Tarle} G, {Thomas} D, {Thomas} RC, {Troxel} MA, {Vikram} V, {Wechsler} RH, {Weller} J, {Dark Energy Survey}, {Dark Energy Camera GW-EM Collaboration}. 2017  {The
  Electromagnetic Counterpart of the Binary Neutron Star Merger LIGO/Virgo GW170817. I. Discovery of the Optical Counterpart Using the Dark Energy Camera}. {\em \apjl} \textbf{848}, L16.
(\href{http://dx.doi.org/10.3847/2041-8213/aa9059}{10.3847/2041-8213/aa9059})

\bibitem{Li2000}
{Li} WD, {Filippenko} AV, {Treffers} RR, {Friedman} A, {Halderson} E, {Johnson} RA, {King} JY, {Modjaz} M, {Papenkova} M, {Sato} Y, {Shefler} T. 2000  {The Lick Observatory Supernova Search}. In {Holt} SS, {Zhang} WW, editors, {\em Cosmic Explosions: Tenth AstroPhysics Conference} vol. 522{\em American Institute of Physics Conference Series} pp. 103--106. AIP.
(\href{http://dx.doi.org/10.1063/1.1291702}{10.1063/1.1291702})

\bibitem{Gehrels2016}
{Gehrels} N, {Cannizzo} JK, {Kanner} J, {Kasliwal} MM, {Nissanke} S, {Singer} LP. 2016  {Galaxy Strategy for LIGO-Virgo Gravitational Wave Counterpart Searches}. {\em \apj} \textbf{820}, 136.
(\href{http://dx.doi.org/10.3847/0004-637X/820/2/136}{10.3847/0004-637X/820/2/136})

\bibitem{Arcavi2017a}
{Arcavi} I, {McCully} C, {Hosseinzadeh} G, {Howell} DA, {Vasylyev} S, {Poznanski} D, {Zaltzman} M, {Maoz} D, {Singer} L, {Valenti} S, {Kasen} D, {Barnes} J, {Piran} T, {Fong} Wf. 2017  {Optical Follow-up of Gravitational-wave Events with Las Cumbres Observatory}. {\em \apjl} \textbf{848}, L33.
(\href{http://dx.doi.org/10.3847/2041-8213/aa910f}{10.3847/2041-8213/aa910f})

\bibitem{Salmon2020}
{Salmon} L, {Hanlon} L, {Jeffrey} RM, {Martin-Carrillo} A. 2020  {Web application for galaxy-targeted follow-up of electromagnetic counterparts to gravitational wave sources}. {\em \aap} \textbf{634}, A32.
(\href{http://dx.doi.org/10.1051/0004-6361/201936573}{10.1051/0004-6361/201936573})

\bibitem{Ducoin2020}
{Ducoin} JG, {Corre} D, {Leroy} N, {Le Floch} E. 2020  {Optimizing gravitational waves follow-up using galaxies stellar mass}. {\em \mnras} \textbf{492}, 4768--4779.
(\href{http://dx.doi.org/10.1093/mnras/staa114}{10.1093/mnras/staa114})

\bibitem{Coulter2024arXiv}
{Coulter} DA, {Kilpatrick} CD, {Jones} DO, {Foley} RJ, {Filippenko} AV, {Zheng} W, {Swift} JJ, {Rahman} GS, {Stacey} HE, {Piro} AL, {Rojas-Bravo} C, {Anais Vilchez} J, {Mu{\~n}oz-Elgueta} N, {Arcavi} I, {Dimitriadis} G, {Siebert} MR, {Bloom} JS, {Bustamante-Rosell} MJ, {Clever} KE, {Davis} KW, {Kutcka} J, {Macias} P, {McGill} P, {Qui{\~n}onez} PJ, {Ramirez-Ruiz} E, {Siellez} K, {Tinyanont} S, {Cenko} SB, {Drout} MR, {Hausen} R, {Jacobson-Gal{\'a}n} WV, {Howell} DA, {Kasen} D, {McCully} C, {Rest} A, {Taggart} K, {Valenti} S. 2024  {The Gravity Collective: A Comprehensive Analysis of the Electromagnetic Search for the Binary Neutron Star Merger GW190425}. {\em arXiv e-prints} p. arXiv:2404.15441.
(\href{http://dx.doi.org/10.48550/arXiv.2404.15441}{10.48550/arXiv.2404.15441})

\bibitem{Abbott2017mma}
{LIGO Scientific Collaboration}, {Virgo Collaboration} et~al.. 2017  {Multi-messenger Observations of a Binary Neutron Star Merger}. {\em \apjl} \textbf{848}, L12.
(\href{http://dx.doi.org/10.3847/2041-8213/aa91c9}{10.3847/2041-8213/aa91c9})

\bibitem{Blanchard2017}
{Blanchard} PK, {Berger} E, {Fong} W, {Nicholl} M, {Leja} J, {Conroy} C, {Alexander} KD, {Margutti} R, {Williams} PKG, {Doctor} Z, {Chornock} R, {Villar} VA, {Cowperthwaite} PS, {Annis} J, {Brout} D, {Brown} DA, {Chen} HY, {Eftekhari} T, {Frieman} JA, {Holz} DE, {Metzger} BD, {Rest} A, {Sako} M, {Soares-Santos} M. 2017  {The Electromagnetic Counterpart of the Binary Neutron Star Merger LIGO/Virgo GW170817. VII. Properties of the Host Galaxy and Constraints on the Merger Timescale}. {\em \apjl} \textbf{848}, L22.
(\href{http://dx.doi.org/10.3847/2041-8213/aa9055}{10.3847/2041-8213/aa9055})

\bibitem{Levan2017}
{Levan} AJ, {Lyman} JD, {Tanvir} NR, {Hjorth} J, {Mandel} I, {Stanway} ER, {Steeghs} D, {Fruchter} AS, {Troja} E, {Schr{\o}der} SL, {Wiersema} K, {Bruun} SH, {Cano} Z, {Cenko} SB, {de Ugarte Postigo} A, {Evans} P, {Fairhurst} S, {Fox} OD, {Fynbo} JPU, {Gompertz} B, {Greiner} J, {Im} M, {Izzo} L, {Jakobsson} P, {Kangas} T, {Khandrika} HG, {Lien} AY, {Malesani} D, {O'Brien} P, {Osborne} JP, {Palazzi} E, {Pian} E, {Perley} DA, {Rosswog} S, {Ryan} RE, {Schulze} S, {Sutton} P, {Th{\"o}ne} CC, {Watson} DJ, {Wijers} RAMJ. 2017  {The Environment of the Binary Neutron Star Merger GW170817}. {\em \apjl} \textbf{848}, L28.
(\href{http://dx.doi.org/10.3847/2041-8213/aa905f}{10.3847/2041-8213/aa905f})

\bibitem{Pan2017}
{Pan} YC, {Kilpatrick} CD, {Simon} JD, {Xhakaj} E, {Boutsia} K, {Coulter} DA, {Drout} MR, {Foley} RJ, {Kasen} D, {Morrell} N, {Murguia-Berthier} A, {Osip} D, {Piro} AL, {Prochaska} JX, {Ramirez-Ruiz} E, {Rest} A, {Rojas-Bravo} C, {Shappee} BJ, {Siebert} MR. 2017  {The Old Host-galaxy Environment of SSS17a, the First Electromagnetic Counterpart to a Gravitational-wave Source}. {\em \apjl} \textbf{848}, L30.
(\href{http://dx.doi.org/10.3847/2041-8213/aa9116}{10.3847/2041-8213/aa9116})

\bibitem{Goldstein2017}
{Goldstein} A, {Veres} P, {Burns} E, {Briggs} MS, {Hamburg} R, {Kocevski} D, {Wilson-Hodge} CA, {Preece} RD, {Poolakkil} S, {Roberts} OJ, {Hui} CM, {Connaughton} V, {Racusin} J, {von Kienlin} A, {Dal Canton} T, {Christensen} N, {Littenberg} T, {Siellez} K, {Blackburn} L, {Broida} J, {Bissaldi} E, {Cleveland} WH, {Gibby} MH, {Giles} MM, {Kippen} RM, {McBreen} S, {McEnery} J, {Meegan} CA, {Paciesas} WS, {Stanbro} M. 2017  {An Ordinary Short Gamma-Ray Burst with Extraordinary Implications: Fermi-GBM Detection of GRB 170817A}. {\em \apjl} \textbf{848}, L14.
(\href{http://dx.doi.org/10.3847/2041-8213/aa8f41}{10.3847/2041-8213/aa8f41})

\bibitem{Savchenko2017}
{Savchenko} V, {Ferrigno} C, {Kuulkers} E, {Bazzano} A, {Bozzo} E, {Brandt} S, {Chenevez} J, {Courvoisier} TJL, {Diehl} R, {Domingo} A, {Hanlon} L, {Jourdain} E, {von Kienlin} A, {Laurent} P, {Lebrun} F, {Lutovinov} A, {Martin-Carrillo} A, {Mereghetti} S, {Natalucci} L, {Rodi} J, {Roques} JP, {Sunyaev} R, {Ubertini} P. 2017  {INTEGRAL Detection of the First Prompt Gamma-Ray Signal Coincident with the Gravitational-wave Event GW170817}. {\em \apjl} \textbf{848}, L15.
(\href{http://dx.doi.org/10.3847/2041-8213/aa8f94}{10.3847/2041-8213/aa8f94})

\bibitem{Abbott2017grb}
{LIGO Scientific Collaboration}, {Virgo Collaboration}, {Fermi Gamma-ray Burst Monitor}, {(INTEGRAL}. 2017  {Gravitational Waves and Gamma-Rays from a Binary Neutron Star Merger: GW170817 and GRB 170817A}. {\em \apjl} \textbf{848}, L13.
(\href{http://dx.doi.org/10.3847/2041-8213/aa920c}{10.3847/2041-8213/aa920c})

\bibitem{Alexander2017}
{Alexander} KD, {Berger} E, {Fong} W, {Williams} PKG, {Guidorzi} C, {Margutti} R, {Metzger} BD, {Annis} J, {Blanchard} PK, {Brout} D, {Brown} DA, {Chen} HY, {Chornock} R, {Cowperthwaite} PS, {Drout} M, {Eftekhari} T, {Frieman} J, {Holz} DE, {Nicholl} M, {Rest} A, {Sako} M, {Soares-Santos} M, {Villar} VA. 2017  {The Electromagnetic Counterpart of the Binary Neutron Star Merger LIGO/Virgo GW170817. VI. Radio Constraints on a Relativistic Jet and Predictions for Late-time Emission from the Kilonova Ejecta}. {\em \apjl} \textbf{848}, L21.
(\href{http://dx.doi.org/10.3847/2041-8213/aa905d}{10.3847/2041-8213/aa905d})

\bibitem{Alexander2018}
{Alexander} KD, {Margutti} R, {Blanchard} PK, {Fong} W, {Berger} E, {Hajela} A, {Eftekhari} T, {Chornock} R, {Cowperthwaite} PS, {Giannios} D, {Guidorzi} C, {Kathirgamaraju} A, {MacFadyen} A, {Metzger} BD, {Nicholl} M, {Sironi} L, {Villar} VA, {Williams} PKG, {Xie} X, {Zrake} J. 2018  {A Decline in the X-Ray through Radio Emission from GW170817 Continues to Support an Off-axis Structured Jet}. {\em \apjl} \textbf{863}, L18.
(\href{http://dx.doi.org/10.3847/2041-8213/aad637}{10.3847/2041-8213/aad637})

\bibitem{Haggard2017}
{Haggard} D, {Nynka} M, {Ruan} JJ, {Kalogera} V, {Cenko} SB, {Evans} P, {Kennea} JA. 2017  {A Deep Chandra X-Ray Study of Neutron Star Coalescence GW170817}. {\em \apjl} \textbf{848}, L25.
(\href{http://dx.doi.org/10.3847/2041-8213/aa8ede}{10.3847/2041-8213/aa8ede})

\bibitem{Hallinan2017}
{Hallinan} G, {Corsi} A, {Mooley} KP, {Hotokezaka} K, {Nakar} E, {Kasliwal} MM, {Kaplan} DL, {Frail} DA, {Myers} ST, {Murphy} T, {De} K, {Dobie} D, {Allison} JR, {Bannister} KW, {Bhalerao} V, {Chandra} P, {Clarke} TE, {Giacintucci} S, {Ho} AYQ, {Horesh} A, {Kassim} NE, {Kulkarni} SR, {Lenc} E, {Lockman} FJ, {Lynch} C, {Nichols} D, {Nissanke} S, {Palliyaguru} N, {Peters} WM, {Piran} T, {Rana} J, {Sadler} EM, {Singer} LP. 2017  {A radio counterpart to a neutron star merger}. {\em Science} \textbf{358}, 1579--1583.
(\href{http://dx.doi.org/10.1126/science.aap9855}{10.1126/science.aap9855})

\bibitem{Margutti2017}
{Margutti} R, {Berger} E, {Fong} W, {Guidorzi} C, {Alexander} KD, {Metzger} BD, {Blanchard} PK, {Cowperthwaite} PS, {Chornock} R, {Eftekhari} T, {Nicholl} M, {Villar} VA, {Williams} PKG, {Annis} J, {Brown} DA, {Chen} H, {Doctor} Z, {Frieman} JA, {Holz} DE, {Sako} M, {Soares-Santos} M. 2017  {The Electromagnetic Counterpart of the Binary Neutron Star Merger LIGO/Virgo GW170817. V. Rising X-Ray Emission from an Off-axis Jet}. {\em \apjl} \textbf{848}, L20.
(\href{http://dx.doi.org/10.3847/2041-8213/aa9057}{10.3847/2041-8213/aa9057})

\bibitem{Margutti2018}
{Margutti} R, {Alexander} KD, {Xie} X, {Sironi} L, {Metzger} BD, {Kathirgamaraju} A, {Fong} W, {Blanchard} PK, {Berger} E, {MacFadyen} A, {Giannios} D, {Guidorzi} C, {Hajela} A, {Chornock} R, {Cowperthwaite} PS, {Eftekhari} T, {Nicholl} M, {Villar} VA, {Williams} PKG, {Zrake} J. 2018  {The Binary Neutron Star Event LIGO/Virgo GW170817 160 Days after Merger: Synchrotron Emission across the Electromagnetic Spectrum}. {\em \apjl} \textbf{856}, L18.
(\href{http://dx.doi.org/10.3847/2041-8213/aab2ad}{10.3847/2041-8213/aab2ad})

\bibitem{Ruan2018}
{Ruan} JJ, {Nynka} M, {Haggard} D, {Kalogera} V, {Evans} P. 2018  {Brightening X-Ray Emission from GW170817/GRB 170817A: Further Evidence for an Outflow}. {\em \apjl} \textbf{853}, L4.
(\href{http://dx.doi.org/10.3847/2041-8213/aaa4f3}{10.3847/2041-8213/aaa4f3})

\bibitem{Troja2017}
{Troja} E, {Piro} L, {van Eerten} H, {Wollaeger} RT, {Im} M, {Fox} OD, {Butler} NR, {Cenko} SB, {Sakamoto} T, {Fryer} CL, {Ricci} R, {Lien} A, {Ryan} RE, {Korobkin} O, {Lee} SK, {Burgess} JM, {Lee} WH, {Watson} AM, {Choi} C, {Covino} S, {D'Avanzo} P, {Fontes} CJ, {Gonz{\'a}lez} JB, {Khandrika} HG, {Kim} J, {Kim} SL, {Lee} CU, {Lee} HM, {Kutyrev} A, {Lim} G, {S{\'a}nchez-Ram{\'{\i}}rez} R, {Veilleux} S, {Wieringa} MH, {Yoon} Y. 2017  {The X-ray counterpart to the gravitational-wave event GW170817}. {\em \nat} \textbf{551}, 71--74.
(\href{http://dx.doi.org/10.1038/nature24290}{10.1038/nature24290})

\bibitem{Troja2018}
{Troja} E, {Piro} L, {Ryan} G, {van Eerten} H, {Ricci} R, {Wieringa} MH, {Lotti} S, {Sakamoto} T, {Cenko} SB. 2018  {The outflow structure of GW170817 from late-time broad-band observations}. {\em \mnras} \textbf{478}, L18--L23.
(\href{http://dx.doi.org/10.1093/mnrasl/sly061}{10.1093/mnrasl/sly061})

\bibitem{Troja2019}
{Troja} E, {van Eerten} H, {Ryan} G, {Ricci} R, {Burgess} JM, {Wieringa} MH, {Piro} L, {Cenko} SB, {Sakamoto} T. 2019  {A year in the life of GW 170817: the rise and fall of a structured jet from a binary neutron star merger}. {\em \mnras} \textbf{489}, 1919--1926.
(\href{http://dx.doi.org/10.1093/mnras/stz2248}{10.1093/mnras/stz2248})

\bibitem{DAvanzo2018}
{D'Avanzo} P, {Campana} S, {Salafia} OS, {Ghirlanda} G, {Ghisellini} G, {Melandri} A, {Bernardini} MG, {Branchesi} M, {Chassande-Mottin} E, {Covino} S, {D'Elia} V, {Nava} L, {Salvaterra} R, {Tagliaferri} G, {Vergani} SD. 2018  {The evolution of the X-ray afterglow emission of GW 170817/ GRB 170817A in XMM-Newton observations}. {\em \aap} \textbf{613}, L1.
(\href{http://dx.doi.org/10.1051/0004-6361/201832664}{10.1051/0004-6361/201832664})

\bibitem{Dobie2018}
{Dobie} D, {Kaplan} DL, {Murphy} T, {Lenc} E, {Mooley} KP, {Lynch} C, {Corsi} A, {Frail} D, {Kasliwal} M, {Hallinan} G. 2018  {A Turnover in the Radio Light Curve of GW170817}. {\em \apjl} \textbf{858}, L15.
(\href{http://dx.doi.org/10.3847/2041-8213/aac105}{10.3847/2041-8213/aac105})

\bibitem{Lyman2018}
{Lyman} JD, {Lamb} GP, {Levan} AJ, {Mandel} I, {Tanvir} NR, {Kobayashi} S, {Gompertz} B, {Hjorth} J, {Fruchter} AS, {Kangas} T, {Steeghs} D, {Steele} IA, {Cano} Z, {Copperwheat} C, {Evans} PA, {Fynbo} JPU, {Gall} C, {Im} M, {Izzo} L, {Jakobsson} P, {Milvang-Jensen} B, {O'Brien} P, {Osborne} JP, {Palazzi} E, {Perley} DA, {Pian} E, {Rosswog} S, {Rowlinson} A, {Schulze} S, {Stanway} ER, {Sutton} P, {Th{\"o}ne} CC, {de Ugarte Postigo} A, {Watson} DJ, {Wiersema} K, {Wijers} RAMJ. 2018  {The optical afterglow of the short gamma-ray burst associated with GW170817}. {\em Nature Astronomy} \textbf{2}, 751--754.
(\href{http://dx.doi.org/10.1038/s41550-018-0511-3}{10.1038/s41550-018-0511-3})

\bibitem{Mooley2018}
{Mooley} KP, {Nakar} E, {Hotokezaka} K, {Hallinan} G, {Corsi} A, {Frail} DA, {Horesh} A, {Murphy} T, {Lenc} E, {Kaplan} DL, {de} K, {Dobie} D, {Chandra} P, {Deller} A, {Gottlieb} O, {Kasliwal} MM, {Kulkarni} SR, {Myers} ST, {Nissanke} S, {Piran} T, {Lynch} C, {Bhalerao} V, {Bourke} S, {Bannister} KW, {Singer} LP. 2018  {A mildly relativistic wide-angle outflow in the neutron-star merger event GW170817}. {\em \nat} \textbf{554}, 207--210.
(\href{http://dx.doi.org/10.1038/nature25452}{10.1038/nature25452})

\bibitem{Fong2019}
{Fong} W, {Blanchard} PK, {Alexander} KD, {Strader} J, {Margutti} R, {Hajela} A, {Villar} VA, {Wu} Y, {Ye} CS, {Berger} E, {Chornock} R, {Coppejans} D, {Cowperthwaite} PS, {Eftekhari} T, {Giannios} D, {Guidorzi} C, {Kathirgamaraju} A, {Laskar} T, {Macfadyen} A, {Metzger} BD, {Nicholl} M, {Paterson} K, {Terreran} G, {Sand} DJ, {Sironi} L, {Williams} PKG, {Xie} X, {Zrake} J. 2019  {The Optical Afterglow of GW170817: An Off-axis Structured Jet and Deep Constraints on a Globular Cluster Origin}. {\em \apjl} \textbf{883}, L1.
(\href{http://dx.doi.org/10.3847/2041-8213/ab3d9e}{10.3847/2041-8213/ab3d9e})

\bibitem{Hajela2019}
{Hajela} A, {Margutti} R, {Alexander} KD, {Kathirgamaraju} A, {Baldeschi} A, {Guidorzi} C, {Giannios} D, {Fong} W, {Wu} Y, {MacFadyen} A, {Paggi} A, {Berger} E, {Blanchard} PK, {Chornock} R, {Coppejans} DL, {Cowperthwaite} PS, {Eftekhari} T, {Gomez} S, {Hosseinzadeh} G, {Laskar} T, {Metzger} BD, {Nicholl} M, {Paterson} K, {Radice} D, {Sironi} L, {Terreran} G, {Villar} VA, {Williams} PKG, {Xie} X, {Zrake} J. 2019  {Two Years of Nonthermal Emission from the Binary Neutron Star Merger GW170817: Rapid Fading of the Jet Afterglow and First Constraints on the Kilonova Fastest Ejecta}. {\em \apjl} \textbf{886}, L17.
(\href{http://dx.doi.org/10.3847/2041-8213/ab5226}{10.3847/2041-8213/ab5226})

\bibitem{Lamb2019}
{Lamb} GP, {Lyman} JD, {Levan} AJ, {Tanvir} NR, {Kangas} T, {Fruchter} AS, {Gompertz} B, {Hjorth} J, {Mandel} I, {Oates} SR, {Steeghs} D, {Wiersema} K. 2019  {The Optical Afterglow of GW170817 at One Year Post-merger}. {\em \apjl} \textbf{870}, L15.
(\href{http://dx.doi.org/10.3847/2041-8213/aaf96b}{10.3847/2041-8213/aaf96b})

\bibitem{Ghirlanda2019}
{Ghirlanda} G, {Salafia} OS, {Paragi} Z, {Giroletti} M, {Yang} J, {Marcote} B, {Blanchard} J, {Agudo} I, {An} T, {Bernardini} MG, {Beswick} R, {Branchesi} M, {Campana} S, {Casadio} C, {Chassande-Mottin} E, {Colpi} M, {Covino} S, {D'Avanzo} P, {D'Elia} V, {Frey} S, {Gawronski} M, {Ghisellini} G, {Gurvits} LI, {Jonker} PG, {van Langevelde} HJ, {Melandri} A, {Moldon} J, {Nava} L, {Perego} A, {Perez-Torres} MA, {Reynolds} C, {Salvaterra} R, {Tagliaferri} G, {Venturi} T, {Vergani} SD, {Zhang} M. 2019  {Compact radio emission indicates a structured jet was produced by a binary neutron star merger}. {\em Science} \textbf{363}, 968--971.
(\href{http://dx.doi.org/10.1126/science.aau8815}{10.1126/science.aau8815})

\bibitem{Coulter2017gcn}
{Coulter} DA, {Kilpatrick} CD, {Siebert} MR, {Foley} RJ, {Shappee} BJ, {Drout} MR, {Simon} JS, {Piro} AL, {Rest} A, {One-Meter Two-Hemisphere (1M2H) Collaboration}. 2017  {LIGO/Virgo G298048: Potential optical counterpart discovered by Swope telescope}. {\em GRB Coordinates Network} \textbf{21529}, 1.

\bibitem{Yang2017gcn1}
{Yang} S, {Valenti} S, {Sand} D, {Tartaglia} L, {Cappellaro} E, {Reichart} D, {Haislip} J, {Vladimir} K. 2017  {LIGO/Virgo G298048: DLT40 optical candidate}. {\em GRB Coordinates Network} \textbf{21531}, 1.

\bibitem{Valenti2017}
{Valenti} S, {David}, {Sand} J, {Yang} S, {Cappellaro} E, {Tartaglia} L, {Corsi} A, {Jha} SW, {Reichart} DE, {Haislip} J, {Kouprianov} V. 2017  {The Discovery of the Electromagnetic Counterpart of GW170817: Kilonova AT 2017gfo/DLT17ck}. {\em \apjl} \textbf{848}, L24.
(\href{http://dx.doi.org/10.3847/2041-8213/aa8edf}{10.3847/2041-8213/aa8edf})

\bibitem{Arcavi2017gcn1}
{Arcavi} I, {Howell} DA, {McCully} C, {Hosseinzadeh} G, {Vasylyev} S, {Zalzman} M, {Poznanski} D, {Singer} LP, {Valenti} S, {Piran} T, {Kasen} D, {Barnes} J, {Fong} Wf. 2017a  {LIGO/Virgo G298048: Las Cumbres Observatory Detection of The Possible Optical Counterpart in NGC 4993}. {\em GRB Coordinates Network} \textbf{21538}, 1.

\bibitem{Arcavi2017}
{Arcavi} I, {Hosseinzadeh} G, {Howell} DA, {McCully} C, {Poznanski} D, {Kasen} D, {Barnes} J, {Zaltzman} M, {Vasylyev} S, {Maoz} D, {Valenti} S. 2017b  {Optical emission from a kilonova following a gravitational-wave-detected neutron-star merger}. {\em \nat} \textbf{551}, 64--66.
(\href{http://dx.doi.org/10.1038/nature24291}{10.1038/nature24291})

\bibitem{Tanvir2017gcn}
{Tanvir} NR, {Levan} AJ, {Vinrouge Collaboration.}. 2017a  {LIGO/Virgo G298048: VISTA/VIRCAM detection of candidate counterpart}. {\em GRB Coordinates Network} \textbf{21544}, 1.

\bibitem{Tanvir2017}
{Tanvir} NR, {Levan} AJ, {Gonz{\'a}lez-Fern{\'a}ndez} C, {Korobkin} O, {Mandel} I, {Rosswog} S, {Hjorth} J, {D'Avanzo} P, {Fruchter} AS, {Fryer} CL, {Kangas} T, {Milvang-Jensen} B, {Rosetti} S, {Steeghs} D, {Wollaeger} RT, {Cano} Z, {Copperwheat} CM, {Covino} S, {D'Elia} V, {de Ugarte Postigo} A, {Evans} PA, {Even} WP, {Fairhurst} S, {Figuera Jaimes} R, {Fontes} CJ, {Fujii} YI, {Fynbo} JPU, {Gompertz} BP, {Greiner} J, {Hodosan} G, {Irwin} MJ, {Jakobsson} P, {J{\o}rgensen} UG, {Kann} DA, {Lyman} JD, {Malesani} D, {McMahon} RG, {Melandri} A, {O'Brien} PT, {Osborne} JP, {Palazzi} E, {Perley} DA, {Pian} E, {Piranomonte} S, {Rabus} M, {Rol} E, {Rowlinson} A, {Schulze} S, {Sutton} P, {Th{\"o}ne} CC, {Ulaczyk} K, {Watson} D, {Wiersema} K, {Wijers} RAMJ. 2017b  {The Emergence of a Lanthanide-rich Kilonova Following the Merger of Two Neutron Stars}. {\em \apjl} \textbf{848}, L27.
(\href{http://dx.doi.org/10.3847/2041-8213/aa90b6}{10.3847/2041-8213/aa90b6})

\bibitem{Lipunov2017gcn}
{Lipunov} VM, {Gorbovskoy} E, {Kornilov} VG, {Tyurina} N, {Shumkov} V, {Kuvshinov} D, {Balanutsa} P, {Gress} O, {Kuznetsov} A, {Panchenko} MI, {Krylov} AV, {Gorbunov} I, {Podesta} R, {Lopez} C, {Podesta} F, {Levato} H, {Saffe} C, {Budnev} NM, {Gress} O, {Ivanov} K, {Yazev} S, {Yurkov} V, {Sergienko} Y, {Gabovich} A, {Buckley} D, {Potter} S, {Kotze} M, {Rebolo} R, {Serra-Ricart} M, {Israelian} G, {Lodieu} N, {Tlatov} A, {Sennik} V. 2017a  {LIGO/Virgo G298048: MASTER observations of the NGC 4993}. {\em GRB Coordinates Network} \textbf{21546}, 1.

\bibitem{Lipunov2017}
{Lipunov} VM, {Gorbovskoy} E, {Kornilov} VG, {.~Tyurina} N, {Balanutsa} P, {Kuznetsov} A, {Vlasenko} D, {Kuvshinov} D, {Gorbunov} I, {Buckley} DAH, {Krylov} AV, {Podesta} R, {Lopez} C, {Podesta} F, {Levato} H, {Saffe} C, {Mallamachi} C, {Potter} S, {Budnev} NM, {Gress} O, {Ishmuhametova} Y, {Vladimirov} V, {Zimnukhov} D, {Yurkov} V, {Sergienko} Y, {Gabovich} A, {Rebolo} R, {Serra-Ricart} M, {Israelyan} G, {Chazov} V, {Wang} X, {Tlatov} A, {Panchenko} MI. 2017b  {MASTER Optical Detection of the First LIGO/Virgo Neutron Star Binary Merger GW170817}. {\em \apjl} \textbf{850}, L1.
(\href{http://dx.doi.org/10.3847/2041-8213/aa92c0}{10.3847/2041-8213/aa92c0})

\bibitem{Allam2017gcn}
{Allam} S, {Annis} J, {Berger} E, {Brout} DJ, {Brown} D, {Butler} RE, {Chen} HY, {Chornock} R, {Cook} E, {Cowperthwaite} P, {Diehl} HT, {Drlica-Wagner} A, {Drout} MR, {Foley} RJ, {Fong} W, {Fox} D, {Frieman} J, {Gruendl} R, {Herner} K, {Holz} D, {Kessler} R, {Margutti} R, {Marshall} J, {Neilsen} E, {Nicholl} M, {Paz-Chincon} F, {Rest} A, {Sako} M, {Scolnic} D, {Smith} N, {Soares-Santos} M, {Tucker} D, {Villar} VA, {Williams} PKG, {Yanny} B. 2017  {LIGO/Virgo G298048: DECam optical candidate}. {\em GRB Coordinates Network} \textbf{21530}, 1.

\bibitem{Andreoni2017}
{Andreoni} I, {Ackley} K, {Cooke} J, {Acharyya} A, {Allison} JR, {Anderson} GE, {Ashley} MCB, {Baade} D, {Bailes} M, {Bannister} K, {Beardsley} A, {Bessell} MS, {Bian} F, {Bland} PA, {Boer} M, {Booler} T, {Brandeker} A, {Brown} IS, {Buckley} DAH, {Chang} SW, {Coward} DM, {Crawford} S, {Crisp} H, {Crosse} B, {Cucchiara} A, {Cup{\'a}k} M, {de Gois} JS, {Deller} A, {Devillepoix} HAR, {Dobie} D, {Elmer} E, {Emrich} D, {Farah} W, {Farrell} TJ, {Franzen} T, {Gaensler} BM, {Galloway} DK, {Gendre} B, {Giblin} T, {Goobar} A, {Green} J, {Hancock} PJ, {Hartig} BAD, {Howell} EJ, {Horsley} L, {Hotan} A, {Howie} RM, {Hu} L, {Hu} Y, {James} CW, {Johnston} S, {Johnston-Hollitt} M, {Kaplan} DL, {Kasliwal} M, {Keane} EF, {Kenney} D, {Klotz} A, {Lau} R, {Laugier} R, {Lenc} E, {Li} X, {Liang} E, {Lidman} C, {Luvaul} LC, {Lynch} C, {Ma} B, {Macpherson} D, {Mao} J, {McClelland} DE, {McCully} C, {M{\"o}ller} A, {Morales} MF, {Morris} D, {Murphy} T, {Noysena} K, {Onken} CA, {Orange} NB, {Os{\l}owski} S, {Pallot} D, {Paxman} J,
  {Potter} SB, {Pritchard} T, {Raja} W, {Ridden-Harper} R, {Romero-Colmenero} E, {Sadler} EM, {Sansom} EK, {Scalzo} RA, {Schmidt} BP, {Scott} SM, {Seghouani} N, {Shang} Z, {Shannon} RM, {Shao} L, {Shara} MM, {Sharp} R, {Sokolowski} M, {Sollerman} J, {Staff} J, {Steele} K, {Sun} T, {Suntzeff} NB, {Tao} C, {Tingay} S, {Towner} MC, {Thierry} P, {Trott} C, {Tucker} BE, {V{\"a}is{\"a}nen} P, {Krishnan} VV, {Walker} M, {Wang} L, {Wang} X, {Wayth} R, {Whiting} M, {Williams} A, {Williams} T, {Wolf} C, {Wu} C, {Wu} X, {Yang} J, {Yuan} X, {Zhang} H, {Zhou} J, {Zovaro} H. 2017  {Follow Up of GW170817 and Its Electromagnetic Counterpart by Australian-Led Observing Programmes}. {\em \pasa} \textbf{34}, e069.
(\href{http://dx.doi.org/10.1017/pasa.2017.65}{10.1017/pasa.2017.65})

\bibitem{Chornock2017}
{Chornock} R, {Berger} E, {Kasen} D, {Cowperthwaite} PS, {Nicholl} M, {Villar} VA, {Alexander} KD, {Blanchard} PK, {Eftekhari} T, {Fong} W, {Margutti} R, {Williams} PKG, {Annis} J, {Brout} D, {Brown} DA, {Chen} HY, {Drout} MR, {Farr} B, {Foley} RJ, {Frieman} JA, {Fryer} CL, {Herner} K, {Holz} DE, {Kessler} R, {Matheson} T, {Metzger} BD, {Quataert} E, {Rest} A, {Sako} M, {Scolnic} DM, {Smith} N, {Soares-Santos} M. 2017  {The Electromagnetic Counterpart of the Binary Neutron Star Merger LIGO/Virgo GW170817. IV. Detection of Near-infrared Signatures of r-process Nucleosynthesis with Gemini-South}. {\em \apjl} \textbf{848}, L19.
(\href{http://dx.doi.org/10.3847/2041-8213/aa905c}{10.3847/2041-8213/aa905c})

\bibitem{Covino2017}
{Covino} S, {Wiersema} K, {Fan} YZ, {Toma} K, {Higgins} AB, {Melandri} A, {D'Avanzo} P, {Mundell} CG, {Palazzi} E, {Tanvir} NR, {Bernardini} MG, {Branchesi} M, {Brocato} E, {Campana} S, {di Serego Alighieri} S, {G{\"o}tz} D, {Fynbo} JPU, {Gao} W, {Gomboc} A, {Gompertz} B, {Greiner} J, {Hjorth} J, {Jin} ZP, {Kaper} L, {Klose} S, {Kobayashi} S, {Kopac} D, {Kouveliotou} C, {Levan} AJ, {Mao} J, {Malesani} D, {Pian} E, {Rossi} A, {Salvaterra} R, {Starling} RLC, {Steele} I, {Tagliaferri} G, {Troja} E, {van der Horst} AJ, {Wijers} RAMJ. 2017  {The unpolarized macronova associated with the gravitational wave event GW 170817}. {\em Nature Astronomy} \textbf{1}, 791--794.
(\href{http://dx.doi.org/10.1038/s41550-017-0285-z}{10.1038/s41550-017-0285-z})

\bibitem{Cowperthwaite2017}
{Cowperthwaite} PS, {Berger} E, {Villar} VA, {Metzger} BD, {Nicholl} M, {Chornock} R, {Blanchard} PK, {Fong} W, {Margutti} R, {Soares-Santos} M, {Alexander} KD, {Allam} S, {Annis} J, {Brout} D, {Brown} DA, {Butler} RE, {Chen} HY, {Diehl} HT, {Doctor} Z, {Drout} MR, {Eftekhari} T, {Farr} B, {Finley} DA, {Foley} RJ, {Frieman} JA, {Fryer} CL, {Garc{\'{\i}}a-Bellido} J, {Gill} MSS, {Guillochon} J, {Herner} K, {Holz} DE, {Kasen} D, {Kessler} R, {Marriner} J, {Matheson} T, {Neilsen}, Jr. EH, {Quataert} E, {Palmese} A, {Rest} A, {Sako} M, {Scolnic} DM, {Smith} N, {Tucker} DL, {Williams} PKG, {Balbinot} E, {Carlin} JL, {Cook} ER, {Durret} F, {Li} TS, {Lopes} PAA, {Louren{\c c}o} ACC, {Marshall} JL, {Medina} GE, {Muir} J, {Mu{\~n}oz} RR, {Sauseda} M, {Schlegel} DJ, {Secco} LF, {Vivas} AK, {Wester} W, {Zenteno} A, {Zhang} Y, {Abbott} TMC, {Banerji} M, {Bechtol} K, {Benoit-L{\'e}vy} A, {Bertin} E, {Buckley-Geer} E, {Burke} DL, {Capozzi} D, {Carnero Rosell} A, {Carrasco Kind} M, {Castander} FJ, {Crocce} M, {Cunha} CE,
  {D'Andrea} CB, {da Costa} LN, {Davis} C, {DePoy} DL, {Desai} S, {Dietrich} JP, {Drlica-Wagner} A, {Eifler} TF, {Evrard} AE, {Fernandez} E, {Flaugher} B, {Fosalba} P, {Gaztanaga} E, {Gerdes} DW, {Giannantonio} T, {Goldstein} DA, {Gruen} D, {Gruendl} RA, {Gutierrez} G, {Honscheid} K, {Jain} B, {James} DJ, {Jeltema} T, {Johnson} MWG, {Johnson} MD, {Kent} S, {Krause} E, {Kron} R, {Kuehn} K, {Nuropatkin} N, {Lahav} O, {Lima} M, {Lin} H, {Maia} MAG, {March} M, {Martini} P, {McMahon} RG, {Menanteau} F, {Miller} CJ, {Miquel} R, {Mohr} JJ, {Neilsen} E, {Nichol} RC, {Ogando} RLC, {Plazas} AA, {Roe} N, {Romer} AK, {Roodman} A, {Rykoff} ES, {Sanchez} E, {Scarpine} V, {Schindler} R, {Schubnell} M, {Sevilla-Noarbe} I, {Smith} M, {Smith} RC, {Sobreira} F, {Suchyta} E, {Swanson} MEC, {Tarle} G, {Thomas} D, {Thomas} RC, {Troxel} MA, {Vikram} V, {Walker} AR, {Wechsler} RH, {Weller} J, {Yanny} B, {Zuntz} J. 2017  {The Electromagnetic Counterpart of the Binary Neutron Star Merger LIGO/Virgo GW170817. II. UV, Optical, and
  Near-infrared Light Curves and Comparison to Kilonova Models}. {\em \apjl} \textbf{848}, L17.
(\href{http://dx.doi.org/10.3847/2041-8213/aa8fc7}{10.3847/2041-8213/aa8fc7})

\bibitem{Diaz2017}
{D{\'{\i}}az} MC, {Macri} LM, {Garcia Lambas} D, {Mendes de Oliveira} C, {Nilo Castell{\'o}n} JL, {Ribeiro} T, {S{\'a}nchez} B, {Schoenell} W, {Abramo} LR, {Akras} S, {Alcaniz} JS, {Artola} R, {Beroiz} M, {Bonoli} S, {Cabral} J, {Camuccio} R, {Castillo} M, {Chavushyan} V, {Coelho} P, {Colazo} C, {Costa-Duarte} MV, {Cuevas Larenas} H, {DePoy} DL, {Dom{\'{\i}}nguez Romero} M, {Dultzin} D, {Fern{\'a}ndez} D, {Garc{\'{\i}}a} J, {Girardini} C, {Gon{\c c}alves} DR, {Gon{\c c}alves} TS, {Gurovich} S, {Jim{\'e}nez-Teja} Y, {Kanaan} A, {Lares} M, {Lopes de Oliveira} R, {L{\'o}pez-Cruz} O, {Marshall} JL, {Melia} R, {Molino} A, {Padilla} N, {Pe{\~n}uela} T, {Placco} VM, {Qui{\~n}ones} C, {Ram{\'{\i}}rez Rivera} A, {Renzi} V, {Riguccini} L, {R{\'{\i}}os-L{\'o}pez} E, {Rodriguez} H, {Sampedro} L, {Schneiter} M, {Sodr{\'e}} L, {Starck} M, {Torres-Flores} S, {Tornatore} M, {Zadro{\.z}ny} A. 2017  {Observations of the First Electromagnetic Counterpart to a Gravitational-wave Source by the TOROS Collaboration}. {\em \apjl}
  \textbf{848}, L29.
(\href{http://dx.doi.org/10.3847/2041-8213/aa9060}{10.3847/2041-8213/aa9060})

\bibitem{Drout2017}
{Drout} MR, {Piro} AL, {Shappee} BJ, {Kilpatrick} CD, {Simon} JD, {Contreras} C, {Coulter} DA, {Foley} RJ, {Siebert} MR, {Morrell} N, {Boutsia} K, {Di Mille} F, {Holoien} TWS, {Kasen} D, {Kollmeier} JA, {Madore} BF, {Monson} AJ, {Murguia-Berthier} A, {Pan} YC, {Prochaska} JX, {Ramirez-Ruiz} E, {Rest} A, {Adams} C, {Alatalo} K, {Ba{\~n}ados} E, {Baughman} J, {Beers} TC, {Bernstein} RA, {Bitsakis} T, {Campillay} A, {Hansen} TT, {Higgs} CR, {Ji} AP, {Maravelias} G, {Marshall} JL, {Moni Bidin} C, {Prieto} JL, {Rasmussen} KC, {Rojas-Bravo} C, {Strom} AL, {Ulloa} N, {Vargas-Gonz{\'a}lez} J, {Wan} Z, {Whitten} DD. 2017  {Light curves of the neutron star merger GW170817/SSS17a: Implications for r-process nucleosynthesis}. {\em Science} \textbf{358}, 1570--1574.
(\href{http://dx.doi.org/10.1126/science.aaq0049}{10.1126/science.aaq0049})

\bibitem{Evans2017}
{Evans} PA, {Cenko} SB, {Kennea} JA, {Emery} SWK, {Kuin} NPM, {Korobkin} O, {Wollaeger} RT, {Fryer} CL, {Madsen} KK, {Harrison} FA, {Xu} Y, {Nakar} E, {Hotokezaka} K, {Lien} A, {Campana} S, {Oates} SR, {Troja} E, {Breeveld} AA, {Marshall} FE, {Barthelmy} SD, {Beardmore} AP, {Burrows} DN, {Cusumano} G, {D'A{\`i}} A, {D'Avanzo} P, {D'Elia} V, {de Pasquale} M, {Even} WP, {Fontes} CJ, {Forster} K, {Garcia} J, {Giommi} P, {Grefenstette} B, {Gronwall} C, {Hartmann} DH, {Heida} M, {Hungerford} AL, {Kasliwal} MM, {Krimm} HA, {Levan} AJ, {Malesani} D, {Melandri} A, {Miyasaka} H, {Nousek} JA, {O'Brien} PT, {Osborne} JP, {Pagani} C, {Page} KL, {Palmer} DM, {Perri} M, {Pike} S, {Racusin} JL, {Rosswog} S, {Siegel} MH, {Sakamoto} T, {Sbarufatti} B, {Tagliaferri} G, {Tanvir} NR, {Tohuvavohu} A. 2017  {Swift and NuSTAR observations of GW170817: Detection of a blue kilonova}. {\em Science} \textbf{358}, 1565--1570.
(\href{http://dx.doi.org/10.1126/science.aap9580}{10.1126/science.aap9580})

\bibitem{Hu2017}
{Hu} L, {Wu} X, {Andreoni} I, {Ashley} MCB, {Cooke} J, {Cui} X, {Du} F, {Dai} Z, {Gu} B, {Hu} Y, {Lu} H, {Li} X, {Li} Z, {Liang} E, {Liu} L, {Ma} B, {Shang} Z, {Sun} T, {Suntzeff} NB, {Tao} C, {Udden} SA, {Wang} L, {Wang} X, {Wen} H, {Xiao} D, {Su} J, {Yang} J, {Yang} S, {Yuan} X, {Zhou} H, {Zhang} H, {Zhou} J, {Zhu} Z. 2017  {Optical observations of LIGO source GW 170817 by the Antarctic Survey Telescopes at Dome A, Antarctica}. {\em Science Bulletin, Vol.~62, No.21, p.1433-1438, 2017} \textbf{62}, 1433--1438.
(\href{http://dx.doi.org/10.1016/j.scib.2017.10.006}{10.1016/j.scib.2017.10.006})

\bibitem{Kasliwal2017}
{Kasliwal} MM, {Nakar} E, {Singer} LP, {Kaplan} DL, {Cook} DO, {Van Sistine} A, {Lau} RM, {Fremling} C, {Gottlieb} O, {Jencson} JE, {Adams} SM, {Feindt} U, {Hotokezaka} K, {Ghosh} S, {Perley} DA, {Yu} PC, {Piran} T, {Allison} JR, {Anupama} GC, {Balasubramanian} A, {Bannister} KW, {Bally} J, {Barnes} J, {Barway} S, {Bellm} E, {Bhalerao} V, {Bhattacharya} D, {Blagorodnova} N, {Bloom} JS, {Brady} PR, {Cannella} C, {Chatterjee} D, {Cenko} SB, {Cobb} BE, {Copperwheat} C, {Corsi} A, {De} K, {Dobie} D, {Emery} SWK, {Evans} PA, {Fox} OD, {Frail} DA, {Frohmaier} C, {Goobar} A, {Hallinan} G, {Harrison} F, {Helou} G, {Hinderer} T, {Ho} AYQ, {Horesh} A, {Ip} WH, {Itoh} R, {Kasen} D, {Kim} H, {Kuin} NPM, {Kupfer} T, {Lynch} C, {Madsen} K, {Mazzali} PA, {Miller} AA, {Mooley} K, {Murphy} T, {Ngeow} CC, {Nichols} D, {Nissanke} S, {Nugent} P, {Ofek} EO, {Qi} H, {Quimby} RM, {Rosswog} S, {Rusu} F, {Sadler} EM, {Schmidt} P, {Sollerman} J, {Steele} I, {Williamson} AR, {Xu} Y, {Yan} L, {Yatsu} Y, {Zhang} C, {Zhao} W. 2017
  {Illuminating gravitational waves: A concordant picture of photons from a neutron star merger}. {\em Science} \textbf{358}, 1559--1565.
(\href{http://dx.doi.org/10.1126/science.aap9455}{10.1126/science.aap9455})

\bibitem{McCully2017}
{McCully} C, {Hiramatsu} D, {Howell} DA, {Hosseinzadeh} G, {Arcavi} I, {Kasen} D, {Barnes} J, {Shara} MM, {Williams} TB, {V{\"a}is{\"a}nen} P, {Potter} SB, {Romero-Colmenero} E, {Crawford} SM, {Buckley} DAH, {Cooke} J, {Andreoni} I, {Pritchard} TA, {Mao} J, {Gromadzki} M, {Burke} J. 2017  {The Rapid Reddening and Featureless Optical Spectra of the Optical Counterpart of GW170817, AT 2017gfo, during the First Four Days}. {\em \apjl} \textbf{848}, L32.
(\href{http://dx.doi.org/10.3847/2041-8213/aa9111}{10.3847/2041-8213/aa9111})

\bibitem{Nicholl2017}
{Nicholl} M, {Berger} E, {Kasen} D, {Metzger} BD, {Elias} J, {Brice{\~n}o} C, {Alexander} KD, {Blanchard} PK, {Chornock} R, {Cowperthwaite} PS, {Eftekhari} T, {Fong} W, {Margutti} R, {Villar} VA, {Williams} PKG, {Brown} W, {Annis} J, {Bahramian} A, {Brout} D, {Brown} DA, {Chen} HY, {Clemens} JC, {Dennihy} E, {Dunlap} B, {Holz} DE, {Marchesini} E, {Massaro} F, {Moskowitz} N, {Pelisoli} I, {Rest} A, {Ricci} F, {Sako} M, {Soares-Santos} M, {Strader} J. 2017  {The Electromagnetic Counterpart of the Binary Neutron Star Merger LIGO/Virgo GW170817. III. Optical and UV Spectra of a Blue Kilonova from Fast Polar Ejecta}. {\em \apjl} \textbf{848}, L18.
(\href{http://dx.doi.org/10.3847/2041-8213/aa9029}{10.3847/2041-8213/aa9029})

\bibitem{Pian2017}
{Pian} E, {D'Avanzo} P, {Benetti} S, {Branchesi} M, {Brocato} E, {Campana} S, {Cappellaro} E, {Covino} S, {D'Elia} V, {Fynbo} JPU, {Getman} F, {Ghirlanda} G, {Ghisellini} G, {Grado} A, {Greco} G, {Hjorth} J, {Kouveliotou} C, {Levan} A, {Limatola} L, {Malesani} D, {Mazzali} PA, {Melandri} A, {M{\o}ller} P, {Nicastro} L, {Palazzi} E, {Piranomonte} S, {Rossi} A, {Salafia} OS, {Selsing} J, {Stratta} G, {Tanaka} M, {Tanvir} NR, {Tomasella} L, {Watson} D, {Yang} S, {Amati} L, {Antonelli} LA, {Ascenzi} S, {Bernardini} MG, {Bo{\"e}r} M, {Bufano} F, {Bulgarelli} A, {Capaccioli} M, {Casella} P, {Castro-Tirado} AJ, {Chassande-Mottin} E, {Ciolfi} R, {Copperwheat} CM, {Dadina} M, {De Cesare} G, {di Paola} A, {Fan} YZ, {Gendre} B, {Giuffrida} G, {Giunta} A, {Hunt} LK, {Israel} GL, {Jin} ZP, {Kasliwal} MM, {Klose} S, {Lisi} M, {Longo} F, {Maiorano} E, {Mapelli} M, {Masetti} N, {Nava} L, {Patricelli} B, {Perley} D, {Pescalli} A, {Piran} T, {Possenti} A, {Pulone} L, {Razzano} M, {Salvaterra} R, {Schipani} P, {Spera} M,
  {Stamerra} A, {Stella} L, {Tagliaferri} G, {Testa} V, {Troja} E, {Turatto} M, {Vergani} SD, {Vergani} D. 2017  {Spectroscopic identification of r-process nucleosynthesis in a double neutron-star merger}. {\em \nat} \textbf{551}, 67--70.
(\href{http://dx.doi.org/10.1038/nature24298}{10.1038/nature24298})

\bibitem{Pozanenko2018}
{Pozanenko} AS, {Barkov} MV, {Minaev} PY, {Volnova} AA, {Mazaeva} ED, {Moskvitin} AS, {Krugov} MA, {Samodurov} VA, {Loznikov} VM, {Lyutikov} M. 2018  {GRB 170817A Associated with GW170817: Multi-frequency Observations and Modeling of Prompt Gamma-Ray Emission}. {\em \apjl} \textbf{852}, L30.
(\href{http://dx.doi.org/10.3847/2041-8213/aaa2f6}{10.3847/2041-8213/aaa2f6})

\bibitem{Shappee2017}
{Shappee} BJ, {Simon} JD, {Drout} MR, {Piro} AL, {Morrell} N, {Prieto} JL, {Kasen} D, {Holoien} TWS, {Kollmeier} JA, {Kelson} DD, {Coulter} DA, {Foley} RJ, {Kilpatrick} CD, {Siebert} MR, {Madore} BF, {Murguia-Berthier} A, {Pan} YC, {Prochaska} JX, {Ramirez-Ruiz} E, {Rest} A, {Adams} C, {Alatalo} K, {Ba{\~n}ados} E, {Baughman} J, {Bernstein} RA, {Bitsakis} T, {Boutsia} K, {Bravo} JR, {Di Mille} F, {Higgs} CR, {Ji} AP, {Maravelias} G, {Marshall} JL, {Placco} VM, {Prieto} G, {Wan} Z. 2017  {Early spectra of the gravitational wave source GW170817: Evolution of a neutron star merger}. {\em Science} \textbf{358}, 1574--1578.
(\href{http://dx.doi.org/10.1126/science.aaq0186}{10.1126/science.aaq0186})

\bibitem{Smartt2017}
{Smartt} SJ, {Chen} TW, {Jerkstrand} A, {Coughlin} M, {Kankare} E, {Sim} SA, {Fraser} M, {Inserra} C, {Maguire} K, {Chambers} KC, {Huber} ME, {Kr{\"u}hler} T, {Leloudas} G, {Magee} M, {Shingles} LJ, {Smith} KW, {Young} DR, {Tonry} J, {Kotak} R, {Gal-Yam} A, {Lyman} JD, {Homan} DS, {Agliozzo} C, {Anderson} JP, {Angus} CR, {Ashall} C, {Barbarino} C, {Bauer} FE, {Berton} M, {Botticella} MT, {Bulla} M, {Bulger} J, {Cannizzaro} G, {Cano} Z, {Cartier} R, {Cikota} A, {Clark} P, {De Cia} A, {Della Valle} M, {Denneau} L, {Dennefeld} M, {Dessart} L, {Dimitriadis} G, {Elias-Rosa} N, {Firth} RE, {Flewelling} H, {Fl{\"o}rs} A, {Franckowiak} A, {Frohmaier} C, {Galbany} L, {Gonz{\'a}lez-Gait{\'a}n} S, {Greiner} J, {Gromadzki} M, {Guelbenzu} AN, {Guti{\'e}rrez} CP, {Hamanowicz} A, {Hanlon} L, {Harmanen} J, {Heintz} KE, {Heinze} A, {Hernandez} MS, {Hodgkin} ST, {Hook} IM, {Izzo} L, {James} PA, {Jonker} PG, {Kerzendorf} WE, {Klose} S, {Kostrzewa-Rutkowska} Z, {Kowalski} M, {Kromer} M, {Kuncarayakti} H, {Lawrence} A, {Lowe} TB,
  {Magnier} EA, {Manulis} I, {Martin-Carrillo} A, {Mattila} S, {McBrien} O, {M{\"u}ller} A, {Nordin} J, {O'Neill} D, {Onori} F, {Palmerio} JT, {Pastorello} A, {Patat} F, {Pignata} G, {Podsiadlowski} P, {Pumo} ML, {Prentice} SJ, {Rau} A, {Razza} A, {Rest} A, {Reynolds} T, {Roy} R, {Ruiter} AJ, {Rybicki} KA, {Salmon} L, {Schady} P, {Schultz} ASB, {Schweyer} T, {Seitenzahl} IR, {Smith} M, {Sollerman} J, {Stalder} B, {Stubbs} CW, {Sullivan} M, {Szegedi} H, {Taddia} F, {Taubenberger} S, {Terreran} G, {van Soelen} B, {Vos} J, {Wainscoat} RJ, {Walton} NA, {Waters} C, {Weiland} H, {Willman} M, {Wiseman} P, {Wright} DE, {Wyrzykowski} {\L}, {Yaron} O. 2017  {A kilonova as the electromagnetic counterpart to a gravitational-wave source}. {\em \nat} \textbf{551}, 75--79.
(\href{http://dx.doi.org/10.1038/nature24303}{10.1038/nature24303})

\bibitem{Utsumi2017}
{Utsumi} Y, {Tanaka} M, {Tominaga} N, {Yoshida} M, {Barway} S, {Nagayama} T, {Zenko} T, {Aoki} K, {Fujiyoshi} T, {Furusawa} H, {Kawabata} KS, {Koshida} S, {Lee} CH, {Morokuma} T, {Motohara} K, {Nakata} F, {Ohsawa} R, {Ohta} K, {Okita} H, {Tajitsu} A, {Tanaka} I, {Terai} T, {Yasuda} N, {Abe} F, {Asakura} Y, {Bond} IA, {Miyazaki} S, {Sumi} T, {Tristram} PJ, {Honda} S, {Itoh} R, {Itoh} Y, {Kawabata} M, {Morihana} K, {Nagashima} H, {Nakaoka} T, {Ohshima} T, {Takahashi} J, {Takayama} M, {Aoki} W, {Baar} S, {Doi} M, {Finet} F, {Kanda} N, {Kawai} N, {Kim} JH, {Kuroda} D, {Liu} W, {Matsubayashi} K, {Murata} KL, {Nagai} H, {Saito} T, {Saito} Y, {Sako} S, {Sekiguchi} Y, {Tamura} Y, {Tanaka} M, {Uemura} M, {Yamaguchi} MS. 2017  {J-GEM observations of an electromagnetic counterpart to the neutron star merger GW170817}. {\em \pasj} \textbf{69}, 101.
(\href{http://dx.doi.org/10.1093/pasj/psx118}{10.1093/pasj/psx118})

\bibitem{Cowperthwaite2017gcn1}
{Cowperthwaite} PS, {Foley} RJ, {Berger} E. 2017  {LIGO/Virgo G298048: Optical Candidate Shows No Evidence for Previous Activity in Archival ASAS-SN Data}. {\em GRB Coordinates Network} \textbf{21533}, 1.

\bibitem{Moller2017gcn}
{M{\"o}ller} A, {Chang} S, {Wolf} C. 2017  {LIGO/Virgo G298048: Optical Candidate Shows No Evidence for Previous Activity in Archival SkyMapper Transient Survey Data}. {\em GRB Coordinates Network} \textbf{21542}, 1.

\bibitem{Foley2017gcn}
{Foley} RJ. 2017  {LIGO/Virgo G298048: Likelihood of SSS17a Being an Unrelated Extragalactic Transient in NGC 4993}. {\em GRB Coordinates Network} \textbf{21557}, 1.

\bibitem{Foley2017gcn2}
{Foley} RJ, {Kilpatrick} CD, {Nicholl} M, {Berger} E. 2017  {LIGO/Virgo G298048: Inspection of archival HST data at the position of the potential optical counterpart}. {\em GRB Coordinates Network} \textbf{21536}, 1.

\bibitem{Cowperthwaite2017gcn2}
{Cowperthwaite} PS, {Nicholl} M, {Berger} E. 2017  {LIGO/VIRGO G298048: Expansion Velocity of SSS17a Consistent With Relativistic Merger Ejecta}. {\em GRB Coordinates Network} \textbf{21578}, 1.

\bibitem{Malesani2017gcn1}
{Malesani} D, {Watson} D, {Hjorth} J. 2017  {LIGO/Virgo G298048: optical spectral energy distribution of SSS17a}. {\em GRB Coordinates Network} \textbf{21577}, 1.

\bibitem{Cenko2017gcn}
{Cenko} SB, {Emery} SWK, {Campana} S, {Evans} PA, {Kennea} JA, {Breeveld} AA, {Troja} EN, {O'Brien} PT, {Barthelmy} SD, {Beardmore} AP, {Burrows} DN, {Cusumano} G, {D'Ai} A, {D'Avanzo} P, {D'Elia} VA, {Giommi} P, {Gronwall} C, {Krimm} HA, {Kuin} NPM, {Lien} AY, {Marshall} FE, {Melandri} A, {Nousek} JA, {Oates} SR, {Osborne} JP, {Pagani} C, {Page} KL, {Palmer} DM, {Perri} M, {Racusin} JL, {Sbarufatti} B, {Siegel} MH, {Tagliaferri} G, {Tohuvavohu} A, {Swift Team}. 2017  {LIGO/Virgo G298048: Continued Swift UV and X-ray Monitoring of SSS17a}. {\em GRB Coordinates Network} \textbf{21572}, 1.

\bibitem{Nicholl2017gcn1}
{Nicholl} M, {Cowperthwaite} PS, {Allam} S, {Annis} J, {Berger} E, {Brout} DJ, {Brown} D, {Butler} RE, {Chen} HY, {Chornock} R, {Cook} E, {Diehl} HT, {Drlica-Wagner} A, {Drout} MR, {Foley} RJ, {Fong} W, {Fox} D, {Frieman} J, {Gill} MSS, {Gruendl} R, {Herner} K, {Holz} D, {Kessler} R, {Margutti} R, {Marshall} J, {Neilsen} E, {Paz-Chincon} F, {Rest} A, {Sako} M, {Smith} N, {Soares-Santos} M, {Tucker} D, {Villar} VA, {Walker} A, {Williams} PKG, {Yanny} B, {Lopes} P, {Durret} F, {Louren{\c{c}}o} A, {Desgw+ Community}. 2017  {LIGO/VIRGO G298048: The Potential Optical Counterpart in NGC 4993 Has Afterglow-Like Colors}. {\em GRB Coordinates Network} \textbf{21541}, 1.

\bibitem{Drout2017gcn}
{Drout} MR, {Simon} JD, {Shappee} BJ, {Boutsia} K, {Bravo} J, {Prieto} G, {Coulter} DA, {Foley} RJ, {Kilpatrick} CD, {Siebert} MR, {Piro} AL. 2017  {LIGO/VIRGO G298048: Magellan Optical Spectrum of the Potential Optical Counterpart Associated with NGC 4993}. {\em GRB Coordinates Network} \textbf{21547}, 1.

\bibitem{Barnes2013}
{Barnes} J, {Kasen} D. 2013  {Effect of a High Opacity on the Light Curves of Radioactively Powered Transients from Compact Object Mergers}. {\em \apj} \textbf{775}, 18.
(\href{http://dx.doi.org/10.1088/0004-637X/775/1/18}{10.1088/0004-637X/775/1/18})

\bibitem{Tanaka2013}
{Tanaka} M, {Hotokezaka} K. 2013  {Radiative Transfer Simulations of Neutron Star Merger Ejecta}. {\em \apj} \textbf{775}, 113.
(\href{http://dx.doi.org/10.1088/0004-637X/775/2/113}{10.1088/0004-637X/775/2/113})

\bibitem{Yang2017gcn2}
{Yang} S, {Valenti} S, {Sand} D, {Tartaglia} L, {Cappellaro} E, {Reichart} D, {Haislip} J, {the Gravitational Wave Follow-Up by DLT40}. 2017  {LIGO/Virgo G298048: Continued observation for DLT17ck}. {\em GRB Coordinates Network} \textbf{21579}, 1.

\bibitem{Nicholl2017gcn2}
{Nicholl} M, {Cowperthwaite} PS, {Berger} E, {Williams} PKG, {Allam} S, {Annis} J, {Garcia Bellido} J, {Brout} DJ, {Brown} D, {Butler} RE, {Chen} HY, {Chornock} R, {Cook} E, {Diehl} HT, {Drlica-Wagner} A, {Doctor} Z, {Drout} MR, {Farr} B, {Foley} RJ, {Fong} W, {Fox} D, {Frieman} J, {Gill} MSS, {Gruendl} R, {Herner} K, {Holz} D, {Kessler} R, {Lin} H, {Marriner} J, {Margutti} R, {Marshall} J, {Neilsen} E, {Paz-Chincon} F, {Rest} A, {Sako} M, {Scolnic} D, {Smith} N, {Soares-Santos} M, {Tucker} D, {Villar} VA, {Walker} A, {Yanny} B, {Lopes} P, {Durret} F, {Louren{\c{c}}o} A. 2017  {LIGO/Virgo G298048: Observed fading of optical counterpart SSS17a}. {\em GRB Coordinates Network} \textbf{21580}, 1.

\bibitem{Chambers2017gcn}
{Chambers} KC, {Smartt} SJ, {Huber} ME, {Smith} KW, {Young} DR, {Coughlin} M, {Chen} TW, {Bulger} J, {Denneau} L, {Flewelling} H, {Heinze} A, {Kankare} E, {Lowe} T, {Magnier} EA, {Rest} A, {Stalder} B, {Schultz} ASB, {Stubbs} CW, {Tonry} J, {Waters} C, {Wainscoat} RJ, {Weiland} H, {Willman} M, {Wright} DE. 2017  {LIGO/Virgo G298048: Further Pan-STARRS izy photometry confirms rapid fading of SSS17a/DLT17ck in i and z bands}. {\em GRB Coordinates Network} \textbf{21590}, 1.

\bibitem{Arcavi2017gcn2}
{Arcavi} I, {Howell} DA, {McCully} C, {Hosseinzadeh} G, {Vasylyev} S, {Poznanski} D, {Zalzman} M, {Singer} LP, {Valenti} S, {Piran} T, {Kasen} D, {Barnes} J, {Fong} W, {Maoz} D. 2017  {LIGO/Virgo G298048: Rapid Evolution of Possible Counterpart}. {\em GRB Coordinates Network} \textbf{21581}, 1.

\bibitem{Lyman2017gcn}
{Lyman} J, {Homan} D, {Maguire} K, {Botticella} MT, {Fraser} M, {Inserra} C, {Kankare} E, {Smartt} SJ, {Smith} KW, {Sullivan} M, {Valenti} S, {Yaron} O, {Manulis} I, {Young} D, {Chen} TW, {Campana} S, {Benetti} S, {Tomasella} L, {Leloudas} G, {Cano} Z. 2017  {LIGO/VIRGO G298048: ePESSTO optical spectra of the candidate optical/NIR counterpart of the gravitational wave G298048 in NGC4993.}. {\em GRB Coordinates Network} \textbf{21582}, 1.

\bibitem{Nicholl2017gcn3}
{Nicholl} M, {Briceno} C, {Cowperthwaite} P, {Berger} E, {Elias} J, {Heathcote} S, {Annis} J, {Tucker} D, {Soares-Santos} M, {Kessler} R, {Sako} M. 2017  {LIGO/Virgo G298048: Possible features in the spectrum of GW counterpart SSS17a}. {\em GRB Coordinates Network} \textbf{21585}, 1.

\bibitem{Wiseman2017gcn}
{Wiseman} P, {Chen} TW, {Greiner} J, {Schady} P, {Grond Team}. 2017  {LIGO/Virgo G298048: GROND photometry of candidate optical counterpart reveals brightening in the NIR}. {\em GRB Coordinates Network} \textbf{21584}, 1.

\bibitem{Malesani2017gcn2}
{Malesani} D, {Pian} E, {Hjorth} J, {Watson} DJ, {Amati} L, {Antonelli} LA, {Ascenzi} S, {Benetti} S, {Botticella} MT, {Branchesi} M, {Campana} S, {Cappellaro} E, {Covino} S, {D'Avanzo} P, {D'Elia} V, {Fugazza} D, {Getman} F, {Grado} A, {Greco} G, {Limatola} L, {Lisi} M, {Melandri} A, {Nicastro} L, {Palazzi} E, {Piranomonte} S, {Pulone} L, {Rossi} A, {Schipani} P, {Stratta} G, {Tagliaferri} G, {Testa} V, {Tomasella} L, {Yang} S, {Brocato} E, {GRavitational Wave Inaf TeAm}. 2017  {LIGO/Virgo G298048: NOT near-infrared observations of SSS17a}. {\em GRB Coordinates Network} \textbf{21591}, 1.

\bibitem{Pian2017gcn}
{Pian} E, {D'Elia} V, {Piranomonte} S, {Branchesi} M, {Campana} S, {Cappellaro} E, {Covino} S, {D'Avanzo} P, {Grado} A, {Greco} G, {Melandri} A, {Palazzi} E, {Stratta} G, {Tomasella} L, {Amati} L, {Antonelli} LA, {Ascenzi} S, {Benetti} S, {Botticella} MT, {Fugazza} D, {Getman} F, {Limatola} L, {Lisi} M, {Nicastro} L, {Pulone} L, {Rossi} A, {Schipani} P, {Tagliaferri} G, {Testa} V, {Yang} S, {Sbordone} L, {Brocato} E, {GRavitational Wave Inaf TeAm}, {Team of the ESO VLT program 099. D-0382(A)}. 2017  {LIGO/Virgo G298048: GRAWITA VLT/X-shooter observations}. {\em GRB Coordinates Network} \textbf{21592}, 1.

\bibitem{Kilpatrick2017}
{Kilpatrick} CD, {Foley} RJ, {Kasen} D, {Murguia-Berthier} A, {Ramirez-Ruiz} E, {Coulter} DA, {Drout} MR, {Piro} AL, {Shappee} BJ, {Boutsia} K, {Contreras} C, {Di Mille} F, {Madore} BF, {Morrell} N, {Pan} YC, {Prochaska} JX, {Rest} A, {Rojas-Bravo} C, {Siebert} MR, {Simon} JD, {Ulloa} N. 2017  {Electromagnetic evidence that SSS17a is the result of a binary neutron star merger}. {\em Science} \textbf{358}, 1583--1587.
(\href{http://dx.doi.org/10.1126/science.aaq0073}{10.1126/science.aaq0073})

\bibitem{Tanaka2017}
{Tanaka} M, {Utsumi} Y, {Mazzali} PA, {Tominaga} N, {Yoshida} M, {Sekiguchi} Y, {Morokuma} T, {Motohara} K, {Ohta} K, {Kawabata} KS, {Abe} F, {Aoki} K, {Asakura} Y, {Baar} S, {Barway} S, {Bond} IA, {Doi} M, {Fujiyoshi} T, {Furusawa} H, {Honda} S, {Itoh} Y, {Kawabata} M, {Kawai} N, {Kim} JH, {Lee} CH, {Miyazaki} S, {Morihana} K, {Nagashima} H, {Nagayama} T, {Nakaoka} T, {Nakata} F, {Ohsawa} R, {Ohshima} T, {Okita} H, {Saito} T, {Sumi} T, {Tajitsu} A, {Takahashi} J, {Takayama} M, {Tamura} Y, {Tanaka} I, {Terai} T, {Tristram} PJ, {Yasuda} N, {Zenko} T. 2017  {Kilonova from post-merger ejecta as an optical and near-Infrared counterpart of GW170817}. {\em \pasj} \textbf{69}, 102.
(\href{http://dx.doi.org/10.1093/pasj/psx121}{10.1093/pasj/psx121})

\bibitem{Kasen2017}
{Kasen} D, {Metzger} B, {Barnes} J, {Quataert} E, {Ramirez-Ruiz} E. 2017  {Origin of the heavy elements in binary neutron-star mergers from a gravitational-wave event}. {\em \nat} \textbf{551}, 80--84.
(\href{http://dx.doi.org/10.1038/nature24453}{10.1038/nature24453})

\bibitem{Villar2017}
{Villar} VA, {Guillochon} J, {Berger} E, {Metzger} BD, {Cowperthwaite} PS, {Nicholl} M, {Alexander} KD, {Blanchard} PK, {Chornock} R, {Eftekhari} T, {Fong} W, {Margutti} R, {Williams} PKG. 2017  {The Combined Ultraviolet, Optical, and Near-infrared Light Curves of the Kilonova Associated with the Binary Neutron Star Merger GW170817: Unified Data Set, Analytic Models, and Physical Implications}. {\em \apjl} \textbf{851}, L21.
(\href{http://dx.doi.org/10.3847/2041-8213/aa9c84}{10.3847/2041-8213/aa9c84})

\bibitem{Hotokezaka2018}
{Hotokezaka} K, {Beniamini} P, {Piran} T. 2018  {Neutron star mergers as sites of r-process nucleosynthesis and short gamma-ray bursts}. {\em International Journal of Modern Physics D} \textbf{27}, 1842005.
(\href{http://dx.doi.org/10.1142/S0218271818420051}{10.1142/S0218271818420051})

\bibitem{margutti2021}
{Margutti} R, {Chornock} R. 2021  {First Multimessenger Observations of a Neutron Star Merger}. {\em \araa} \textbf{59}, 155--202.
(\href{http://dx.doi.org/10.1146/annurev-astro-112420-030742}{10.1146/annurev-astro-112420-030742})

\bibitem{Sekiguchi2015}
{Sekiguchi} Y, {Kiuchi} K, {Kyutoku} K, {Shibata} M. 2015  {Dynamical mass ejection from binary neutron star mergers: Radiation-hydrodynamics study in general relativity}. {\em \prd} \textbf{91}, 064059.
(\href{http://dx.doi.org/10.1103/PhysRevD.91.064059}{10.1103/PhysRevD.91.064059})

\bibitem{Hotokezaka2013}
{Hotokezaka} K, {Kiuchi} K, {Kyutoku} K, {Okawa} H, {Sekiguchi} Yi, {Shibata} M, {Taniguchi} K. 2013  {Mass ejection from the merger of binary neutron stars}. {\em \prd} \textbf{87}, 024001.
(\href{http://dx.doi.org/10.1103/PhysRevD.87.024001}{10.1103/PhysRevD.87.024001})

\bibitem{Bauswein2013}
{Bauswein} A, {Goriely} S, {Janka} HT. 2013  {Systematics of Dynamical Mass Ejection, Nucleosynthesis, and Radioactively Powered Electromagnetic Signals from Neutron-star Mergers}. {\em \apj} \textbf{773}, 78.
(\href{http://dx.doi.org/10.1088/0004-637X/773/1/78}{10.1088/0004-637X/773/1/78})

\bibitem{Wanajo2014}
{Wanajo} S, {Sekiguchi} Y, {Nishimura} N, {Kiuchi} K, {Kyutoku} K, {Shibata} M. 2014  {Production of All the r-process Nuclides in the Dynamical Ejecta of Neutron Star Mergers}. {\em \apjl} \textbf{789}, L39.
(\href{http://dx.doi.org/10.1088/2041-8205/789/2/L39}{10.1088/2041-8205/789/2/L39})

\bibitem{Goriely2015}
{Goriely} S, {Bauswein} A, {Just} O, {Pllumbi} E, {Janka} HT. 2015  {Impact of weak interactions of free nucleons on the r-process in dynamical ejecta from neutron star mergers}. {\em \mnras} \textbf{452}, 3894--3904.
(\href{http://dx.doi.org/10.1093/mnras/stv1526}{10.1093/mnras/stv1526})

\bibitem{Foucart2016}
{Foucart} F, {O'Connor} E, {Roberts} L, {Kidder} LE, {Pfeiffer} HP, {Scheel} MA. 2016  {Impact of an improved neutrino energy estimate on outflows in neutron star merger simulations}. {\em \prd} \textbf{94}, 123016.
(\href{http://dx.doi.org/10.1103/PhysRevD.94.123016}{10.1103/PhysRevD.94.123016})

\bibitem{Metzger2014}
{Metzger} BD, {Fern{\'a}ndez} R. 2014  {Red or blue? A potential kilonova imprint of the delay until black hole formation following a neutron star merger}. {\em \mnras} \textbf{441}, 3444--3453.
(\href{http://dx.doi.org/10.1093/mnras/stu802}{10.1093/mnras/stu802})

\bibitem{Fernandez2016}
{Fern{\'a}ndez} R, {Metzger} BD. 2016  {Electromagnetic Signatures of Neutron Star Mergers in the Advanced LIGO Era}. {\em Annual Review of Nuclear and Particle Science} \textbf{66}, 23--45.
(\href{http://dx.doi.org/10.1146/annurev-nucl-102115-044819}{10.1146/annurev-nucl-102115-044819})

\bibitem{Margalit2017}
{Margalit} B, {Metzger} BD. 2017  {Constraining the Maximum Mass of Neutron Stars from Multi-messenger Observations of GW170817}. {\em \apjl} \textbf{850}, L19.
(\href{http://dx.doi.org/10.3847/2041-8213/aa991c}{10.3847/2041-8213/aa991c})

\bibitem{Nicholl2021}
{Nicholl} M, {Margalit} B, {Schmidt} P, {Smith} GP, {Ridley} EJ, {Nuttall} J. 2021  {Tight multimessenger constraints on the neutron star equation of state from GW170817 and a forward model for kilonova light-curve synthesis}. {\em \mnras} \textbf{505}, 3016--3032.
(\href{http://dx.doi.org/10.1093/mnras/stab1523}{10.1093/mnras/stab1523})

\bibitem{Arnett1982}
{Arnett} WD. 1982  {Type I supernovae. I - Analytic solutions for the early part of the light curve}. {\em \apj} \textbf{253}, 785--797.
(\href{http://dx.doi.org/10.1086/159681}{10.1086/159681})

\bibitem{Tanaka2020}
{Tanaka} M, {Kato} D, {Gaigalas} G, {Kawaguchi} K. 2020  {Systematic opacity calculations for kilonovae}. {\em \mnras} \textbf{496}, 1369--1392.
(\href{http://dx.doi.org/10.1093/mnras/staa1576}{10.1093/mnras/staa1576})

\bibitem{Arcavi2018}
{Arcavi} I. 2018  {The First Hours of the GW170817 Kilonova and the Importance of Early Optical and Ultraviolet Observations for Constraining Emission Models}. {\em \apjl} \textbf{855}, L23.
(\href{http://dx.doi.org/10.3847/2041-8213/aab267}{10.3847/2041-8213/aab267})

\bibitem{Bulla2019}
{Bulla} M. 2019  {POSSIS: predicting spectra, light curves, and polarization for multidimensional models of supernovae and kilonovae}. {\em \mnras} \textbf{489}, 5037--5045.
(\href{http://dx.doi.org/10.1093/mnras/stz2495}{10.1093/mnras/stz2495})

\bibitem{Piro2018}
{Piro} AL, {Kollmeier} JA. 2018  {Evidence for Cocoon Emission from the Early Light Curve of SSS17a}. {\em \apj} \textbf{855}, 103.
(\href{http://dx.doi.org/10.3847/1538-4357/aaaab3}{10.3847/1538-4357/aaaab3})

\bibitem{Gottlieb2018}
{Gottlieb} O, {Nakar} E, {Piran} T. 2018  {The cocoon emission - an electromagnetic counterpart to gravitational waves from neutron star mergers}. {\em \mnras} \textbf{473}, 576--584.
(\href{http://dx.doi.org/10.1093/mnras/stx2357}{10.1093/mnras/stx2357})

\bibitem{Metzger2015}
{Metzger} BD, {Bauswein} A, {Goriely} S, {Kasen} D. 2015  {Neutron-powered precursors of kilonovae}. {\em \mnras} \textbf{446}, 1115--1120.
(\href{http://dx.doi.org/10.1093/mnras/stu2225}{10.1093/mnras/stu2225})

\bibitem{Coughlin2019}
{Coughlin} MW, {Antier} S, {Corre} D, {Alqassimi} K, {Anand} S, {Christensen} N, {Coulter} DA, {Foley} RJ, {Guessoum} N, {Mikulski} TM, {Mualla} MA, {Reed} D, {Tao} D. 2019  {Optimizing multitelescope observations of gravitational-wave counterparts}. {\em \mnras} \textbf{489}, 5775--5783.
(\href{http://dx.doi.org/10.1093/mnras/stz2485}{10.1093/mnras/stz2485})

\bibitem{Dietrich2020}
{Dietrich} T, {Coughlin} MW, {Pang} PTH, {Bulla} M, {Heinzel} J, {Issa} L, {Tews} I, {Antier} S. 2020  {Multimessenger constraints on the neutron-star equation of state and the Hubble constant}. {\em Science} \textbf{370}, 1450--1453.
(\href{http://dx.doi.org/10.1126/science.abb4317}{10.1126/science.abb4317})

\bibitem{Breschi2021}
{Breschi} M, {Perego} A, {Bernuzzi} S, {Del Pozzo} W, {Nedora} V, {Radice} D, {Vescovi} D. 2021  {AT2017gfo: Bayesian inference and model selection of multicomponent kilonovae and constraints on the neutron star equation of state}. {\em \mnras} \textbf{505}, 1661--1677.
(\href{http://dx.doi.org/10.1093/mnras/stab1287}{10.1093/mnras/stab1287})

\bibitem{Raaijmakers2021}
{Raaijmakers} G, {Greif} SK, {Hebeler} K, {Hinderer} T, {Nissanke} S, {Schwenk} A, {Riley} TE, {Watts} AL, {Lattimer} JM, {Ho} WCG. 2021  {Constraints on the Dense Matter Equation of State and Neutron Star Properties from NICER's Mass-Radius Estimate of PSR J0740+6620 and Multimessenger Observations}. {\em \apjl} \textbf{918}, L29.
(\href{http://dx.doi.org/10.3847/2041-8213/ac089a}{10.3847/2041-8213/ac089a})

\bibitem{Schutz1986}
{Schutz} BF. 1986  {Determining the Hubble constant from gravitational wave observations}. {\em \nat} \textbf{323}, 310--311.
(\href{http://dx.doi.org/10.1038/323310a0}{10.1038/323310a0})

\bibitem{Abbott2017H0}
{LIGO Scientific Collaboration}, {Virgo Collaboration} et~al.. 2017  {A gravitational-wave standard siren measurement of the Hubble constant}. {\em \nat} \textbf{551}, 85--88.
(\href{http://dx.doi.org/10.1038/nature24471}{10.1038/nature24471})

\bibitem{Riess2016}
{Riess} AG, {Macri} LM, {Hoffmann} SL, {Scolnic} D, {Casertano} S, {Filippenko} AV, {Tucker} BE, {Reid} MJ, {Jones} DO, {Silverman} JM, {Chornock} R, {Challis} P, {Yuan} W, {Brown} PJ, {Foley} RJ. 2016  {A 2.4\% Determination of the Local Value of the Hubble Constant}. {\em \apj} \textbf{826}, 56.
(\href{http://dx.doi.org/10.3847/0004-637X/826/1/56}{10.3847/0004-637X/826/1/56})

\bibitem{Planck2016}
{Planck Collaboration}. 2016  {Planck 2015 results. XIII. Cosmological parameters}. {\em \aap} \textbf{594}, A13.
(\href{http://dx.doi.org/10.1051/0004-6361/201525830}{10.1051/0004-6361/201525830})

\bibitem{Guidorzi2017}
{Guidorzi} C, {Margutti} R, {Brout} D, {Scolnic} D, {Fong} W, {Alexander} KD, {Cowperthwaite} PS, {Annis} J, {Berger} E, {Blanchard} PK, {Chornock} R, {Coppejans} DL, {Eftekhari} T, {Frieman} JA, {Huterer} D, {Nicholl} M, {Soares-Santos} M, {Terreran} G, {Villar} VA, {Williams} PKG. 2017  {Improved Constraints on H $_{0}$ from a Combined Analysis of Gravitational-wave and Electromagnetic Emission from GW170817}. {\em \apjl} \textbf{851}, L36.
(\href{http://dx.doi.org/10.3847/2041-8213/aaa009}{10.3847/2041-8213/aaa009})

\bibitem{Palmese2024}
{Palmese} A, {Kaur} R, {Hajela} A, {Margutti} R, {McDowell} A, {MacFadyen} A. 2024  {Standard siren measurement of the Hubble constant using GW170817 and the latest observations of the electromagnetic counterpart afterglow}. {\em \prd} \textbf{109}, 063508.
(\href{http://dx.doi.org/10.1103/PhysRevD.109.063508}{10.1103/PhysRevD.109.063508})

\bibitem{Dhawan2020}
{Dhawan} S, {Bulla} M, {Goobar} A, {Sagu{\'e}s Carracedo} A, {Setzer} CN. 2020  {Constraining the Observer Angle of the Kilonova AT2017gfo Associated with GW170817: Implications for the Hubble Constant}. {\em \apj} \textbf{888}, 67.
(\href{http://dx.doi.org/10.3847/1538-4357/ab5799}{10.3847/1538-4357/ab5799})

\bibitem{Bulla2022}
{Bulla} M, {Coughlin} MW, {Dhawan} S, {Dietrich} T. 2022  {Multi-Messenger Constraints on the Hubble Constant Through Combination of Gravitational Waves, Gamma-Ray Bursts and Kilonovae from Neutron Star Mergers}. {\em Universe} \textbf{8}, 289.
(\href{http://dx.doi.org/10.3390/universe8050289}{10.3390/universe8050289})

\bibitem{Mooley2018a}
{Mooley} KP, {Deller} AT, {Gottlieb} O, {Nakar} E, {Hallinan} G, {Bourke} S, {Frail} DA, {Horesh} A, {Corsi} A, {Hotokezaka} K. 2018  {Superluminal motion of a relativistic jet in the neutron-star merger GW170817}. {\em \nat} \textbf{561}, 355--359.
(\href{http://dx.doi.org/10.1038/s41586-018-0486-3}{10.1038/s41586-018-0486-3})

\bibitem{Hotokezaka2019}
{Hotokezaka} K, {Nakar} E, {Gottlieb} O, {Nissanke} S, {Masuda} K, {Hallinan} G, {Mooley} KP, {Deller} AT. 2019  {A Hubble constant measurement from superluminal motion of the jet in GW170817}. {\em Nature Astronomy} \textbf{3}, 940--944.
(\href{http://dx.doi.org/10.1038/s41550-019-0820-1}{10.1038/s41550-019-0820-1})

\bibitem{Gianfagna2024}
{Gianfagna} G, {Piro} L, {Pannarale} F, {Van Eerten} H, {Ricci} F, {Ryan} G. 2024  {Potential biases and prospects for the Hubble constant estimation via electromagnetic and gravitational-wave joint analyses}. {\em \mnras} \textbf{528}, 2600--2613.
(\href{http://dx.doi.org/10.1093/mnras/stae198}{10.1093/mnras/stae198})

\bibitem{Mooley2022}
{Mooley} KP, {Anderson} J, {Lu} W. 2022  {Optical superluminal motion measurement in the neutron-star merger GW170817}. {\em \nat} \textbf{610}, 273--276.
(\href{http://dx.doi.org/10.1038/s41586-022-05145-7}{10.1038/s41586-022-05145-7})

\bibitem{Doctor2020}
{Doctor} Z. 2020  {Thunder and Lightning: Using Neutron-star Mergers as Simultaneous Standard Candles and Sirens to Measure Cosmological Parameters}. {\em \apjl} \textbf{892}, L16.
(\href{http://dx.doi.org/10.3847/2041-8213/ab7cd8}{10.3847/2041-8213/ab7cd8})

\bibitem{Coughlin2020PhRvR}
{Coughlin} MW, {Dietrich} T, {Heinzel} J, {Khetan} N, {Antier} S, {Bulla} M, {Christensen} N, {Coulter} DA, {Foley} RJ. 2020a  {Standardizing kilonovae and their use as standard candles to measure the Hubble constant}. {\em Physical Review Research} \textbf{2}, 022006.
(\href{http://dx.doi.org/10.1103/PhysRevResearch.2.022006}{10.1103/PhysRevResearch.2.022006})

\bibitem{Coughlin2020NatCo}
{Coughlin} MW, {Antier} S, {Dietrich} T, {Foley} RJ, {Heinzel} J, {Bulla} M, {Christensen} N, {Coulter} DA, {Issa} L, {Khetan} N. 2020b  {Measuring the Hubble constant with a sample of kilonovae}. {\em Nature Communications} \textbf{11}, 4129.
(\href{http://dx.doi.org/10.1038/s41467-020-17998-5}{10.1038/s41467-020-17998-5})

\bibitem{Abbott2020b}
{LIGO Scientific Collaboration}, {Virgo Collaboration}. 2020  {GW190814: Gravitational Waves from the Coalescence of a 23 Solar Mass Black Hole with a 2.6 Solar Mass Compact Object}. {\em \apjl} \textbf{896}, L44.
(\href{http://dx.doi.org/10.3847/2041-8213/ab960f}{10.3847/2041-8213/ab960f})

\bibitem{Singer2016}
{Singer} LP, {Price} LR. 2016  {Rapid Bayesian position reconstruction for gravitational-wave transients}. {\em \prd} \textbf{93}, 024013.
(\href{http://dx.doi.org/10.1103/PhysRevD.93.024013}{10.1103/PhysRevD.93.024013})

\bibitem{Berry2015}
{Berry} CPL, {Mandel} I, {Middleton} H, {Singer} LP, {Urban} AL, {Vecchio} A, {Vitale} S, {Cannon} K, {Farr} B, {Farr} WM, {Graff} PB, {Hanna} C, {Haster} CJ, {Mohapatra} S, {Pankow} C, {Price} LR, {Sidery} T, {Veitch} J. 2015  {Parameter Estimation for Binary Neutron-star Coalescences with Realistic Noise during the Advanced LIGO Era}. {\em \apj} \textbf{804}, 114.
(\href{http://dx.doi.org/10.1088/0004-637X/804/2/114}{10.1088/0004-637X/804/2/114})

\bibitem{Kiendrebeogo2023}
{Kiendrebeogo} RW, {Farah} AM, {Foley} EM, {Gray} A, {Kunert} N, {Puecher} A, {Toivonen} A, {VandenBerg} RO, {Anand} S, {Ahumada} T, {Karambelkar} V, {Coughlin} MW, {Dietrich} T, {Kam} SZ, {Pang} PTH, {Singer} LP, {Sravan} N. 2023  {Updated Observing Scenarios and Multimessenger Implications for the International Gravitational-wave Networks O4 and O5}. {\em \apj} \textbf{958}, 158.
(\href{http://dx.doi.org/10.3847/1538-4357/acfcb1}{10.3847/1538-4357/acfcb1})

\bibitem{Cabero2020}
{Cabero} M, {Mahabal} A, {McIver} J. 2020  {GWSkyNet: A Real-time Classifier for Public Gravitational-wave Candidates}. {\em \apjl} \textbf{904}, L9.
(\href{http://dx.doi.org/10.3847/2041-8213/abc5b5}{10.3847/2041-8213/abc5b5})

\bibitem{Hosseinzadeh2019}
{Hosseinzadeh} G, {Cowperthwaite} PS, {Gomez} S, {Villar} VA, {Nicholl} M, {Margutti} R, {Berger} E, {Chornock} R, {Paterson} K, {Fong} W, {Savchenko} V, {Short} P, {Alexander} KD, {Blanchard} PK, {Braga} J, {Calkins} ML, {Cartier} R, {Coppejans} DL, {Eftekhari} T, {Laskar} T, {Ly} C, {Patton} L, {Pelisoli} I, {Reichart} DE, {Terreran} G, {Williams} PKG. 2019  {Follow-up of the Neutron Star Bearing Gravitational-wave Candidate Events S190425z and S190426c with MMT and SOAR}. {\em \apjl} \textbf{880}, L4.
(\href{http://dx.doi.org/10.3847/2041-8213/ab271c}{10.3847/2041-8213/ab271c})

\bibitem{Lundquist2019}
{Lundquist} MJ, {Paterson} K, {Fong} W, {Sand} DJ, {Andrews} JE, {Shivaei} I, {Daly} PN, {Valenti} S, {Yang} S, {Christensen} E, {Gibbs} AR, {Shelly} F, {Wyatt} S, {Eskandari} O, {Kuhn} O, {Amaro} RC, {Arcavi} I, {Behroozi} P, {Butler} N, {Chomiuk} L, {Corsi} A, {Drout} MR, {Egami} E, {Fan} X, {Foley} RJ, {Frye} B, {Gabor} P, {Green} EM, {Grier} CJ, {Guzman} F, {Hamden} E, {Howell} DA, {Jannuzi} BT, {Kelly} P, {Milne} P, {Moe} M, {Nugent} A, {Olszewski} E, {Palazzi} E, {Paschalidis} V, {Psaltis} D, {Reichart} D, {Rest} A, {Rossi} A, {Schroeder} G, {Smith} PS, {Smith} N, {Spekkens} K, {Strader} J, {Stark} DP, {Trilling} D, {Veillet} C, {Wagner} M, {Weiner} B, {Wheeler} JC, {Williams} GG, {Zabludoff} A. 2019  {Searches after Gravitational Waves Using ARizona Observatories (SAGUARO): System Overview and First Results from Advanced LIGO/Virgo{\textquoteright}s Third Observing Run}. {\em \apjl} \textbf{881}, L26.
(\href{http://dx.doi.org/10.3847/2041-8213/ab32f2}{10.3847/2041-8213/ab32f2})

\bibitem{Coughlin2019b}
{Coughlin} MW, {Ahumada} T, {Anand} S, {De} K, {Hankins} MJ, {Kasliwal} MM, {Singer} LP, {Bellm} EC, {Andreoni} I, {Cenko} SB, {Cooke} J, {Copperwheat} CM, {Dugas} AM, {Jencson} JE, {Perley} DA, {Yu} PC, {Bhalerao} V, {Kumar} H, {Bloom} JS, {Anupama} GC, {Ashley} MCB, {Bagdasaryan} A, {Biswas} R, {Buckley} DAH, {Burdge} KB, {Cook} DO, {Cromer} J, {Cunningham} V, {D'A{\`\i}} A, {Dekany} RG, {Delacroix} Ar, {Dichiara} S, {Duev} DA, {Dutta} A, {Feeney} M, {Frederick} S, {Gatkine} P, {Ghosh} S, {Goldstein} DA, {Golkhou} VZ, {Goobar} A, {Graham} MJ, {Hanayama} H, {Horiuchi} T, {Hung} T, {Jha} SW, {Kong} AKH, {Giomi} M, {Kaplan} DL, {Karambelkar} VR, {Kowalski} M, {Kulkarni} SR, {Kupfer} T, {Masci} FJ, {Mazzali} P, {Moore} AM, {Mogotsi} M, {Neill} JD, {Ngeow} CC, {Mart{\'\i}nez-Palomera} J, {La Parola} V, {Pavana} M, {Ofek} EO, {Patil} AS, {Riddle} R, {Rigault} M, {Rusholme} B, {Serabyn} E, {Shupe} DL, {Sharma} Y, {Singh} A, {Sollerman} J, {Soon} J, {Staats} K, {Taggart} K, {Tan} H, {Travouillon} T, {Troja} E,
  {Waratkar} G, {Yatsu} Y. 2019  {GROWTH on S190425z: Searching Thousands of Square Degrees to Identify an Optical or Infrared Counterpart to a Binary Neutron Star Merger with the Zwicky Transient Facility and Palomar Gattini-IR}. {\em \apjl} \textbf{885}, L19.
(\href{http://dx.doi.org/10.3847/2041-8213/ab4ad8}{10.3847/2041-8213/ab4ad8})

\bibitem{Antier2020}
{Antier} S, {Agayeva} S, {Aivazyan} V, {Alishov} S, {Arbouch} E, {Baransky} A, {Barynova} K, {Bai} JM, {Basa} S, {Beradze} S, {Bertin} E, {Berthier} J, {Bla{\v{z}}ek} M, {Bo{\"e}r} M, {Burkhonov} O, {Burrell} A, {Cailleau} A, {Chabert} B, {Chen} JC, {Christensen} N, {Coleiro} A, {Cordier} B, {Corre} D, {Coughlin} MW, {Coward} D, {Crisp} H, {Delattre} C, {Dietrich} T, {Ducoin} JG, {Duverne} PA, {Marchal-Duval} G, {Gendre} B, {Eymar} L, {Fock-Hang} P, {Han} X, {Hello} P, {Howell} EJ, {Inasaridze} R, {Ismailov} N, {Kann} DA, {Kapanadze} G, {Klotz} A, {Kochiashvili} N, {Lachaud} C, {Leroy} N, {Le Van Su} A, {Lin} WL, {Li} WX, {Lognone} P, {Marron} R, {Mo} J, {Moore} J, {Natsvlishvili} R, {Noysena} K, {Perrigault} S, {Peyrot} A, {Samadov} D, {Sadibekova} T, {Simon} A, {Stachie} C, {Teng} JP, {Thierry} P, {Th{\"o}ne} CC, {Tillayev} Y, {Turpin} D, {de Ugarte Postigo} A, {Vachier} F, {Vardosanidze} M, {Vasylenko} V, {Vidadi} Z, {Wang} XF, {Wang} CJ, {Wei} J, {Yan} SY, {Zhang} JC, {Zhang} JJ, {Zhang} XH. 2020  {The
  first six months of the Advanced LIGO's and Advanced Virgo's third observing run with GRANDMA}. {\em \mnras} \textbf{492}, 3904--3927.
(\href{http://dx.doi.org/10.1093/mnras/stz3142}{10.1093/mnras/stz3142})

\bibitem{Gompertz2020}
{Gompertz} BP, {Cutter} R, {Steeghs} D, {Galloway} DK, {Lyman} J, {Ulaczyk} K, {Dyer} MJ, {Ackley} K, {Dhillon} VS, {O'Brien} PT, {Ramsay} G, {Poshyachinda} S, {Kotak} R, {Nuttall} L, {Breton} RP, {Pall{\'e}} E, {Pollacco} D, {Thrane} E, {Aukkaravittayapun} S, {Awiphan} S, {Brown} MJI, {Burhanudin} U, {Chote} P, {Chrimes} AA, {Daw} E, {Duffy} C, {Eyles-Ferris} RAJ, {Heikkil{\"a}} T, {Irawati} P, {Kennedy} MR, {Killestein} T, {Levan} AJ, {Littlefair} S, {Makrygianni} L, {Marsh} T, {Mata S{\'a}nchez} D, {Mattila} S, {Maund} J, {McCormac} J, {Mkrtichian} D, {Mong} YL, {Mullaney} J, {M{\"u}ller} B, {Obradovic} A, {Rol} E, {Sawangwit} U, {Stanway} ER, {Starling} RLC, {Str{\o}m} PA, {Tooke} S, {West} R, {Wiersema} K. 2020  {Searching for electromagnetic counterparts to gravitational-wave merger events with the prototype Gravitational-Wave Optical Transient Observer (GOTO-4)}. {\em \mnras} \textbf{497}, 726--738.
(\href{http://dx.doi.org/10.1093/mnras/staa1845}{10.1093/mnras/staa1845})

\bibitem{Paterson2020}
{Paterson} K, {Lundquist} MJ, {Rastinejad} JC, {Fong} W, {Sand} DJ, {Andrews} JE, {Amaro} RC, {Eskandari} O, {Wyatt} S, {Daly} PN, {Bradley} H, {Zhou-Wright} S, {Valenti} S, {Yang} S, {Christensen} E, {Gibbs} AR, {Shelly} F, {Bilinski} C, {Chomiuk} L, {Corsi} A, {Drout} MR, {Foley} RJ, {Gabor} P, {Garnavich} P, {Grier} CJ, {Hamden} E, {Krantz} H, {Olszewski} E, {Paschalidis} V, {Reichart} D, {Rest} A, {Smith} N, {Strader} J, {Trilling} D, {Veillet} C, {Wagner} RM, {Zabludoff} A. 2020  {Searches after Gravitational Waves Using ARizona Observatories (SAGUARO): Observations and Analysis from Advanced LIGO/Virgo's Third Observing Run}. {\em arXiv e-prints} p. arXiv:2012.11700.

\bibitem{Oates2021}
{Oates} SR, {Marshall} FE, {Breeveld} AA, {Kuin} NPM, {Brown} PJ, {De Pasquale} M, {Evans} PA, {Fenney} AJ, {Gronwall} C, {Kennea} JA, {Klingler} NJ, {Page} MJ, {Siegel} MH, {Tohuvavohu} A, {Ambrosi} E, {Barthelmy} SD, {Beardmore} AP, {Bernardini} MG, {Campana} S, {Caputo} R, {Cenko} SB, {Cusumano} G, {D'A{\`\i}} A, {D'Avanzo} P, {D'Elia} V, {Giommi} P, {Hartmann} DH, {Krimm} HA, {Laha} S, {Malesani} DB, {Melandri} A, {Nousek} JA, {O'Brien} PT, {Osborne} JP, {Pagani} C, {Page} KL, {Palmer} DM, {Perri} M, {Racusin} JL, {Sakamoto} T, {Sbarufatti} B, {Schlieder} JE, {Tagliaferri} G, {Troja} E. 2021  {Swift/UVOT follow-up of gravitational wave alerts in the O3 era}. {\em \mnras} \textbf{507}, 1296--1317.
(\href{http://dx.doi.org/10.1093/mnras/stab2189}{10.1093/mnras/stab2189})

\bibitem{Smartt2024}
{Smartt} SJ, {Nicholl} M, {Srivastav} S, {Huber} ME, {Chambers} KC, {Smith} KW, {Young} DR, {Fulton} MD, {Tonry} JL, {Stubbs} CW, {Denneau} L, {Cooper} AJ, {Aamer} A, {Anderson} JP, {Andersson} A, {Bulger} J, {Chen} TW, {Clark} P, {de Boer} T, {Gao} H, {Gillanders} JH, {Lawrence} A, {Lin} CC, {Lowe} TB, {Magnier} EA, {Minguez} P, {Moore} T, {Rest} A, {Shingles} L, {Siverd} R, {Smith} IA, {Stalder} B, {Stevance} HF, {Wainscoat} R, {Williams} R. 2024  {GW190425: Pan-STARRS and ATLAS coverage of the skymap and limits on optical emission associated with FRB 20190425A}. {\em \mnras} \textbf{528}, 2299--2307.
(\href{http://dx.doi.org/10.1093/mnras/stae100}{10.1093/mnras/stae100})

\bibitem{Rastinejad2022b}
{Rastinejad} JC, {Paterson} K, {Fong} W, {Sand} DJ, {Lundquist} MJ, {Hosseinzadeh} G, {Christensen} E, {Daly} PN, {Gibbs} AR, {Hall} S, {Shelly} F, {Yang} S. 2022  {A Systematic Exploration of Kilonova Candidates from Neutron Star Mergers during the Third Gravitational-wave Observing Run}. {\em \apj} \textbf{927}, 50.
(\href{http://dx.doi.org/10.3847/1538-4357/ac4d34}{10.3847/1538-4357/ac4d34})

\bibitem{gcn24204}
{Perley} DA, {Copperwheat} CM, {Taggart} KL. 2019  {LIGO/Virgo S190425z: Liverpool Telescope spectroscopy of ZTF19aarykkb.}. {\em GCN} \textbf{24204}, 1.

\bibitem{gcn24205}
{Buckley} DAH, {Jha} SW, {Cooke} J, {Mogotsi} M. 2019  {LIGO/Virgo S190425z: SALT spectroscopy of ZTF19aarzaod as a likely type II supernova.}. {\em GCN} \textbf{24205}, 1.

\bibitem{gcn24215}
{Short} P, {Nicholl} M, {Smartt} SJ, {Smith} KW, {Chambers} K, {Huber} M, {Anderson} J, {Chen} TW, {Inserra} C, {Yaron} O. 2019  {LIGO/Virgo S190425z - ePESSTO+ NTT spectrum of candidate PS19qo.}. {\em GCN} \textbf{24215}, 1.

\bibitem{gcn24230}
{Morokuma} T, {Ohta} K, {Yoshida} M, {Aoki} K, {Tanaka} M, {Sasada} M, {Nakaoka} T, {Akitaya} H, {Kawabata} KS, {Itoh} R. 2019  {LIGO/Virgo S190425z: J-GEM spectroscopic observations of AT2019ebq/PS19qp with Subaru/FOCAS.}. {\em GCN} \textbf{24230}, 1.

\bibitem{gcn24233}
{Jencson} J, {De} K, {Anand} S, {Kasliwal} MM, {Andreoni} I, {Ahumada} T, {Perley} DA. 2019  {LIGO/Virgo S190425z: Keck NIR spectroscopy shows AT2019ebq is a supernova.}. {\em GCN} \textbf{24233}, 1.

\bibitem{gcn24269}
{Short} P, {Nicholl} M, {Anderson} J, {Chen} TW, {Inserra} C, {Yaron} O, {Young} DR, {Angus} C, {Pursiainen} M, {Wiseman} P. 2019  {LIGO/Virgo S190426c - ePESSTO+ spectrum of candidate PS19qu.}. {\em GCN} \textbf{24269}, 1.

\bibitem{gcn24321}
{Nicholl} M, {Cartier} R, {Pelisoli} I, {Berger} E, {Blanchard} P, {Eftekhari} T, {Gomez} S, {Hosseinzadeh} G, {Villar} A, {Williams} P. 2019  {LIGO/Virgo S190425z: Spectroscopic observations of two ZTF candidates with SOAR.}. {\em GCN} \textbf{24321}, 1.

\bibitem{Barbieri2019}
{Barbieri} C, {Salafia} OS, {Perego} A, {Colpi} M, {Ghirlanda} G. 2019  {Light-curve models of black hole - neutron star mergers: steps towards a multi-messenger parameter estimation}. {\em \aap} \textbf{625}, A152.
(\href{http://dx.doi.org/10.1051/0004-6361/201935443}{10.1051/0004-6361/201935443})

\bibitem{Kruger2020}
{Kr{\"u}ger} CJ, {Foucart} F. 2020  {Estimates for disk and ejecta masses produced in compact binary mergers}. {\em \prd} \textbf{101}, 103002.
(\href{http://dx.doi.org/10.1103/PhysRevD.101.103002}{10.1103/PhysRevD.101.103002})

\bibitem{Mochkovitch2021}
{Mochkovitch} R, {Daigne} F, {Duque} R, {Zitouni} H. 2021  {Prospects for kilonova signals in the gravitational-wave era}. {\em \aap} \textbf{651}, A83.
(\href{http://dx.doi.org/10.1051/0004-6361/202140689}{10.1051/0004-6361/202140689})

\bibitem{Nedora2022}
{Nedora} V, {Schianchi} F, {Bernuzzi} S, {Radice} D, {Daszuta} B, {Endrizzi} A, {Perego} A, {Prakash} A, {Zappa} F. 2022  {Mapping dynamical ejecta and disk masses from numerical relativity simulations of neutron star mergers}. {\em Classical and Quantum Gravity} \textbf{39}, 015008.
(\href{http://dx.doi.org/10.1088/1361-6382/ac35a8}{10.1088/1361-6382/ac35a8})

\bibitem{Gompertz2023a}
{Gompertz} BP, {Ravasio} ME, {Nicholl} M, {Levan} AJ, {Metzger} BD, {Oates} SR, {Lamb} GP, {Fong} Wf, {Malesani} DB, {Rastinejad} JC, {Tanvir} NR, {Evans} PA, {Jonker} PG, {Page} KL, {Pe'er} A. 2023  {The case for a minute-long merger-driven gamma-ray burst from fast-cooling synchrotron emission}. {\em Nature Astronomy} \textbf{7}, 67--79.
(\href{http://dx.doi.org/10.1038/s41550-022-01819-4}{10.1038/s41550-022-01819-4})

\bibitem{Setzer2023}
{Setzer} CN, {Peiris} HV, {Korobkin} O, {Rosswog} S. 2023  {Modelling populations of kilonovae}. {\em \mnras} \textbf{520}, 2829--2842.
(\href{http://dx.doi.org/10.1093/mnras/stad257}{10.1093/mnras/stad257})

\bibitem{Foley2020}
{Foley} RJ, {Coulter} DA, {Kilpatrick} CD, {Piro} AL, {Ramirez-Ruiz} E, {Schwab} J. 2020  {Updated parameter estimates for GW190425 using astrophysical arguments and implications for the electromagnetic counterpart}. {\em \mnras} \textbf{494}, 190--198.
(\href{http://dx.doi.org/10.1093/mnras/staa725}{10.1093/mnras/staa725})

\bibitem{Sagues2021}
{Sagu{\'e}s Carracedo} A, {Bulla} M, {Feindt} U, {Goobar} A. 2021  {Detectability of kilonovae in optical surveys: post-mortem examination of the LVC O3 run follow-up}. {\em \mnras} \textbf{504}, 1294--1303.
(\href{http://dx.doi.org/10.1093/mnras/stab872}{10.1093/mnras/stab872})

\bibitem{Raaijmakers2021a}
{Raaijmakers} G, {Nissanke} S, {Foucart} F, {Kasliwal} MM, {Bulla} M, {Fern{\'a}ndez} R, {Henkel} A, {Hinderer} T, {Hotokezaka} K, {Luko{\v{s}}i{\={u}}t{\.{e}}} K, {Venumadhav} T, {Antier} S, {Coughlin} MW, {Dietrich} T, {Edwards} TDP. 2021  {The Challenges Ahead for Multimessenger Analyses of Gravitational Waves and Kilonova: A Case Study on GW190425}. {\em \apj} \textbf{922}, 269.
(\href{http://dx.doi.org/10.3847/1538-4357/ac222d}{10.3847/1538-4357/ac222d})

\bibitem{Barbieri2021}
{Barbieri} C, {Salafia} OS, {Colpi} M, {Ghirlanda} G, {Perego} A. 2021  {Exploring the nature of ambiguous merging systems: GW190425 in low latency}. {\em \aap} \textbf{654}, A12.
(\href{http://dx.doi.org/10.1051/0004-6361/202037778}{10.1051/0004-6361/202037778})

\bibitem{Bulla2023}
{Bulla} M. 2023  {The critical role of nuclear heating rates, thermalization efficiencies, and opacities for kilonova modelling and parameter inference}. {\em \mnras} \textbf{520}, 2558--2570.
(\href{http://dx.doi.org/10.1093/mnras/stad232}{10.1093/mnras/stad232})

\bibitem{Smith2023}
{Smith} GP, {Robertson} A, {Mahler} G, {Nicholl} M, {Ryczanowski} D, {Bianconi} M, {Sharon} K, {Massey} R, {Richard} J, {Jauzac} M. 2023  {Discovering gravitationally lensed gravitational waves: predicted rates, candidate selection, and localization with the Vera Rubin Observatory}. {\em \mnras} \textbf{520}, 702--721.
(\href{http://dx.doi.org/10.1093/mnras/stad140}{10.1093/mnras/stad140})

\bibitem{Bianconi2023}
{Bianconi} M, {Smith} GP, {Nicholl} M, {Ryczanowski} D, {Richard} J, {Jauzac} M, {Massey} R, {Robertson} A, {Sharon} K, {Ridley} E. 2023  {On the gravitational lensing interpretation of three gravitational wave detections in the mass gap by LIGO and Virgo}. {\em \mnras} \textbf{521}, 3421--3430.
(\href{http://dx.doi.org/10.1093/mnras/stad673}{10.1093/mnras/stad673})

\bibitem{Moroianu2023}
{Moroianu} A, {Wen} L, {James} CW, {Ai} S, {Kovalam} M, {Panther} FH, {Zhang} B. 2023  {An assessment of the association between a fast radio burst and binary neutron star merger}. {\em Nature Astronomy} \textbf{7}, 579--589.
(\href{http://dx.doi.org/10.1038/s41550-023-01917-x}{10.1038/s41550-023-01917-x})

\bibitem{Panther2023}
{Panther} FH, {Anderson} GE, {Bhandari} S, {Goodwin} AJ, {Hurley-Walker} N, {James} CW, {Kawka} A, {Ai} S, {Kovalam} M, {Moroianu} A, {Wen} L, {Zhang} B. 2023  {The most probable host of CHIME FRB 190425A, associated with binary neutron star merger GW190425, and a late-time transient search}. {\em \mnras} \textbf{519}, 2235--2250.
(\href{http://dx.doi.org/10.1093/mnras/stac3597}{10.1093/mnras/stac3597})

\bibitem{Price2006}
{Price} DJ, {Rosswog} S. 2006  {Producing Ultrastrong Magnetic Fields in Neutron Star Mergers}. {\em Science} \textbf{312}, 719--722.
(\href{http://dx.doi.org/10.1126/science.1125201}{10.1126/science.1125201})

\bibitem{Yu2013}
{Yu} YW, {Zhang} B, {Gao} H. 2013  {Bright ``Merger-nova'' from the Remnant of a Neutron Star Binary Merger: A Signature of a Newly Born, Massive, Millisecond Magnetar}. {\em \apjl} \textbf{776}, L40.
(\href{http://dx.doi.org/10.1088/2041-8205/776/2/L40}{10.1088/2041-8205/776/2/L40})

\bibitem{Zrake2013}
{Zrake} J, {MacFadyen} AI. 2013  {Magnetic Energy Production by Turbulence in Binary Neutron Star Mergers}. {\em \apjl} \textbf{769}, L29.
(\href{http://dx.doi.org/10.1088/2041-8205/769/2/L29}{10.1088/2041-8205/769/2/L29})

\bibitem{Metzger2019}
{Metzger} BD. 2019  {Kilonovae}. {\em Living Reviews in Relativity} \textbf{23}, 1.
(\href{http://dx.doi.org/10.1007/s41114-019-0024-0}{10.1007/s41114-019-0024-0})

\bibitem{Sarin2022}
{Sarin} N, {Omand} CMB, {Margalit} B, {Jones} DI. 2022  {On the diversity of magnetar-driven kilonovae}. {\em \mnras} \textbf{516}, 4949--4962.
(\href{http://dx.doi.org/10.1093/mnras/stac2609}{10.1093/mnras/stac2609})

\bibitem{Kiuchi2023}
{Kiuchi} K, {Reboul-Salze} A, {Shibata} M, {Sekiguchi} Y. 2023  {A large-scale magnetic field via $\alpha\Omega$ dynamo in binary neutron star mergers}. {\em arXiv e-prints} p. arXiv:2306.15721.
(\href{http://dx.doi.org/10.48550/arXiv.2306.15721}{10.48550/arXiv.2306.15721})

\bibitem{MaganaHernandez2024}
{Maga{\~n}a Hernandez} I, {d'Emilio} V, {Morisaki} S, {Bhardwaj} M, {Palmese} A. 2024  {On the Association of GW190425 with Its Potential Electromagnetic Counterpart FRB 20190425A}. {\em \apjl} \textbf{971}, L5.
(\href{http://dx.doi.org/10.3847/2041-8213/ad5b4c}{10.3847/2041-8213/ad5b4c})

\bibitem{Radice2024}
{Radice} D, {Ricigliano} G, {Bhattacharya} M, {Perego} A, {Fattoyev} FJ, {Murase} K. 2024  {What if GW190425 did not produce a black hole promptly?}. {\em \mnras} \textbf{528}, 5836--5844.
(\href{http://dx.doi.org/10.1093/mnras/stae400}{10.1093/mnras/stae400})

\bibitem{Fattoyev2020}
{Fattoyev} FJ, {Horowitz} CJ, {Piekarewicz} J, {Reed} B. 2020  {GW190814: Impact of a 2.6 solar mass neutron star on the nucleonic equations of state}. {\em \prc} \textbf{102}, 065805.
(\href{http://dx.doi.org/10.1103/PhysRevC.102.065805}{10.1103/PhysRevC.102.065805})

\bibitem{Tsokaros2020}
{Tsokaros} A, {Ruiz} M, {Shapiro} SL. 2020  {GW190814: Spin and Equation of State of a Neutron Star Companion}. {\em \apj} \textbf{905}, 48.
(\href{http://dx.doi.org/10.3847/1538-4357/abc421}{10.3847/1538-4357/abc421})

\bibitem{Palmese2020}
{Palmese} A, {deVicente} J, {Pereira} MES, {Annis} J, {Hartley} W, {Herner} K, {Soares-Santos} M, {Crocce} M, {Huterer} D, {Maga{\~n}a Hernandez} I, {Garcia} A, {Garcia-Bellido} J, {Gschwend} J, {Holz} DE, {Kessler} R, {Lahav} O, {Morgan} R, {Nicolaou} C, {Conselice} C, {Foley} RJ, {Gill} MSS, {Abbott} TMC, {Aguena} M, {Allam} S, {Avila} S, {Bechtol} K, {Bertin} E, {Bhargava} S, {Brooks} D, {Buckley-Geer} E, {Burke} DL, {Carrasco Kind} M, {Carretero} J, {Castander} FJ, {Chang} C, {Costanzi} M, {da Costa} LN, {Davis} TM, {Desai} S, {Diehl} HT, {Doel} P, {Drlica-Wagner} A, {Estrada} J, {Everett} S, {Evrard} AE, {Fernandez} E, {Finley} DA, {Flaugher} B, {Fosalba} P, {Frieman} J, {Gaztanaga} E, {Gerdes} DW, {Gruen} D, {Gruendl} RA, {Gutierrez} G, {Hinton} SR, {Hollowood} DL, {Honscheid} K, {James} DJ, {Kent} S, {Krause} E, {Kuehn} K, {Lin} H, {Maia} MAG, {March} M, {Marshall} JL, {Melchior} P, {Menanteau} F, {Miquel} R, {Ogando} RLC, {Paz-Chinch{\'o}n} F, {Plazas} AA, {Roodman} A, {Sako} M, {Sanchez} E,
  {Scarpine} V, {Schubnell} M, {Serrano} S, {Sevilla-Noarbe} I, {Smith} JA, {Smith} M, {Suchyta} E, {Tarle} G, {Troxel} MA, {Tucker} DL, {Walker} AR, {Wester} W, {Wilkinson} RD, {Zuntz} J, {DES Collaboration}. 2020  {A Statistical Standard Siren Measurement of the Hubble Constant from the LIGO/Virgo Gravitational Wave Compact Object Merger GW190814 and Dark Energy Survey Galaxies}. {\em \apjl} \textbf{900}, L33.
(\href{http://dx.doi.org/10.3847/2041-8213/abaeff}{10.3847/2041-8213/abaeff})

\bibitem{Vasylyev2020}
{Vasylyev} SS, {Filippenko} AV. 2020  {A Measurement of the Hubble Constant Using Gravitational Waves from the Binary Merger GW190814}. {\em \apj} \textbf{902}, 149.
(\href{http://dx.doi.org/10.3847/1538-4357/abb5f9}{10.3847/1538-4357/abb5f9})

\bibitem{Dobie2019}
{Dobie} D, {Stewart} A, {Murphy} T, {Lenc} E, {Wang} Z, {Kaplan} DL, {Andreoni} I, {Banfield} J, {Brown} I, {Corsi} A, {De} K, {Goldstein} DA, {Hallinan} G, {Hotan} A, {Hotokezaka} K, {Jaodand} AD, {Karambelkar} V, {Kasliwal} MM, {McConnell} D, {Mooley} K, {Moss} VA, {Newman} JA, {Perley} DA, {Prakash} A, {Pritchard} J, {Sadler} EM, {Sharma} Y, {Ward} C, {Whiting} M, {Zhou} R. 2019  {An ASKAP Search for a Radio Counterpart to the First High-significance Neutron Star-Black Hole Merger LIGO/Virgo S190814bv}. {\em \apjl} \textbf{887}, L13.
(\href{http://dx.doi.org/10.3847/2041-8213/ab59db}{10.3847/2041-8213/ab59db})

\bibitem{Andreoni2020bv}
{Andreoni} I, {Goldstein} DA, {Kasliwal} MM, {Nugent} PE, {Zhou} R, {Newman} JA, {Bulla} M, {Foucart} F, {Hotokezaka} K, {Nakar} E, {Nissanke} S, {Raaijmakers} G, {Bloom} JS, {De} K, {Jencson} JE, {Ward} C, {Ahumada} T, {Anand} S, {Buckley} DAH, {Caballero-Garc{\'\i}a} MD, {Castro-Tirado} AJ, {Copperwheat} CM, {Coughlin} MW, {Cenko} SB, {Gromadzki} M, {Hu} Y, {Karambelkar} VR, {Perley} DA, {Sharma} Y, {Valeev} AF, {Cook} DO, {Fremling} UC, {Kumar} H, {Taggart} K, {Bagdasaryan} A, {Cooke} J, {Dahiwale} A, {Dhawan} S, {Dobie} D, {Gatkine} P, {Golkhou} VZ, {Goobar} A, {Chaves} AG, {Hankins} M, {Kaplan} DL, {Kong} AKH, {Kool} EC, {Mohite} S, {Sollerman} J, {Tzanidakis} A, {Webb} S, {Zhang} K. 2020  {GROWTH on S190814bv: Deep Synoptic Limits on the Optical/Near-infrared Counterpart to a Neutron Star-Black Hole Merger}. {\em \apj} \textbf{890}, 131.
(\href{http://dx.doi.org/10.3847/1538-4357/ab6a1b}{10.3847/1538-4357/ab6a1b})

\bibitem{Morgan2020}
{Morgan} R, {Soares-Santos} M, {Annis} J, {Herner} K, {Garcia} A, {Palmese} A, {Drlica-Wagner} A, {Kessler} R, {Garc{\'\i}a-Bellido} J, {Bachmann} TG, {Sherman} N, {Allam} S, {Bechtol} K, {Bom} CR, {Brout} D, {Butler} RE, {Butner} M, {Cartier} R, {Chen} H, {Conselice} C, {Cook} E, {Davis} TM, {Doctor} Z, {Farr} B, {Figueiredo} AL, {Finley} DA, {Foley} RJ, {Galarza} JY, {Gill} MSS, {Gruendl} RA, {Holz} DE, {Kuropatkin} N, {Lidman} C, {Lin} H, {Malik} U, {Mann} AW, {Marriner} J, {Marshall} JL, {Mart{\'\i}nez-V{\'a}zquez} CE, {Meza} N, {Neilsen} E, {Nicolaou} C, {Olivares E.} F, {Paz-Chinch{\'o}n} F, {Points} S, {Quirola-V{\'a}squez} J, {Rodriguez} O, {Sako} M, {Scolnic} D, {Smith} M, {Sobreira} F, {Tucker} DL, {Vivas} AK, {Wiesner} M, {Wood} ML, {Yanny} B, {Zenteno} A, {Abbott} TMC, {Aguena} M, {Avila} S, {Bertin} E, {Bhargava} S, {Brooks} D, {Burke} DL, {Rosell} AC, {Kind} MC, {Carretero} J, {Costa} LNd, {Costanzi} M, {De Vicente} J, {Desai} S, {Diehl} HT, {Doel} P, {Eifler} TF, {Everett} S, {Flaugher} B,
  {Frieman} J, {Gaztanaga} E, {Gerdes} DW, {Gruen} D, {Gschwend} J, {Gutierrez} G, {Hartley} WG, {Hinton} SR, {Hollowood} DL, {Honscheid} K, {James} DJ, {Kuehn} K, {Lahav} O, {Lima} M, {Maia} MAG, {March} M, {Miquel} R, {Ogando} RLC, {Plazas} AA, {Roodman} A, {Sanchez} E, {Scarpine} V, {Schubnell} M, {Serrano} S, {Sevilla-Noarbe} I, {Suchyta} E, {Tarle} G. 2020  {Constraints on the Physical Properties of GW190814 through Simulations Based on DECam Follow-up Observations by the Dark Energy Survey}. {\em \apj} \textbf{901}, 83.
(\href{http://dx.doi.org/10.3847/1538-4357/abafaa}{10.3847/1538-4357/abafaa})

\bibitem{Kasliwal2020}
{Kasliwal} MM, {Anand} S, {Ahumada} T, {Stein} R, {Carracedo} AS, {Andreoni} I, {Coughlin} MW, {Singer} LP, {Kool} EC, {De} K, {Kumar} H, {AlMualla} M, {Yao} Y, {Bulla} M, {Dobie} D, {Reusch} S, {Perley} DA, {Cenko} SB, {Bhalerao} V, {Kaplan} DL, {Sollerman} J, {Goobar} A, {Copperwheat} CM, {Bellm} EC, {Anupama} GC, {Corsi} A, {Nissanke} S, {Agudo} I, {Bagdasaryan} A, {Barway} S, {Belicki} J, {Bloom} JS, {Bolin} B, {Buckley} DAH, {Burdge} KB, {Burruss} R, {Caballero-Garc{\'\i}a} MD, {Cannella} C, {Castro-Tirado} AJ, {Cook} DO, {Cooke} J, {Cunningham} V, {Dahiwale} A, {Deshmukh} K, {Dichiara} S, {Duev} DA, {Dutta} A, {Feeney} M, {Franckowiak} A, {Frederick} S, {Fremling} C, {Gal-Yam} A, {Gatkine} P, {Ghosh} S, {Goldstein} DA, {Golkhou} VZ, {Graham} MJ, {Graham} ML, {Hankins} MJ, {Helou} G, {Hu} Y, {Ip} WH, {Jaodand} A, {Karambelkar} V, {Kong} AKH, {Kowalski} M, {Khandagale} M, {Kulkarni} SR, {Kumar} B, {Laher} RR, {Li} KL, {Mahabal} A, {Masci} FJ, {Miller} AA, {Mogotsi} M, {Mohite} S, {Mooley} K, {Mroz} P,
  {Newman} JA, {Ngeow} CC, {Oates} SR, {Patil} AS, {Pandey} SB, {Pavana} M, {Pian} E, {Riddle} R, {S{\'a}nchez-Ram{\'\i}rez} R, {Sharma} Y, {Singh} A, {Smith} R, {Soumagnac} MT, {Taggart} K, {Tan} H, {Tzanidakis} A, {Troja} E, {Valeev} AF, {Walters} R, {Waratkar} G, {Webb} S, {Yu} PC, {Zhang} BB, {Zhou} R, {Zolkower} J. 2020  {Kilonova Luminosity Function Constraints Based on Zwicky Transient Facility Searches for 13 Neutron Star Merger Triggers during O3}. {\em \apj} \textbf{905}, 145.
(\href{http://dx.doi.org/10.3847/1538-4357/abc335}{10.3847/1538-4357/abc335})

\bibitem{Vieira2020}
{Vieira} N, {Ruan} JJ, {Haggard} D, {Drout} MR, {Nynka} MC, {Boyce} H, {Spekkens} K, {Safi-Harb} S, {Carlberg} RG, {Fern{\'a}ndez} R, {Piro} AL, {Afsariardchi} N, {Moon} DS. 2020  {A Deep CFHT Optical Search for a Counterpart to the Possible Neutron Star-Black Hole Merger GW190814}. {\em \apj} \textbf{895}, 96.
(\href{http://dx.doi.org/10.3847/1538-4357/ab917d}{10.3847/1538-4357/ab917d})

\bibitem{Watson2020}
{Watson} AM, {Butler} NR, {Lee} WH, {Becerra} RL, {Pereyra} M, {Angeles} F, {Farah} A, {Figueroa} L, {G{\'o}nzalez-Buitrago} D, {Quir{\'o}s} F, {Ru{\'\i}z-D{\'\i}az-Soto} J, {Tejada de Vargas} C, {Tinoco} SJ, {Wolfram} T. 2020  {Limits on the electromagnetic counterpart to S190814bv}. {\em \mnras} \textbf{492}, 5916--5921.
(\href{http://dx.doi.org/10.1093/mnras/staa161}{10.1093/mnras/staa161})

\bibitem{Becerra2021MNRAS}
{Becerra} RL, {Dichiara} S, {Watson} AM, {Troja} E, {Butler} NR, {Pereyra} M, {Moreno M{\'e}ndez} E, {De Colle} F, {Lee} WH, {Kutyrev} AS, {L{\'o}pez} KOC. 2021  {DDOTI observations of gravitational-wave sources discovered in O3}. {\em \mnras} \textbf{507}, 1401--1420.
(\href{http://dx.doi.org/10.1093/mnras/stab2086}{10.1093/mnras/stab2086})

\bibitem{Chang2021}
{Chang} SW, {Onken} CA, {Wolf} C, {Luvaul} L, {M{\"o}ller} A, {Scalzo} R, {Schmidt} BP, {Scott} SM, {Sura} N, {Yuan} F. 2021  {SkyMapper optical follow-up of gravitational wave triggers: Alert science data pipeline and LIGO/Virgo O3 run}. {\em \pasa} \textbf{38}, e024.
(\href{http://dx.doi.org/10.1017/pasa.2021.17}{10.1017/pasa.2021.17})

\bibitem{deWet2021}
{de Wet} S, {Groot} PJ, {Bloemen} S, {Le Poole} R, {Klein-Wolt} M, {K{\"o}rding} E, {McBride} V, {Paterson} K, {Pieterse} DLA, {Vreeswijk} PM, {Woudt} P. 2021  {GW190814 follow-up with the optical telescope MeerLICHT}. {\em \aap} \textbf{649}, A72.
(\href{http://dx.doi.org/10.1051/0004-6361/202040231}{10.1051/0004-6361/202040231})

\bibitem{Dobie2022}
{Dobie} D, {Stewart} A, {Hotokezaka} K, {Murphy} T, {Kaplan} DL, {Buckley} DAH, {Cooke} J, {Ho} AYQ, {Lenc} E, {Leung} JK, {Gromadzki} M, {O'Brien} A, {Pintaldi} S, {Pritchard} J, {Wang} Y, {Wang} Z. 2022  {A comprehensive search for the radio counterpart of GW190814 with the Australian Square Kilometre Array Pathfinder}. {\em \mnras} \textbf{510}, 3794--3805.
(\href{http://dx.doi.org/10.1093/mnras/stab3628}{10.1093/mnras/stab3628})

\bibitem{Gomez2019}
{Gomez} S, {Hosseinzadeh} G, {Cowperthwaite} PS, {Villar} VA, {Berger} E, {Gardner} T, {Alexander} KD, {Blanchard} PK, {Chornock} R, {Drout} MR, {Eftekhari} T, {Fong} W, {Gill} K, {Margutti} R, {Nicholl} M, {Paterson} K, {Williams} PKG. 2019  {A Galaxy-targeted Search for the Optical Counterpart of the Candidate NS-BH Merger S190814bv with Magellan}. {\em \apjl} \textbf{884}, L55.
(\href{http://dx.doi.org/10.3847/2041-8213/ab4ad5}{10.3847/2041-8213/ab4ad5})

\bibitem{Page2020}
{Page} KL, {Evans} PA, {Tohuvavohu} A, {Kennea} JA, {Klingler} NJ, {Cenko} SB, {Oates} SR, {Ambrosi} E, {Barthelmy} SD, {Beardmore} AP, {Bernardini} MG, {Breeveld} AA, {Brown} PJ, {Burrows} DN, {Campana} S, {Caputo} R, {Cusumano} G, {D'A{\`\i}} A, {D'Avanzo} P, {D'Elia} V, {De Pasquale} M, {Emery} SWK, {Giommi} P, {Gronwall} C, {Hartmann} DH, {Krimm} HA, {Kuin} NPM, {Malesani} DB, {Marshall} FE, {Melandri} A, {Nousek} JA, {O'Brien} PT, {Osborne} JP, {Pagani} C, {Page} MJ, {Palmer} DM, {Perri} M, {Racusin} JL, {Sakamoto} T, {Sbarufatti} B, {Schlieder} JE, {Siegel} MH, {Tagliaferri} G, {Troja} E. 2020  {Swift-XRT follow-up of gravitational wave triggers during the third aLIGO/Virgo observing run}. {\em \mnras} \textbf{499}, 3459--3480.
(\href{http://dx.doi.org/10.1093/mnras/staa3032}{10.1093/mnras/staa3032})

\bibitem{Alexander2021}
{Alexander} KD, {Schroeder} G, {Paterson} K, {Fong} W, {Cowperthwaite} P, {Gomez} S, {Margalit} B, {Margutti} R, {Berger} E, {Blanchard} P, {Chornock} R, {Eftekhari} T, {Laskar} T, {Metzger} BD, {Nicholl} M, {Villar} VA, {Williams} PKG. 2021  {A Late-time Galaxy-targeted Search for the Radio Counterpart of GW190814}. {\em \apj} \textbf{923}, 66.
(\href{http://dx.doi.org/10.3847/1538-4357/ac281a}{10.3847/1538-4357/ac281a})

\bibitem{Ackley2020}
{Ackley} K, {Amati} L, {Barbieri} C, {Bauer} FE, {Benetti} S, {Bernardini} MG, {Bhirombhakdi} K, {Botticella} MT, {Branchesi} M, {Brocato} E, {Bruun} SH, {Bulla} M, {Campana} S, {Cappellaro} E, {Castro-Tirado} AJ, {Chambers} KC, {Chaty} S, {Chen} TW, {Ciolfi} R, {Coleiro} A, {Copperwheat} CM, {Covino} S, {Cutter} R, {D'Ammando} F, {D'Avanzo} P, {De Cesare} G, {D'Elia} V, {Della Valle} M, {Denneau} L, {De Pasquale} M, {Dhillon} VS, {Dyer} MJ, {Elias-Rosa} N, {Evans} PA, {Eyles-Ferris} RAJ, {Fiore} A, {Fraser} M, {Fruchter} AS, {Fynbo} JPU, {Galbany} L, {Gall} C, {Galloway} DK, {Getman} FI, {Ghirlanda} G, {Gillanders} JH, {Gomboc} A, {Gompertz} BP, {Gonz{\'a}lez-Fern{\'a}ndez} C, {Gonz{\'a}lez-Gait{\'a}n} S, {Grado} A, {Greco} G, {Gromadzki} M, {Groot} PJ, {Guti{\'e}rrez} CP, {Heikkil{\"a}} T, {Heintz} KE, {Hjorth} J, {Hu} YD, {Huber} ME, {Inserra} C, {Izzo} L, {Japelj} J, {Jerkstrand} A, {Jin} ZP, {Jonker} PG, {Kankare} E, {Kann} DA, {Kennedy} M, {Kim} S, {Klose} S, {Kool} EC, {Kotak} R, {Kuncarayakti} H,
  {Lamb} GP, {Leloudas} G, {Levan} AJ, {Longo} F, {Lowe} TB, {Lyman} JD, {Magnier} E, {Maguire} K, {Maiorano} E, {Mandel} I, {Mapelli} M, {Mattila} S, {McBrien} OR, {Melandri} A, {Micha{\l}owski} MJ, {Milvang-Jensen} B, {Moran} S, {Nicastro} L, {Nicholl} M, {Nicuesa Guelbenzu} A, {Nuttal} L, {Oates} SR, {O'Brien} PT, {Onori} F, {Palazzi} E, {Patricelli} B, {Perego} A, {Torres} MAP, {Perley} DA, {Pian} E, {Pignata} G, {Piranomonte} S, {Poshyachinda} S, {Possenti} A, {Pumo} ML, {Quirola-V{\'a}squez} J, {Ragosta} F, {Ramsay} G, {Rau} A, {Rest} A, {Reynolds} TM, {Rosetti} SS, {Rossi} A, {Rosswog} S, {Sabha} NB, {Sagu{\'e}s Carracedo} A, {Salafia} OS, {Salmon} L, {Salvaterra} R, {Savaglio} S, {Sbordone} L, {Schady} P, {Schipani} P, {Schultz} ASB, {Schweyer} T, {Smartt} SJ, {Smith} KW, {Smith} M, {Sollerman} J, {Srivastav} S, {Stanway} ER, {Starling} RLC, {Steeghs} D, {Stratta} G, {Stubbs} CW, {Tanvir} NR, {Testa} V, {Thrane} E, {Tonry} JL, {Turatto} M, {Ulaczyk} K, {van der Horst} AJ, {Vergani} SD, {Walton} NA,
  {Watson} D, {Wiersema} K, {Wiik} K, {Wyrzykowski} {\L}, {Yang} S, {Yi} SX, {Young} DR. 2020  {Observational constraints on the optical and near-infrared emission from the neutron star-black hole binary merger candidate S190814bv}. {\em \aap} \textbf{643}, A113.
(\href{http://dx.doi.org/10.1051/0004-6361/202037669}{10.1051/0004-6361/202037669})

\bibitem{Thakur2020}
{Thakur} AL, {Dichiara} S, {Troja} E, {Chase} EA, {S{\'a}nchez-Ram{\'\i}rez} R, {Piro} L, {Fryer} CL, {Butler} NR, {Watson} AM, {Wollaeger} RT, {Ambrosi} E, {Becerra Gonz{\'a}lez} J, {Becerra} RL, {Bruni} G, {Cenko} SB, {Cusumano} G, {D'A{\`\i}} A, {Durbak} J, {Fontes} CJ, {Gatkine} P, {Hungerford} AL, {Korobkin} O, {Kutyrev} AS, {Lee} WH, {Lotti} S, {Minervini} G, {Novara} G, {Parola} VL, {Pereyra} M, {Ricci} R, {Tiengo} A, {Veilleux} S. 2020  {A search for optical and near-infrared counterparts of the compact binary merger GW190814}. {\em \mnras} \textbf{499}, 3868--3883.
(\href{http://dx.doi.org/10.1093/mnras/staa2798}{10.1093/mnras/staa2798})

\bibitem{Kilpatrick2021}
{Kilpatrick} CD, {Coulter} DA, {Arcavi} I, {Brink} TG, {Dimitriadis} G, {Filippenko} AV, {Foley} RJ, {Howell} DA, {Jones} DO, {Kasen} D, {Makler} M, {Piro} AL, {Rojas-Bravo} C, {Sand} DJ, {Swift} JJ, {Tucker} D, {Zheng} W, {Allam} SS, {Annis} JT, {Antilen} J, {Bachmann} TG, {Bloom} JS, {Bom} CR, {Bostroem} KA, {Brout} D, {Burke} J, {Butler} RE, {Butner} M, {Campillay} A, {Clever} KE, {Conselice} CJ, {Cooke} J, {Dage} KC, {de Carvalho} RR, {de Jaeger} T, {Desai} S, {Garcia} A, {Garcia-Bellido} J, {Gill} MSS, {Girish} N, {Hallakoun} N, {Herner} K, {Hiramatsu} D, {Holz} DE, {Huber} G, {Kawash} AM, {McCully} C, {Medallon} SA, {Metzger} BD, {Modak} S, {Morgan} R, {Mu{\~n}oz} RR, {Mu{\~n}oz-Elgueta} N, {Murakami} YS, {Felipe Olivares} E, {Palmese} A, {Patra} KC, {Pereira} MES, {Pessi} TL, {Pineda-Garcia} J, {Quirola-V{\'a}squez} J, {Ramirez-Ruiz} E, {Rembold} SB, {Rest} A, {Rodr{\'\i}guez} {\'O}, {Santana-Silva} L, {Sherman} NF, {Siebert} MR, {Smith} C, {Smith} JA, {Soares-Santos} M, {Stacey} H, {Stahl} BE,
  {Strader} J, {Strasburger} E, {Sunseri} J, {Tinyanont} S, {Tucker} BE, {Ulloa} N, {Valenti} S, {Vasylyev} SS, {Wiesner} MP, {Zhang} KD. 2021  {The Gravity Collective: A Search for the Electromagnetic Counterpart to the Neutron Star-Black Hole Merger GW190814}. {\em \apj} \textbf{923}, 258.
(\href{http://dx.doi.org/10.3847/1538-4357/ac23c6}{10.3847/1538-4357/ac23c6})

\bibitem{Tews2021}
{Tews} I, {Pang} PTH, {Dietrich} T, {Coughlin} MW, {Antier} S, {Bulla} M, {Heinzel} J, {Issa} L. 2021  {On the Nature of GW190814 and Its Impact on the Understanding of Supranuclear Matter}. {\em \apjl} \textbf{908}, L1.
(\href{http://dx.doi.org/10.3847/2041-8213/abdaae}{10.3847/2041-8213/abdaae})

\bibitem{Foucart2012}
{Foucart} F. 2012  {Black-hole-neutron-star mergers: Disk mass predictions}. {\em \prd} \textbf{86}, 124007.
(\href{http://dx.doi.org/10.1103/PhysRevD.86.124007}{10.1103/PhysRevD.86.124007})

\bibitem{Kawaguchi2015}
{Kawaguchi} K, {Kyutoku} K, {Nakano} H, {Okawa} H, {Shibata} M, {Taniguchi} K. 2015  {Black hole-neutron star binary merger: Dependence on black hole spin orientation and equation of state}. {\em \prd} \textbf{92}, 024014.
(\href{http://dx.doi.org/10.1103/PhysRevD.92.024014}{10.1103/PhysRevD.92.024014})

\bibitem{Zhou2021}
{Zhou} R, {Newman} JA, {Mao} YY, {Meisner} A, {Moustakas} J, {Myers} AD, {Prakash} A, {Zentner} AR, {Brooks} D, {Duan} Y, {Landriau} M, {Levi} ME, {Prada} F, {Tarle} G. 2021  {The clustering of DESI-like luminous red galaxies using photometric redshifts}. {\em \mnras} \textbf{501}, 3309--3331.
(\href{http://dx.doi.org/10.1093/mnras/staa3764}{10.1093/mnras/staa3764})

\bibitem{Bilicki2014}
{Bilicki} M, {Jarrett} TH, {Peacock} JA, {Cluver} ME, {Steward} L. 2014  {Two Micron All Sky Survey Photometric Redshift Catalog: A Comprehensive Three-dimensional Census of the Whole Sky}. {\em \apjs} \textbf{210}, 9.
(\href{http://dx.doi.org/10.1088/0067-0049/210/1/9}{10.1088/0067-0049/210/1/9})

\bibitem{Dey2019}
{Dey} A, {Schlegel} DJ, {Lang} D, {Blum} R, {Burleigh} K, {Fan} X, {Findlay} JR, {Finkbeiner} D, {Herrera} D, {Juneau} S, {Landriau} M, {Levi} M, {McGreer} I, {Meisner} A, {Myers} AD, {Moustakas} J, {Nugent} P, {Patej} A, {Schlafly} EF, {Walker} AR, {Valdes} F, {Weaver} BA, {Y{\`e}che} C, {Zou} H, {Zhou} X, {Abareshi} B, {Abbott} TMC, {Abolfathi} B, {Aguilera} C, {Alam} S, {Allen} L, {Alvarez} A, {Annis} J, {Ansarinejad} B, {Aubert} M, {Beechert} J, {Bell} EF, {BenZvi} SY, {Beutler} F, {Bielby} RM, {Bolton} AS, {Brice{\~n}o} C, {Buckley-Geer} EJ, {Butler} K, {Calamida} A, {Carlberg} RG, {Carter} P, {Casas} R, {Castander} FJ, {Choi} Y, {Comparat} J, {Cukanovaite} E, {Delubac} T, {DeVries} K, {Dey} S, {Dhungana} G, {Dickinson} M, {Ding} Z, {Donaldson} JB, {Duan} Y, {Duckworth} CJ, {Eftekharzadeh} S, {Eisenstein} DJ, {Etourneau} T, {Fagrelius} PA, {Farihi} J, {Fitzpatrick} M, {Font-Ribera} A, {Fulmer} L, {G{\"a}nsicke} BT, {Gaztanaga} E, {George} K, {Gerdes} DW, {Gontcho} SGA, {Gorgoni} C, {Green} G, {Guy} J,
  {Harmer} D, {Hernandez} M, {Honscheid} K, {Huang} LW, {James} DJ, {Jannuzi} BT, {Jiang} L, {Joyce} R, {Karcher} A, {Karkar} S, {Kehoe} R, {Kneib} JP, {Kueter-Young} A, {Lan} TW, {Lauer} TR, {Le Guillou} L, {Le Van Suu} A, {Lee} JH, {Lesser} M, {Perreault Levasseur} L, {Li} TS, {Mann} JL, {Marshall} R, {Mart{\'\i}nez-V{\'a}zquez} CE, {Martini} P, {du Mas des Bourboux} H, {McManus} S, {Meier} TG, {M{\'e}nard} B, {Metcalfe} N, {Mu{\~n}oz-Guti{\'e}rrez} A, {Najita} J, {Napier} K, {Narayan} G, {Newman} JA, {Nie} J, {Nord} B, {Norman} DJ, {Olsen} KAG, {Paat} A, {Palanque-Delabrouille} N, {Peng} X, {Poppett} CL, {Poremba} MR, {Prakash} A, {Rabinowitz} D, {Raichoor} A, {Rezaie} M, {Robertson} AN, {Roe} NA, {Ross} AJ, {Ross} NP, {Rudnick} G, {Safonova} S, {Saha} A, {S{\'a}nchez} FJ, {Savary} E, {Schweiker} H, {Scott} A, {Seo} HJ, {Shan} H, {Silva} DR, {Slepian} Z, {Soto} C, {Sprayberry} D, {Staten} R, {Stillman} CM, {Stupak} RJ, {Summers} DL, {Sien Tie} S, {Tirado} H, {Vargas-Maga{\~n}a} M, {Vivas} AK, {Wechsler}
  RH, {Williams} D, {Yang} J, {Yang} Q, {Yapici} T, {Zaritsky} D, {Zenteno} A, {Zhang} K, {Zhang} T, {Zhou} R, {Zhou} Z. 2019  {Overview of the DESI Legacy Imaging Surveys}. {\em \aj} \textbf{157}, 168.
(\href{http://dx.doi.org/10.3847/1538-3881/ab089d}{10.3847/1538-3881/ab089d})

\bibitem{Beck2021}
{Beck} R, {Szapudi} I, {Flewelling} H, {Holmberg} C, {Magnier} E, {Chambers} KC. 2021  {PS1-STRM: neural network source classification and photometric redshift catalogue for PS1 3{\ensuremath{\pi}} DR1}. {\em \mnras} \textbf{500}, 1633--1644.
(\href{http://dx.doi.org/10.1093/mnras/staa2587}{10.1093/mnras/staa2587})

\bibitem{Colless2001}
{Colless} M, {Dalton} G, {Maddox} S, {Sutherland} W, {Norberg} P, {Cole} S, {Bland-Hawthorn} J, {Bridges} T, {Cannon} R, {Collins} C, {Couch} W, {Cross} N, {Deeley} K, {De Propris} R, {Driver} SP, {Efstathiou} G, {Ellis} RS, {Frenk} CS, {Glazebrook} K, {Jackson} C, {Lahav} O, {Lewis} I, {Lumsden} S, {Madgwick} D, {Peacock} JA, {Peterson} BA, {Price} I, {Seaborne} M, {Taylor} K. 2001  {The 2dF Galaxy Redshift Survey: spectra and redshifts}. {\em \mnras} \textbf{328}, 1039--1063.
(\href{http://dx.doi.org/10.1046/j.1365-8711.2001.04902.x}{10.1046/j.1365-8711.2001.04902.x})

\bibitem{Nissanke2013}
{Nissanke} S, {Kasliwal} M, {Georgieva} A. 2013  {Identifying Elusive Electromagnetic Counterparts to Gravitational Wave Mergers: An End-to-end Simulation}. {\em \apj} \textbf{767}, 124.
(\href{http://dx.doi.org/10.1088/0004-637X/767/2/124}{10.1088/0004-637X/767/2/124})

\bibitem{DESI2016}
{DESI Collaboration}, {Aghamousa} A, {Aguilar} J, {Ahlen} S, {Alam} S, {Allen} LE, {Allende Prieto} C, {Annis} J, {Bailey} S, {Balland} C, {Ballester} O, {Baltay} C, {Beaufore} L, {Bebek} C, {Beers} TC, {Bell} EF, {Bernal} JL, {Besuner} R, {Beutler} F, {Blake} C, {Bleuler} H, {Blomqvist} M, {Blum} R, {Bolton} AS, {Briceno} C, {Brooks} D, {Brownstein} JR, {Buckley-Geer} E, {Burden} A, {Burtin} E, {Busca} NG, {Cahn} RN, {Cai} YC, {Cardiel-Sas} L, {Carlberg} RG, {Carton} PH, {Casas} R, {Castander} FJ, {Cervantes-Cota} JL, {Claybaugh} TM, {Close} M, {Coker} CT, {Cole} S, {Comparat} J, {Cooper} AP, {Cousinou} MC, {Crocce} M, {Cuby} JG, {Cunningham} DP, {Davis} TM, {Dawson} KS, {de la Macorra} A, {De Vicente} J, {Delubac} T, {Derwent} M, {Dey} A, {Dhungana} G, {Ding} Z, {Doel} P, {Duan} YT, {Ealet} A, {Edelstein} J, {Eftekharzadeh} S, {Eisenstein} DJ, {Elliott} A, {Escoffier} S, {Evatt} M, {Fagrelius} P, {Fan} X, {Fanning} K, {Farahi} A, {Farihi} J, {Favole} G, {Feng} Y, {Fernandez} E, {Findlay} JR, {Finkbeiner}
  DP, {Fitzpatrick} MJ, {Flaugher} B, {Flender} S, {Font-Ribera} A, {Forero-Romero} JE, {Fosalba} P, {Frenk} CS, {Fumagalli} M, {Gaensicke} BT, {Gallo} G, {Garcia-Bellido} J, {Gaztanaga} E, {Pietro Gentile Fusillo} N, {Gerard} T, {Gershkovich} I, {Giannantonio} T, {Gillet} D, {Gonzalez-de-Rivera} G, {Gonzalez-Perez} V, {Gott} S, {Graur} O, {Gutierrez} G, {Guy} J, {Habib} S, {Heetderks} H, {Heetderks} I, {Heitmann} K, {Hellwing} WA, {Herrera} DA, {Ho} S, {Holland} S, {Honscheid} K, {Huff} E, {Hutchinson} TA, {Huterer} D, {Hwang} HS, {Illa Laguna} JM, {Ishikawa} Y, {Jacobs} D, {Jeffrey} N, {Jelinsky} P, {Jennings} E, {Jiang} L, {Jimenez} J, {Johnson} J, {Joyce} R, {Jullo} E, {Juneau} S, {Kama} S, {Karcher} A, {Karkar} S, {Kehoe} R, {Kennamer} N, {Kent} S, {Kilbinger} M, {Kim} AG, {Kirkby} D, {Kisner} T, {Kitanidis} E, {Kneib} JP, {Koposov} S, {Kovacs} E, {Koyama} K, {Kremin} A, {Kron} R, {Kronig} L, {Kueter-Young} A, {Lacey} CG, {Lafever} R, {Lahav} O, {Lambert} A, {Lampton} M, {Landriau} M, {Lang} D, {Lauer}
  TR, {Le Goff} JM, {Le Guillou} L, {Le Van Suu} A, {Lee} JH, {Lee} SJ, {Leitner} D, {Lesser} M, {Levi} ME, {L'Huillier} B, {Li} B, {Liang} M, {Lin} H, {Linder} E, {Loebman} SR, {Luki{\'c}} Z, {Ma} J, {MacCrann} N, {Magneville} C, {Makarem} L, {Manera} M, {Manser} CJ, {Marshall} R, {Martini} P, {Massey} R, {Matheson} T, {McCauley} J, {McDonald} P, {McGreer} ID, {Meisner} A, {Metcalfe} N, {Miller} TN, {Miquel} R, {Moustakas} J, {Myers} A, {Naik} M, {Newman} JA, {Nichol} RC, {Nicola} A, {Nicolati da Costa} L, {Nie} J, {Niz} G, {Norberg} P, {Nord} B, {Norman} D, {Nugent} P, {O'Brien} T, {Oh} M, {Olsen} KAG, {Padilla} C, {Padmanabhan} H, {Padmanabhan} N, {Palanque-Delabrouille} N, {Palmese} A, {Pappalardo} D, {P{\^a}ris} I, {Park} C, {Patej} A, {Peacock} JA, {Peiris} HV, {Peng} X, {Percival} WJ, {Perruchot} S, {Pieri} MM, {Pogge} R, {Pollack} JE, {Poppett} C, {Prada} F, {Prakash} A, {Probst} RG, {Rabinowitz} D, {Raichoor} A, {Ree} CH, {Refregier} A, {Regal} X, {Reid} B, {Reil} K, {Rezaie} M, {Rockosi} CM, {Roe}
  N, {Ronayette} S, {Roodman} A, {Ross} AJ, {Ross} NP, {Rossi} G, {Rozo} E, {Ruhlmann-Kleider} V, {Rykoff} ES, {Sabiu} C, {Samushia} L, {Sanchez} E, {Sanchez} J, {Schlegel} DJ, {Schneider} M, {Schubnell} M, {Secroun} A, {Seljak} U, {Seo} HJ, {Serrano} S, {Shafieloo} A, {Shan} H, {Sharples} R, {Sholl} MJ, {Shourt} WV, {Silber} JH, {Silva} DR, {Sirk} MM, {Slosar} A, {Smith} A, {Smoot} GF, {Som} D, {Song} YS, {Sprayberry} D, {Staten} R, {Stefanik} A, {Tarle} G, {Sien Tie} S, {Tinker} JL, {Tojeiro} R, {Valdes} F, {Valenzuela} O, {Valluri} M, {Vargas-Magana} M, {Verde} L, {Walker} AR, {Wang} J, {Wang} Y, {Weaver} BA, {Weaverdyck} C, {Wechsler} RH, {Weinberg} DH, {White} M, {Yang} Q, {Yeche} C, {Zhang} T, {Zhao} GB, {Zheng} Y, {Zhou} X, {Zhou} Z, {Zhu} Y, {Zou} H, {Zu} Y. 2016  {The DESI Experiment Part I: Science,Targeting, and Survey Design}. {\em arXiv e-prints} p. arXiv:1611.00036.
(\href{http://dx.doi.org/10.48550/arXiv.1611.00036}{10.48550/arXiv.1611.00036})

\bibitem{deJong2019}
{de Jong} RS, {Agertz} O, {Berbel} AA, {Aird} J, {Alexander} DA, {Amarsi} A, {Anders} F, {Andrae} R, {Ansarinejad} B, {Ansorge} W, {Antilogus} P, {Anwand-Heerwart} H, {Arentsen} A, {Arnadottir} A, {Asplund} M, {Auger} M, {Azais} N, {Baade} D, {Baker} G, {Baker} S, {Balbinot} E, {Baldry} IK, {Banerji} M, {Barden} S, {Barklem} P, {Barth{\'e}l{\'e}my-Mazot} E, {Battistini} C, {Bauer} S, {Bell} CPM, {Bellido-Tirado} O, {Bellstedt} S, {Belokurov} V, {Bensby} T, {Bergemann} M, {Bestenlehner} JM, {Bielby} R, {Bilicki} M, {Blake} C, {Bland-Hawthorn} J, {Boeche} C, {Boland} W, {Boller} T, {Bongard} S, {Bongiorno} A, {Bonifacio} P, {Boudon} D, {Brooks} D, {Brown} MJI, {Brown} R, {Br{\"u}ggen} M, {Brynnel} J, {Brzeski} J, {Buchert} T, {Buschkamp} P, {Caffau} E, {Caillier} P, {Carrick} J, {Casagrande} L, {Case} S, {Casey} A, {Cesarini} I, {Cescutti} G, {Chapuis} D, {Chiappini} C, {Childress} M, {Christlieb} N, {Church} R, {Cioni} MRL, {Cluver} M, {Colless} M, {Collett} T, {Comparat} J, {Cooper} A, {Couch} W, {Courbin} F,
  {Croom} S, {Croton} D, {Daguis{\'e}} E, {Dalton} G, {Davies} LJM, {Davis} T, {de Laverny} P, {Deason} A, {Dionies} F, {Disseau} K, {Doel} P, {D{\"o}scher} D, {Driver} SP, {Dwelly} T, {Eckert} D, {Edge} A, {Edvardsson} B, {Youssoufi} DE, {Elhaddad} A, {Enke} H, {Erfanianfar} G, {Farrell} T, {Fechner} T, {Feiz} C, {Feltzing} S, {Ferreras} I, {Feuerstein} D, {Feuillet} D, {Finoguenov} A, {Ford} D, {Fotopoulou} S, {Fouesneau} M, {Frenk} C, {Frey} S, {Gaessler} W, {Geier} S, {Gentile Fusillo} N, {Gerhard} O, {Giannantonio} T, {Giannone} D, {Gibson} B, {Gillingham} P, {Gonz{\'a}lez-Fern{\'a}ndez} C, {Gonzalez-Solares} E, {Gottloeber} S, {Gould} A, {Grebel} EK, {Gueguen} A, {Guiglion} G, {Haehnelt} M, {Hahn} T, {Hansen} CJ, {Hartman} H, {Hauptner} K, {Hawkins} K, {Haynes} D, {Haynes} R, {Heiter} U, {Helmi} A, {Aguayo} CH, {Hewett} P, {Hinton} S, {Hobbs} D, {Hoenig} S, {Hofman} D, {Hook} I, {Hopgood} J, {Hopkins} A, {Hourihane} A, {Howes} L, {Howlett} C, {Huet} T, {Irwin} M, {Iwert} O, {Jablonka} P, {Jahn} T,
  {Jahnke} K, {Jarno} A, {Jin} S, {Jofre} P, {Johl} D, {Jones} D, {J{\"o}nsson} H, {Jordan} C, {Karovicova} I, {Khalatyan} A, {Kelz} A, {Kennicutt} R, {King} D, {Kitaura} F, {Klar} J, {Klauser} U, {Kneib} JP, {Koch} A, {Koposov} S, {Kordopatis} G, {Korn} A, {Kosmalski} J, {Kotak} R, {Kovalev} M, {Kreckel} K, {Kripak} Y, {Krumpe} M, {Kuijken} K, {Kunder} A, {Kushniruk} I, {Lam} MI, {Lamer} G, {Laurent} F, {Lawrence} J, {Lehmitz} M, {Lemasle} B, {Lewis} J, {Li} B, {Lidman} C, {Lind} K, {Liske} J, {Lizon} JL, {Loveday} J, {Ludwig} HG, {McDermid} RM, {Maguire} K, {Mainieri} V, {Mali} S, {Mandel} H, {Mandel} K, {Mannering} L, {Martell} S, {Martinez Delgado} D, {Matijevic} G, {McGregor} H, {McMahon} R, {McMillan} P, {Mena} O, {Merloni} A, {Meyer} MJ, {Michel} C, {Micheva} G, {Migniau} JE, {Minchev} I, {Monari} G, {Muller} R, {Murphy} D, {Muthukrishna} D, {Nandra} K, {Navarro} R, {Ness} M, {Nichani} V, {Nichol} R, {Nicklas} H, {Niederhofer} F, {Norberg} P, {Obreschkow} D, {Oliver} S, {Owers} M, {Pai} N, {Pankratow}
  S, {Parkinson} D, {Paschke} J, {Paterson} R, {Pecontal} A, {Parry} I, {Phillips} D, {Pillepich} A, {Pinard} L, {Pirard} J, {Piskunov} N, {Plank} V, {Pl{\"u}schke} D, {Pons} E, {Popesso} P, {Power} C, {Pragt} J, {Pramskiy} A, {Pryer} D, {Quattri} M, {Queiroz} ABdA, {Quirrenbach} A, {Rahurkar} S, {Raichoor} A, {Ramstedt} S, {Rau} A, {Recio-Blanco} A, {Reiss} R, {Renaud} F, {Revaz} Y, {Rhode} P, {Richard} J, {Richter} AD, {Rix} HW, {Robotham} ASG, {Roelfsema} R, {Romaniello} M, {Rosario} D, {Rothmaier} F, {Roukema} B, {Ruchti} G, {Rupprecht} G, {Rybizki} J, {Ryde} N, {Saar} A, {Sadler} E, {Sahl{\'e}n} M, {Salvato} M, {Sassolas} B, {Saunders} W, {Saviauk} A, {Sbordone} L, {Schmidt} T, {Schnurr} O, {Scholz} RD, {Schwope} A, {Seifert} W, {Shanks} T, {Sheinis} A, {Sivov} T, {Sk{\'u}lad{\'o}ttir} {\'A}, {Smartt} S, {Smedley} S, {Smith} G, {Smith} R, {Sorce} J, {Spitler} L, {Starkenburg} E, {Steinmetz} M, {Stilz} I, {Storm} J, {Sullivan} M, {Sutherland} W, {Swann} E, {Tamone} A, {Taylor} EN, {Teillon} J, {Tempel} E,
  {ter Horst} R, {Thi} WF, {Tolstoy} E, {Trager} S, {Traven} G, {Tremblay} PE, {Tresse} L, {Valentini} M, {van de Weygaert} R, {van den Ancker} M, {Veljanoski} J, {Venkatesan} S, {Wagner} L, {Wagner} K, {Walcher} CJ, {Waller} L, {Walton} N, {Wang} L, {Winkler} R, {Wisotzki} L, {Worley} CC, {Worseck} G, {Xiang} M, {Xu} W, {Yong} D, {Zhao} C, {Zheng} J, {Zscheyge} F, {Zucker} D. 2019  {4MOST: Project overview and information for the First Call for Proposals}. {\em The Messenger} \textbf{175}, 3--11.
(\href{http://dx.doi.org/10.18727/0722-6691/5117}{10.18727/0722-6691/5117})

\bibitem{Takada2014}
{Takada} M, {Ellis} RS, {Chiba} M, {Greene} JE, {Aihara} H, {Arimoto} N, {Bundy} K, {Cohen} J, {Dor{\'e}} O, {Graves} G, {Gunn} JE, {Heckman} T, {Hirata} CM, {Ho} P, {Kneib} JP, {Le F{\`e}vre} O, {Lin} L, {More} S, {Murayama} H, {Nagao} T, {Ouchi} M, {Seiffert} M, {Silverman} JD, {Sodr{\'e}} L, {Spergel} DN, {Strauss} MA, {Sugai} H, {Suto} Y, {Takami} H, {Wyse} R. 2014  {Extragalactic science, cosmology, and Galactic archaeology with the Subaru Prime Focus Spectrograph}. {\em \pasj} \textbf{66}, R1.
(\href{http://dx.doi.org/10.1093/pasj/pst019}{10.1093/pasj/pst019})

\bibitem{Cook2023}
{Cook} DO, {Mazzarella} JM, {Helou} G, {Alcala} A, {Chen} TX, {Ebert} R, {Frayer} C, {Kim} J, {Lo} T, {Madore} BF, {Ogle} PM, {Schmitz} M, {Singer} LP, {Terek} S, {Valladon} J, {Wu} X. 2023  {Completeness of the NASA/IPAC Extragalactic Database (NED) Local Volume Sample}. {\em \apjs} \textbf{268}, 14.
(\href{http://dx.doi.org/10.3847/1538-4365/acdd06}{10.3847/1538-4365/acdd06})

\bibitem{Antoniadis2013}
{Antoniadis} J, {Freire} PCC, {Wex} N, {Tauris} TM, {Lynch} RS, {van Kerkwijk} MH, {Kramer} M, {Bassa} C, {Dhillon} VS, {Driebe} T, {Hessels} JWT, {Kaspi} VM, {Kondratiev} VI, {Langer} N, {Marsh} TR, {McLaughlin} MA, {Pennucci} TT, {Ransom} SM, {Stairs} IH, {van Leeuwen} J, {Verbiest} JPW, {Whelan} DG. 2013  {A Massive Pulsar in a Compact Relativistic Binary}. {\em Science} \textbf{340}, 448.
(\href{http://dx.doi.org/10.1126/science.1233232}{10.1126/science.1233232})

\bibitem{Cromartie2020}
{Cromartie} HT, {Fonseca} E, {Ransom} SM, {Demorest} PB, {Arzoumanian} Z, {Blumer} H, {Brook} PR, {DeCesar} ME, {Dolch} T, {Ellis} JA, {Ferdman} RD, {Ferrara} EC, {Garver-Daniels} N, {Gentile} PA, {Jones} ML, {Lam} MT, {Lorimer} DR, {Lynch} RS, {McLaughlin} MA, {Ng} C, {Nice} DJ, {Pennucci} TT, {Spiewak} R, {Stairs} IH, {Stovall} K, {Swiggum} JK, {Zhu} WW. 2020  {Relativistic Shapiro delay measurements of an extremely massive millisecond pulsar}. {\em Nature Astronomy} \textbf{4}, 72--76.
(\href{http://dx.doi.org/10.1038/s41550-019-0880-2}{10.1038/s41550-019-0880-2})

\bibitem{Fonseca2021}
{Fonseca} E, {Cromartie} HT, {Pennucci} TT, {Ray} PS, {Kirichenko} AY, {Ransom} SM, {Demorest} PB, {Stairs} IH, {Arzoumanian} Z, {Guillemot} L, {Parthasarathy} A, {Kerr} M, {Cognard} I, {Baker} PT, {Blumer} H, {Brook} PR, {DeCesar} M, {Dolch} T, {Dong} FA, {Ferrara} EC, {Fiore} W, {Garver-Daniels} N, {Good} DC, {Jennings} R, {Jones} ML, {Kaspi} VM, {Lam} MT, {Lorimer} DR, {Luo} J, {McEwen} A, {McKee} JW, {McLaughlin} MA, {McMann} N, {Meyers} BW, {Naidu} A, {Ng} C, {Nice} DJ, {Pol} N, {Radovan} HA, {Shapiro-Albert} B, {Tan} CM, {Tendulkar} SP, {Swiggum} JK, {Wahl} HM, {Zhu} WW. 2021  {Refined Mass and Geometric Measurements of the High-mass PSR J0740+6620}. {\em \apjl} \textbf{915}, L12.
(\href{http://dx.doi.org/10.3847/2041-8213/ac03b8}{10.3847/2041-8213/ac03b8})

\bibitem{Goldstein2019}
{Goldstein} DA, {Andreoni} I, {Nugent} PE, {Kasliwal} MM, {Coughlin} MW, {Anand} S, {Bloom} JS, {Mart{\'\i}nez-Palomera} J, {Zhang} K, {Ahumada} T, {Bagdasaryan} A, {Cooke} J, {De} K, {Duev} DA, {Fremling} UC, {Gatkine} P, {Graham} M, {Ofek} EO, {Singer} LP, {Yan} L. 2019  {GROWTH on S190426c: Real-time Search for a Counterpart to the Probable Neutron Star-Black Hole Merger using an Automated Difference Imaging Pipeline for DECam}. {\em \apjl} \textbf{881}, L7.
(\href{http://dx.doi.org/10.3847/2041-8213/ab3046}{10.3847/2041-8213/ab3046})

\bibitem{Kumar2022}
{Kumar} H, {Bhalerao} V, {Anupama} GC, {Barway} S, {Coughlin} MW, {De} K, {Deshmukh} K, {Dutta} A, {Goldstein} DA, {Jassani} A, {Joharle} S, {Karambelker} V, {Khandagale} M, {Kumar} B, {Saraogi} D, {Sharma} Y, {Shenoy} V, {singer} L, {Singh} A, {Waratkar} G. 2022  {GROWTH on S190426c II: GROWTH-India Telescope search for an optical counterpart with a custom image reduction and candidate vetting pipeline}. {\em \mnras} \textbf{516}, 4517--4528.
(\href{http://dx.doi.org/10.1093/mnras/stac2516}{10.1093/mnras/stac2516})

\bibitem{Gourdji2023}
{Gourdji} K, {Rowlinson} A, {Wijers} RAMJ, {Broderick} JW, {Shulevski} A. 2023  {LOFAR observations of gravitational wave merger events: O3 results and O4 strategy}. {\em \mnras} \textbf{523}, 4748--4755.
(\href{http://dx.doi.org/10.1093/mnras/stad1714}{10.1093/mnras/stad1714})

\bibitem{Anand2021}
{Anand} S, {Coughlin} MW, {Kasliwal} MM, {Bulla} M, {Ahumada} T, {Sagu{\'e}s Carracedo} A, {Almualla} M, {Andreoni} I, {Stein} R, {Foucart} F, {Singer} LP, {Sollerman} J, {Bellm} EC, {Bolin} B, {Caballero-Garc{\'\i}a} MD, {Castro-Tirado} AJ, {Cenko} SB, {De} K, {Dekany} RG, {Duev} DA, {Feeney} M, {Fremling} C, {Goldstein} DA, {Golkhou} VZ, {Graham} MJ, {Guessoum} N, {Hankins} MJ, {Hu} Y, {Kong} AKH, {Kool} EC, {Kulkarni} SR, {Kumar} H, {Laher} RR, {Masci} FJ, {Mr{\'o}z} P, {Nissanke} S, {Porter} M, {Reusch} S, {Riddle} R, {Rosnet} P, {Rusholme} B, {Serabyn} E, {S{\'a}nchez-Ram{\'\i}rez} R, {Rigault} M, {Shupe} DL, {Smith} R, {Soumagnac} MT, {Walters} R, {Valeev} AF. 2021  {Optical follow-up of the neutron star-black hole mergers S200105ae and S200115j}. {\em Nature Astronomy} \textbf{5}, 46--53.
(\href{http://dx.doi.org/10.1038/s41550-020-1183-3}{10.1038/s41550-020-1183-3})

\bibitem{Dichiara2021}
{Dichiara} S, {Becerra} RL, {Chase} EA, {Troja} E, {Lee} WH, {Watson} AM, {Butler} NR, {O'Connor} B, {Pereyra} M, {L{\'o}pez} KOC, {Lien} AY, {Gottlieb} A, {Kutyrev} AS. 2021  {Constraints on the Electromagnetic Counterpart of the Neutron-star-Black-hole Merger GW200115}. {\em \apjl} \textbf{923}, L32.
(\href{http://dx.doi.org/10.3847/2041-8213/ac4259}{10.3847/2041-8213/ac4259})

\bibitem{Coughlin2020c}
{Coughlin} MW, {Dietrich} T, {Antier} S, {Almualla} M, {Anand} S, {Bulla} M, {Foucart} F, {Guessoum} N, {Hotokezaka} K, {Kumar} V, {Raaijmakers} G, {Nissanke} S. 2020  {Implications of the search for optical counterparts during the second part of the Advanced LIGO's and Advanced Virgo's third observing run: lessons learned for future follow-up observations}. {\em \mnras} \textbf{497}, 1181--1196.
(\href{http://dx.doi.org/10.1093/mnras/staa1925}{10.1093/mnras/staa1925})

\bibitem{Broekgaarden2021}
{Broekgaarden} FS, {Berger} E. 2021  {Formation of the First Two Black Hole-Neutron Star Mergers (GW200115 and GW200105) from Isolated Binary Evolution}. {\em \apjl} \textbf{920}, L13.
(\href{http://dx.doi.org/10.3847/2041-8213/ac2832}{10.3847/2041-8213/ac2832})

\bibitem{Chattopadhyay2022}
{Chattopadhyay} D, {Stevenson} S, {Broekgaarden} F, {Antonini} F, {Belczynski} K. 2022  {Modelling the formation of the first two neutron star-black hole mergers, GW200105 and GW200115: metallicity, chirp masses, and merger remnant spins}. {\em \mnras} \textbf{513}, 5780--5789.
(\href{http://dx.doi.org/10.1093/mnras/stac1283}{10.1093/mnras/stac1283})

\bibitem{Mandel2022}
{Mandel} I, {Broekgaarden} FS. 2022  {Rates of compact object coalescences}. {\em Living Reviews in Relativity} \textbf{25}, 1.
(\href{http://dx.doi.org/10.1007/s41114-021-00034-3}{10.1007/s41114-021-00034-3})

\bibitem{Gompertz2022}
{Gompertz} BP, {Nicholl} M, {Schmidt} P, {Pratten} G, {Vecchio} A. 2022  {Constraints on compact binary merger evolution from spin-orbit misalignment in gravitational-wave observations}. {\em \mnras} \textbf{511}, 1454--1461.
(\href{http://dx.doi.org/10.1093/mnras/stac029}{10.1093/mnras/stac029})

\bibitem{Fragione2021}
{Fragione} G, {Loeb} A, {Rasio} FA. 2021  {Impact of Natal Kicks on Merger Rates and Spin-Orbit Misalignments of Black Hole-Neutron Star Mergers}. {\em \apjl} \textbf{918}, L38.
(\href{http://dx.doi.org/10.3847/2041-8213/ac225a}{10.3847/2041-8213/ac225a})

\bibitem{Mandel2021}
{Mandel} I, {Smith} RJE. 2021  {GW200115: A Nonspinning Black Hole-Neutron Star Merger}. {\em \apjl} \textbf{922}, L14.
(\href{http://dx.doi.org/10.3847/2041-8213/ac35dd}{10.3847/2041-8213/ac35dd})

\bibitem{Foucart2018}
{Foucart} F, {Hinderer} T, {Nissanke} S. 2018  {Remnant baryon mass in neutron star-black hole mergers: Predictions for binary neutron star mimickers and rapidly spinning black holes}. {\em \prd} \textbf{98}, 081501.
(\href{http://dx.doi.org/10.1103/PhysRevD.98.081501}{10.1103/PhysRevD.98.081501})

\bibitem{Zhu2021}
{Zhu} JP, {Wu} S, {Yang} YP, {Zhang} B, {Yu} YW, {Gao} H, {Cao} Z, {Liu} LD. 2021  {No Detectable Kilonova Counterpart is Expected for O3 Neutron Star-Black Hole Candidates}. {\em \apj} \textbf{921}, 156.
(\href{http://dx.doi.org/10.3847/1538-4357/ac19a7}{10.3847/1538-4357/ac19a7})

\bibitem{Fragione2021a}
{Fragione} G. 2021  {Black-hole-Neutron-star Mergers Are Unlikely Multimessenger Sources}. {\em \apjl} \textbf{923}, L2.
(\href{http://dx.doi.org/10.3847/2041-8213/ac3bcd}{10.3847/2041-8213/ac3bcd})

\bibitem{Drozda2022}
{Drozda} P, {Belczynski} K, {O'Shaughnessy} R, {Bulik} T, {Fryer} CL. 2022  {Black hole-neutron star mergers: The first mass gap and kilonovae}. {\em \aap} \textbf{667}, A126.
(\href{http://dx.doi.org/10.1051/0004-6361/202039418}{10.1051/0004-6361/202039418})

\bibitem{Biscoveanu2023}
{Biscoveanu} S, {Landry} P, {Vitale} S. 2023  {Population properties and multimessenger prospects of neutron star-black hole mergers following GWTC-3}. {\em \mnras} \textbf{518}, 5298--5312.
(\href{http://dx.doi.org/10.1093/mnras/stac3052}{10.1093/mnras/stac3052})

\bibitem{Colombo2023}
{Colombo} A, {Duqu{\'e}} R, {Sharan Salafia} O, {Broekgaarden} FS, {Iacovelli} F, {Mancarella} M, {Andreoni} I, {Gabrielli} F, {Ragosta} F, {Ghirlanda} G, {Fragos} T, {Levan} AJ, {Piranomonte} S, {Melandri} A, {Giacomazzo} B, {Colpi} M. 2023  {Multi-messenger prospects for black hole - neutron star mergers in the O4 and O5 runs}. {\em arXiv e-prints} p. arXiv:2310.16894.
(\href{http://dx.doi.org/10.48550/arXiv.2310.16894}{10.48550/arXiv.2310.16894})

\bibitem{Fuller2019}
{Fuller} J, {Ma} L. 2019  {Most Black Holes Are Born Very Slowly Rotating}. {\em \apjl} \textbf{881}, L1.
(\href{http://dx.doi.org/10.3847/2041-8213/ab339b}{10.3847/2041-8213/ab339b})

\bibitem{Hu2022}
{Hu} RC, {Zhu} JP, {Qin} Y, {Zhang} B, {Liang} EW, {Shao} Y. 2022  {A Channel to Form Fast-spinning Black Hole-Neutron Star Binary Mergers as Multimessenger Sources}. {\em \apj} \textbf{928}, 163.
(\href{http://dx.doi.org/10.3847/1538-4357/ac573f}{10.3847/1538-4357/ac573f})

\bibitem{Steinle2023}
{Steinle} N, {Gompertz} BP, {Nicholl} M. 2023  {Mechanisms for high spin in black-hole neutron-star binaries and kilonova emission: inheritance and accretion}. {\em \mnras} \textbf{519}, 891--901.
(\href{http://dx.doi.org/10.1093/mnras/stac3626}{10.1093/mnras/stac3626})

\bibitem{Wang2024}
{Wang} ZHT, {Hu} RC, {Qin} Y, {Zhu} JP, {Zhang} B, {Yi} SX, {Tang} QW, {Shu} XW, {Lyu} F, {Liang} EW. 2024  {A Channel to Form Fast-spinning Black Hole{\textendash}Neutron Star Binary Mergers as Multimessenger Sources. II. Accretion-induced Spin-up}. {\em \apj} \textbf{965}, 177.
(\href{http://dx.doi.org/10.3847/1538-4357/ad2fc1}{10.3847/1538-4357/ad2fc1})

\bibitem{Tanaka2014}
{Tanaka} M, {Hotokezaka} K, {Kyutoku} K, {Wanajo} S, {Kiuchi} K, {Sekiguchi} Y, {Shibata} M. 2014  {Radioactively Powered Emission from Black Hole-Neutron Star Mergers}. {\em \apj} \textbf{780}, 31.
(\href{http://dx.doi.org/10.1088/0004-637X/780/1/31}{10.1088/0004-637X/780/1/31})

\bibitem{Kyutoku2015}
{Kyutoku} K, {Ioka} K, {Okawa} H, {Shibata} M, {Taniguchi} K. 2015  {Dynamical mass ejection from black hole-neutron star binaries}. {\em \prd} \textbf{92}, 044028.
(\href{http://dx.doi.org/10.1103/PhysRevD.92.044028}{10.1103/PhysRevD.92.044028})

\bibitem{Kawaguchi2016}
{Kawaguchi} K, {Kyutoku} K, {Shibata} M, {Tanaka} M. 2016  {Models of Kilonova/Macronova Emission from Black Hole-Neutron Star Mergers}. {\em \apj} \textbf{825}, 52.
(\href{http://dx.doi.org/10.3847/0004-637X/825/1/52}{10.3847/0004-637X/825/1/52})

\bibitem{Fernandez2017}
{Fern{\'a}ndez} R, {Foucart} F, {Kasen} D, {Lippuner} J, {Desai} D, {Roberts} LF. 2017  {Dynamics, nucleosynthesis, and kilonova signature of black hole{\textemdash}neutron star merger ejecta}. {\em Classical and Quantum Gravity} \textbf{34}, 154001.
(\href{http://dx.doi.org/10.1088/1361-6382/aa7a77}{10.1088/1361-6382/aa7a77})

\bibitem{Kyutoku2018}
{Kyutoku} K, {Kiuchi} K, {Sekiguchi} Y, {Shibata} M, {Taniguchi} K. 2018  {Neutrino transport in black hole-neutron star binaries: Neutrino emission and dynamical mass ejection}. {\em \prd} \textbf{97}, 023009.
(\href{http://dx.doi.org/10.1103/PhysRevD.97.023009}{10.1103/PhysRevD.97.023009})

\bibitem{Bartos2017}
Bartos I, Kocsis B, Haiman Z, M{\'a}rka S. 2017  Rapid and {{Bright Stellar-mass Binary Black Hole Mergers}} in {{Active Galactic Nuclei}}. {\em The Astrophysical Journal} \textbf{835}, 165.
(\href{http://dx.doi.org/10.3847/1538-4357/835/2/165}{10.3847/1538-4357/835/2/165})

\bibitem{Tagawa2023}
{Tagawa} H, {Kimura} SS, {Haiman} Z, {Perna} R, {Bartos} I. 2023  {Observable Signature of Merging Stellar-mass Black Holes in Active Galactic Nuclei}. {\em \apj} \textbf{950}, 13.
(\href{http://dx.doi.org/10.3847/1538-4357/acc4bb}{10.3847/1538-4357/acc4bb})

\bibitem{McKernan2012}
{McKernan} B, {Ford} KES, {Lyra} W, {Perets} HB. 2012  {Intermediate mass black holes in AGN discs - I. Production and growth}. {\em \mnras} \textbf{425}, 460--469.
(\href{http://dx.doi.org/10.1111/j.1365-2966.2012.21486.x}{10.1111/j.1365-2966.2012.21486.x})

\bibitem{McKernan2014}
{McKernan} B, {Ford} KES, {Kocsis} B, {Lyra} W, {Winter} LM. 2014  {Intermediate-mass black holes in AGN discs - II. Model predictions and observational constraints}. {\em \mnras} \textbf{441}, 900--909.
(\href{http://dx.doi.org/10.1093/mnras/stu553}{10.1093/mnras/stu553})

\bibitem{Gerosa2019}
{Gerosa} D, {Berti} E. 2019  {Escape speed of stellar clusters from multiple-generation black-hole mergers in the upper mass gap}. {\em \prd} \textbf{100}, 041301.
(\href{http://dx.doi.org/10.1103/PhysRevD.100.041301}{10.1103/PhysRevD.100.041301})

\bibitem{Graham2023}
{Graham} MJ, {McKernan} B, {Ford} KES, {Stern} D, {Djorgovski} SG, {Coughlin} M, {Burdge} KB, {Bellm} EC, {Helou} G, {Mahabal} AA, {Masci} FJ, {Purdum} J, {Rosnet} P, {Rusholme} B. 2023  {A Light in the Dark: Searching for Electromagnetic Counterparts to Black Hole-Black Hole Mergers in LIGO/Virgo O3 with the Zwicky Transient Facility}. {\em \apj} \textbf{942}, 99.
(\href{http://dx.doi.org/10.3847/1538-4357/aca480}{10.3847/1538-4357/aca480})

\bibitem{2023GCN.33813_S230518h_disc}
{Ligo Scientific Collaboration}, {VIRGO Collaboration}, {Kagra Collaboration}. 2023a  {LIGO/Virgo/KAGRA S230518h: Identification of a GW compact binary merger candidate}. {\em GRB Coordinates Network} \textbf{33813}, 1.

\bibitem{2023GCN.33816_S230518h_update}
{Ligo Scientific Collaboration}, {VIRGO Collaboration}, {Kagra Collaboration}. 2023b  {LIGO/Virgo/KAGRA S230518h: Updated Sky localization and EM Bright Classification}. {\em GRB Coordinates Network} \textbf{33816}, 1.

\bibitem{2023GCN.33884_S230518h_update2}
{Ligo Scientific Collaboration}, {VIRGO Collaboration}, {Kagra Collaboration}. 2023c  {LIGO/Virgo/KAGRA S230518h: Updated Sky localization}. {\em GRB Coordinates Network} \textbf{33884}, 1.

\bibitem{2023GCN.33891_S230529ay_update}
{Ligo Scientific Collaboration}, {VIRGO Collaboration}, {Kagra Collaboration}. 2023d  {LIGO/Virgo/KAGRA S230529ay: Updated Sky localization and EM Bright Classification}. {\em GRB Coordinates Network} \textbf{33891}, 1.

\bibitem{2023GCN.34148_S230529ay_update2}
{Ligo Scientific Collaboration}, {VIRGO Collaboration}, {Kagra Collaboration}. 2023e  {LIGO/Virgo/KAGRA S230529ay: Updated Sky localization and EM Bright Classification}. {\em GRB Coordinates Network} \textbf{34148}, 1.

\bibitem{2024GCN.36236_S240422ed_disc}
{Ligo Scientific Collaboration}, {VIRGO Collaboration}, {Kagra Collaboration}. 2024a  {LIGO/Virgo/KAGRA S240422ed: Identification of a GW compact binary merger candidate}. {\em GRB Coordinates Network} \textbf{36236}, 1.

\bibitem{2024GCN.36240_S240422ed_update}
{Ligo Scientific Collaboration}, {VIRGO Collaboration}, {Kagra Collaboration}. 2024b  {LIGO/Virgo/KAGRA S240422ed: Updated Sky localization and EM Bright Classification}. {\em GRB Coordinates Network} \textbf{36240}, 1.

\bibitem{Ashton2019bilby}
{Ashton} G, {H{\"u}bner} M, {Lasky} PD, {Talbot} C, {Ackley} K, {Biscoveanu} S, {Chu} Q, {Divakarla} A, {Easter} PJ, {Goncharov} B, {Hernandez Vivanco} F, {Harms} J, {Lower} ME, {Meadors} GD, {Melchor} D, {Payne} E, {Pitkin} MD, {Powell} J, {Sarin} N, {Smith} RJE, {Thrane} E. 2019  {BILBY: A User-friendly Bayesian Inference Library for Gravitational-wave Astronomy}. {\em \apjs} \textbf{241}, 27.
(\href{http://dx.doi.org/10.3847/1538-4365/ab06fc}{10.3847/1538-4365/ab06fc})

\bibitem{Chatterjee2020}
{Chatterjee} D, {Ghosh} S, {Brady} PR, {Kapadia} SJ, {Miller} AL, {Nissanke} S, {Pannarale} F. 2020  {A Machine Learning-based Source Property Inference for Compact Binary Mergers}. {\em \apj} \textbf{896}, 54.
(\href{http://dx.doi.org/10.3847/1538-4357/ab8dbe}{10.3847/1538-4357/ab8dbe})

\bibitem{2024GCN.36236....1L}
{Ligo Scientific Collaboration}, {VIRGO Collaboration}, {Kagra Collaboration}. 2024a  {LIGO/Virgo/KAGRA S240422ed: Identification of a GW compact binary merger candidate}. {\em GRB Coordinates Network} \textbf{36236}, 1.

\bibitem{2024GCN.36812....1L}
{Ligo Scientific Collaboration}, {VIRGO Collaboration}, {Kagra Collaboration}. 2024b  {LIGO/Virgo/KAGRA S240422ed: Updated significance estimate}. {\em GRB Coordinates Network} \textbf{36812}, 1.

\bibitem{Abac2024}
{Ligo Scientific Collaboration}, {VIRGO Collaboration}, {Kagra Collaboration}. 2024c  {Observation of Gravitational Waves from the Coalescence of a 2.5{\textendash}4.5 M $_{{\ensuremath{\odot}}}$ Compact Object and a Neutron Star}. {\em \apjl} \textbf{970}, L34.
(\href{http://dx.doi.org/10.3847/2041-8213/ad5beb}{10.3847/2041-8213/ad5beb})

\bibitem{Sherman2024GCN.36288}
{Sherman} N, {McMahon} I, {Kesler} E, {MacBride} S, {Santos} A, {Kaur} S, {Santana-Silva} L, {Soares-Santos} M, {de Bom} C, {Herner} K, {Vivas} K, {Desgw Team}. 2024  {LIGO/Virgo/KAGRA S240413p: DESGW DECam Follow-Up}. {\em GRB Coordinates Network} \textbf{36288}, 1.

\bibitem{Mohite2022}
{Mohite} SR, {Rajkumar} P, {Anand} S, {Kaplan} DL, {Coughlin} MW, {Sagu{\'e}s-Carracedo} A, {Saleem} M, {Creighton} J, {Brady} PR, {Ahumada} T, {Almualla} M, {Andreoni} I, {Bulla} M, {Graham} MJ, {Kasliwal} MM, {Kaye} S, {Laher} RR, {Shin} KM, {Shupe} DL, {Singer} LP. 2022  {Inferring Kilonova Population Properties with a Hierarchical Bayesian Framework. I. Nondetection Methodology and Single-event Analyses}. {\em \apj} \textbf{925}, 58.
(\href{http://dx.doi.org/10.3847/1538-4357/ac3981}{10.3847/1538-4357/ac3981})

\bibitem{Ahumada2024arXiv}
{Ahumada} T, {Anand} S, {Coughlin} MW, {Gupta} V, {Kasliwal} MM, {Karambelkar} VR, {Stein} RD, {Waratkar} G, {Swain} V, {Jegou du Laz} T, {Anumarlapudi} A, {Andreoni} I, {Bulla} M, {Srinivasaragavan} GP, {Toivonen} A, {Wold} A, {Bellm} EC, {Cenko} SB, {Kaplan} DL, {Sollerman} J, {Bhalerao} V, {Perley} D, {Salgundi} A, {Suresh} A, {Hinds} KR, {Reusch} S, {Necker} J, {Cook} DO, {Pletskova} N, {Singer} LP, {Banerjee} S, {Barna} T, {Copperwheat} CM, {Healy} B, {Weizmann Kiendrebeogo} R, {Kumar} H, {Kumar} R, {Pezzella} M, {Sagues-Carracedo} A, {Sravan} N, {Bloom} JS, {Chen} TX, {Graham} M, {Helou} G, {Laher} RR, {Mahabal} AA, {Purdum} J, {Anupama} GC, {Barway} S, {Basu} J, {Raman} D, {Roychowdhury} T. 2024  {Searching for gravitational wave optical counterparts with the Zwicky Transient Facility: summary of O4a}. {\em arXiv e-prints} p. arXiv:2405.12403.
(\href{http://dx.doi.org/10.48550/arXiv.2405.12403}{10.48550/arXiv.2405.12403})

\bibitem{Ascenzi2019}
{Ascenzi} S, {Coughlin} MW, {Dietrich} T, {Foley} RJ, {Ramirez-Ruiz} E, {Piranomonte} S, {Mockler} B, {Murguia-Berthier} A, {Fryer} CL, {Lloyd-Ronning} NM, {Rosswog} S. 2019  {A luminosity distribution for kilonovae based on short gamma-ray burst afterglows}. {\em \mnras} \textbf{486}, 672--690.
(\href{http://dx.doi.org/10.1093/mnras/stz891}{10.1093/mnras/stz891})

\bibitem{Rastinejad2022}
{Rastinejad} JC, {Gompertz} BP, {Levan} AJ, {Fong} Wf, {Nicholl} M, {Lamb} GP, {Malesani} DB, {Nugent} AE, {Oates} SR, {Tanvir} NR, {de Ugarte Postigo} A, {Kilpatrick} CD, {Moore} CJ, {Metzger} BD, {Ravasio} ME, {Rossi} A, {Schroeder} G, {Jencson} J, {Sand} DJ, {Smith} N, {Ag{\"u}{\'\i} Fern{\'a}ndez} JF, {Berger} E, {Blanchard} PK, {Chornock} R, {Cobb} BE, {De Pasquale} M, {Fynbo} JPU, {Izzo} L, {Kann} DA, {Laskar} T, {Marini} E, {Paterson} K, {Escorial} AR, {Sears} HM, {Th{\"o}ne} CC. 2022  {A kilonova following a long-duration gamma-ray burst at 350 Mpc}. {\em \nat} \textbf{612}, 223--227.
(\href{http://dx.doi.org/10.1038/s41586-022-05390-w}{10.1038/s41586-022-05390-w})

\bibitem{Tanvir2013}
{Tanvir} NR, {Levan} AJ, {Fruchter} AS, {Hjorth} J, {Hounsell} RA, {Wiersema} K, {Tunnicliffe} RL. 2013  {A `kilonova' associated with the short-duration {\ensuremath{\gamma}}-ray burst GRB 130603B}. {\em \nat} \textbf{500}, 547--549.
(\href{http://dx.doi.org/10.1038/nature12505}{10.1038/nature12505})

\bibitem{Berger2013}
{Berger} E, {Fong} W, {Chornock} R. 2013  {An r-process Kilonova Associated with the Short-hard GRB 130603B}. {\em \apjl} \textbf{774}, L23.
(\href{http://dx.doi.org/10.1088/2041-8205/774/2/L23}{10.1088/2041-8205/774/2/L23})

\bibitem{Yang2015}
{Yang} B, {Jin} ZP, {Li} X, {Covino} S, {Zheng} XZ, {Hotokezaka} K, {Fan} YZ, {Piran} T, {Wei} DM. 2015  {A possible macronova in the late afterglow of the long-short burst GRB 060614}. {\em Nature Communications} \textbf{6}, 7323.
(\href{http://dx.doi.org/10.1038/ncomms8323}{10.1038/ncomms8323})

\bibitem{Jin2015}
{Jin} ZP, {Li} X, {Cano} Z, {Covino} S, {Fan} YZ, {Wei} DM. 2015  {The Light Curve of the Macronova Associated with the Long-Short Burst GRB 060614}. {\em \apjl} \textbf{811}, L22.
(\href{http://dx.doi.org/10.1088/2041-8205/811/2/L22}{10.1088/2041-8205/811/2/L22})

\bibitem{Jin2016}
{Jin} ZP, {Hotokezaka} K, {Li} X, {Tanaka} M, {D'Avanzo} P, {Fan} YZ, {Covino} S, {Wei} DM, {Piran} T. 2016  {The Macronova in GRB 050709 and the GRB-macronova connection}. {\em Nature Communications} \textbf{7}, 12898.
(\href{http://dx.doi.org/10.1038/ncomms12898}{10.1038/ncomms12898})

\bibitem{Troja2018a}
{Troja} E, {Ryan} G, {Piro} L, {van Eerten} H, {Cenko} SB, {Yoon} Y, {Lee} SK, {Im} M, {Sakamoto} T, {Gatkine} P, {Kutyrev} A, {Veilleux} S. 2018  {A luminous blue kilonova and an off-axis jet from a compact binary merger at z = 0.1341}. {\em Nature Communications} \textbf{9}, 4089.
(\href{http://dx.doi.org/10.1038/s41467-018-06558-7}{10.1038/s41467-018-06558-7})

\bibitem{Zhu2023}
{Zhu} YM, {Zhou} H, {Wang} Y, {Liao} NH, {Jin} ZP, {Wei} DM. 2023  {The afterglow of GRB 070707 and a possible kilonova component}. {\em \mnras} \textbf{521}, 269--277.
(\href{http://dx.doi.org/10.1093/mnras/stad541}{10.1093/mnras/stad541})

\bibitem{Gao2015}
{Gao} H, {Ding} X, {Wu} XF, {Dai} ZG, {Zhang} B. 2015  {GRB 080503 Late Afterglow Re-brightening: Signature of a Magnetar-powered Merger-nova}. {\em \apj} \textbf{807}, 163.
(\href{http://dx.doi.org/10.1088/0004-637X/807/2/163}{10.1088/0004-637X/807/2/163})

\bibitem{Zhou2023}
{Zhou} H, {Jin} ZP, {Covino} S, {Lei} L, {An} Y, {Gong} HY, {Fan} YZ, {Wei} DM. 2023  {GRB 080503: A Very Early Blue Kilonova and an Adjacent Nonthermal Radiation Component}. {\em \apj} \textbf{943}, 104.
(\href{http://dx.doi.org/10.3847/1538-4357/acac9b}{10.3847/1538-4357/acac9b})

\bibitem{Gao2017}
{Gao} H, {Zhang} B, {L{\"u}} HJ, {Li} Y. 2017  {Searching for Magnetar-powered Merger-novae from Short GRBS}. {\em \apj} \textbf{837}, 50.
(\href{http://dx.doi.org/10.3847/1538-4357/aa5be3}{10.3847/1538-4357/aa5be3})

\bibitem{Fong2020}
{Fong} W, {Laskar} T, {Rastinejad} J, {Rouco Escorial} A, {Schroeder} G, {Barnes} J, {Kilpatrick} CD, {Paterson} K, {Berger} E, {Metzger} BD, {Dong} Y, {Nugent} AE, {Strausbaugh} R, {Blanchard} PK, {Goyal} A, {Cucchiara} A, {Terreran} G, {Alexander} KD, {Eftekhari} T, {Fryer} C, {Margalit} B, {Margutti} R, {Nicholl} M. 2020  {The Broad-band Counterpart of the Short GRB 200522A at $z=0.5536$: A Luminous Kilonova or a Collimated Outflow with a Reverse Shock?}. {\em arXiv e-prints} p. arXiv:2008.08593.

\bibitem{OConnor2021}
{O'Connor} B, {Troja} E, {Dichiara} S, {Chase} EA, {Ryan} G, {Cenko} SB, {Fryer} CL, {Ricci} R, {Marshall} F, {Kouveliotou} C, {Wollaeger} RT, {Fontes} CJ, {Korobkin} O, {Gatkine} P, {Kutyrev} A, {Veilleux} S, {Kawai} N, {Sakamoto} T. 2021  {A tale of two mergers: constraints on kilonova detection in two short GRBs at z {\ensuremath{\sim}} 0.5}. {\em \mnras} \textbf{502}, 1279--1298.
(\href{http://dx.doi.org/10.1093/mnras/stab132}{10.1093/mnras/stab132})

\bibitem{Levan2024}
{Levan} AJ, {Gompertz} BP, {Salafia} OS, {Bulla} M, {Burns} E, {Hotokezaka} K, {Izzo} L, {Lamb} GP, {Malesani} DB, {Oates} SR, {Ravasio} ME, {Rouco Escorial} A, {Schneider} B, {Sarin} N, {Schulze} S, {Tanvir} NR, {Ackley} K, {Anderson} G, {Brammer} GB, {Christensen} L, {Dhillon} VS, {Evans} PA, {Fausnaugh} M, {Fong} Wf, {Fruchter} AS, {Fryer} C, {Fynbo} JPU, {Gaspari} N, {Heintz} KE, {Hjorth} J, {Kennea} JA, {Kennedy} MR, {Laskar} T, {Leloudas} G, {Mandel} I, {Martin-Carrillo} A, {Metzger} BD, {Nicholl} M, {Nugent} A, {Palmerio} JT, {Pugliese} G, {Rastinejad} J, {Rhodes} L, {Rossi} A, {Saccardi} A, {Smartt} SJ, {Stevance} HF, {Tohuvavohu} A, {van der Horst} A, {Vergani} SD, {Watson} D, {Barclay} T, {Bhirombhakdi} K, {Breedt} E, {Breeveld} AA, {Brown} AJ, {Campana} S, {Chrimes} AA, {D'Avanzo} P, {D'Elia} V, {De Pasquale} M, {Dyer} MJ, {Galloway} DK, {Garbutt} JA, {Green} MJ, {Hartmann} DH, {Jakobsson} P, {Kerry} P, {Kouveliotou} C, {Langeroodi} D, {Le Floc'h} E, {Leung} JK, {Littlefair} SP, {Munday} J,
  {O'Brien} P, {Parsons} SG, {Pelisoli} I, {Sahman} DI, {Salvaterra} R, {Sbarufatti} B, {Steeghs} D, {Tagliaferri} G, {Th{\"o}ne} CC, {de Ugarte Postigo} A, {Kann} DA. 2024  {Heavy-element production in a compact object merger observed by JWST}. {\em \nat} \textbf{626}, 737--741.
(\href{http://dx.doi.org/10.1038/s41586-023-06759-1}{10.1038/s41586-023-06759-1})

\bibitem{Troja2019a}
{Troja} E, {Castro-Tirado} AJ, {Becerra Gonz{\'a}lez} J, {Hu} Y, {Ryan} GS, {Cenko} SB, {Ricci} R, {Novara} G, {S{\'a}nchez-R{\'a}mirez} R, {Acosta-Pulido} JA, {Ackley} KD, {Caballero Garc{\'\i}a} MD, {Eikenberry} SS, {Guziy} S, {Jeong} S, {Lien} AY, {M{\'a}rquez} I, {Pand ey} SB, {Park} IH, {Sakamoto} T, {Tello} JC, {Sokolov} IV, {Sokolov} VV, {Tiengo} A, {Valeev} AF, {Zhang} BB, {Veilleux} S. 2019  {The afterglow and kilonova of the short GRB 160821B}. {\em \mnras} \textbf{489}, 2104--2116.
(\href{http://dx.doi.org/10.1093/mnras/stz2255}{10.1093/mnras/stz2255})

\bibitem{Yang2024}
{Yang} YH, {Troja} E, {O'Connor} B, {Fryer} CL, {Im} M, {Durbak} J, {Paek} GSH, {Ricci} R, {Bom} CR, {Gillanders} JH, {Castro-Tirado} AJ, {Peng} ZK, {Dichiara} S, {Ryan} G, {van Eerten} H, {Dai} ZG, {Chang} SW, {Choi} H, {De} K, {Hu} Y, {Kilpatrick} CD, {Kutyrev} A, {Jeong} M, {Lee} CU, {Makler} M, {Navarete} F, {P{\'e}rez-Garc{\'\i}a} I. 2024  {A lanthanide-rich kilonova in the aftermath of a long gamma-ray burst}. {\em \nat} \textbf{626}, 742--745.
(\href{http://dx.doi.org/10.1038/s41586-023-06979-5}{10.1038/s41586-023-06979-5})

\bibitem{Gillanders2023}
{Gillanders} JH, {Troja} E, {Fryer} CL, {Ristic} M, {O'Connor} B, {Fontes} CJ, {Yang} YH, {Domoto} N, {Rahmouni} S, {Tanaka} M, {Fox} OD, {Dichiara} S. 2023  {Heavy element nucleosynthesis associated with a gamma-ray burst}. {\em arXiv e-prints} p. arXiv:2308.00633.
(\href{http://dx.doi.org/10.48550/arXiv.2308.00633}{10.48550/arXiv.2308.00633})

\bibitem{Hotokezaka2023}
{Hotokezaka} K, {Tanaka} M, {Kato} D, {Gaigalas} G. 2023  {Tellurium emission line in kilonova AT 2017gfo}. {\em \mnras} \textbf{526}, L155--L159.
(\href{http://dx.doi.org/10.1093/mnrasl/slad128}{10.1093/mnrasl/slad128})

\bibitem{Norris2006}
{Norris} JP, {Bonnell} JT. 2006  {Short Gamma-Ray Bursts with Extended Emission}. {\em \apj} \textbf{643}, 266--275.
(\href{http://dx.doi.org/10.1086/502796}{10.1086/502796})

\bibitem{Norris2010}
{Norris} JP, {Gehrels} N, {Scargle} JD. 2010  {Threshold for Extended Emission in Short Gamma-ray Bursts}. {\em \apj} \textbf{717}, 411--419.
(\href{http://dx.doi.org/10.1088/0004-637X/717/1/411}{10.1088/0004-637X/717/1/411})

\bibitem{Metzger2008}
{Metzger} BD, {Quataert} E, {Thompson} TA. 2008  {Short-duration gamma-ray bursts with extended emission from protomagnetar spin-down}. {\em \mnras} \textbf{385}, 1455--1460.
(\href{http://dx.doi.org/10.1111/j.1365-2966.2008.12923.x}{10.1111/j.1365-2966.2008.12923.x})

\bibitem{Bucciantini2012}
{Bucciantini} N, {Metzger} BD, {Thompson} TA, {Quataert} E. 2012  {Short gamma-ray bursts with extended emission from magnetar birth: jet formation and collimation}. {\em \mnras} \textbf{419}, 1537--1545.
(\href{http://dx.doi.org/10.1111/j.1365-2966.2011.19810.x}{10.1111/j.1365-2966.2011.19810.x})

\bibitem{Giacomazzo2013}
{Giacomazzo} B, {Perna} R. 2013  {Formation of Stable Magnetars from Binary Neutron Star Mergers}. {\em \apjl} \textbf{771}, L26.
(\href{http://dx.doi.org/10.1088/2041-8205/771/2/L26}{10.1088/2041-8205/771/2/L26})

\bibitem{Gao2022}
{Gao} H, {Lei} WH, {Zhu} ZP. 2022  {GRB 211211A: a Prolonged Central Engine under a Strong Magnetic Field Environment}. {\em \apjl} \textbf{934}, L12.
(\href{http://dx.doi.org/10.3847/2041-8213/ac80c7}{10.3847/2041-8213/ac80c7})

\bibitem{Rosswog2007}
{Rosswog} S. 2007  {Fallback accretion in the aftermath of a compact binary merger}. {\em \mnras} \textbf{376}, L48--L51.
(\href{http://dx.doi.org/10.1111/j.1745-3933.2007.00284.x}{10.1111/j.1745-3933.2007.00284.x})

\bibitem{Desai2019}
{Desai} D, {Metzger} BD, {Foucart} F. 2019  {Imprints of r-process heating on fall-back accretion: distinguishing black hole-neutron star from double neutron star mergers}. {\em \mnras} \textbf{485}, 4404--4412.
(\href{http://dx.doi.org/10.1093/mnras/stz644}{10.1093/mnras/stz644})

\bibitem{Gompertz2020a}
{Gompertz} BP, {Levan} AJ, {Tanvir} NR. 2020  {A Search for Neutron Star-Black Hole Binary Mergers in the Short Gamma-Ray Burst Population}. {\em \apj} \textbf{895}, 58.
(\href{http://dx.doi.org/10.3847/1538-4357/ab8d24}{10.3847/1538-4357/ab8d24})

\bibitem{Zhu2022}
{Zhu} JP, {Wang} XI, {Sun} H, {Yang} YP, {Li} Z, {Hu} RC, {Qin} Y, {Wu} S. 2022  {Long-duration Gamma-Ray Burst and Associated Kilonova Emission from Fast-spinning Black Hole-Neutron Star Mergers}. {\em \apjl} \textbf{936}, L10.
(\href{http://dx.doi.org/10.3847/2041-8213/ac85ad}{10.3847/2041-8213/ac85ad})

\bibitem{Yang2022}
{Yang} J, {Ai} S, {Zhang} BB, {Zhang} B, {Liu} ZK, {Wang} XI, {Yang} YH, {Yin} YH, {Li} Y, {L{\"u}} HJ. 2022  {A long-duration gamma-ray burst with a peculiar origin}. {\em \nat} \textbf{612}, 232--235.
(\href{http://dx.doi.org/10.1038/s41586-022-05403-8}{10.1038/s41586-022-05403-8})

\bibitem{Gompertz2018}
{Gompertz} BP, {Levan} AJ, {Tanvir} NR, {Hjorth} J, {Covino} S, {Evans} PA, {Fruchter} AS, {Gonz{\'a}lez-Fern{\'a}ndez} C, {Jin} ZP, {Lyman} JD, {Oates} SR, {O'Brien} PT, {Wiersema} K. 2018  {The Diversity of Kilonova Emission in Short Gamma-Ray Bursts}. {\em \apj} \textbf{860}, 62.
(\href{http://dx.doi.org/10.3847/1538-4357/aac206}{10.3847/1538-4357/aac206})

\bibitem{Rossi2020}
{Rossi} A, {Stratta} G, {Maiorano} E, {Spighi} D, {Masetti} N, {Palazzi} E, {Gardini} A, {Melandri} A, {Nicastro} L, {Pian} E, {Branchesi} M, {Dadina} M, {Testa} V, {Brocato} E, {Benetti} S, {Ciolfi} R, {Covino} S, {D'Elia} V, {Grado} A, {Izzo} L, {Perego} A, {Piranomonte} S, {Salvaterra} R, {Selsing} J, {Tomasella} L, {Yang} S, {Vergani} D, {Amati} L, {Stephen} JB. 2020  {A comparison between short GRB afterglows and kilonova AT2017gfo: shedding light on kilonovae properties}. {\em \mnras} \textbf{493}, 3379--3397.
(\href{http://dx.doi.org/10.1093/mnras/staa479}{10.1093/mnras/staa479})

\bibitem{Rastinejad2021}
{Rastinejad} JC, {Fong} W, {Kilpatrick} CD, {Paterson} K, {Tanvir} NR, {Levan} AJ, {Metzger} BD, {Berger} E, {Chornock} R, {Cobb} BE, {Laskar} T, {Milne} P, {Nugent} AE, {Smith} N. 2021  {Probing Kilonova Ejecta Properties Using a Catalog of Short Gamma-Ray Burst Observations}. {\em \apj} \textbf{916}, 89.
(\href{http://dx.doi.org/10.3847/1538-4357/ac04b4}{10.3847/1538-4357/ac04b4})

\bibitem{Kasen2015}
{Kasen} D, {Fern{\'a}ndez} R, {Metzger} BD. 2015  {Kilonova light curves from the disc wind outflows of compact object mergers}. {\em \mnras} \textbf{450}, 1777--1786.
(\href{http://dx.doi.org/10.1093/mnras/stv721}{10.1093/mnras/stv721})

\bibitem{Kawaguchi2018}
{Kawaguchi} K, {Shibata} M, {Tanaka} M. 2018  {Radiative Transfer Simulation for the Optical and Near-infrared Electromagnetic Counterparts to GW170817}. {\em \apjl} \textbf{865}, L21.
(\href{http://dx.doi.org/10.3847/2041-8213/aade02}{10.3847/2041-8213/aade02})

\bibitem{Wollaeger2018}
{Wollaeger} RT, {Korobkin} O, {Fontes} CJ, {Rosswog} SK, {Even} WP, {Fryer} CL, {Sollerman} J, {Hungerford} AL, {van Rossum} DR, {Wollaber} AB. 2018  {Impact of ejecta morphology and composition on the electromagnetic signatures of neutron star mergers}. {\em \mnras} \textbf{478}, 3298--3334.
(\href{http://dx.doi.org/10.1093/mnras/sty1018}{10.1093/mnras/sty1018})

\bibitem{Korobkin2020}
{Korobkin} O, {Wollaeger} R, {Fryer} C, {Hungerford} AL, {Rosswog} S, {Fontes} C, {Mumpower} M, {Chase} E, {Even} W, {Miller} JM, {Misch} GW, {Lippuner} J. 2020  {Axisymmetric Radiative Transfer Models of Kilonovae}. {\em arXiv e-prints} p. arXiv:2004.00102.

\bibitem{Nativi2021}
{Nativi} L, {Bulla} M, {Rosswog} S, {Lundman} C, {Kowal} G, {Gizzi} D, {Lamb} GP, {Perego} A. 2021  {Can jets make the radioactively powered emission from neutron star mergers bluer?}. {\em \mnras} \textbf{500}, 1772--1783.
(\href{http://dx.doi.org/10.1093/mnras/staa3337}{10.1093/mnras/staa3337})

\bibitem{Klion2021}
{Klion} H, {Duffell} PC, {Kasen} D, {Quataert} E. 2021  {The Effect of Jet-Ejecta Interaction on the Viewing Angle Dependence of Kilonova Light Curves}. {\em \mnras}.
(\href{http://dx.doi.org/10.1093/mnras/stab042}{10.1093/mnras/stab042})

\bibitem{Doctor2017}
{Doctor} Z, {Kessler} R, {Chen} HY, {Farr} B, {Finley} DA, {Foley} RJ, {Goldstein} DA, {Holz} DE, {Kim} AG, {Morganson} E, {Sako} M, {Scolnic} D, {Smith} M, {Soares-Santos} M, {Spinka} H, {Abbott} TMC, {Abdalla} FB, {Allam} S, {Annis} J, {Bechtol} K, {Benoit-L{\'e}vy} A, {Bertin} E, {Brooks} D, {Buckley-Geer} E, {Burke} DL, {Carnero Rosell} A, {Carrasco Kind} M, {Carretero} J, {Cunha} CE, {D'Andrea} CB, {da Costa} LN, {DePoy} DL, {Desai} S, {Diehl} HT, {Drlica-Wagner} A, {Eifler} TF, {Frieman} J, {Garc{\'\i}a-Bellido} J, {Gaztanaga} E, {Gerdes} DW, {Gruendl} RA, {Gschwend} J, {Gutierrez} G, {James} DJ, {Krause} E, {Kuehn} K, {Kuropatkin} N, {Lahav} O, {Li} TS, {Lima} M, {Maia} MAG, {March} M, {Marshall} JL, {Menanteau} F, {Miquel} R, {Neilsen} E, {Nichol} RC, {Nord} B, {Plazas} AA, {Romer} AK, {Sanchez} E, {Scarpine} V, {Schubnell} M, {Sevilla-Noarbe} I, {Smith} RC, {Sobreira} F, {Suchyta} E, {Swanson} MEC, {Tarle} G, {Walker} AR, {Wester} W, {DES Collaboration}. 2017  {A Search for Kilonovae in the Dark
  Energy Survey}. {\em \apj} \textbf{837}, 57.
(\href{http://dx.doi.org/10.3847/1538-4357/aa5d09}{10.3847/1538-4357/aa5d09})

\bibitem{Yang2017}
{Yang} S, {Valenti} S, {Cappellaro} E, {Sand} DJ, {Tartaglia} L, {Corsi} A, {Reichart} DE, {Haislip} J, {Kouprianov} V. 2017  {An Empirical Limit on the Kilonova Rate from the DLT40 One Day Cadence Supernova Survey}. {\em \apjl} \textbf{851}, L48.
(\href{http://dx.doi.org/10.3847/2041-8213/aaa07d}{10.3847/2041-8213/aaa07d})

\bibitem{Andreoni2020}
{Andreoni} I, {Kool} EC, {Sagu{\'e}s Carracedo} A, {Kasliwal} MM, {Bulla} M, {Ahumada} T, {Coughlin} MW, {Anand} S, {Sollerman} J, {Goobar} A, {Kaplan} DL, {Loveridge} TT, {Karambelkar} V, {Cooke} J, {Bagdasaryan} A, {Bellm} EC, {Cenko} SB, {Cook} DO, {De} K, {Dekany} R, {Delacroix} A, {Drake} A, {Duev} DA, {Fremling} C, {Golkhou} VZ, {Graham} MJ, {Hale} D, {Kulkarni} SR, {Kupfer} T, {Laher} RR, {Mahabal} AA, {Masci} FJ, {Rusholme} B, {Smith} RM, {Tzanidakis} A, {Van Sistine} A, {Yao} Y. 2020  {Constraining the Kilonova Rate with Zwicky Transient Facility Searches Independent of Gravitational Wave and Short Gamma-Ray Burst Triggers}. {\em \apj} \textbf{904}, 155.
(\href{http://dx.doi.org/10.3847/1538-4357/abbf4c}{10.3847/1538-4357/abbf4c})

\bibitem{Andreoni2021ztfrest}
{Andreoni} I, {Coughlin} MW, {Kool} EC, {Kasliwal} MM, {Kumar} H, {Bhalerao} V, {Carracedo} AS, {Ho} AYQ, {Pang} PTH, {Saraogi} D, {Sharma} K, {Shenoy} V, {Burns} E, {Ahumada} T, {Anand} S, {Singer} LP, {Perley} DA, {De} K, {Fremling} UC, {Bellm} EC, {Bulla} M, {Crellin-Quick} A, {Dietrich} T, {Drake} A, {Duev} DA, {Goobar} A, {Graham} MJ, {Kaplan} DL, {Kulkarni} SR, {Laher} RR, {Mahabal} AA, {Shupe} DL, {Sollerman} J, {Walters} R, {Yao} Y. 2021  {Fast-transient Searches in Real Time with ZTFReST: Identification of Three Optically Discovered Gamma-Ray Burst Afterglows and New Constraints on the Kilonova Rate}. {\em \apj} \textbf{918}, 63.
(\href{http://dx.doi.org/10.3847/1538-4357/ac0bc7}{10.3847/1538-4357/ac0bc7})

\bibitem{McBrien2021}
{McBrien} OR, {Smartt} SJ, {Huber} ME, {Rest} A, {Chambers} KC, {Barbieri} C, {Bulla} M, {Jha} S, {Gromadzki} M, {Srivastav} S, {Smith} KW, {Young} DR, {McLaughlin} S, {Inserra} C, {Nicholl} M, {Fraser} M, {Maguire} K, {Chen} TW, {Wevers} T, {Anderson} JP, {M{\"u}ller-Bravo} TE, {Olivares E.} F, {Kankare} E, {Gal-Yam} A, {Waters} C. 2021  {PS15cey and PS17cke: prospective candidates from the Pan-STARRS Search for kilonovae}. {\em \mnras} \textbf{500}, 4213--4228.
(\href{http://dx.doi.org/10.1093/mnras/staa3361}{10.1093/mnras/staa3361})

\bibitem{Li2023}
{Li} W, {Arcavi} I, {Nakar} E, {Filippenko} AV, {Brink} TG, {Zheng} W, {Yang} Y, {Lam} MC, {Keinan} I, {Brennan} SJ, {Shitrit} N. 2023  {Rapidly Evolving Transients in Archival ZTF Public Alerts}. {\em \apj} \textbf{955}, 144.
(\href{http://dx.doi.org/10.3847/1538-4357/ace4bc}{10.3847/1538-4357/ace4bc})

\bibitem{Mo2023}
{Mo} G, {Jayaraman} R, {Fausnaugh} M, {Katsavounidis} E, {Ricker} GR, {Vanderspek} R. 2023  {Searching for Gravitational-wave Counterparts Using the Transiting Exoplanet Survey Satellite}. {\em \apjl} \textbf{948}, L3.
(\href{http://dx.doi.org/10.3847/2041-8213/acca70}{10.3847/2041-8213/acca70})

\bibitem{Flewelling2020}
{Flewelling} HA, {Magnier} EA, {Chambers} KC, {Heasley} JN, {Holmberg} C, {Huber} ME, {Sweeney} W, {Waters} CZ, {Calamida} A, {Casertano} S, {Chen} X, {Farrow} D, {Hasinger} G, {Henderson} R, {Long} KS, {Metcalfe} N, {Narayan} G, {Nieto-Santisteban} MA, {Norberg} P, {Rest} A, {Saglia} RP, {Szalay} A, {Thakar} AR, {Tonry} JL, {Valenti} J, {Werner} S, {White} R, {Denneau} L, {Draper} PW, {Hodapp} KW, {Jedicke} R, {Kaiser} N, {Kudritzki} RP, {Price} PA, {Wainscoat} RJ, {Chastel} S, {McLean} B, {Postman} M, {Shiao} B. 2020  {The Pan-STARRS1 Database and Data Products}. {\em \apjs} \textbf{251}, 7.
(\href{http://dx.doi.org/10.3847/1538-4365/abb82d}{10.3847/1538-4365/abb82d})

\bibitem{Andreoni2019PASP}
{Andreoni} I, {Anand} S, {Bianco} FB, {Cenko} SB, {Cowperthwaite} PS, {Coughlin} MW, {Drout} M, {Golkhou} VZ, {Kaplan} DL, {Mooley} KP, {Pritchard} TA, {Singer} LP, {Webb} S, {LSST Transient} wsot, {Variable Stars Collaboration}. 2019  {A Strategy for LSST to Unveil a Population of Kilonovae without Gravitational-wave Triggers}. {\em \pasp} \textbf{131}, 068004.
(\href{http://dx.doi.org/10.1088/1538-3873/ab1531}{10.1088/1538-3873/ab1531})

\bibitem{Setzer2019MNRAS}
{Setzer} CN, {Biswas} R, {Peiris} HV, {Rosswog} S, {Korobkin} O, {Wollaeger} RT, {LSST Dark Energy Science Collaboration}. 2019  {Serendipitous discoveries of kilonovae in the LSST main survey: maximizing detections of sub-threshold gravitational wave events}. {\em \mnras} \textbf{485}, 4260--4273.
(\href{http://dx.doi.org/10.1093/mnras/stz506}{10.1093/mnras/stz506})

\bibitem{Andreoni2022serendip}
{Andreoni} I, {Coughlin} MW, {Almualla} M, {Bellm} EC, {Bianco} FB, {Bulla} M, {Cucchiara} A, {Dietrich} T, {Goobar} A, {Kool} EC, {Li} X, {Ragosta} F, {Sagu{\'e}s-Carracedo} A, {Singer} LP. 2022  {Optimizing Cadences with Realistic Light-curve Filtering for Serendipitous Kilonova Discovery with Vera Rubin Observatory}. {\em \apjs} \textbf{258}, 5.
(\href{http://dx.doi.org/10.3847/1538-4365/ac3bae}{10.3847/1538-4365/ac3bae})

\bibitem{SCOCtoo}
{Survey Cadence Optimization Committee for Vera Rubin Observatory}. 2023  Survey Cadence Optimization Committee’s Phase 2 Recommendations. .

\bibitem{rubintoo}
Andreoni I, Margutti R et~al.. 2024  Rubin ToO 2024: Envisioning the Vera C. Rubin Observatory LSST Target of Opportunity program. .

\bibitem{Andreoni2022ToO}
{Andreoni} I, {Margutti} R, {Salafia} OS, {Parazin} B, {Villar} VA, {Coughlin} MW, {Yoachim} P, {Mortensen} K, {Brethauer} D, {Smartt} SJ, {Kasliwal} MM, {Alexander} KD, {Anand} S, {Berger} E, {Bernardini} MG, {Bianco} FB, {Blanchard} PK, {Bloom} JS, {Brocato} E, {Bulla} M, {Cartier} R, {Cenko} SB, {Chornock} R, {Copperwheat} CM, {Corsi} A, {D'Ammando} F, {D'Avanzo} P, {H{\'e}l{\`e}ne Datrier} L{\'E}, {Foley} RJ, {Ghirlanda} G, {Goobar} A, {Grindlay} J, {Hajela} A, {Holz} DE, {Karambelkar} V, {Kool} EC, {Lamb} GP, {Laskar} T, {Levan} A, {Maguire} K, {May} M, {Melandri} A, {Milisavljevic} D, {Miller} AA, {Nicholl} M, {Nissanke} SM, {Palmese} A, {Piranomonte} S, {Rest} A, {Sagu{\'e}s-Carracedo} A, {Siellez} K, {Singer} LP, {Smith} M, {Steeghs} D, {Tanvir} N. 2022  {Target-of-opportunity Observations of Gravitational-wave Events with Vera C. Rubin Observatory}. {\em \apjs} \textbf{260}, 18.
(\href{http://dx.doi.org/10.3847/1538-4365/ac617c}{10.3847/1538-4365/ac617c})

\bibitem{Andreoni2024APh}
{Andreoni} I, {Coughlin} MW, {Criswell} AW, {Bulla} M, {Toivonen} A, {Singer} LP, {Palmese} A, {Burns} E, {Gezari} S, {Kasliwal} MM, {Kiendrebeogo} RW, {Mahabal} A, {Moriya} TJ, {Rest} A, {Scolnic} D, {Simcoe} RA, {Soon} J, {Stein} R, {Travouillon} T. 2024  {Enabling kilonova science with Nancy Grace Roman Space Telescope}. {\em Astroparticle Physics} \textbf{155}, 102904.
(\href{http://dx.doi.org/10.1016/j.astropartphys.2023.102904}{10.1016/j.astropartphys.2023.102904})

\bibitem{Margutti2018WP}
{Margutti} R, {Cowperthwaite} P, {Doctor} Z, {Mortensen} K, {Pankow} CP, {Salafia} O, {Villar} VA, {Alexander} K, {Annis} J, {Andreoni} I, {Baldeschi} A, {Balmaverde} B, {Berger} E, {Bernardini} MG, {Berry} CPL, {Bianco} F, {Blanchard} PK, {Brocato} E, {Carnerero} MI, {Cartier} R, {Cenko} SB, {Chornock} R, {Chomiuk} L, {Copperwheat} CM, {Coughlin} MW, {Coppejans} DL, {Corsi} A, {D'Ammando} F, {Datrier} L, {D'Avanzo} P, {Dimitriadis} G, {Drout} MR, {Foley} RJ, {Fong} W, {Fox} O, {Ghirlanda} G, {Goldstein} D, {Grindlay} J, {Guidorzi} C, {Haiman} Z, {Hendry} M, {Holz} D, {Hung} T, {Inserra} C, {Jones} DO, {Kalogera} V, {Kilpatrick} CD, {Lamb} G, {Laskar} T, {Levan} A, {Mason} E, {Maguire} K, {Melandri} A, {Milisavljevic} D, {Miller} A, {Narayan} G, {Nielsen} E, {Nicholl} M, {Nissanke} S, {Nugent} P, {Pan} YC, {Pasham} D, {Paterson} K, {Piranomonte} S, {Racusin} J, {Rest} A, {Righi} C, {Sand} D, {Seaman} R, {Scolnic} D, {Siellez} K, {Singer} L, {Szkody} P, {Smith} M, {Steeghs} D, {Sullivan} M, {Tanvir} N,
  {Terreran} G, {Trimble} V, {Valenti} S, {LSST Transient} wtsot, {Variable Stars Collaboration}. 2018  {Target of Opportunity Observations of Gravitational Wave Events with LSST}. {\em arXiv e-prints} p. arXiv:1812.04051.

\bibitem{Cowperthwaite2019}
{Cowperthwaite} PS, {Villar} VA, {Scolnic} DM, {Berger} E. 2019  {LSST Target-of-opportunity Observations of Gravitational-wave Events: Essential and Efficient}. {\em \apj} \textbf{874}, 88.
(\href{http://dx.doi.org/10.3847/1538-4357/ab07b6}{10.3847/1538-4357/ab07b6})

\bibitem{Scolnic2018}
{Scolnic} D, {Kessler} R, {Brout} D, {Cowperthwaite} PS, {Soares-Santos} M, {Annis} J, {Herner} K, {Chen} HY, {Sako} M, {Doctor} Z, {Butler} RE, {Palmese} A, {Diehl} HT, {Frieman} J, {Holz} DE, {Berger} E, {Chornock} R, {Villar} VA, {Nicholl} M, {Biswas} R, {Hounsell} R, {Foley} RJ, {Metzger} J, {Rest} A, {Garc{\'\i}a-Bellido} J, {M{\"o}ller} A, {Nugent} P, {Abbott} TMC, {Abdalla} FB, {Allam} S, {Bechtol} K, {Benoit-L{\'e}vy} A, {Bertin} E, {Brooks} D, {Buckley-Geer} E, {Carnero Rosell} A, {Carrasco Kind} M, {Carretero} J, {Castander} FJ, {Cunha} CE, {D'Andrea} CB, {da Costa} LN, {Davis} C, {Doel} P, {Drlica-Wagner} A, {Eifler} TF, {Flaugher} B, {Fosalba} P, {Gaztanaga} E, {Gerdes} DW, {Gruen} D, {Gruendl} RA, {Gschwend} J, {Gutierrez} G, {Hartley} WG, {Honscheid} K, {James} DJ, {Johnson} MWG, {Johnson} MD, {Krause} E, {Kuehn} K, {Kuhlmann} S, {Lahav} O, {Li} TS, {Lima} M, {Maia} MAG, {March} M, {Marshall} JL, {Menanteau} F, {Miquel} R, {Neilsen} E, {Plazas} AA, {Sanchez} E, {Scarpine} V, {Schubnell} M,
  {Sevilla-Noarbe} I, {Smith} M, {Smith} RC, {Sobreira} F, {Suchyta} E, {Swanson} MEC, {Tarle} G, {Thomas} RC, {Tucker} DL, {Walker} AR, {DES Collaboration}. 2018  {How Many Kilonovae Can Be Found in Past, Present, and Future Survey Data Sets?}. {\em \apjl} \textbf{852}, L3.
(\href{http://dx.doi.org/10.3847/2041-8213/aa9d82}{10.3847/2041-8213/aa9d82})

\bibitem{Bianco2019}
{Bianco} FB, {Drout} MR, {Graham} ML, {Pritchard} TA, {Biswas} R, {Narayan} G, {Andreoni} I, {Cowperthwaite} PS, {Ribeiro} T, {LSST Transient} WSot, {Variable Stars Collaboration}. 2019  {Presto-Color: A Photometric Survey Cadence for Explosive Physics and Fast Transients}. {\em \pasp} \textbf{131}, 068002.
(\href{http://dx.doi.org/10.1088/1538-3873/ab121a}{10.1088/1538-3873/ab121a})

\bibitem{Almualla2021}
{Almualla} M, {Anand} S, {Coughlin} MW, {Dietrich} T, {Guessoum} N, {Sagu{\'e}s Carracedo} A, {Ahumada} T, {Andreoni} I, {Antier} S, {Bellm} EC, {Bulla} M, {Singer} LP. 2021  {Optimizing serendipitous detections of kilonovae: cadence and filter selection}. {\em \mnras} \textbf{504}, 2822--2831.
(\href{http://dx.doi.org/10.1093/mnras/stab1090}{10.1093/mnras/stab1090})

\bibitem{Gompertz2023}
{Gompertz} BP, {Nicholl} M, {Smith} JC, {Harisankar} S, {Pratten} G, {Schmidt} P, {Smith} GP. 2023  {A multimessenger model for neutron star-black hole mergers}. {\em \mnras} \textbf{526}, 4585--4598.
(\href{http://dx.doi.org/10.1093/mnras/stad2990}{10.1093/mnras/stad2990})

\bibitem{Hlozek2023}
{Hlo{\v{z}}ek} R, {Malz} AI, {Ponder} KA, {Dai} M, {Narayan} G, {Ishida} EEO, {Allam}, T. J, {Bahmanyar} A, {Bi} X, {Biswas} R, {Boone} K, {Chen} S, {Du} N, {Erdem} A, {Galbany} L, {Garreta} A, {Jha} SW, {Jones} DO, {Kessler} R, {Lin} M, {Liu} J, {Lochner} M, {Mahabal} AA, {Mandel} KS, {Margolis} P, {Mart{\'\i}nez-Galarza} JR, {McEwen} JD, {Muthukrishna} D, {Nakatsuka} Y, {Noumi} T, {Oya} T, {Peiris} HV, {Peters} CM, {Puget} JF, {Setzer} CN, {Siddhartha}, {Stefanov} S, {Xie} T, {Yan} L, {Yeh} KH, {Zuo} W. 2023  {Results of the Photometric LSST Astronomical Time-series Classification Challenge (PLAsTiCC)}. {\em \apjs} \textbf{267}, 25.
(\href{http://dx.doi.org/10.3847/1538-4365/accd6a}{10.3847/1538-4365/accd6a})

\bibitem{Ragosta2024}
{Ragosta} F, {Ahumada} T, {Piranomonte} S, {Andreoni} I, {Melandri} A, {Colombo} A, {Coughlin} MW. 2024  {Kilonova Parameter Estimation with LSST at Vera C. Rubin Observatory}. {\em \apj} \textbf{966}, 214.
(\href{http://dx.doi.org/10.3847/1538-4357/ad35c1}{10.3847/1538-4357/ad35c1})

\bibitem{Chase2022}
{Chase} EA, {O'Connor} B, {Fryer} CL, {Troja} E, {Korobkin} O, {Wollaeger} RT, {Ristic} M, {Fontes} CJ, {Hungerford} AL, {Herring} AM. 2022  {Kilonova Detectability with Wide-field Instruments}. {\em \apj} \textbf{927}, 163.
(\href{http://dx.doi.org/10.3847/1538-4357/ac3d25}{10.3847/1538-4357/ac3d25})

\bibitem{Rest2024}
{Rest} A, {others}. 2024  Transient Exploration in the high latitude Wide Area Survey for Roman. {\em Submitted}.

\bibitem{Ma2023}
{Ma} H, {Lu} Y, {Guo} X, {Zhang} S, {Chu} Q. 2023  {On the detection of the electromagnetic counterparts from lensed gravitational wave events by binary neutron star mergers}. {\em \mnras} \textbf{518}, 6183--6198.
(\href{http://dx.doi.org/10.1093/mnras/stac3418}{10.1093/mnras/stac3418})

\bibitem{Agudo2023}
{Agudo} I, {Amati} L, {An} T, {Bauer} FE, {Benetti} S, {Bernardini} MG, {Beswick} R, {Bhirombhakdi} K, {de Boer} T, {Branchesi} M, {Brennan} SJ, {Brocato} E, {Caballero-Garc{\'\i}a} MD, {Cappellaro} E, {Castro Rodr{\'\i}guez} N, {Castro-Tirado} AJ, {Chambers} KC, {Chassande-Mottin} E, {Chaty} S, {Chen} TW, {Coleiro} A, {Covino} S, {D'Ammando} F, {D'Avanzo} P, {D'Elia} V, {Fiore} A, {Fl{\"o}rs} A, {Fraser} M, {Frey} S, {Frohmaier} C, {Fulton} M, {Galbany} L, {Gall} C, {Gao} H, {Garc{\'\i}a-Rojas} J, {Ghirlanda} G, {Giarratana} S, {Gillanders} JH, {Giroletti} M, {Gompertz} BP, {Gromadzki} M, {Heintz} KE, {Hjorth} J, {Hu} YD, {Huber} ME, {Inkenhaag} A, {Izzo} L, {Jin} ZP, {Jonker} PG, {Kann} DA, {Kool} EC, {Kotak} R, {Leloudas} G, {Levan} AJ, {Lin} CC, {Lyman} JD, {Magnier} EA, {Maguire} K, {Mandel} I, {Marcote} B, {Mata S{\'a}nchez} D, {Mattila} S, {Melandri} A, {Micha{\l}owski} MJ, {Moldon} J, {Nicholl} M, {Nicuesa Guelbenzu} A, {Oates} SR, {Onori} F, {Orienti} M, {Paladino} R, {Paragi} Z, {Perez-Torres} M,
  {Pian} E, {Pignata} G, {Piranomonte} S, {Quirola-V{\'a}squez} J, {Ragosta} F, {Rau} A, {Ronchini} S, {Rossi} A, {S{\'a}nchez-Ram{\'\i}rez} R, {Salafia} OS, {Schulze} S, {Smartt} SJ, {Smith} KW, {Sollerman} J, {Srivastav} S, {Starling} RLC, {Steeghs} D, {Stevance} HF, {Tanvir} NR, {Testa} V, {Torres} MAP, {Valeev} A, {Vergani} SD, {Vescovi} D, {Wainscost} R, {Watson} D, {Wiersema} K, {Wyrzykowski} {\L}, {Yang} J, {Yang} S, {Young} DR. 2023  {Panning for gold, but finding helium: Discovery of the ultra-stripped supernova SN 2019wxt from gravitational-wave follow-up observations}. {\em \aap} \textbf{675}, A201.
(\href{http://dx.doi.org/10.1051/0004-6361/202244751}{10.1051/0004-6361/202244751})

\bibitem{Wyatt2020}
{Wyatt} SD, {Tohuvavohu} A, {Arcavi} I, {Lundquist} MJ, {Howell} DA, {Sand} DJ. 2020  {The Gravitational Wave Treasure Map: A Tool to Coordinate, Visualize, and Assess the Electromagnetic Follow-up of Gravitational-wave Events}. {\em \apj} \textbf{894}, 127.
(\href{http://dx.doi.org/10.3847/1538-4357/ab855e}{10.3847/1538-4357/ab855e})

\bibitem{Keinan2024arXiv}
{Keinan} I, {Arcavi} I. 2024  {Coordinated Followup Could Have Enabled the Discovery of the GW190425 Kilonova}. {\em arXiv e-prints} p. arXiv:2405.17558.
(\href{http://dx.doi.org/10.48550/arXiv.2405.17558}{10.48550/arXiv.2405.17558})

\bibitem{Frohmaier2017}
{Frohmaier} C, {Sullivan} M, {Nugent} PE, {Goldstein} DA, {DeRose} J. 2017  {Real-time Recovery Efficiencies and Performance of the Palomar Transient Factory{\textquoteright}s Transient Discovery Pipeline}. {\em \apjs} \textbf{230}, 4.
(\href{http://dx.doi.org/10.3847/1538-4365/aa6d70}{10.3847/1538-4365/aa6d70})

\bibitem{Carrasco-Davis2021}
{Carrasco-Davis} R, {Reyes} E, {Valenzuela} C, {F{\"o}rster} F, {Est{\'e}vez} PA, {Pignata} G, {Bauer} FE, {Reyes} I, {S{\'a}nchez-S{\'a}ez} P, {Cabrera-Vives} G, {Eyheramendy} S, {Catelan} M, {Arredondo} J, {Castillo-Navarrete} E, {Rodr{\'\i}guez-Mancini} D, {Ruz-Mieres} D, {Moya} A, {Sabatini-Gacit{\'u}a} L, {Sep{\'u}lveda-Cobo} C, {Mahabal} AA, {Silva-Farf{\'a}n} J, {Camacho-I{\~n}iguez} E, {Galbany} L. 2021  {Alert Classification for the ALeRCE Broker System: The Real-time Stamp Classifier}. {\em \aj} \textbf{162}, 231.
(\href{http://dx.doi.org/10.3847/1538-3881/ac0ef1}{10.3847/1538-3881/ac0ef1})

\end{thebibliography}

\end{document}